\documentclass[10pt]{article}

\usepackage{graphicx}  
\graphicspath{{figures/}}
\usepackage{amsmath, mathtools, eqparbox}   
\usepackage[compress]{cite}
\usepackage{amssymb}   
\usepackage{bm} 
\usepackage{dcolumn}
\usepackage{color}
\usepackage[dvipsnames]{xcolor}
\usepackage{mathrsfs}
\usepackage{gensymb}
\usepackage{float}
\usepackage{amsfonts}
\usepackage{varioref}
\usepackage{textcomp} 
\usepackage{enumerate}
\usepackage{tikz}
\usepackage{multicol}
\usepackage{notoccite}
\usepackage{graphicx}
\usepackage{mathtools}
\usepackage{mathrsfs}
\usepackage{booktabs}
\usepackage{url}

\usepackage[utf8]{inputenc}
\usepackage[T1]{fontenc}

\newcommand\numberthis{\addtocounter{equation}{1}\tag{\theequation}}
\newcommand{\meqbox}[2]{\eqmakebox[#1][r]{$\displaystyle#2$}}
\usepackage[caption=true]{subfig}
\RequirePackage[colorlinks,citecolor=Cyan,urlcolor=Purple,linkcolor=Blue]{hyperref}

\allowdisplaybreaks
\usepackage{mathpazo,palatino}
\renewcommand{\bfseries}{\fontfamily{ppl}\fontseries{bx}\selectfont}
\DeclareMathAlphabet{\mathbf} {OT1}{cmbr}{bx}{n}
\DeclareMathAlphabet{\mathbold}{OML}{cmbrm}{b}{it}
\def\RN{Reissner-Nordstr\"{o}m }

\newcommand{\bigO}{\mathcal{O}}

\RequirePackage{parskip}
\usepackage{parskip}
\usepackage{array,multirow}
\usepackage{etoolbox}
\makeatletter
\patchcmd{\@sect}{#8}{\boldmath #8}{}{}
\let\ori@section\@section
\def\@section[#1]#2{\ori@section[\boldmath#1]{\boldmath#2}}
\makeatother

\usepackage{float}
\usepackage[caption=true]{subfig}
\usepackage[subfigure]{tocloft}
\usepackage{titlesec}

\usepackage{sectsty}
\sectionfont{\centering\fontsize{14}{17}\selectfont}
\titleformat{\section}[block]
  {\titlerule\addvspace{2pt}\titlerule\addvspace{7pt}\centering\normalfont\scshape\fontsize{14}{17}\bfseries}
  {\thesection\enspace}{0pt}{}[\vspace{7pt}\titlerule\addvspace{2pt}\titlerule\vspace{10pt}]
  
  \usepackage{tabularx}
\usepackage{adjustbox}
\usepackage{array}
\newcommand{\PreserveBackslash}[1]{\let\temp=\\#1\let\\=\temp}
\newcolumntype{C}[1]{>{\PreserveBackslash\centering}p{#1}}
\newcolumntype{R}[1]{>{\PreserveBackslash\raggedleft}p{#1}}
\newcolumntype{L}[1]{>{\PreserveBackslash\raggedright}p{#1}}
\usepackage{array, makecell}
\newcolumntype{P}[1]{>{\centering\arraybackslash}p{#1}}

\renewcommand*{\arraystretch}{1.1}
\newcommand*{\mline}[1]{%
\begingroup
    \renewcommand*{\arraystretch}{1.1}%
   \begin{tabular}[c]{@{}>{\raggedright\arraybackslash}p{.2\textwidth}@{}}#1\end{tabular}%
  \endgroup
}
\newcommand*{\Mline}[1]{%
\begingroup
    \renewcommand*{\arraystretch}{1.1}%
   \begin{tabular}[c]{@{}>{\raggedright\arraybackslash}p{.2\textwidth}@{}}#1\end{tabular}%
  \endgroup
}

\newcommand*{\Nline}[1]{%
\begingroup
    \renewcommand*{\arraystretch}{1.1}%
   \begin{tabular}[c]{@{}>{\raggedright\arraybackslash}p{.3\textwidth}@{}}#1\end{tabular}%
  \endgroup
}

\newcommand*{\qline}[1]{%
\begingroup
    \renewcommand*{\arraystretch}{1.1}%
   \begin{tabular}[c]{@{}>{\raggedright\arraybackslash}p{.085\textwidth}@{}}#1\end{tabular}%
  \endgroup
}

\labelformat{section}{section #1}
\labelformat{subsection}{section #1}
\labelformat{subsubsection}{section #1}
\labelformat{subsubsubsection}{section #1}
\labelformat{equation}{Eq.~(#1)}
\labelformat{figure}{fig.~[#1]}
\labelformat{subfigure}{fig.~[\thefigure#1]}
\labelformat{table}{table.~#1}
\labelformat{appendix}{Appendix #1}
\definecolor{lime}{HTML}{A6CE39}
\DeclareRobustCommand{\orcidicon}{
	\begin{tikzpicture}
		\draw[lime, fill=lime] (0,0)
		circle [radius=0.16]
		node[white] {{\fontfamily{qag}\selectfont \tiny ID}};
		\draw[white, fill=white] (-0.0625,0.095)
		circle [radius=0.007];
	\end{tikzpicture}
	\hspace{-2mm}
}
\foreach \x in {A, ..., Z}{\expandafter\xdef\csname orcid\x\endcsname{\noexpand\href{https://orcid.org/\csname orcidauthor\x\endcsname}
		{\noexpand\orcidicon}}
}

\title{Gravitational Light Deflection:Geometrical Methods and Non-Linear Electrodynamics}

\author{ Partha Pratim Basumallick\footnote{\color{Blue}basuparth314@gmail.com}~$^{1}$\orcidP{}, 
Saheb Das\footnote{\color{Blue}sahebdasjgm3@gmail.com}~$^{2,3}$\orcidQ{},Bhaswati Mandal\footnote{\color{Blue}bhaswatimandaliitm92@gmail.com}~$^{4}$\orcidB{} and Subhadip Sau\footnote{\color{Blue}subhadipsau2@gmail.com}~$^{5}$\orcidA{}\\
         {$^{1}$\small{M. P, Birla Institute of Fundamental Research,  Kolkata-700071}}\\
          {$^{2}$\small{Saha Institute of Nuclear Physics,AF Block, Sector 1, Bidhannagar, Kolkata, West Bengal 700064}}\\ {$^{3}$\small{Homi Bhabha National Institute ,Training School Complex, Central Ave Rd, Anushakti Nagar, Mumbai, Maharashtra 400094}}\\ 
{$^{4}$\small{Physics and Applied Mathematics Unit, Indian Statistical Institute, Baranagar, West Bengal 700108}}\\
 {$^{5}$\small{Department of Physics, Jhargram Raj College, Jhargram, West Bengal-721507}}\\
	}

\makeatletter
\newif\ifGBNeedInitialContents
\newcommand{\GBInitialContents}{%
\contentsline {section}{\numberline {1}Introduction}{5}{section.1}%
\contentsline {section}{\numberline {2}Rudiments of Gravitational Lensing}{8}{section.2}%
\contentsline {subsection}{\numberline {2.1}Observational regimes of gravitational lensing}{9}{subsection.2.1}%
\contentsline {section}{\numberline {3}Early Beginnings}{11}{section.3}%
\contentsline {section}{\numberline {4}Bending Angle of Light in Newtonian Mechanics}{14}{section.4}%
\contentsline {section}{\numberline {5}The Lens Equation}{18}{section.5}%
\contentsline {section}{\numberline {6}Lensing Potential and Lensing Distance}{21}{section.6}%
\contentsline {subsection}{\numberline {6.1}Fermat's principle -- A general relativistic adaptation}{22}{subsection.6.1}%
\contentsline {subsubsection}{\numberline {6.1.1}Effective refractive index in conformally stationary spacetime through Fermat's principle}{23}{subsubsection.6.1.1}%
\contentsline {subsubsection}{\numberline {6.1.2}Effective refractive index in conformally static spacetime through Fermat's principle}{24}{subsubsection.6.1.2}%
\contentsline {section}{\numberline {7}Deflection of Light in a Static, Spherically Symmetric Spacetime}{24}{section.7}%
\contentsline {subsection}{\numberline {7.1}Photon Trajectories and Light Deflection in Static Spherically Symmetric Spacetimes}{25}{subsection.7.1}%
\contentsline {section}{\numberline {8}Deflection Angle of Light in the Equatorial Plane of a Kerr Black Hole}{27}{section.8}%
\contentsline {subsection}{\numberline {8.1}Kerr Geometry and Geodesic Framework}{28}{subsection.8.1}%
\contentsline {subsection}{\numberline {8.2}Photon Orbits in Reciprocal Coordinates and the Analytical Structure of the Deflection Angle}{30}{subsection.8.2}%
\contentsline {subsection}{\numberline {8.3}Determination of the Distance of Closest Approach and Photon Sphere Radius}{31}{subsection.8.3}%
\contentsline {subsection}{\numberline {8.4}Analytic Determination of Critical Parameters for Prograde and Retrograde Orbits}{32}{subsection.8.4}%
\contentsline {paragraph}{Critical Impact Parameter for Prograde and Retrograde Orbits:}{33}{section*.15}%
\contentsline {paragraph}{Radius of the Circular Photon Orbit:}{33}{section*.16}%
\contentsline {paragraph}{Geometric Visualization:}{33}{section*.17}%
\contentsline {paragraph}{Comparison and Interpretation:}{34}{section*.19}%
\contentsline {paragraph}{Numerical Profiles:}{34}{section*.20}%
\contentsline {paragraph}{Significance:}{34}{section*.23}%
\contentsline {section}{\numberline {9}Rindler--Ishak Method}{36}{section.9}%
\contentsline {subsection}{\numberline {9.1}Application to Schwarzschild de Sitter spacetime}{39}{subsection.9.1}%
\contentsline {subsection}{\numberline {9.2}Application to Weyl conformal gravity: sign of the linear potential term from the invariant angle}{40}{subsection.9.2}%
\contentsline {subsection}{\numberline {9.3}Application to 4D Einstein--Gauss--Bonnet--de Sitter black holes}{41}{subsection.9.3}%
\contentsline {subsection}{\numberline {9.4}Charged massive particle bending in a charged galactic wormhole}{43}{subsection.9.4}%
\contentsline {section}{\numberline {10}Bending Angle of Light and the Gauss-Bonnet Theorem}{44}{section.10}%
\contentsline {subsection}{\numberline {10.1}The Gauss-Bonnet Theorem}{45}{subsection.10.1}%
\contentsline {subsection}{\numberline {10.2}Gravitational Lens, Photon trajectory and Geometry}{45}{subsection.10.2}%
\contentsline {subsection}{\numberline {10.3}Mathematical Formulation of Deflection Angle from the Gauss-Bonnet Theorem}{47}{subsection.10.3}%
\contentsline {subsubsection}{\numberline {10.3.1}Gauss-Bonnet theorem in the region $^{\rm V}\nabla _{\rm L}^{\rm S}$}{47}{subsubsection.10.3.1}%
\contentsline {subsubsection}{\numberline {10.3.2}Gauss-Bonnet theorem in the region $^\infty _{\rm V}\square ^\infty _{\rm S}$}{48}{subsubsection.10.3.2}%
\contentsline {subsection}{\numberline {10.4}Implications for Direct Astronomical Observations}{51}{subsection.10.4}%
\contentsline {section}{\numberline {11}Generalized Expression of Bending Angle in Axisymmetric Spacetime by OIA Formalism}{52}{section.11}%
\contentsline {subsection}{\numberline {11.1}Extracting the Fermat Metric from the Randers-Finsler Form}{53}{subsection.11.1}%
\contentsline {subsubsection}{\numberline {11.1.1}Fermat metric --- Why?}{54}{subsubsection.11.1.1}%
\contentsline {subsubsection}{\numberline {11.1.2}Affine parameter and the Asada--Kasai versus Fermat spatial metrics}{55}{subsubsection.11.1.2}%
\contentsline {subsection}{\numberline {11.2}Determination of the Angles $\psi _{\rm V}$, $\psi _{\rm S}$ \& $\phi _{\rm VS}$}{56}{subsection.11.2}%
\contentsline {subsubsection}{\numberline {11.2.1}Defining the unit tangent vector and unit radial vector}{56}{subsubsection.11.2.1}%
\contentsline {subsubsection}{\numberline {11.2.2}Orbit equation in terms of impact parameter}{57}{subsubsection.11.2.2}%
\contentsline {subsubsection}{\numberline {11.2.3}Deriving the expression of the angles $\psi _{\rm V}$ \& $\psi _{\rm S}$}{59}{subsubsection.11.2.3}%
\contentsline {subsubsection}{\numberline {11.2.4}Deriving the expression of the angle $\phi _{\rm VS}$}{60}{subsubsection.11.2.4}%
\contentsline {subsection}{\numberline {11.3}Applying OIA Formalism to Kerr metric}{61}{subsection.11.3}%
\contentsline {subsubsection}{\numberline {11.3.1}Evaluating $\psi _{\rm V}$ \& $\psi _{\rm S}$ for Kerr spacetime}{61}{subsubsection.11.3.1}%
\contentsline {subsubsection}{\numberline {11.3.2}Evaluating $\phi _{\rm VS}$ for Kerr spacetime}{63}{subsubsection.11.3.2}%
\contentsline {subsubsection}{\numberline {11.3.3}Bending angle of a photon trajectory in Kerr metric from OIA formalism}{66}{subsubsection.11.3.3}%
\contentsline {section}{\numberline {12}Generalized Expression of Bending Angle in Axisymmetric Spacetime by GW-OIA Formalism}{68}{section.12}%
\contentsline {subsection}{\numberline {12.1}Gaussian curvature \& Geodesic curvature}{68}{subsection.12.1}%
\contentsline {subsection}{\numberline {12.2}Solving the Orbit Equation to Evaluate $r_{\rm VS}$}{69}{subsection.12.2}%
\contentsline {subsection}{\numberline {12.3}Surface Integral of the Gaussian Curvature for Kerr metric (in equatorial plane)}{75}{subsection.12.3}%
\contentsline {subsection}{\numberline {12.4}Evaluating the Path Integral of Geodesic Curvature (in equatorial plane) }{77}{subsection.12.4}%
\contentsline {section}{\numberline {13}Generalized Expression of Bending Angle in Static Spherically Symmetric Spacetime by OIA Formalism}{88}{section.13}%
\contentsline {section}{\numberline {14}Nonlinear electrodynamics in finite-distance light deflection}{92}{section.14}%
\contentsline {subsection}{\numberline {14.1}Where NED enters a lensing calculation}{94}{subsection.14.1}%
\contentsline {subsubsection}{\numberline {14.1.1}Action normalization and background field equations}{94}{subsubsection.14.1.1}%
\contentsline {subsubsection}{\numberline {14.1.2}Dual descriptions and branch control}{95}{subsubsection.14.1.2}%
\contentsline {subsubsection}{\numberline {14.1.3}The principal symbol and its polarization factors}{96}{subsubsection.14.1.3}%
\contentsline {subsection}{\numberline {14.2}From characteristic covectors to finite-distance observables}{97}{subsection.14.2}%
\contentsline {subsubsection}{\numberline {14.2.1}Hamiltonian turning points and critical curves}{97}{subsubsection.14.2.1}%
\contentsline {subsubsection}{\numberline {14.2.2}Two local directions: phase normal and energy ray}{98}{subsubsection.14.2.2}%
\contentsline {subsubsection}{\numberline {14.2.3}Optical curvature and endpoint geometry}{99}{subsubsection.14.2.3}%
\contentsline {subsection}{\numberline {14.3}Representative NED models and optical regimes}{100}{subsection.14.3}%
\contentsline {subsubsection}{\numberline {14.3.1}Regular background with a nontrivial cone: the Dymnikova spacetime}{100}{subsubsection.14.3.1}%
\contentsline {subsubsection}{\numberline {14.3.2}Background-metric optics in the Hayward spacetime}{103}{subsubsection.14.3.2}%
\contentsline {subsubsection}{\numberline {14.3.3}Suppressed far-field corrections and frequency shifts: the $h$-family}{104}{subsubsection.14.3.3}%
\contentsline {paragraph}{Observer-frequency relation.}{106}{section*.69}%
\contentsline {subsubsection}{\numberline {14.3.4}Branch-resolved birefringence in a quadratic two-invariant model}{108}{subsubsection.14.3.4}%
\contentsline {subsection}{\numberline {14.4}Physical synthesis: which observable sees which nonlinear layer?}{110}{subsection.14.4}%
\contentsline {subsubsection}{\numberline {14.4.1}Far-field bending and strong-field capture are different filters}{112}{subsubsection.14.4.1}%
\contentsline {subsubsection}{\numberline {14.4.2}The exterior branch must exist before it can cast a shadow}{112}{subsubsection.14.4.2}%
\contentsline {subsubsection}{\numberline {14.4.3}Finite distance changes the observable, not the critical orbit}{113}{subsubsection.14.4.3}%
\contentsline {subsubsection}{\numberline {14.4.4}Critical-size degeneracies are broken by other channels}{114}{subsubsection.14.4.4}%
\contentsline {subsubsection}{\numberline {14.4.5}Matched parameters and observational scope}{114}{subsubsection.14.4.5}%
\contentsline {subsection}{\numberline {14.5}Conventions and consistency across the literature}{115}{subsection.14.5}%
\contentsline {subsection}{\numberline {14.6}From optical calculations to observational claims}{116}{subsection.14.6}%
\contentsline {subsection}{\numberline {14.7}Conclusions}{117}{subsection.14.7}%
\contentsline {section}{\numberline {15}Light deflection across methods and spacetimes}{118}{section.15}%
\contentsline {subsection}{\numberline {15.1}Schwarzschild: from Newtonian bending to optical curvature}{119}{subsection.15.1}%
\contentsline {subsubsection}{\numberline {15.1.1}The mass contribution}{119}{subsubsection.15.1.1}%
\contentsline {subsubsection}{\numberline {15.1.2}From weak bending to smaller impact parameters}{121}{subsubsection.15.1.2}%
\contentsline {subsection}{\numberline {15.2}Kerr: mass bending and frame dragging}{121}{subsection.15.2}%
\contentsline {subsubsection}{\numberline {15.2.1}The two orbital orientations}{121}{subsubsection.15.2.1}%
\contentsline {subsubsection}{\numberline {15.2.2}Separating the mass and spin responses}{122}{subsubsection.15.2.2}%
\contentsline {subsection}{\numberline {15.3}Kottler: cosmological curvature and the measured direction}{124}{subsection.15.3}%
\contentsline {subsection}{\numberline {15.4}Charge and nonlinear electrodynamics: geometry and propagation}{125}{subsection.15.4}%
\contentsline {subsubsection}{\numberline {15.4.1}Background geometry and photon propagation}{125}{subsubsection.15.4.1}%
\contentsline {subsubsection}{\numberline {15.4.2}Charge at fixed mass: RN and Dymnikova lenses}{126}{subsubsection.15.4.2}%
\contentsline {subsubsection}{\numberline {15.4.3}From the bending angle to the concentration of a beam}{128}{subsubsection.15.4.3}%
\contentsline {subsubsection}{\numberline {15.4.4}Polarization and the relation between bending and travel time}{129}{subsubsection.15.4.4}%
\contentsline {subsection}{\numberline {15.5}Frequency-dependent bending in a plasma}{132}{subsection.15.5}%
\contentsline {subsection}{\numberline {15.6}From weak bending to rays near capture}{134}{subsection.15.6}%
\contentsline {subsubsection}{\numberline {15.6.1}Lenses with the same capture scale}{134}{subsubsection.15.6.1}%
\contentsline {subsubsection}{\numberline {15.6.2}The growth of the angle near trapping}{135}{subsubsection.15.6.2}%
\contentsline {subsubsection}{\numberline {15.6.3}The spacing of successive winding rays}{136}{subsubsection.15.6.3}%
\contentsline {subsection}{\numberline {15.7}Concluding discussion}{137}{subsection.15.7}%
}
\newcommand{\GBTableOfContents}{%
  \begingroup
  \IfFileExists{\jobname.toc}{\GBNeedInitialContentsfalse}{\GBNeedInitialContentstrue}%
  \ifGBNeedInitialContents
    \let\GBOriginalStartToc\@starttoc
    \def\@starttoc##1{\GBInitialContents\GBOriginalStartToc{##1}}%
  \fi
  \tableofcontents
  \endgroup
}
\makeatother

\makeatletter
\AtBeginDocument{%
\@ifundefined{b@carroll1997lecturenotesgeneralrelativity}{\bibcite{carroll1997lecturenotesgeneralrelativity}{1}}{}%
\@ifundefined{b@zaveri2025periodicrelativitytheorygravity}{\bibcite{zaveri2025periodicrelativitytheorygravity}{2}}{}%
\@ifundefined{b@Narayan1997}{\bibcite{Narayan1997}{3}}{}%
\@ifundefined{b@Dodelson_2017}{\bibcite{Dodelson_2017}{4}}{}%
\@ifundefined{b@Bartelmann_2010}{\bibcite{Bartelmann_2010}{5}}{}%
\@ifundefined{b@Jalilvand_2020}{\bibcite{Jalilvand_2020}{6}}{}%
\@ifundefined{b@Torres_Ballesteros_2022}{\bibcite{Torres_Ballesteros_2022}{7}}{}%
\@ifundefined{b@Natarajan_2024}{\bibcite{Natarajan_2024}{8}}{}%
\@ifundefined{b@Ferrami_2024}{\bibcite{Ferrami_2024}{9}}{}%
\@ifundefined{b@Zamani_2024}{\bibcite{Zamani_2024}{10}}{}%
\@ifundefined{b@Buncher_2025}{\bibcite{Buncher_2025}{11}}{}%
\@ifundefined{b@Massey_2010}{\bibcite{Massey_2010}{12}}{}%
\@ifundefined{b@Ellis_2010}{\bibcite{Ellis_2010}{13}}{}%
\@ifundefined{b@Karamazov_2021}{\bibcite{Karamazov_2021}{14}}{}%
\@ifundefined{b@eng_l_2022}{\bibcite{eng_l_2022}{15}}{}%
\@ifundefined{b@Vegetti_2023}{\bibcite{Vegetti_2023}{16}}{}%
\@ifundefined{b@Vachher_2024}{\bibcite{Vachher_2024}{17}}{}%
\@ifundefined{b@Gannon_2025}{\bibcite{Gannon_2025}{18}}{}%
\@ifundefined{b@Tsapras_2018}{\bibcite{Tsapras_2018}{19}}{}%
\@ifundefined{b@Turyshev_2020}{\bibcite{Turyshev_2020}{20}}{}%
\@ifundefined{b@Turyshev_2022}{\bibcite{Turyshev_2022}{21}}{}%
\@ifundefined{b@Toth_2023}{\bibcite{Toth_2023}{22}}{}%
\@ifundefined{b@Mr_z_2024}{\bibcite{Mr_z_2024}{23}}{}%
\@ifundefined{b@Li_2018}{\bibcite{Li_2018}{24}}{}%
\@ifundefined{b@Diego_2019}{\bibcite{Diego_2019}{25}}{}%
\@ifundefined{b@Pagano_2020}{\bibcite{Pagano_2020}{26}}{}%
\@ifundefined{b@Bulashenko_2022}{\bibcite{Bulashenko_2022}{27}}{}%
\@ifundefined{b@Tambalo_2023}{\bibcite{Tambalo_2023}{28}}{}%
\@ifundefined{b@Ashish_2024}{\bibcite{Ashish_2024}{29}}{}%
\@ifundefined{b@Shan2025}{\bibcite{Shan2025}{30}}{}%
\@ifundefined{b@Bolton_2006}{\bibcite{Bolton_2006}{31}}{}%
\@ifundefined{b@etherington2023}{\bibcite{etherington2023}{32}}{}%
\@ifundefined{b@kumar2023}{\bibcite{kumar2023}{33}}{}%
\@ifundefined{b@lemon_2023}{\bibcite{lemon_2023}{34}}{}%
\@ifundefined{b@saha2024}{\bibcite{saha2024}{35}}{}%
\@ifundefined{b@mccarty2024}{\bibcite{mccarty2024}{36}}{}%
\@ifundefined{b@Press1973}{\bibcite{Press1973}{37}}{}%
\@ifundefined{b@Treu2010}{\bibcite{Treu2010}{38}}{}%
\@ifundefined{b@Jouvel2014}{\bibcite{Jouvel2014}{39}}{}%
\@ifundefined{b@Molino2017}{\bibcite{Molino2017}{40}}{}%
\@ifundefined{b@Gonzalez_2018}{\bibcite{Gonzalez_2018}{41}}{}%
\@ifundefined{b@Narlikar1996}{\bibcite{Narlikar1996}{42}}{}%
\@ifundefined{b@Bernal_2016}{\bibcite{Bernal_2016}{43}}{}%
\@ifundefined{b@Chwolson_1924}{\bibcite{Chwolson_1924}{44}}{}%
\@ifundefined{b@Einstein_1936}{\bibcite{Einstein_1936}{45}}{}%
\@ifundefined{b@Fort_1988}{\bibcite{Fort_1988}{46}}{}%
\@ifundefined{b@Tyson_1990}{\bibcite{Tyson_1990}{47}}{}%
\@ifundefined{b@Bartelmann_2001}{\bibcite{Bartelmann_2001}{48}}{}%
\@ifundefined{b@Refregier_2003}{\bibcite{Refregier_2003}{49}}{}%
\@ifundefined{b@hoekstra2013}{\bibcite{hoekstra2013}{50}}{}%
\@ifundefined{b@Giblin_2021}{\bibcite{Giblin_2021}{51}}{}%
\@ifundefined{b@Li_2022}{\bibcite{Li_2022}{52}}{}%
\@ifundefined{b@Zhang_2024}{\bibcite{Zhang_2024}{53}}{}%
\@ifundefined{b@Adam_2025}{\bibcite{Adam_2025}{54}}{}%
\@ifundefined{b@LEWIS20061}{\bibcite{LEWIS20061}{55}}{}%
\@ifundefined{b@Kaiser1992}{\bibcite{Kaiser1992}{56}}{}%
\@ifundefined{b@Kaiser1993}{\bibcite{Kaiser1993}{57}}{}%
\@ifundefined{b@Kohlinger2017}{\bibcite{Kohlinger2017}{58}}{}%
\@ifundefined{b@Congdon2018}{\bibcite{Congdon2018}{59}}{}%
\@ifundefined{b@PDG2022}{\bibcite{PDG2022}{60}}{}%
\@ifundefined{b@troja2022}{\bibcite{troja2022}{61}}{}%
\@ifundefined{b@Euclid:2024yrr}{\bibcite{Euclid:2024yrr}{62}}{}%
\@ifundefined{b@Mao_2012}{\bibcite{Mao_2012}{63}}{}%
\@ifundefined{b@Gaudi2012}{\bibcite{Gaudi2012}{64}}{}%
\@ifundefined{b@Suyu2024}{\bibcite{Suyu2024}{65}}{}%
\@ifundefined{b@Michell_1784}{\bibcite{Michell_1784}{66}}{}%
\@ifundefined{b@Laplace_2009}{\bibcite{Laplace_2009}{67}}{}%
\@ifundefined{b@Jaki_1978}{\bibcite{Jaki_1978}{68}}{}%
\@ifundefined{b@Treder}{\bibcite{Treder}{69}}{}%
\@ifundefined{b@Einstein_1911}{\bibcite{Einstein_1911}{70}}{}%
\@ifundefined{b@Einstein_1915}{\bibcite{Einstein_1915}{71}}{}%
\@ifundefined{b@Dyson_1920}{\bibcite{Dyson_1920}{72}}{}%
\@ifundefined{b@Will2014}{\bibcite{Will2014}{73}}{}%
\@ifundefined{b@Meneghetti_2022}{\bibcite{Meneghetti_2022}{74}}{}%
\@ifundefined{b@Strutt_2009}{\bibcite{Strutt_2009}{75}}{}%
\@ifundefined{b@Yarimoto_2025}{\bibcite{Yarimoto_2025}{76}}{}%
\@ifundefined{b@Misner1973}{\bibcite{Misner1973}{77}}{}%
\@ifundefined{b@carroll2003spacetime}{\bibcite{carroll2003spacetime}{78}}{}%
\@ifundefined{b@Schneider}{\bibcite{Schneider}{79}}{}%
\@ifundefined{b@Perlick1_1990}{\bibcite{Perlick1_1990}{80}}{}%
\@ifundefined{b@Perlick2_1990}{\bibcite{Perlick2_1990}{81}}{}%
\@ifundefined{b@Faraoni_1992}{\bibcite{Faraoni_1992}{82}}{}%
\@ifundefined{b@Perlick1998}{\bibcite{Perlick1998}{83}}{}%
\@ifundefined{b@Perlick2000}{\bibcite{Perlick2000}{84}}{}%
\@ifundefined{b@Giannoni_2002}{\bibcite{Giannoni_2002}{85}}{}%
\@ifundefined{b@Perlick_2006}{\bibcite{Perlick_2006}{86}}{}%
\@ifundefined{b@Frolov_2013}{\bibcite{Frolov_2013}{87}}{}%
\@ifundefined{b@Weinberg_Steven1972-07}{\bibcite{Weinberg_Steven1972-07}{88}}{}%
\@ifundefined{b@Congdon_Iyer_Keeton_2022}{\bibcite{Congdon_Iyer_Keeton_2022}{89}}{}%
\@ifundefined{b@RoySenKerrMaterial2015}{\bibcite{RoySenKerrMaterial2015}{90}}{}%
\@ifundefined{b@1959RSPSA.249..180D}{\bibcite{1959RSPSA.249..180D}{91}}{}%
\@ifundefined{b@1967JMP.....8..265B}{\bibcite{1967JMP.....8..265B}{92}}{}%
\@ifundefined{b@Chandrasekhar}{\bibcite{Chandrasekhar}{93}}{}%
\@ifundefined{b@Iyer1}{\bibcite{Iyer1}{94}}{}%
\@ifundefined{b@Iyer2009Strong}{\bibcite{Iyer2009Strong}{95}}{}%
\@ifundefined{b@Rindler:2007zz}{\bibcite{Rindler:2007zz}{96}}{}%
\@ifundefined{b@1989ApJ...342..635M}{\bibcite{1989ApJ...342..635M}{97}}{}%
\@ifundefined{b@2026IJGMM..2350151H}{\bibcite{2026IJGMM..2350151H}{98}}{}%
\@ifundefined{b@PhysRevD.63.064014}{\bibcite{PhysRevD.63.064014}{99}}{}%
\@ifundefined{b@Gibbons_2008}{\bibcite{Gibbons_2008}{100}}{}%
\@ifundefined{b@Ishihara_2016}{\bibcite{Ishihara_2016}{101}}{}%
\@ifundefined{b@Ono_2017}{\bibcite{Ono_2017}{102}}{}%
\@ifundefined{b@Ishihara_2017}{\bibcite{Ishihara_2017}{103}}{}%
\@ifundefined{b@Ono_2018}{\bibcite{Ono_2018}{104}}{}%
\@ifundefined{b@Avgun_2018}{\bibcite{Avgun_2018}{105}}{}%
\@ifundefined{b@Ono_2019}{\bibcite{Ono_2019}{106}}{}%
\@ifundefined{b@Ono_2019b}{\bibcite{Ono_2019b}{107}}{}%
\@ifundefined{b@Avgun_2019}{\bibcite{Avgun_2019}{108}}{}%
\@ifundefined{b@Carmo}{\bibcite{Carmo}{109}}{}%
\@ifundefined{b@John_Oprea2019-02-06}{\bibcite{John_Oprea2019-02-06}{110}}{}%
\@ifundefined{b@Takizawa_2020}{\bibcite{Takizawa_2020}{111}}{}%
\@ifundefined{b@Huang_2024}{\bibcite{Huang_2024}{112}}{}%
\@ifundefined{b@takahashi2023equivalence}{\bibcite{takahashi2023equivalence}{113}}{}%
\@ifundefined{b@Madison_2013}{\bibcite{Madison_2013}{114}}{}%
\@ifundefined{b@Wang_2017}{\bibcite{Wang_2017}{115}}{}%
\@ifundefined{b@Deller_2019}{\bibcite{Deller_2019}{116}}{}%
\@ifundefined{b@Liu_2023}{\bibcite{Liu_2023}{117}}{}%
\@ifundefined{b@Ding_2024}{\bibcite{Ding_2024}{118}}{}%
\@ifundefined{b@Pearlman_2024}{\bibcite{Pearlman_2024}{119}}{}%
\@ifundefined{b@Walker_1996}{\bibcite{Walker_1996}{120}}{}%
\@ifundefined{b@Lai:2004ue}{\bibcite{Lai:2004ue}{121}}{}%
\@ifundefined{b@Hobill2008}{\bibcite{Hobill2008}{122}}{}%
\@ifundefined{b@Chanda_2019}{\bibcite{Chanda_2019}{123}}{}%
\@ifundefined{b@Werner_2012}{\bibcite{Werner_2012}{124}}{}%
\@ifundefined{b@Gibbons_2009}{\bibcite{Gibbons_2009}{125}}{}%
\@ifundefined{b@Landau1980}{\bibcite{Landau1980}{126}}{}%
\@ifundefined{b@Asada_2000}{\bibcite{Asada_2000}{127}}{}%
\@ifundefined{b@Bao_D_2012-10-03}{\bibcite{Bao_D_2012-10-03}{128}}{}%
\@ifundefined{b@Teukolsky_2015}{\bibcite{Teukolsky_2015}{129}}{}%
\@ifundefined{b@Epstein}{\bibcite{Epstein}{130}}{}%
\@ifundefined{b@Ibanez}{\bibcite{Ibanez}{131}}{}%
\@ifundefined{b@Caponio_2011}{\bibcite{Caponio_2011}{132}}{}%
\@ifundefined{b@Novello2000}{\bibcite{Novello2000}{133}}{}%
\@ifundefined{b@DeLorenci2000}{\bibcite{DeLorenci2000}{134}}{}%
\@ifundefined{b@ObukhovRubilar2002}{\bibcite{ObukhovRubilar2002}{135}}{}%
\@ifundefined{b@RoySenChargedMedium2017}{\bibcite{RoySenChargedMedium2017}{136}}{}%
\@ifundefined{b@RoyKalaGhoshKerrNewman2025}{\bibcite{RoyKalaGhoshKerrNewman2025}{137}}{}%
\@ifundefined{b@RoyKalaSinghaKerrSen2025}{\bibcite{RoyKalaSinghaKerrSen2025}{138}}{}%
\@ifundefined{b@dePaulaDymnikova2023}{\bibcite{dePaulaDymnikova2023}{139}}{}%
\@ifundefined{b@dePaulaAzreg2025}{\bibcite{dePaulaAzreg2025}{140}}{}%
\@ifundefined{b@dePaulaBirefringence2026}{\bibcite{dePaulaBirefringence2026}{141}}{}%
\@ifundefined{b@dePaulaHayward2023}{\bibcite{dePaulaHayward2023}{142}}{}%
\@ifundefined{b@Allahyari2020}{\bibcite{Allahyari2020}{143}}{}%
\@ifundefined{b@Kruglov2020}{\bibcite{Kruglov2020}{144}}{}%
\@ifundefined{b@Abalos2015}{\bibcite{Abalos2015}{145}}{}%
\@ifundefined{b@Oancea2020}{\bibcite{Oancea2020}{146}}{}%
\@ifundefined{b@Dymnikova2004}{\bibcite{Dymnikova2004}{147}}{}%
\@ifundefined{b@Hayward2006}{\bibcite{Hayward2006}{148}}{}%
\@ifundefined{b@HeisenbergEuler1936}{\bibcite{HeisenbergEuler1936}{149}}{}%
\@ifundefined{b@Dunne2004}{\bibcite{Dunne2004}{150}}{}%
\@ifundefined{b@c15LebedevLake2013}{\bibcite{c15LebedevLake2013}{151}}{}%
\@ifundefined{b@c15Oliveira2018}{\bibcite{c15Oliveira2018}{152}}{}%
\@ifundefined{b@c15CrisnejoGallo2018}{\bibcite{c15CrisnejoGallo2018}{153}}{}%
\@ifundefined{b@c15Crisnejo2019}{\bibcite{c15Crisnejo2019}{154}}{}%
\@ifundefined{b@c15Cardoso2009}{\bibcite{c15Cardoso2009}{155}}{}%
\@ifundefined{b@c15Bozza2002}{\bibcite{c15Bozza2002}{156}}{}%
\@ifundefined{b@c15BozzaMancini2004}{\bibcite{c15BozzaMancini2004}{157}}{}%
}
\makeatother
\date{September 29, 2026}
\begin{document}
	\maketitle
	\GBTableOfContents
	\pagebreak 
	\begin{abstract}
	Gravitational lensing constitutes one of the most direct observational manifestations of spacetime curvature and provides a powerful probe of compact astrophysical objects. In this work, we present a comprehensive analysis of the bending of light in curved spacetime, beginning with the fundamental aspects of gravitational lensing and the Newtonian approximation to light deflection. The relativistic formulation of lensing is then developed through the lens equation, lensing potential, and a geometrical interpretation of Fermat’s principle using an effective refractive index in curved spacetime. Photon trajectories and light deflection are subsequently investigated in static, spherically symmetric geometries, followed by a detailed study of photon motion in the equatorial plane of the Kerr spacetime. Analytical expressions for the closest approach distance and critical parameters governing photon orbits are derived. Furthermore, the bending angle is examined using the Rindler-Ishak method and the Gauss–Bonnet theorem within the optical geometry. Finally, the analysis is extended to axisymmetric spacetimes using the OIA and GW–OIA formalisms, providing a unified geometrical framework for computing light deflection in both static and rotating gravitational fields. The review also examines nonlinear electrodynamics (NED), distinguishing changes in the gravitational background from changes in the characteristic propagation of electromagnetic waves. Representative models are used to relate these effects to finite-distance deflection, polarization-dependent optical observables, and comparisons between weak bending and strong-field capture.

	\end{abstract}
	
	

\section{Introduction}{\label{s1}}

Gravitational lensing is one of the most remarkable predictions of 
\textit{General Relativity} (GR)\cite{carroll1997lecturenotesgeneralrelativity}, where the presence of mass and energy curves spacetime and consequently influences the propagation of light. In a curved spacetime geometry, photons do not travel along straight lines in the Euclidean sense\cite{zaveri2025periodicrelativitytheorygravity}; instead, they follow \textit{null geodesics} determined by the gravitational field of intervening matter. As a result, when light emitted from a distant astrophysical source propagates through the universe, its trajectory can be deflected by the gravitational potential of massive objects such as stars, galaxies, galaxy clusters, or even 
large-scale cosmic structures. This phenomenon is known as \textit{gravitational lensing}.

The basic physical principle underlying gravitational lensing is closely analogous to the refraction of light in classical optics. 
In optical systems, variations in the refractive index of a medium bend the path of light rays. Similarly, in gravitational lensing, the curvature of spacetime induced by mass distributions alters the propagation of electromagnetic radiation. Because of this strong conceptual similarity, gravitational lensing is often 
referred to as \textit{gravitational optics}. 
However, unlike conventional optical systems, the ``lens'' in gravitational lensing is not a physical medium but rather the spacetime curvature generated by matter and energy.

Historically, the deflection of light by gravity was first predicted by Einstein shortly after the formulation of GR in 1915. 
The earliest observational confirmation of this effect came during the 
1919 total solar eclipse expedition led by Sir Arthur Eddington, which 
measured the apparent displacement of background stars near the Sun. 
The observed deflection was consistent with Einstein’s theoretical prediction, providing one of the first empirical tests of GR and marking the beginning of gravitational lensing as a physical phenomenon of observational relevance. Since then, improvements in observational technology and astronomical surveys have transformed gravitational lensing from a theoretical curiosity into a 
central tool of modern astrophysics and cosmology.

Gravitational lensing phenomena are commonly classified into three regimes:\textbf{strong lensing, weak lensing, and microlensing}. 
These regimes differ primarily in the strength of the gravitational field and the geometric alignment between the source, lens, and observer. Strong gravitational lensing occurs when the gravitational potential of the lens is sufficiently large and the alignment between the observer, lens, and source is close to perfect. In such situations, multiple images of the background source may be formed. Depending on the symmetry of the system, these images can appear as arcs, Einstein rings, or multiple distinct images. 
Strong lensing is frequently observed in systems involving massive galaxies or galaxy clusters acting as lenses for distant quasars or galaxies.

Weak gravitational lensing, on the other hand, arises when the gravitational potential of the lens is relatively small or the alignment is less precise. 
In this regime, the deflection of light does not produce multiple images but instead induces subtle distortions in the shapes and orientations of background galaxies. These distortions, often referred to as \textit{cosmic shear}, are extremely small and cannot be detected for individual sources. Instead, they are measured statistically over large ensembles of galaxies. Weak lensing has become an indispensable observational probe for studying the 
large-scale structure of the universe and mapping the distribution of dark matter.

Microlensing represents another important regime of gravitational lensing. It occurs when the lensing mass is relatively small, typically of stellar or planetary scale. In such cases, the angular separation between the multiple images is extremely small, usually of the order of microarcseconds, making them impossible to resolve directly with current telescopes. Instead, microlensing events are detected through characteristic variations in the brightness of the background source as the lensing object moves across the 
line of sight. Microlensing has proven to be a powerful method for detecting exoplanets, compact stellar remnants, and other faint astrophysical objects.

From a theoretical perspective, gravitational lensing describes the deflection of null geodesics corresponding to massless waves such as electromagnetic radiation and gravitational waves. As these waves propagate through curved spacetime generated by a massive 
object, their trajectories are bent by the gravitational potential of the intervening mass distribution. The magnitude of this bending is quantified by the \textit{deflection angle}, which depends on the mass, geometry, and spatial distribution of the lens. 
In simple cases, such as a point mass lens, the deflection angle can be calculated analytically, while more realistic astrophysical systems require numerical modeling and sophisticated lens reconstruction techniques.

One of the most important outcomes of gravitational lensing is the formation of distinct image configurations, magnification patterns, and time delays between multiple images of the same source. 
These observational signatures encode valuable information about the geometry of the lensing system as well as the underlying gravitational field. By analyzing these effects, astronomers can reconstruct the mass distribution of the lens, even when that mass is not directly visible.

A particularly significant application of gravitational lensing lies in the study of dark matter. Although dark matter does not emit, absorb, or scatter electromagnetic radiation, it contributes to the gravitational potential that bends light. Consequently, gravitational lensing provides a unique and direct method for mapping the distribution of dark matter in galaxies and galaxy clusters. 
Weak lensing surveys, in particular, allow researchers to reconstruct the large-scale distribution of dark matter across cosmological distances, offering important insights into the formation and evolution of cosmic structures.

In addition to dark matter studies, gravitational lensing has become an essential tool for investigating dark energy, measuring cosmological parameters, and probing the expansion history of the universe. Time delays between multiple images of strongly lensed quasars can be used to estimate the Hubble constant, while large weak-lensing surveys provide constraints on the growth of cosmic structure and the dynamics of cosmic acceleration.

In recent years, gravitational lensing has also gained renewed attention in the context of multi-messenger astronomy. With the detection of gravitational waves from compact binary mergers, it has 
become possible to investigate the lensing of gravitational waves themselves. Just like electromagnetic radiation, gravitational waves propagate along null geodesics and can therefore be deflected, magnified, or multiply imaged by massive objects. Studying such lensing effects opens new avenues for testing gravitational 
theories and probing the matter distribution of the universe.

Given its wide range of theoretical implications and observational 
applications, gravitational lensing has emerged as one of the most powerful tools in modern astrophysics and cosmology. 
It provides a natural laboratory for testing the predictions of GR, exploring the properties of dark matter and dark energy, and studying distant astrophysical sources that would otherwise remain inaccessible.

The purpose of this review is to provide a comprehensive overview of the theoretical foundations and astrophysical applications of gravitational lensing. We aim to synthesize key developments in the literature, highlighting both the mathematical framework underlying lensing phenomena and the diverse range of observational contexts in which these effects play a crucial role. Through this discussion, we seek to emphasize the importance of gravitational lensing as a bridge between fundamental gravitational physics and observational cosmology.

We further consider nonlinear electrodynamics (NED), in which electromagnetic self-interactions can modify both the spacetime geometry and the characteristic cones governing photon propagation. Through the Dymnikova and Hayward examples, the fully regular $h$-family, and a quadratic two-invariant model, we distinguish background-null calculations from electromagnetic characteristic calculations, including birefringence where present. Particular attention is given to finite source and observer distances, the distinction between phase-normal and energy-transport directions, and comparisons of deflection, capture scales, and polarization-dependent travel times.

\section{Rudiments of Gravitational Lensing}{\label{s1.1}}

Gravitational lensing, like everything else associated with General Relativity, is corrupted with the seductive grace and elegance of mathematical rigor that lends itself readily to the explanation of breathtaking physical phenomena. This is perhaps one of the strongest reasons why gravitational lensing has emerged as a major field of research in the domain of modern astrophysics. Simply put the deflection of light rays from a far-off source as they approach a massive object, like a galaxy or a black hole, is known as gravitational lensing \cite{{Narayan1997}, {Dodelson_2017}}. The massive object's gravitational field causes spacetime to curve, which results in this bending of light. The deflection angle, a crucial parameter that is essential to both theoretical astrophysics and observational cosmology, describes the degree of this bending. This review attempts to provide a detailed explanation of the importance of light bending and how the deflection angle is measured with strong emphasis on the effects of finite distance corrections based on novel mathematical techniques. Far from being an affair of theoretical curiosities lensing is very much an asset for experimental verification and/or designing new analytical frameworks for current and upcoming observations. Ranging from fields such as structural mapping of galaxies and galaxy clusters \cite{{Bartelmann_2010}, {Jalilvand_2020}, {Torres_Ballesteros_2022}, {Natarajan_2024}, {Ferrami_2024}, {Zamani_2024}, {Buncher_2025}} to investigating the elusive properties of dark matter and dark energy \cite{{Massey_2010}, {Ellis_2010}, {Karamazov_2021}, {eng_l_2022}, {Vegetti_2023}, {Vachher_2024}, {Gannon_2025}}, gravitational lensing encompasses a vast spread of research topics. Although it does not directly interact with light, dark matter, which makes up a sizable amount of the universe's mass, gravitationally affects passing photons. In order to reconstruct the mass distribution of dark matter halos, scientists examine the deflection angles of light from distant background galaxies. Astronomers can gain a deeper understanding of the enigmatic nature of dark matter by using methods like weak gravitational lensing to infer the large-scale structure of dark matter in cosmic voids and galaxy clusters. In fact gravitational lensing has been successfully employed in studies related to exoplanets as well. The lensing effects of a planet that orbits the lensing star can also be measured. This method is employed to find planets that are far from Earth. \cite{Tsapras_2018, Turyshev_2020, Turyshev_2022, Toth_2023, Mr_z_2024}. 

\par The deflection of null geodesics—trajectories followed by massless waves, such as electromagnetic waves (EMW) and gravitational waves (GWs)—as they propagate through the curved spacetime surrounding a massive object falls very much within the purview of gravitational lensing as well \cite{{Li_2018}, {Diego_2019}, {Pagano_2020},  {Bulashenko_2022}, {Tambalo_2023},  {Ashish_2024}, {Shan2025}}. Needless to say the gravitational potential of the intervening mass acts as the primary agent causing the said deflection, in accordance with the predictions of general relativity. The resulting lensing effect can generate intricate interference patterns and distinct image configurations or \emph{\emph{caustics}}, which depend on the mass distribution of the lens and the geometric alignment between the source, lens, and observer. Hence, we can claim that via gravitational lensing, we gain a deeper understanding of how both electromagnetic radiation and gravitational waves are detected and interpreted from distant astrophysical sources. Moreover, gravitational lensing also plays a fundamental role as a cosmological probe. But before we delve into that discussion it is essential to gain some fundamental understanding of the different regimes of gravitational lensing.

\subsection{Observational regimes of gravitational lensing}{\label{s1.1.1}}
\par Gravitational lensing can be classified into three distinct categories — strong lensing, weak lensing, and micro lensing — depending on how much the lensing object influences the light rays coming from the source towards the observer. A cursory description of these phenomena is given below with relevant images illustrating the observations made and in some cases the corresponding geometric alignment.
\begin{figure}[hbt!]
\begin{center}\includegraphics[scale=0.15]{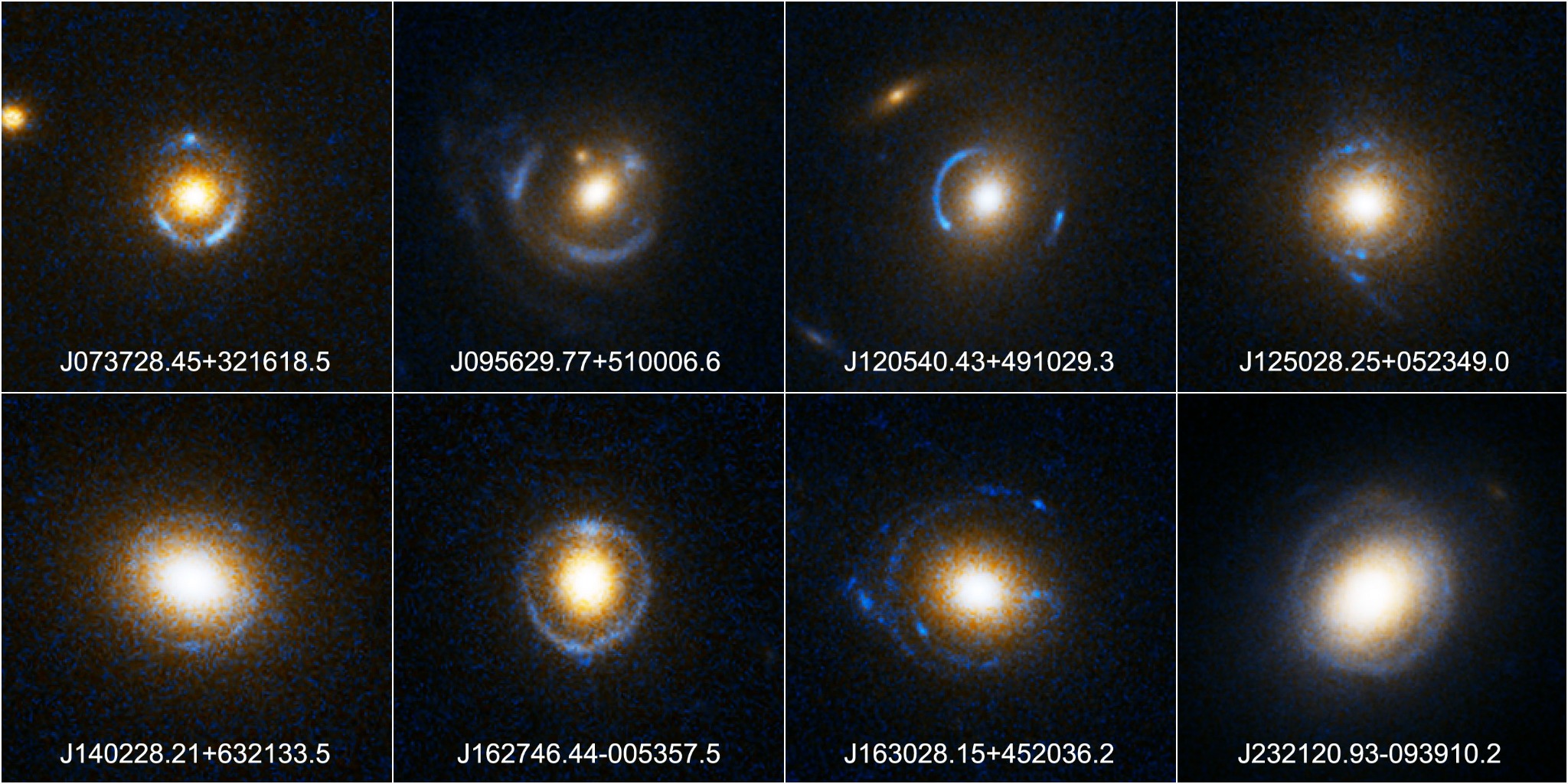}
\end{center}
\caption{An assortment of Einstein rings photographed by Hubble Space Telescope in 2005 as part of the Sloan Lens Arc Survey. \cite{Bolton_2006}}
\label{Einstein_rings}
\end{figure}
\begin{itemize}
\item[\textbf{(a)}] \textbf{Strong lensing} -- In scenarios where the mass of the lensing object is sufficiently large and can generate a strong gravitational field strong lensing effects may be induced. Of course the geometry involving the trinity of source, lens \& observer must conspire accordingly in order for this effect to be observed. In cases of strong lensing there are usually a multitude of images formed of the source and the deflection angle of the light rays are comparatively larger. Detailed discussions on the subject may be found in \cite{etherington2023,kumar2023,lemon_2023,saha2024,mccarty2024}. Since it only affects a small percentage of distant sources, multiple-imaging caused by strong lensing is a relatively uncommon phenomenon \cite{Press1973}. Subarcsecond resolutions and, depending on the application, knowledge of the source and deflector's redshifts are typical requirements for the measurements. Photometric redshifts are frequently used as estimates when redshift information is required and cannot be obtained through spectroscopy. However, because light from the background source and the foreground object mix, these are challenging to compute in crowded fields \cite{{Treu2010}, {Jouvel2014}, {Molino2017}}. Furthermore, they may occasionally turn out to be insufficiently accurate for the purposes in question \cite{Gonzalez_2018}. Strong lensing may also be utilised for determining cosmological parameters like the Hubble-Lemaitre parameter $H_0$ \cite{{Narlikar1996}, {Bernal_2016}}. Distortions of background galaxies or other light sources are highly noticeable in strong gravitational lensing, producing arcs, multiple images, or so called Einstein rings \ref{Einstein_rings} first proposed in \cite{Chwolson_1924} and later by \cite{Einstein_1936}.
\begin{figure}[hbt!]
\begin{center}\includegraphics[width=\linewidth,keepaspectratio]{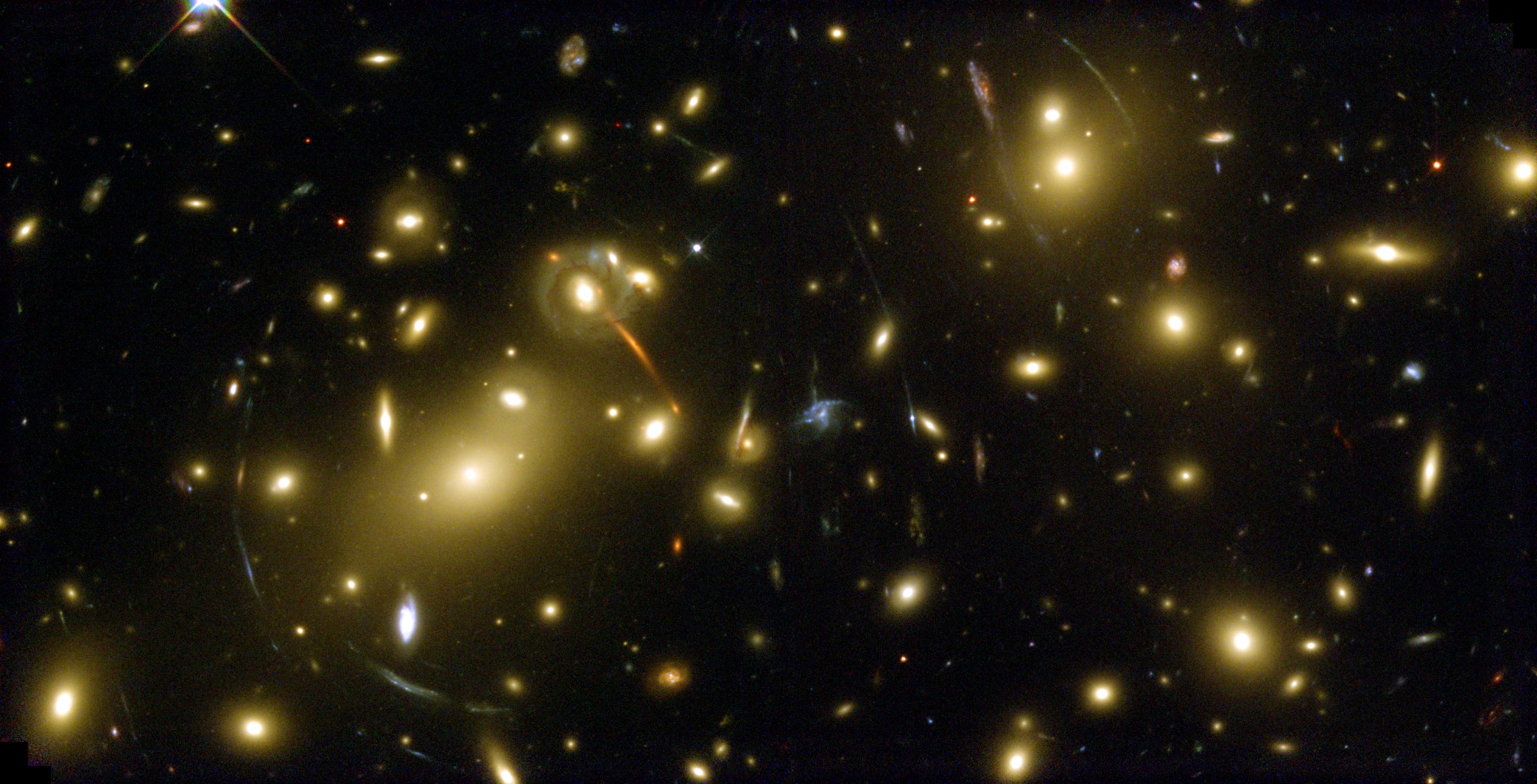}
\end{center}
\caption[Lensing of distant galaxy by the cluster Abell 2218 as an instance of weak lensing]{Lensing of distant galaxy by the cluster Abell 2218 as an instance of weak lensing. In the image provided above there are quite a few instances of multiple images or arcs bearing testimony for strong lensing. However, the background galaxies tell a different story with their distorted images. Although these distortions/magnifications are quite difficult to measure [\emph{as shown in later sections}].}
\label{abell_weak_lens}
\end{figure}
\item[\textbf{(b)}] \textbf{Weak lensing} -- In spite of the lensing object having a significant mass(either concentrated locally or distributed on large scales) the geometrical alignment of the aforementioned trinity may be less than ideal. As a consequence of this the image of the source suffers distortion to some extent (via magnification and shearing far from \emph{critical lines}) and in some cases mild arcing (or arclet formation) as well \cite{Fort_1988,Tyson_1990}. This effect is known as \emph{shear} in the literature of gravitational lensing. The existence of the shearing effect implies that the alignment of the objects in the background appear to be non-random. As a result of this the shearing effect can be measured statistically even if the distortions of individual objects are too insignificant for direct identification. In general formation of small arclets detectable directly is quite rare. The measurement of the slight distortions that lensing causes in the form of background galaxies when photons pass through large-scale structures is the basis of the cosmic shear technique. \cite{{Bartelmann_2001}, {Refregier_2003}, {hoekstra2013}, {Giblin_2021}, {Li_2022}, {Zhang_2024}, {Adam_2025}}. It is worth mentioning in this context that CMB lensing is classified as weak lensing and can accommodate the formation of multiple images for very small deflections provided the source is also very small \cite{LEWIS20061}.  A further sub-division of weak lensing is \textbf{statistical weak lensing} for cases which rely solely on statistical analysis and is devoid other observable characteristics. Foreground galaxy clusters and cosmic web filaments, as well as dark matter along the photon path, are examples of large-scale matter inhomogeneities that weakly lens high-redshift background galaxies. In these cases the lensing induced shear is observed statistically via averaging over a large number of galaxies \cite{{Kaiser1992}, {Kaiser1993}, {Kohlinger2017}, {Congdon2018}, {PDG2022}}. 
\par Because there are more dark-matter structures acting as lenses between us and the light sources, the distortions caused by weak gravitational lensing become more noticeable the farther we look. A three-dimensional layout of the distribution of dark matter in our universe will be provided by Euclid, which will measure the warped forms of billions of galaxies over ten billion years of cosmic history \cite{{troja2022}, {Euclid:2024yrr}}. This will clarify the nature of the  exotic dark-matter element. As such weak lensing has emerged as a potent tool in modern cosmology.

\item[\textbf{(c)}] \textbf{Microlensing} -- Microlensing is typically associated with lensing objects belonging to the category of stellar mass objects(\emph{e.g.}  normal stars, brown dwarfs or stellar remnants [white dwarfs, neutron stars and black holes])  . In this scenario the geometrical alignment is extremely favourable, but the image splitting is too small (of the order of milliarcseconds in the local group) to be resolved by ground-based telescopes. The effects that we can register are changes in magnification as a function of time. For the duration of the lensing event the image of the source appears to brighten. Note that a specific reference to \emph{time duration} is mentioned in the context of microlensing. This is due to the fact that the trinity of source, lens \& observer all have relative motions, and the alignment for which microlensing takes place is temporary. Microlensing has been employed as a tool for searching MACHO'S in the halo of the galaxy. For further discussions the reader is referred to the following works \cite{{Mao_2012}, {Gaudi2012}, {Suyu2024}}. In fact, signatures of microlensing in gravitational waves are also an active frontier of research \cite{Diego_2019}. 
\end{itemize} 

\par Regarding the formation of Einstein rings and the importance of geometrical alignment in gravitational lensing a detailed description is provided with appropriate schematics in \ref{lensing_geometry}.

\begin{table}[htbp]
\centering
\caption[Effect of lensing geometry on the observed image formation.]{The table provided below presents a series of schematics depicting the formation of Einstein rings and arcs of varying dimensions. The source (S), lensing object (L) and Viewer (V) are denoted by the colours orange, black and blue respectively. The corresponding alignment scenario with their respective astronomical observations allow for a readymade conceptualisation of the mechanics of image formation by gravitational lensing. A comparative study with the panel given in \ref{Einstein_rings} should help us understand the several arcs present alongside the actual rings and how the geometrical alignment of the source, lens and observer must have been for those images.\\ Scenario III presents the most ideal perfectly symmetrical alignment of the source, lens and viewer giving rise to the Einstein ring. {\color{black} For an ideal isolated point-mass lens and a source at any finite nonzero angular offset, the point-lens equation has two mathematical image solutions, one on either side of the lens.  In scenarios I and V the secondary image is highly demagnified and lies close to the lens; the schematic should therefore be read as showing the image that is likely to be detectable, not the total number of mathematical images.} Depending on the direction in which the source shifts from the perfect alignment the corresponding image also shifts accordingly. Scenario II \& IV presents only a slight shifting of the source from the perfect alignment. In this case there are two images formed. Depending on the direction of the shift of the source the locations of the elongated and diminished images are fixed accordingly. Note that whenever the source is offset from the ideal position the elongated image is formed slightly outside the Einstein radius while the smaller image is formed slightly within. }
\label{lensing_geometry}
\small
\begin{tabular}{|c|c|c|c|c|c|}
\hline
\bf Scenario & \bf I & \bf II & \bf III & \bf IV & \bf V \\
\hline  
 \multirow{-4.5}{*}{\textbf{\makecell{Lensing\\ geometry}}}  & \includegraphics[scale=0.185]{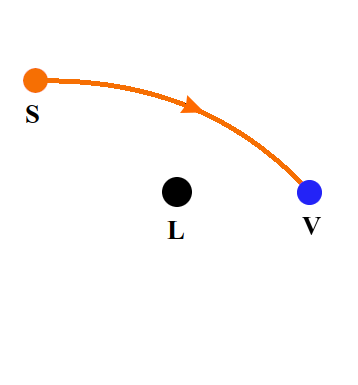} & \includegraphics[scale=0.185]{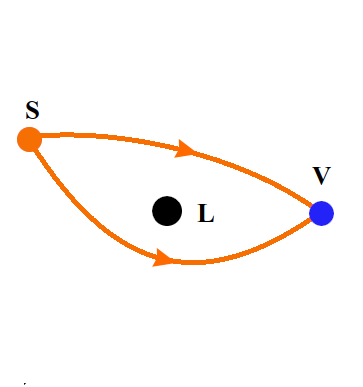} & \includegraphics[scale=0.185]{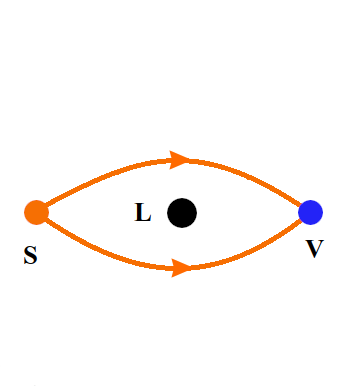} & \includegraphics[scale=0.185]{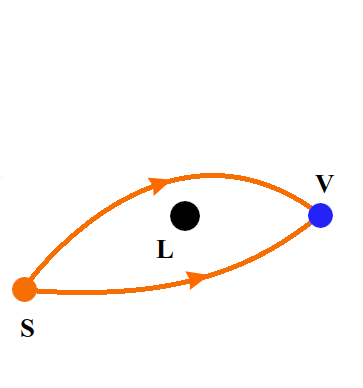} & \includegraphics[scale=0.185]{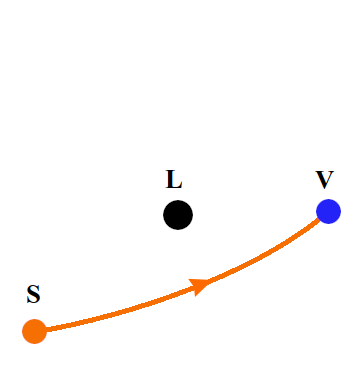} \\
\hline
\multirow{-4.5}{*}{\textbf{\makecell{Observed\\ image}}}  & \includegraphics[scale=0.185]{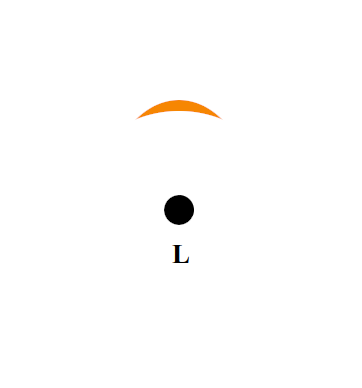} & \includegraphics[scale=0.185]{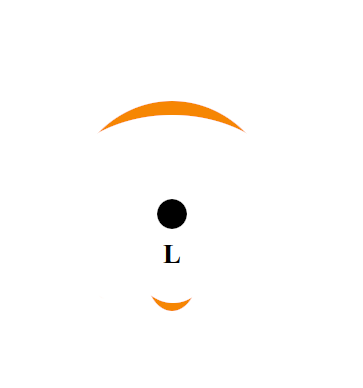} & \includegraphics[scale=0.185]{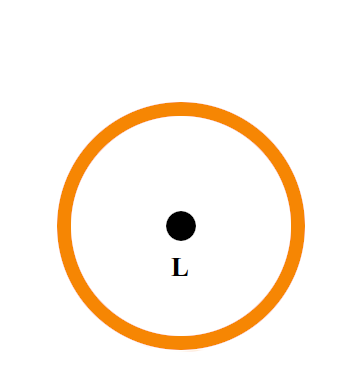} & \includegraphics[scale=0.185]{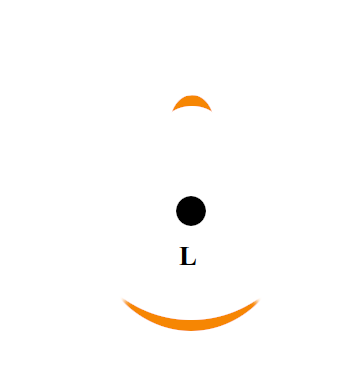} & \includegraphics[scale=0.185]{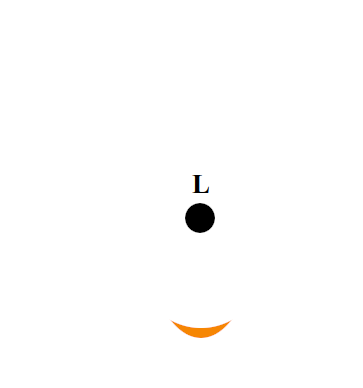} \\
\hline
\end{tabular}
\end{table}

\section{Early Beginnings}{\label{s1.2}}
Like several fundamental physical concepts that are the cornerstones of modern physics, the idea of light rays being influenced by a gravitational field can be traced back to Isaac Newton. However, it seems there was a debt owed to time and it was only in 1784 that Henry Cavendish computed the deflection angle of light due to a point mass. The two major assumptions behind his derivation were Newton's theory of gravitation and Newton's corpuscular theory of light. John Michell a friend and colleague of Cavendish had a pivotal role to play in this chapter of history as well. Michell in a 1783 correspondence to Cavendish laid down a proposal to determine the mass of stars by measuring the reduction in speed of the light corpuscles traveling from the star to Earth \cite{Michell_1784}. This letter turned out to be a key motivator for Cavendish's subsequent work on the deflection of light in a gravitational field. Strangely enough Cavendish was reluctant to publish his calculations and his work gained recognition only after he had passed away. 
The 1783 letter from Michell reserves its place in the history of science for another reason as well. It was here that he ventured the idea that the universe might host a certain variety of "dark stars" commanding such an enormous gravitational field that not even light can escape from their surface. This is now widely accepted as the very first instance of anyone proposing the concept of a black hole. Pierre-Simon Laplace made independent contributions to the subject over a decade later and ascertained the conditions that would render the escape velocity from the surface of an object  greater than the velocity of light \cite{Laplace_2009}.   
\par Laplace's work, it seems, had a strong influence on Johann Georg von Soldner, which motivated him to compute detailed calculation of the deflection of light under the gravitational influence of a massive object \cite{{Jaki_1978}, {Treder}}. von Soldner's results showed that for an object of mass $M$ a ray of light passing close to the object is deflected by
\begin{equation}
\alpha_{\rm Newton} = \dfrac{2GM}{c^2 b}\label{e_1.1}
\end{equation} 
where symbols having their usual meaning. Einstein in 1911 came up with the same results as Soldner employing the equivalence principle of special theory of relativity \cite{Einstein_1911}. Following this Einstein persuaded his friend, the German astronomer Erwin Freundlich to lead an expedition to the Crimean peninsula with the objective of measuring the deflection of light during a solar eclipse. Unfortunately or (perhaps fortunately?), the onset of WWI prevented Freundlich's team from fulfilling their undertaking. This setback in the meantime allowed Einstein to perfect his \emph{magnum opus}-- The General Theory of relativity. Armed with the tools of general relativity he revisited the problem of deflection of light in a gravitational field and found that the bending angle was 
\begin{equation}
\alpha_{\rm Einstein} = \dfrac{4GM}{c^2 b}
\label{e_1.2}
\end{equation} 
\emph{i.e.,} twice the value in Newtonian regime \cite{Einstein_1915}. While the discrepancy in Mercury's orbit was successfully addressed  by Einstein's theory, the experimental verification of bending angle of light near a massive object was to be the acid test that would establish general relativity as an accurate physical theory. The stage was set. 
\par A total solar eclipse was due to occur in May, 1919. Frank Dyson was put in charge of organizing a British expedition. The decision was made to send two groups of two observers each. One team traveled to Sobral, Brazil, and included employees from the Royal Observatory in Greenwich. 220 kilometers off the coast of West Africa, the Island of Principe is a small volcanic island that was visited by the second group from the Cambridge Observatory. Andrew Crommelin and Charles Davidson, his less experienced colleague, made up the Greenwich group. Edwin Cottingham, who maintained the Observatory's clocks, and Arthur Eddington, the director of the Observatory, made up the Cambridge group. The joint standing eclipse committee of the Royal Society and Royal Astronomical Society provided support and some funding for the expeditions.
It is interesting to note that the previous recorded solar eclipse by Dyson was in 1905 and the few stars that were observed during that event were insufficient for any conclusive evidence either in favour or against general relativity. As far as other attempts at experimental verification are concerned the attempt by a group of astronomers at Lick observatory during the total eclipse of June 1918 had also been unsuccessful.    
 
\par Eddington's experiment not only cemented Einstein's theory as the most beautiful and precise theory of gravitation but it also set a precedent of international collaboration amongst scientists rising above political animosity \cite{Dyson_1920,Will2014}. After all, the times were that of WWI and it was a British astronomer toiling after the predictions of a German theoretician. Needless to say, the successful union of an elegant mathematical structure with clinical experimental validation at the solar eclipse of May 1919 ensured the passing over of the baton from Newton to Einstein. Ironically, it took an eclipse for physics to come out of the shadow of an intellectual giant by the name of Newton.{\color{black} It should however be noted that the historical plates had different uncertainties and their interpretation has subsequently been re-examined; nevertheless the result was important evidence for the relativistic prediction, but neither a unique nor a ``clinical'' validation of the whole theory.}

\section{Bending Angle of Light in Newtonian Mechanics}{\label{s1.3}}
Although we now know that as far as the bending of light is concerned, Newton's theory is off by a factor of $2$ it is still instructive to follow the actual derivation in Newtonian mechanics. {\color{black} The additional factor of two in general relativity is not obtained by a Newtonian adjustment: at leading post-Newtonian order one contribution is associated with the time-time part of the metric and an equal contribution with spatial curvature.  A relativistic derivation, or an explicitly stated post-Newtonian optical metric, is therefore required \cite{Congdon2018,Meneghetti_2022}.}.
\begin{figure}[hbt!]
\begin{center}
\includegraphics[scale=.25, angle=0]{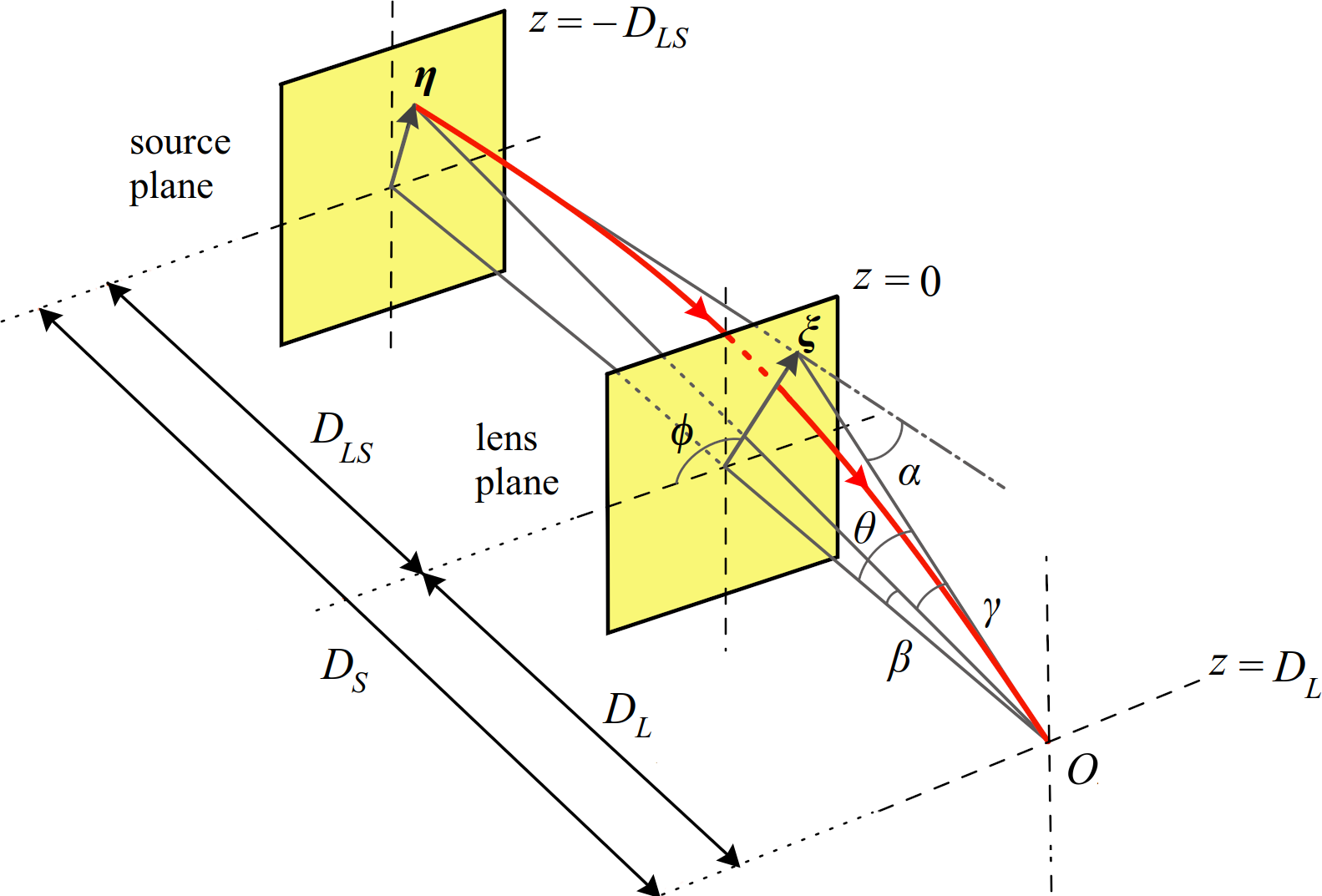}
\end{center}
\caption[Illustration depicting the bending of light under the influence of gravitational field.]{Illustration depicting the bending of light under the influence of gravitational field. The actual path of the light ray is traced in red. The lensing object is an object of mass $M$ at $z=0$. The deflection angle is given by the angle $\alpha=\left(\overrightarrow{\xi},\; \phi  \right)$ where $\overrightarrow{\xi}$ denotes the distance of closest approach for the light ray. The other symbols depicted here will come in useful while deriving the lens equation and other essential quantities. But for the time being we will concentrate on the variables necessary for deriving the expression for deflection angle. \textbf{Note:} The mass $M$ may be a point mass as well as an extended distribution of mass in which case $\overrightarrow{\xi}$ denotes the radius of the extended mass.}
\label{lens_schematic1}
\end{figure} 
\par As depicted in \ref{lens_schematic1} the lens is located at  $z=0$ and with respect to it the observer and the source are located at $z=D_L$ and $z=-D_{LS}$ respectively (the negative sign indicates distance being measured in the opposite sense from that of the observer). The distance between the source and the observer is $D_{S}$. In this particular scenario we have chosen the cylindrical coordinates $\left(\xi, \phi, z \right)$ where the $z$-axis is the central axis passing through the different planes shown in \ref{lens_schematic1}. Let us now assume that the velocity of light ray, traveling initially parallel to the $z$-axis is given as $c\hat{e}_z\equiv \overrightarrow{v}_{\rm st}$ where $\hat{e_z}$ is the unit vector along the $z$-axis and $c$ of course is the speed of light. It is important to note in this context that $\overrightarrow{\xi}$ is the coordinate along the axis perpendicular to $z$-axis. Given that we are working with the underlying framework of corpuscular theory and Newtonian gravitation, we can stipulate that the lens induces an acceleration of $\overrightarrow{a}=-\overrightarrow{\nabla} \Phi(\overrightarrow{\xi})$. $\Phi$ represents the gravitational potential at location $\overrightarrow{\xi}$. Therefore, the corresponding change in velocity of the light corpuscle along its trajectory from the source to the observer is,
\begin{equation}
\Delta \overrightarrow{v} =\dfrac{1}{c} \int\limits_{-D_{LS}}^{D_L} \overrightarrow{a} {\rm d}z=-\dfrac{1}{c} \int\limits_{-D_{LS}}^{D_L} \overrightarrow{\nabla}\Phi {\rm d}z
\label{e1.3}
\end{equation}  
{\color{black} The minus sign follows directly from $\overrightarrow a=-\overrightarrow\nabla\Phi$.  With the conventional negative Newtonian potential, the transverse acceleration points toward the lens.}
Note, that we have simply swapped out the integration w.r.t time coordinate $t$ by using the relation $z=ct$. Furthermore, a crucial point to notice is that the limits of integration in \ref{e1.3} actually mean that the integration is not performed along the light trajectory but along the straight line connecting the source to the observer. However, this issue regarding the approximate nature of the integral can be circumvented by considering the first order Born approximation \cite{Strutt_2009}. The validity and accuracy of the Born approximation for a single lens plane is dealt with rigorously in \cite{Yarimoto_2025}. This essentially means that the light trajectory in the absence of gravitational field can be taken as zeroeth-order path from which we can determine the first order correction which can be fed into a second-order correction in turn. 

\par The gradient of the gravitational potential can be decomposed into the components along $z$ and $\xi$ as,
\begin{equation}
\overrightarrow{\nabla} \Phi = \dfrac{{\rm d} \Phi}{{\rm d}z}\hat{e}_z + \dfrac{{\rm d} \Phi}{{\rm d}\xi}\hat{e}_\xi = \overrightarrow{\nabla}_\parallel \Phi \hat{e}_z + \overrightarrow{\nabla}_\perp \Phi \hat{e}_\xi
\label{e1.4}
\end{equation}
Similarly, the change in velocity of the light corpuscle will have two components along $\hat{e}_z$ and $\hat{e}_\xi$ as well.
\begin{equation}
\Delta \overrightarrow{v} = \Delta v_\parallel \hat{e}_z + \Delta v_\perp \hat{e}_\xi
\label{e1.5}
\end{equation}
For the perpendicular component of change in velocity \ref{e1.3} yields,
\begin{equation}
\Delta v_\perp = -\dfrac{1}{c} \int\limits_{-D_{LS}}^{D_L} \dfrac{{\rm d}\Phi\left(\xi \right)}{{\rm d}\xi}{\rm d}z
\label{e1.6}
\end{equation}
As for the parallel component (\emph{i.e.,} along the line of sight) the amount of change in velocity is given by
\begin{align*}
\Delta v_\parallel &= -\dfrac{1}{c} \int\limits_{-D_{LS}}^{D_L} \dfrac{{\rm d}\Phi\left(\xi \right)}{{\rm d}z}{\rm d}z\\
&= \dfrac{1}{c} \left[ \Phi\left(\xi, -D_{LS}  \right)  - \Phi\left(\xi, D_L  \right)  \right] \numberthis \label{e1.7}
\end{align*}
Now, we may consider that the distance between the lens and source as well as the distance between lens and observer are both extremely large. As such $|-D_{LS}|$ \& $|D_L|$ may be considered to be $\infty$. Consequently as $\lim _{|z|\rightarrow \infty} \Phi = 0$, which yields $\Delta v_{\parallel}=0$. This means that no additional velocity is acquired by the light corpuscle along the direction of motion \emph{i.e.,} the velocity along the $z$-axis remains as $c$. 
\par Therefore, the velocity of light after getting deviated due to the gravitational field of the lens will be given by,
\begin{equation}
\overrightarrow{v}_{\rm dev} = \overrightarrow{v}_{\rm st} + \Delta \overrightarrow{v} = \Delta v_\perp \hat{e}_\xi + c \hat{e}_z
\label{e1.8}
\end{equation}
where $\overrightarrow{v}_{\rm st}$ represents the initial velocity of light.
The potential $\Phi$ has the well known form $-\dfrac{GM}{\sqrt{\xi ^2 + z^2}}$. Substituting, this in \ref{e1.6} with appropriate limits we get
\begin{equation}
\Delta v_\perp = -\dfrac{GM \xi}{c}\int\limits_{-\infty}^{\infty} \left(\xi ^2 + z^2  \right)^{-\frac{3}{2}} {\rm d}z
\label{e1.9}
\end{equation} 
We can now easily solve this integral by means of a variable change $\dfrac{z}{\xi}=\tan \theta$.
\begin{equation}
\Delta v_\perp = -\dfrac{GM}{c\,\xi} \int \limits_{-\frac{\pi}{2}}^{\frac{\pi}{2}} \cos \theta {\rm d}\theta = -\dfrac{2GM}{c\, \xi}
\label{e1.10}
\end{equation} 
This when substituted in \ref{e1.8} yields
\begin{equation}
\overrightarrow{v}_{\rm dev} = c\hat{e}_z - \dfrac{2GM}{c\, \xi}\hat{e}_\xi
\label{e1.11}
\end{equation}
Therefore, the angle of deflection is given by, 
\begin{equation}
\alpha_{\rm N}\left(\overrightarrow{\xi } \right) = \hat{e}_{\rm st} - \hat{e}_{\rm dev} = \dfrac{2GM}{c^2\xi}\hat{e}_\xi
\label{e1.12}
\end{equation}
Here, we have used the magnitude of $|v_{\rm dev}| $ as
\begin{equation}
|v_{\rm dev}| = \sqrt{c^2 + \dfrac{4G^2M^2}{c^2 \xi ^2}} \approx c
\label{e1.13}
\end{equation}
 This is precisely half the value of the result obtained from general relativistic calculations \cite{{Misner1973}, {carroll2003spacetime}}. 
\begin{equation}
\alpha_{\rm E}\left(\overrightarrow{\xi}  \right) = \hat{e}_{\rm st} - \hat{e}_{\rm dev} = \dfrac{4GM}{c^2\xi}\hat{e}_\xi
\label{e1.14}
\end{equation} 
If we substitute the value of mass as solar mass and the radius as solar radius in \ref{e1.12}, then we have a scenario where the light ray gets deflected while grazing the surface of the Sun. \\
\textbf{For Newtonian case---}
\begin{equation}
\alpha_{\rm N_{\odot}} = \dfrac{2G\times 1.989\times 10^{30}}{c^2\times 6.96 \times 10^8} \approx 0.875^{''}
\label{e1.15}
\end{equation}
\textbf{For Einsteinian case---}
\begin{equation}
\alpha_{\rm E_{\odot}} = \dfrac{4G\times 1.989\times 10^{30}}{c^2\times 6.96 \times 10^8} \approx 1.75^{''}
\label{e1.16}
\end{equation}
Another interesting case to consider is the case for an extended mass distribution. In particular the introduction of terms such as surface mass density etc will prove extremely essential in defining some of the key concepts related to gravitational lensing.
\par For any arbitrary 2-dimensional surface mass distribution, the mass term in the expression for deflection angle of light can be expressed as, ${\rm d}M = \Sigma \left(\overrightarrow{\xi}   \right) {\rm d}^2 \xi$. {\color{black} Thus $\Sigma$ is mass per unit area on the lens plane, while $\Sigma\,{\rm d}^{2}\xi$ is the mass in the area element ${\rm d}^{2}\xi$.}. Introducing the correction due to general relativity the deflection angle for this case is given by
\begin{equation}
\alpha\left(\overrightarrow{\xi}  \right) = \dfrac{4G}{c^2}\int\limits_{\mathcal{R}^2} {\rm d}^2 \xi ^\prime \Sigma\left(\overrightarrow{\xi ^\prime}   \right) \dfrac{\overrightarrow{\xi} - \overrightarrow{\xi ^\prime}}{|\overrightarrow{\xi}- \overrightarrow{\xi ^\prime}|^2}
\label{e1.17}
\end{equation}
where the limits of integration are so chosen as to encompass the the entire mass distribution in the lens plane. {\color{black} Equation~\ref{e1.17} is the standard weak-field, thin-lens expression under the following conditions:}
\begin{itemize}
\item[\textbf{(a)}] The gravitational fields are weak, \emph{i.e.,} the deflection angle is small.
\item[\textbf{(b)}] {\color{black} The characteristic velocities of the lens matter satisfy $|\boldsymbol v_{\rm lens}|\ll c$, unless gravitomagnetic corrections are included explicitly.}
\end{itemize} 
In general for almost every conceivable astrophysical scenario these conditions are applicable.

\par The following sections will essentially provide derivations of some of the key concepts/relations regarding --- lens geometry, multiple imaging, magnification ratio and time delay. 

\section{The Lens Equation}{\label{s1.4}}
In order to derive the lens equation let us first define some of the essential elements of lensing geometry.
\begin{itemize}
\item[] $D_L = $ {\color{black} observer-lens angular-diameter distance}
\item[] $D_S = $ {\color{black} observer-source angular-diameter distance}
\item[] $D_{LS} = $ {\color{black} lens-source angular-diameter distance}
\item[] $\overrightarrow{\beta} = $ actual angle between the lens and the source
\item[] $\overrightarrow{\theta} = $ observed angle between the lens and the source \emph{i.e.,} angle between the observer and the image
\item[] $\overrightarrow{\xi} = $ distance from the lens to photon trajectory at the lens plane
\item[] $\overrightarrow{\alpha} = $ the angle of deflection 
\end{itemize}
Here, $\overrightarrow{\beta}$, $\overrightarrow{\theta}$, $\overrightarrow{\alpha}$, and $\overrightarrow{\xi}$ are all vector quantities and our derivations should include both negative as well as positive angles. In case of simple point source lens objects (needless to say an approximation) the vectorial aspects of the angles don't play any significant role. However, for complex lensing systems like galaxies or clusters vectorial treatment becomes necessary.

\par {\color{black} In cosmological lensing the three distances above are angular-diameter distances and do not, in general, obey the Euclidean identity $D_S=D_L+D_{LS}$.  Equations~\ref{e1.18}--\ref{e1.22} use the small-angle, thin-lens geometry with these distance factors; $\overrightarrow\xi=D_L\overrightarrow\theta$ is a physical transverse vector in the lens plane.}

\par
\begin{figure}[htbp!]
    \centering
    \subfloat[]{{\includegraphics[scale=0.2]{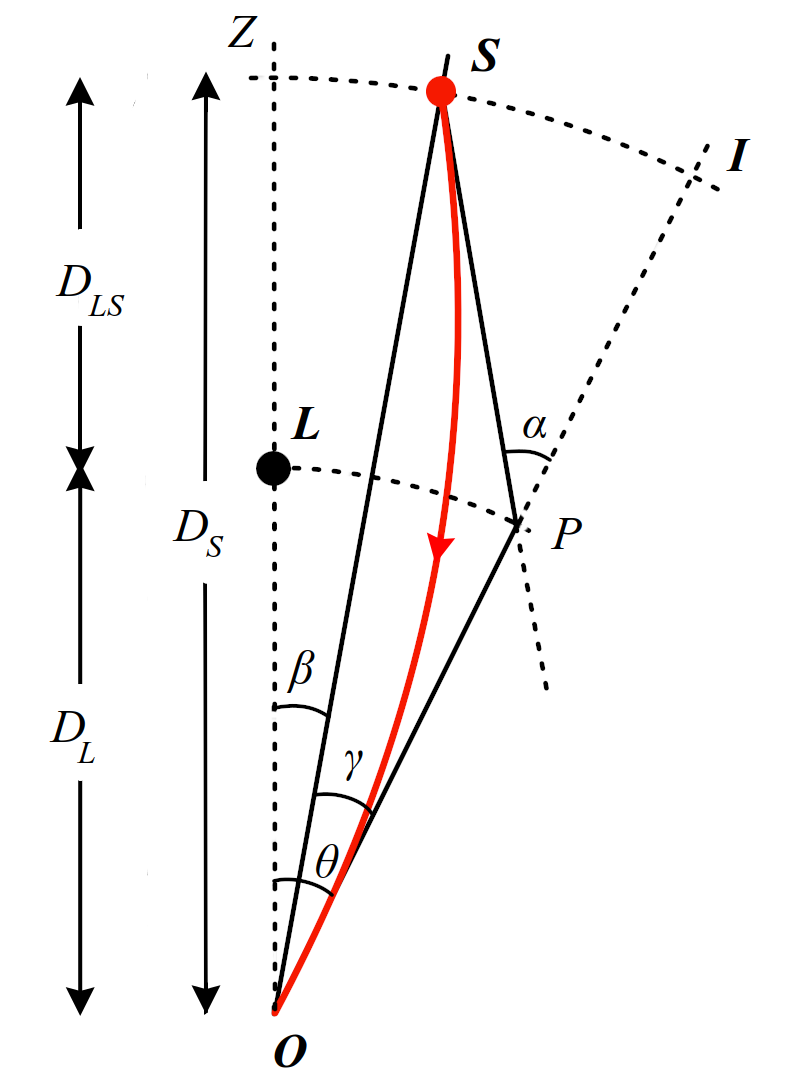} }}%
    \hfill
    \subfloat[]{{\includegraphics[scale=0.2, angle=0]{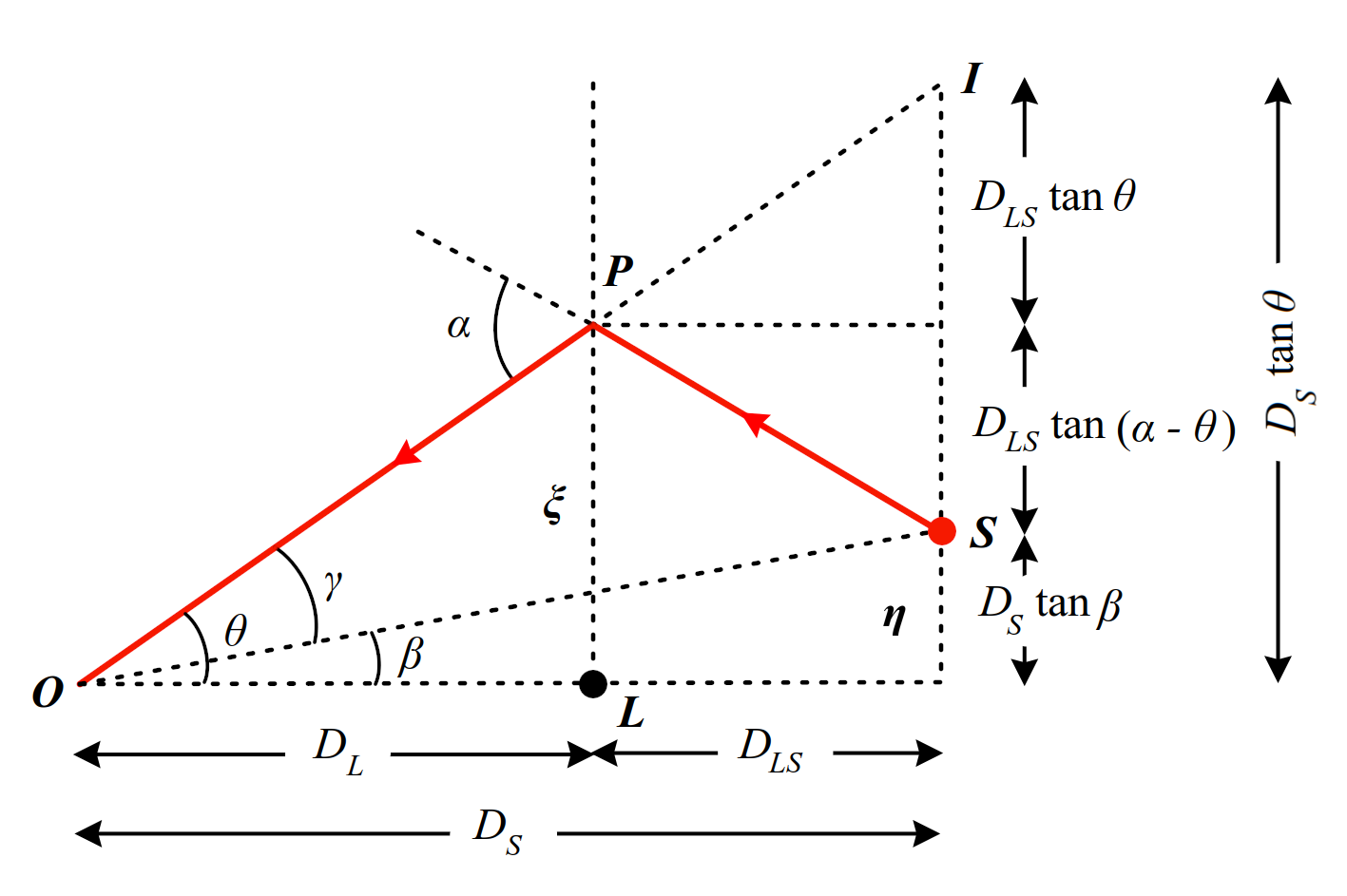} }}%
    \caption[Illustration depicting the projection of the lensing geometry onto a 2-D spatial slice.]{\textbf{(a)} Illustration depicting the projection of the lensing geometry onto a 2-D spatial slice. Here we consider the spacetime to be Friedmann-Lemaitre-Robertson-Walker (FLRW) and the coordinates to be comoving. As such we are at liberty to project the spacetime diagram onto a spatial slice. The source \textbf{S} emits a photon which follows the path (spatial projection) indicated by the red solid line and reaches the observer \textbf{O}. The lens is marked by the black circle at \textbf{L}. The source is at an angular position $\beta$ w.r.t the observer whereas the image \textbf{I} is at an angular position $\theta$. As we consider the flat sky approximation the spheres at the source and lens can be replaced by their tangent planes, \emph{i.e.,} the arcs LP and SI can be replaced by their tangents at L and S respectively. \\ \textbf{(b)} For thin lens approximation the actual spatial path is replaced by the piecewise spatial path given by SPO. The magnitude of the deflection angle is given by the angle between SP and PI.}%
    \label{lens_schematic2}%
\end{figure}

 We can invoke the small angle approximation as $\beta$, $\theta$ and $\alpha$ are all very small. Furthermore, since $D_S$, $D_{LS},$ and $D_L$ are all considerably larger than $\xi$ this approximation holds good [\ref{lens_schematic2}]. 
 
\par With reference to \ref{lens_schematic2} we can write the geometric relation
\begin{equation}
D_S \tan \beta = D_S \tan \theta - D_{LS} \tan \alpha
\label{e1.18}
\end{equation}
As a consequence of the small angle approximation we have $\left(\tan \theta \approx \theta  \right)$. Therefore, we can write \ref{e1.18} as
\begin{align*}
& D_S \beta = D_S \theta - D_{LS} \alpha\\
\Rightarrow & \beta = \theta - \dfrac{D_{LS}}{D_S} \alpha \numberthis
\label{e1.19}
\end{align*}
In vector notation which is
\begin{equation}
\overrightarrow{\beta} = \overrightarrow{\theta} - \dfrac{D_{LS}}{D_S}\alpha \left(  \overrightarrow{\xi}\right)
\label{e1.20}
\end{equation}
We can define a scaled deflection angle $\alpha\left( \overrightarrow{\theta} \right) = \dfrac{D_{LS}}{D_S}\alpha\left(\overrightarrow{\xi}   \right)$ and express the lens equation given by \ref{e1.20} as
\begin{equation}
\overrightarrow{\beta} = \overrightarrow{\theta} - \alpha\left(\overrightarrow{\theta}  \right)
\label{e1.21}
\end{equation}
Similarly, if we want to express the lens equation in terms of displacement vectors $\overrightarrow{\xi}$ and $\overrightarrow{\eta}$ then we have
\begin{equation}
\overrightarrow{\eta} = \dfrac{D_S}{D_L}\overrightarrow{\xi} - D_{LS}\alpha\left(\overrightarrow{\xi}  \right)
\label{e1.22}
\end{equation}
In \ref{e1.21} the dependent variable has been changed from $\xi$  to $\theta$. The physical interpretation of the lens equation here is that an image of a source at true position $\overrightarrow{\beta}$ will be observed at angular positions $\overrightarrow{\theta}$ so long as \ref{e1.21} is satisfied. Multiple images are produced for cases where the lens equation has multiple solutions. 
\par A common practice in gravitational lensing is to write the lens equation in {\color{black}its dimensionless scaled form} by defining a length scale $\xi_0$ on the lens plane and a corresponding length scale $\eta_0 = \dfrac{\xi_0 D_S}{D_L}$ on the source plane and using them to define the dimensionless (scaled) vectors
\begin{equation}
\overrightarrow{x}\equiv \dfrac{\overrightarrow{\xi}}{\xi_0},\;\;\;\;\; \overrightarrow{y}\equiv \dfrac{\overrightarrow{\eta}}{\eta_0}
\label{e1.23}
\end{equation}
Hence, the dimensionless lens equation is,
\begin{equation}
\overrightarrow{y} = \overrightarrow{x} - \alpha\left(\overrightarrow{x}  \right)
\label{e1.24}
\end{equation}
where
\begin{equation}
\alpha\left(\overrightarrow{x}  \right) = \dfrac{1}{\pi}\int\limits_{\mathcal{R}^2} {\rm d}^2x^{\prime}\kappa \left( \overrightarrow{x}^\prime \right) \dfrac{\overrightarrow{x} - \overrightarrow{x}^\prime}{|\overrightarrow{x} - \overrightarrow{x}^\prime|^2} = \dfrac{D_L D_{LS}}{\xi_0 D_S} \alpha \left(\xi_0\overrightarrow{x}  \right)
\label{e1.25}
\end{equation}
is the scaled deflection angle, and
\begin{equation}
\kappa \left(\overrightarrow{x}  \right) = \dfrac{\Sigma \left(\xi_0 \overrightarrow{x}  \right)}{\Sigma_{cr}}
\label{e1.26}
\end{equation}
is the dimensionless surface mass density also known as the convergence. The term $\Sigma_{cr}$ is the critical surface mass density expressed as
\begin{equation}
\Sigma_{cr} = \dfrac{c^2 D_S}{4\pi G D_L D_{LS}}
\label{e1.27}
\end{equation}
{\color{black} $\Sigma_{\rm cr}$ sets the natural normalization of the projected density, so that $\kappa=\Sigma/\Sigma_{\rm cr}$.  It is not, for a general lens, a universal pointwise threshold that by itself separates strong from weak lensing.  The local mapping is governed by the Jacobian
\begin{equation}
\mathcal A_{ij}\equiv\frac{\partial y_i}{\partial x_j}
=\delta_{ij}-\psi_{,ij},
\label{eq:lens_jacobian}
\end{equation}
whose determinant may be written
\begin{equation}
\det\mathcal A=(1-\kappa)^2-|\gamma|^2 .
\label{eq:critical_curve_condition}
\end{equation}
Here $\gamma$ is the two-component shear.  Critical curves occur where $\det\mathcal A=0$, and their images under the lens mapping are caustics in the source plane.  Multiple imaging is a global property of this mapping and depends on the full convergence and shear fields.  A region with $\kappa<1$ can therefore participate in strong lensing when shear is appreciable, while $\kappa>1$ at one point is not, by itself, a complete classification of an observed lensing regime.}

\section{Lensing Potential and Lensing Distance}{\label{s1.5}}
In case of an extended distribution of matter the deflection tendencies can be defined by its effective lensing potential.
\begin{equation}
\alpha \left(\overrightarrow{x}  \right) = \nabla \psi \left(\overrightarrow{x}  \right)
\label{e1.28}
\end{equation}
where
\begin{equation}
\psi \left(\overrightarrow{x}  \right) = \dfrac{1}{\pi} \int \limits_{\mathcal{R}^2} {\rm d}^2 x^\prime \kappa \left(\overrightarrow{x}^\prime  \right) {\rm ln} |\overrightarrow{x} - \overrightarrow{x}^\prime|
\label{e1.29}
\end{equation}
The mapping of $\overrightarrow{x} \rightarrow \overrightarrow{y}$ is a gradient mapping,
\begin{equation}
\overrightarrow{y} = \nabla \left(\dfrac{1}{2} \overrightarrow{x}^2 - \psi \left(\overrightarrow{x}  \right)  \right)
\label{e1.30}
\end{equation}
or,
\begin{equation}
\nabla \phi \left(\overrightarrow{x}, \overrightarrow{y}   \right) = 0
\label{e1.31}
\end{equation}
where,
\begin{equation}
\phi \left(\overrightarrow{x}, \overrightarrow{y}  \right) = \dfrac{1}{2}\left(\overrightarrow{x} - \overrightarrow{y} \right)^2 - \psi \left(\overrightarrow{x}  \right)
\label{e1.32}
\end{equation}
is called the Fermat potential.
\par The Laplacian of the lensing potential, with the help of the identity $\nabla ^2 {\rm ln} |\overrightarrow{x} - \overrightarrow{x}^\prime| = 2 \pi \delta ^2 \left(\overrightarrow{x} - \overrightarrow{x}^\prime   \right)$ is
\begin{equation}
\nabla ^2 \psi = 2 \kappa
\label{e1.33}
\end{equation}
\emph{i.e.,} twice the convergence.
\par Lensing potential, deflection angle, critical surface density all strongly depend on distances between the trinity of source, lens and observer. This in turn depends on the source and lens redshifts. The ratio of the distances as given below
\begin{equation}
\dfrac{D_L D_{LS}}{D_S}
\label{e1.34}
\end{equation} 
is called lensing distance. {\color{black} For a fixed source redshift and a specified cosmological model, the combination $D_LD_{LS}/D_S$ vanishes when the lens approaches either endpoint and commonly reaches a maximum at an intermediate lens redshift.  Its dependence on source redshift is not universally monotonic, because every factor is an angular-diameter distance and therefore depends on the background cosmology.  Moreover, a larger value of this distance combination increases the geometric lensing efficiency for a fixed physical surface-mass distribution; it does not by itself imply a larger convergence unless the surface density and the distance convention are also specified.}
\par It is instructive to notice the schematic panel given in \ref{lensing_geometry}. As the relative position of the source keeps changing the corresponding nature and positioning of images also keeps changing. The following section will firmly establish the mathematical relation defining the exact image positioning based on the lens equation.
 
\subsection{Fermat's principle -- A general relativistic adaptation}{\label{s1.9.1}}
In the classical context {\color{black} Fermat's principle states that the optical travel time is stationary under admissible first-order variations; the stationary path may be a minimum, a maximum, or a saddle.} Now, in general relativity the path of light curves are represented by null geodesics \emph{i.e.,} the line element satisfies ${\rm d}s^2=0$ for any null curve between two events. As such it is tricky to lay down the Fermat's principle for null geodesics which are already a byproduct of extremalisation. Following in the footsteps of \cite{Schneider}, we can lay down the theorem as
\begin{itemize}
\item[] For any event $S$ (emission of light from a source) let the worldline $\mathcal{L}$ for an observer be connected by a null curve $\gamma$. Such a null curve will represent a photon trajectory \emph{iff} its time of arrival $\tau$ on $\mathcal{L}$ is stationary under first order variations of $\gamma$, where the entire family of smooth null curves from $S$ to $\mathcal{L}$ are taken into account.
\begin{equation}
\delta \tau=0
\label{e1.59}
\end{equation}
\end{itemize}   
\begin{figure}[t!]
\begin{center}\includegraphics[scale=0.25, angle=0]{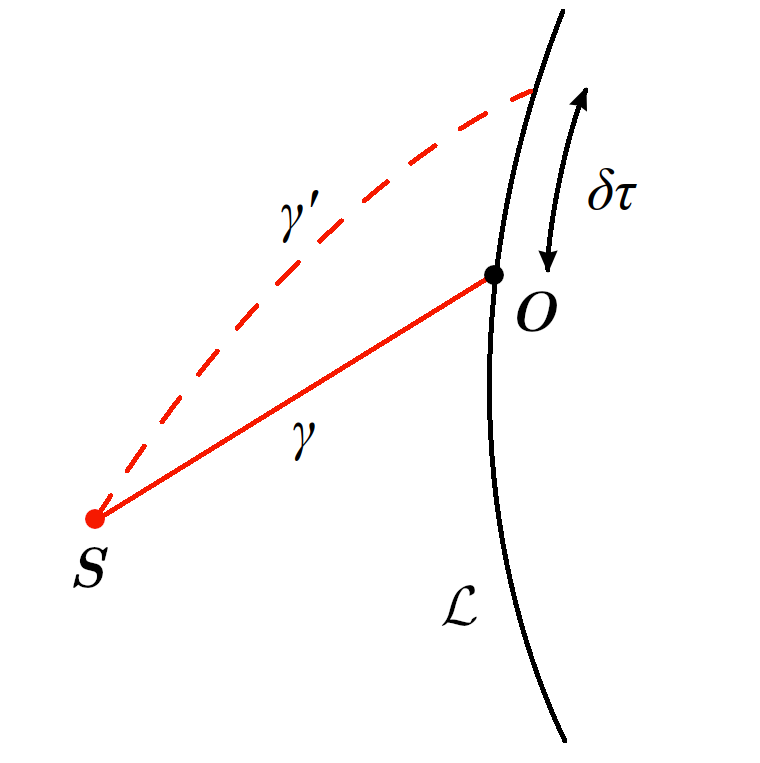}
\end{center}
\caption[Illustration depicting the formulation of relativistic Fermat's principle]{Illustration depicting the formulation of relativistic Fermat's principle}
\label{fermat}
\end{figure}
The path $\gamma^\prime$ is a slightly different path than $\gamma$ so that the both of these paths emerging from $S$ arrive at the worldline $\mathcal{L}$ with a relative time delay of $\delta\tau$ [\ref{fermat}]. {\color{black} The condition $\delta\tau=0$ holds at the physical null geodesic: the first variation vanishes there for every admissible variation, not for every neighbouring trial curve separately.} Comprehensive treatment of Fermat's principle in general relativity may be found in the following literature \cite{{Perlick1_1990}, {Perlick2_1990}, {Faraoni_1992}, {Perlick1998}, {Perlick2000}, {Giannoni_2002}, {Perlick_2006}, {Frolov_2013}}. 
\par Fermat's principle will have a strong bearing on the mathematical treatments in \ref{s5}. As a precursor it will be instructive to delve into the mathematical formulations regarding Fermat's principle for two specific spacetimes, namely (a) conformally stationary and (b) conformally static.
\subsubsection{Effective refractive index in conformally stationary spacetime through Fermat's principle}{\label{s1.9.1.1}}
Any spacetime whose geometrical identity does not change with respect to time is termed as a stationary spacetime. The line element of a stationary spacetime has the form 
\begin{equation}
{\rm d}s^2 = e^{2\Phi}\left({\rm d}t -\omega_i{\rm d}x^i  \right)^2 - e^{-2\Phi}{\rm d}l^2
\label{e1.60}
\end{equation}
where ${\rm d}l^2 = \gamma_{ij}{\rm d}x^i{\rm d}x^j$. Note that $\Phi$, $\omega_i$ and $\gamma_{ij}$ depend on the spatial coordinates $x_i$ alone. The term $\omega_i$ determines rotational effects in the spacetime geometry. Abiding by the criteria for null geodesics, ${\rm d}s^2=0$ and applying it to \ref{e1.60} we get
\begin{align*}
&{\rm d}t = \omega_i{\rm d}x^i + e^{-2\Phi}{\rm d}l\\
\Rightarrow & t = \int\limits_\gamma \omega_i{\rm d}x^i + e^{-2\Phi}{\rm d}l\numberthis
\label{e1.61}
\end{align*} 
where we integrate over the null geodesic $\gamma$. Applying Fermat's principle to \ref{e1.61} we get
\begin{equation}
\delta \int\limits_\gamma \omega_i{\rm d}x^i + e^{-2\Phi}{\rm d}l = \delta \int\limits_\gamma \left(\omega_i\dfrac{{\rm d}x^i}{{\rm d}l} + e^{-2\Phi}\right){\rm d}l = 0
\label{e1.62}
\end{equation}
Comparing with the Fermat's principle in the classical domain {\color{black} the integrand in \ref{e1.62} defines a direction-dependent optical functional rather than, in general, a scalar refractive index:}
\begin{equation}
\left(\omega_i\dfrac{{\rm d}x^i}{{\rm d}l} + e^{-2\Phi}\right) = n_{\rm stationary}
\label{e1.63}
\end{equation}
{\color{black} The quantity $n_{\rm stationary}$ depends on both position and propagation direction and is the Randers-type Fermat integrand in the chosen stationary chart.} As \ref{e1.63} reveals the effective refractive index is dependent on both position and direction. Thus we can conclude that the spatial paths of photons are geodesics for the Finsler metric $\left(\omega_i{\rm d}x^i + e^{-2\Phi}{\rm d}l  \right)$ which reduces to the Riemannian metric for $\omega_i=0$ [\ref{s5.1}].
 
\subsubsection{Effective refractive index in conformally static spacetime through Fermat's principle}{\label{s1.9.1.2}}
For special cases where the spacetime geometry is irrotational we have $\omega_i=0$. In instances such as this \ref{e1.63} reduces down to 
\begin{equation}
n_{\rm static} = e^{-2\Phi}
\label{e1.64}
\end{equation}
Physically speaking we can attribute the effective refractive index as a property of isotropic, non-dispersive medium.

\section{Deflection of Light in a Static, Spherically Symmetric Spacetime}
 
 We begin by reviewing the derivation of the light deflection angle in a general static, spherically symmetric gravitational field, following the formalism outlined by Weinberg~\cite{Weinberg_Steven1972-07}. In such spacetimes, one may introduce a set of quasi-Minkowskian coordinates $x^\mu = (t, r, \theta, \phi)$ in which the line element takes the standard form
 \begin{equation}
 	ds^2 = -f(r)\,dt^2 + g(r)\,dr^2 + r^2\,d\theta^2 + r^2\sin^2\theta\,d\phi^2,
 	\label{eq:metric}
 \end{equation}
 where the metric functions $f(r)$ and $g(r)$ depend only on the radial coordinate due to the imposed static and spherical symmetry.
 
 The trajectory of a photon is determined by the null geodesic condition $ds^2 = 0$ along with the geodesic equations
 \begin{equation}
 	\frac{d^2 x^\mu}{dp^2} + \Gamma^\mu_{\nu\lambda} \frac{dx^\nu}{dp} \frac{dx^\lambda}{dp} = 0,
 	\label{eq:geodesic}
 \end{equation}
 where $p$ is an affine parameter along the photon's path. The Christoffel symbols $\Gamma^\mu_{\nu\lambda}$, required for these equations, are computed from the metric via
 \begin{equation}
 	\Gamma^\mu_{\nu\lambda} = \frac{1}{2}g^{\mu\eta} \left( \partial_\lambda g_{\eta\nu} + \partial_\nu g_{\eta\lambda} - \partial_\eta g_{\nu\lambda} \right).
 	\label{eq:christoffel}
 \end{equation}
 
 For the line element in ~\ref{eq:metric}, the non-vanishing Christoffel symbols relevant to the photon's motion are
 \begin{equation}
 	\begin{aligned}
 		&\Gamma^r_{tt} = \frac{1}{2g(r)} \frac{df}{dr}, \qquad 
 		\Gamma^r_{rr} = \frac{1}{2g(r)} \frac{dg}{dr}, \qquad
 		\Gamma^r_{\theta\theta} = -\frac{r}{g(r)}, \\
 		&\Gamma^r_{\phi\phi} = -\frac{r \sin^2\theta}{g(r)}, \qquad
 		\Gamma^\theta_{r\theta} = \Gamma^\theta_{\theta r} = \frac{1}{r}, \qquad
 		\Gamma^\theta_{\phi\phi} = -\sin\theta\cos\theta, \\
 		&\Gamma^\phi_{\phi r} = \Gamma^\phi_{r\phi} = \frac{1}{r}, \qquad
 		\Gamma^\phi_{\phi\theta} = \Gamma^\phi_{\theta\phi} = \cot\theta, \qquad
 		\Gamma^t_{tr} = \Gamma^t_{rt} = \frac{1}{2f(r)} \frac{df}{dr}.
 	\end{aligned}
 	\label{eq:christoffel_list}
 \end{equation}
 
 Due to the spherical symmetry of the system, one may restrict the motion to the equatorial plane $\theta = \pi/2$ without loss of generality. Substituting the relevant components into ~\ref{eq:geodesic} yields the equations of motion for each coordinate. The resulting system governs the trajectory of a light ray in the curved spacetime and ultimately leads to the expression for the deflection angle, which we derive in the subsequent subsection.

 \subsection{Photon Trajectories and Light Deflection in Static Spherically Symmetric Spacetimes}
 
 The deflection of light by gravity is one of the most striking predictions of general relativity. In a static, spherically symmetric spacetime described by the general metric
 \begin{equation}
 	ds^2 = -f(r) \, dt^2 + g(r) \, dr^2 + r^2 d\theta^2 + r^2 \sin^2\theta \, d\phi^2,
 	\label{eq:metric_static}
 \end{equation}
 null geodesics capture the path of light rays in the curved geometry. Here, $f(r)$ and $g(r)$ encode the gravitational potential, and their precise form depends on the underlying gravitational theory or solution (e.g., Schwarzschild, Reissner–Nordström, or regular black holes).
 
 Due to spherical symmetry, the motion can be confined to the equatorial plane, $\theta = \pi/2$, without loss of generality. The symmetries of the spacetime ensure the conservation of two quantities along geodesics: {\color{black} the photon energy parameter $E=-p_t$ and axial angular-momentum parameter $J=p_\phi$.  Since a photon has no rest mass, these constants are not ``per unit mass''; their common normalization changes under an affine rescaling, whereas the ratio $b=J/E$ is invariant under that rescaling.} These are associated with the timelike and rotational Killing vectors, respectively, and yield the first integrals:
 \begin{align}
 	\frac{dt}{dp} &= \frac{E}{f(r)}, \label{eq:t_eom_phys} \\
 	\frac{d\phi}{dp} &= \frac{J}{r^2}, \label{eq:phi_eom_phys}
 \end{align}
 where $p$ is an affine parameter along the null trajectory. 
 
 The condition for null motion, $ds^2 = 0$, translates into a radial equation for the light ray:
 \begin{equation}
 	g(r) \left( \frac{dr}{dp} \right)^2 = \frac{E^2}{f(r)} - \frac{J^2}{r^2}.
 	\label{eq:radial_null_phys}
 \end{equation}
 This relation governs the radial evolution of the photon and sets the foundation for determining the deflection angle.
 
 To understand the bending of light, it is more insightful to consider the photon's trajectory $r(\phi)$, obtained by eliminating the affine parameter $p$. Combining Eqs.~\ref{eq:phi_eom_phys} and \ref{eq:radial_null_phys} yields:
 \begin{equation}
 	\left( \frac{dr}{d\phi} \right)^2 = \frac{r^4}{g(r)} \left[ \frac{E^2}{J^2 f(r)} - \frac{1}{r^2} \right].
 	\label{eq:orbit_eq_phys}
 \end{equation}
 
 The light ray reaches its closest approach at $r = r_0$, where $\left( dr/d\phi \right) = 0$. This condition defines the turning point of the trajectory and leads to:
 \begin{equation}
 	\frac{E^2}{J^2} = \frac{f(r_0)}{r_0^2}.
 	\label{eq:turning_point_phys}
 \end{equation}
 Equivalently, the impact parameter obeys $b^2\equiv J^2/E^2=r_0^2/f(r_0)$.
 \par This identity is fundamental: it encodes how the curvature of spacetime modifies the effective centrifugal barrier for photons, displacing the turning point from the flat-space prediction.
 
 Substituting \ref{eq:turning_point_phys} back into \ref{eq:orbit_eq_phys}, we arrive at the differential equation that governs the angular evolution of the light ray:
 \begin{equation}
 	\left|\frac{d\phi}{dr}\right|
 	= \frac{\sqrt{g(r)}}{r}
 	\left[ \frac{f(r_0)}{f(r)} \left( \frac{r}{r_0} \right)^2 - 1 \right]^{-1/2}.
 	\label{eq:phidr_phys}
 \end{equation}
 
 Integrating from the point of closest approach to infinity yields the total change in the azimuthal angle:
 \begin{equation}
 	\left|\phi_\infty-\phi(r)\right|
 	= \int_r^\infty \frac{\sqrt{g(r')}}{r'}
 	\left[ \frac{f(r_0)}{f(r')} \left( \frac{r'}{r_0} \right)^2 - 1 \right]^{-1/2}{\rm d}r' .
 	\label{eq:phi_integral_phys}
 \end{equation}
 {\color{black}The absolute value avoids assigning an orientation before the incoming or outgoing branch is specified.}
 Physically, this integral measures how the presence of spacetime curvature alters the straight-line path of light, causing it to bend. 
 
 The total bending angle is then defined as the angular excess over the Euclidean prediction:
 \begin{equation}
 	\Delta\phi
 	=2\int_{r_0}^\infty \frac{\sqrt{g(r)}}{r}
 	\left[ \frac{f(r_0)}{f(r)} \left( \frac{r}{r_0} \right)^2 - 1 \right]^{-1/2}{\rm d}r-\pi .
 	\label{eq:delta_phi_phys}
 \end{equation}
 
 This expression, valid for {\color{black}static, spherically symmetric spacetimes written in the areal-radius gauge of \ref{eq:metric_static}}. For instance, in the weak-field approximation around a Schwarzschild black hole, this recovers Einstein’s celebrated result $\Delta\phi \approx 4GM/b$, where $b$ represents the impact parameter.
 \begin{figure}[hbt!]
\begin{center}
\includegraphics[scale=0.35]{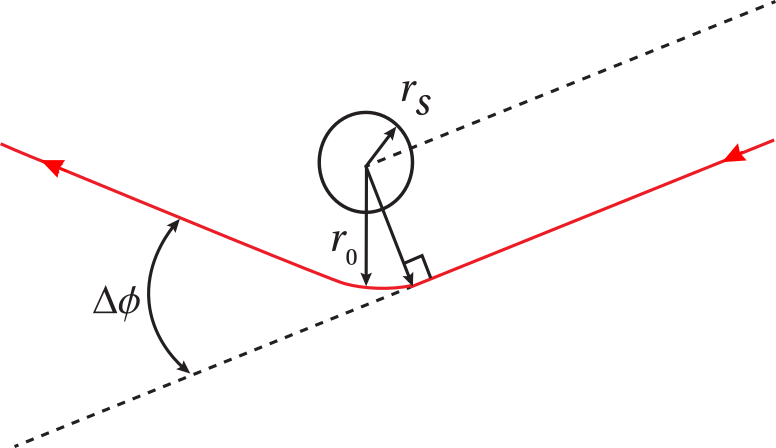}
\end{center}
\caption{Classical depiction of bending of light under the influence of a gravitational field}
\label{fig0}
\end{figure}



{

\section{Deflection Angle of Light in the Equatorial Plane of a Kerr Black Hole}\label{s3}

The deflection of light in curved spacetime lies at the heart of gravitational lensing, a phenomenon whose most striking realizations occur near compact astrophysical objects such as black holes. For a Schwarzschild black hole, which is non-rotating and spherically symmetric, the bending of null geodesics is independent of the direction of approach. However, the situation changes drastically when the central object possesses angular momentum. In the case of a Kerr black hole, the deflection angle becomes a directional quantity, sensitive to whether the photon's trajectory is aligned with or opposed to the black hole's spin. 

{\color{black}This spin-dependent asymmetry is produced by frame dragging.  Fix $a>0$ and let $b>0$ denote the magnitude of the equatorial impact parameter.  With $s=+1$ for a prograde ray and $s=-1$ for a retrograde ray, the leading weak-deflection result is
\begin{equation*}
\hat\alpha_s=\frac{4M}{b}-\frac{4s\,aM}{b^2}
+\mathcal O\!\left(\frac{M^2}{b^2},\frac{a^2M}{b^3}\right).
\end{equation*}
At fixed $b$, the prograde correction is therefore negative and the retrograde correction positive.  This statement concerns the weak bending at a fixed impact-parameter magnitude; the prograde critical orbit can nevertheless lie closer to the horizon and has a smaller capture threshold \cite{Congdon_Iyer_Keeton_2022}.}

This breaking of spherical symmetry leads to the loss of reflection symmetry about the spin axis in observational features, such as black hole shadows and the location of relativistic images. The greater the spin, the more asymmetric the bending becomes.
`
A complementary formulation uses an equivalent material-medium description of the Kerr geometry: Roy and Sen evaluated its frame-dragging-dependent refractive index and far-field light bending in this approach \cite{RoySenKerrMaterial2015}.  This is an alternative optical description of propagation in a fixed Kerr spacetime, not by itself a nonlinear-electrodynamics correction to the photon characteristic cone.

In what follows, we provide a detailed analytical formulation of this phenomenon in the equatorial plane of a Kerr black hole. Our treatment draws inspiration from the foundational work of Darwin \cite{1959RSPSA.249..180D}, Bardeen \cite{1967JMP.....8..265B}, and Chandrasekhar \cite{Chandrasekhar}, and we adopt the approach developed by Iyer and Hansen \cite{Iyer1}, \cite{Iyer2009Strong}, which yields exact expressions for the bending angle in both prograde and retrograde regimes.

\vspace{1em}
\subsection{Kerr Geometry and Geodesic Framework}

We restrict our analysis to the equatorial plane of the Kerr spacetime (\( \theta = \pi/2 \)), where the metric in Boyer-Lindquist coordinates takes the form:
\begin{equation}
ds^2 = g_{tt} dt^2 + g_{rr} dr^2 + g_{\phi\phi} d\phi^2 + 2g_{t\phi} dt d\phi,
\label{e3.1}
\end{equation}
with the non-vanishing metric components given by:
\begin{subequations}
\begin{align*}
g_{tt} &= -\left(1 - \frac{2M}{r}\right)\\
g_{rr} &= \left(1 - \frac{2M}{r} + \frac{a^2}{r^2}\right)^{-1}\\
g_{\phi\phi} &= r^2 + a^2 + \frac{2a^2 M}{r}\\
g_{t\phi} &= -\frac{2aM}{r}
\end{align*}
\end{subequations}

Photons follow null geodesics, characterized by \( ds^2 = 0 \), and their four-momentum \( p^\mu = \dot{x}^\mu = dx^\mu/d\lambda \) is conserved along the trajectory due to the stationarity and axial symmetry of the Kerr spacetime. These symmetries lead to two conserved quantities:
\begin{itemize}
    \item  {\color{black}Conserved energy parameter: \( E = -p_t \)},
    \item  {\color{black}Conserved axial angular-momentum parameter: \( J_\phi = p_\phi \)}.
\end{itemize}

Exploiting these conserved quantities and the form of the metric, one obtains the following expressions for the time and azimuthal components of the motion:
\begin{align}
\dot{t} &= \frac{E r(r^2 + a^2) - 2aM(J_\phi - aE)}{r \Delta}, \label{e3.5} \\
\dot{\phi} &= \frac{J_\phi\left(1 - \frac{2M}{r}\right) + \frac{2aM}{r} E}{\Delta}, \label{e3.4}
\end{align}
where \( \Delta = r^2 - 2Mr + a^2 \) governs the horizon structure of the black hole.

The radial equation of motion is obtained by inserting these results into the null condition \( p^\mu p_\mu = 0 \). After simplification, the radial velocity takes the form:
\begin{equation}
\dot{r}^2 = E^2 + \frac{a^2 E^2}{r^2} + \frac{2Ma^2 E^2}{r^3} - \frac{4MaE J_\phi}{r^3} - \frac{J_\phi^2}{r^2} + \frac{2MJ_\phi^2}{r^3}.
\label{e3.12}
\end{equation}

This expression reveals the non-trivial coupling between spin \( a \), energy \( E \), and angular momentum \( J_\phi \). It also displays the effective potential barrier structure that governs the turning points of photon trajectories.

To express the dynamics more intuitively, {\color{black} we define the positive impact-parameter magnitude \(b=|J_\phi/E|>0\) and the signed impact parameter \(b_s=J_\phi/E=s\,b\)}, where \( s = +1 \) for prograde and \( s = -1 \) for retrograde motion. In terms of these, the radial equation becomes:
\begin{equation}
\dot{r}^2 = J_\phi^2 \left[ \frac{1}{b^2} + \frac{a^2}{b^2 r^2} + \frac{2Ma^2}{b^2 r^3} - \frac{4Ma}{b_s r^3} - \frac{1}{r^2} + \frac{2M}{r^3} \right].
\label{e3.13}
\end{equation}

 \ref{e3.13} is pivotal -- it encodes the entire influence of rotation on the radial trajectory. The term \( -\frac{4Ma}{b_s r^3} \) distinguishes prograde and retrograde motion through its sign and modifies the centrifugal barrier. This spin-induced shift in effective potential has direct consequences for the location of turning points, the critical impact parameter, and ultimately the deflection angle.

{\color{black} Physically, frame dragging permits the prograde critical photon orbit to occur at a smaller radius and the retrograde one at a larger radius.  For scattering rays compared at the same positive magnitude \(b\), however, the leading spin term decreases the prograde bending and increases the retrograde bending, as shown above.  Statements about fixed \(b\), fixed closest approach, and proximity to the capture threshold must not be interchanged.} Pictorial depiction has been shown in \ref{Fig:Kerr} for better visualisation.

 \begin{figure}[hbt!]
\begin{center}
\includegraphics[scale=0.3]{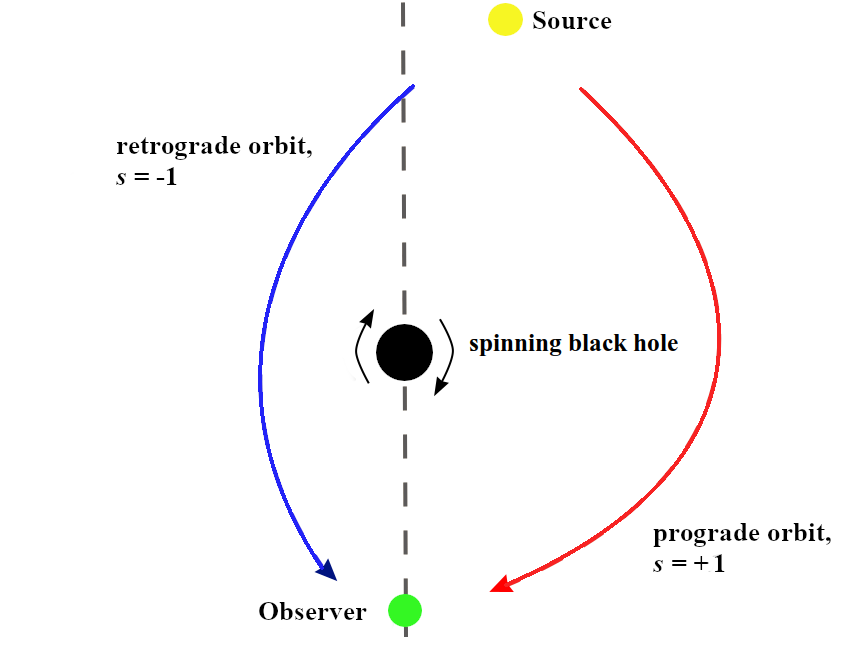}
\end{center}
\caption[Schematic diagram illustrating the prograde and retrograde orbits (on the equatorial plane) in the vicinity of a spinning black hole.]{Schematic diagram illustrating the prograde and retrograde orbits (on the equatorial plane) in the vicinity of a spinning black hole. In the above scenario the axis of rotation of the black hole points into the page.}
\label{Fig:Kerr}
\end{figure}
This difference, albeit subtle in weak field regimes, becomes dramatically pronounced in the strong gravity domain near the photon sphere. The next sections develop these features in detail, constructing the bending angle as an integral over the radial coordinate, identifying the critical impact parameter \( b_{\rm sc} \), and demonstrating how the deflection diverges logarithmically as the turning point approaches the photon capture radius.

\subsection{Photon Orbits in Reciprocal Coordinates and the Analytical Structure of the Deflection Angle}

To understand the bending of light in a Kerr spacetime, it is convenient to transition from the radial coordinate \( r \) to its reciprocal \( u = 1/r \). This transformation simplifies the analysis of geodesics, particularly for light rays deflected in the equatorial plane. The radial part of the null geodesic equation, when recast in terms of \( u(\phi) \), links the rate of change of the inverse radius with respect to the azimuthal angle to the fundamental characteristics of the geometry and the motion constants:
\begin{equation}
\left( \frac{{\rm d}u}{{\rm d}\phi} \right)^2 = u^4 \frac{\dot{r}^2}{\dot{\phi}^2}.
\end{equation}

Using earlier expressions for \( \dot{r}^2 \) and \( \dot{\phi} \) in terms of the conserved angular momentum \( J_\phi \), spin parameter \( a \), and impact parameter \( b_s \), the right-hand side simplifies into a form that transparently encodes spin effects:
\begin{equation}
\left( \frac{{\rm d}u}{{\rm d}\phi} \right)^2 = \frac{[1 - 2Mu + a^2u^2]^2}{\left[1 - 2Mu\left(1 - \frac{a}{b_s} \right)\right]^2} \left[ 2M\left(1 - \frac{a}{b_s} \right)^2 u^3 - \left(1 - \frac{a^2}{b^2}\right) u^2 + \frac{1}{b^2} \right].
\label{eq:du_dphi_squared}
\end{equation}

Equation~\eqref{eq:du_dphi_squared} captures how frame dragging reshapes the geometry of photon orbits. The numerator encodes the modified centrifugal barrier due to spin, while the denominator reflects the reduced effective radial force for prograde orbits and enhanced force for retrograde ones.

We define the deflection angle \( \hat{\alpha}_{\rm Kerr} \) as the angular deviation of the light ray from a straight path. Assuming that the light originates and terminates at spatial infinity, and bends to a minimum radius \( r_0 \), the deflection is calculated by integrating the azimuthal change from the closest approach back to infinity and subtracting the asymptotic straight-line contribution \( \pi \):
\begin{equation}
\hat{\alpha}_{\rm Kerr} = -\pi + 2 \int_0^{1/r_0} \frac{1 - 2Mu\left(1 - \frac{a}{b_s} \right)}{[1 - 2Mu + a^2u^2] \sqrt{2M\left(1 - \frac{a}{b_s} \right)^2 u^3 - \left(1 - \frac{a^2}{b^2}\right) u^2 + \frac{1}{b^2}}} \, {\rm d}u.
\label{eq:deflection_integral}
\end{equation}

{\color{black} The integral can be reduced to combinations of elliptic integrals; Iyer and Hansen give an exact closed representation in that sense \cite{Iyer1}.  It is generally not an elementary antiderivative.}. We denote this polynomial in the denominator as \( B(u) \), and it plays a central role in determining the nature of the orbit:
\[
B(u) = 2M\left(1 - \frac{a}{b_s} \right)^2 u^3 - \left(1 - \frac{a^2}{b^2}\right) u^2 + \frac{1}{b^2}.
\]

{\color{black} The real roots delimit the intervals in which the radial polynomial is non-negative.  For a scattering ray arriving from infinity, the physical closest approach is the largest accessible positive root reached from the exterior branch.  The remaining roots are algebraic roots of the radial potential and should not generically be called the periapsis and apoapsis of a bound photon orbit.  At the capture threshold two relevant roots coalesce, producing the unstable circular equatorial photon orbit.} In the high bending limit, the light path approaches a circular photon orbit (CPO), where the closest approach radius coincides with the location of a light ring.

\subsection{Determination of the Distance of Closest Approach and Photon Sphere Radius}

The condition \( \dot{r} = 0 \) at \( r = r_0 \) defines the turning point of the trajectory. Substituting into the radial equation of motion, we obtain a cubic relation for \( r_0 \):
\begin{equation}
r_0^3 - b^2 \left(1 - \frac{a^2}{b^2} \right) r_0 + 2M b^2 \left(1 - \frac{a}{b_s} \right)^2 = 0.
\label{eq:cubic_r0}
\end{equation}
This expression reflects how both the magnitude and direction of the black hole’s spin alter the photon’s radial dynamics. The above equation can be solved using trigonometric methods for cubics, yielding an analytic expression for \( r_0 \) in terms of \( b, a, \) and \( b_s \):
\begin{equation}
r_0=\frac{2b}{\sqrt3}\sqrt{1-\frac{a^2}{b^2}}\,
\cos\!\left[
\frac13\cos^{-1}\!\left(
-\frac{3\sqrt3 M}{b}
\frac{\left(1-a/b_s\right)^2}
{\left(1-a^2/b^2\right)^{3/2}}
\right)\right] ,
\label{eq:kerr_r0_solution}
\end{equation}
{\color{black} where the displayed branch is the largest accessible positive root for the scattering orbit in its domain of validity.}

{\color{black} For \(a=0\), the turning-point relation becomes \(b^2=r_0^3/(r_0-2M)\); thus \(r_0\) depends on \(b\).  The special value \(r_0=3M\) occurs only at the critical Schwarzschild impact parameter \(b=3\sqrt3\,M\).}

The effective potential governing photon motion in this geometry is given by:
\begin{equation}
W_{\rm eff}(r) = \frac{1}{r^2}\left(1 - \frac{a^2}{b^2} \right) - \frac{2M}{r^3} \left(1 - \frac{a}{b_s} \right)^2.
\label{eq:weff}
\end{equation}
Its extrema define the circular photon orbits, with the radius of the stable or unstable orbit \( r_{\rm sc} \) found by setting \(  \dfrac{{\rm d}W_{\rm eff}}{{\rm d}r} = 0 \). Solving this yields:
\begin{equation}
\dfrac{r_{\rm sc}}{3M} = {\left(1 - \dfrac{a}{b_s}\right)}{\left(1 + \dfrac{a}{b_s}\right)}^{-1}.
\label{eq:rsc}
\end{equation}
This elegant result confirms that frame-dragging reduces the critical radius for prograde orbits and increases it for retrograde ones.  Using the critical circular orbit radius \( r_{\rm sc} \), we can further determine the corresponding impact parameter \( b_{\rm sc} \), which represents the threshold for photon capture. It satisfies the cubic identity:
\begin{equation}
\left(b_{\rm sc} + a\right)^3 = 27M^2 \left(b_{\rm sc} - a\right).
\label{eq:bsc}
\end{equation}
This relation governs the asymmetry in capture cross-sections between prograde and retrograde orbits, with the effective photon cross-section being reduced for co-rotating light and enlarged for counter-rotating ones.

{\color{black} The closest approach \( r_0 \) and the equatorial critical parameters \( b_{\rm sc}, r_{\rm sc} \) are foundational for strong-deflection expansions and the equatorial edges of capture. They do not by themselves generate the full two-dimensional Kerr shadow: generic non-equatorial null geodesics also require the Carter constant, and the observed shadow depends on the observer's inclination.}

\subsection{Analytic Determination of Critical Parameters for Prograde and Retrograde Orbits}

The deflection of null geodesics around a Kerr black hole is fundamentally altered by the inclusion of angular momentum. Unlike the symmetric case of Schwarzschild geometry, where the bending of light is independent of direction due to spherical symmetry, the Kerr spacetime breaks this symmetry due to its axisymmetric and rotating nature. This distinction is most prominently reflected in the impact parameter and the radius of closest approach for light rays undergoing prograde and retrograde motion.

\paragraph{Critical Impact Parameter for Prograde and Retrograde Orbits:}

The critical impact parameter $b_{\rm sc}$ corresponds to the boundary between photon trajectories that are scattered away to infinity and those that plunge into the black hole. To derive its form analytically, we begin with the cubic relation governing its dynamics:
\begin{equation}
\left(b_{\rm sc} + a \right)^3 = 27M^2\left(b_{\rm sc} - a \right),
\label{e3.32_re}
\end{equation}
which can be transformed to yield a more general, unified solution valid for both prograde and retrograde motion:
\begin{equation}
b_{\rm sc} = -a + s\, 6M\, \cos\left[\frac{1}{3}\cos^{-1}\left(-s \frac{a}{M}\right)\right],
\label{e3.38_re}
\end{equation}
where the spin-dependent sign $s = +1$ corresponds to prograde motion (co-rotating photons), while $s = -1$ refers to retrograde motion (counter-rotating photons). This compact expression, adapted from the results in \cite{Chandrasekhar}, elegantly incorporates the rotational influence of the black hole on the deflection dynamics.

{\color{black} With the signed convention of \ref{e3.38_re}, the Schwarzschild limit is $b_{\rm sc}=s\,3\sqrt3\,M$; the corresponding magnitude is $|b_{\rm sc}|=3\sqrt3\,M$.}. As the spin increases positively (prograde), the effective potential well becomes deeper, allowing light to orbit closer to the black hole, thus lowering the critical impact parameter. Conversely, for retrograde orbits, the effective potential becomes more repulsive, and $b_{\rm sc}$ increases, pushing the photon sphere outward.

\paragraph{Radius of the Circular Photon Orbit:}

Closely related to $b_{\rm sc}$ is the radius $r_{\rm sc}$ of the unstable circular photon orbit, often referred to as the photon sphere radius in the Schwarzschild context. For Kerr geometry, this radius is spin-dependent and given by:
\begin{equation}
r_{\rm sc} = 2M\left[1 + \cos\left(\frac{2}{3}\cos^{-1}\left(-s \frac{a}{M}\right)\right)\right].
\label{e3.40_re}
\end{equation}
This relation captures how frame dragging, encoded in the sign and magnitude of $a$, affects the effective potential landscape. For prograde trajectories ($s=+1$), $r_{\rm sc}$ lies inside the Schwarzschild photon sphere ($r_c = 3M$), while for retrograde ones ($s=-1$), the radius shifts outward, increasing the minimum allowed distance for circular photon motion.

\paragraph{Geometric Visualization:}

To visualize these effects, ~\ref{Kerr2} presents an overhead view of the equatorial plane around a spinning black hole. The dashed-dotted lines indicate the closest approach trajectories -- blue for prograde and red for retrograde. The solid curves denote the corresponding circular photon orbits -- cyan for prograde and orange for retrograde. The rotation axis is oriented perpendicular to the plane, consistent with the convention adopted in ~\ref{Fig:Kerr}.

\begin{figure}[hbt!]
\begin{center}
\includegraphics[scale=0.275]{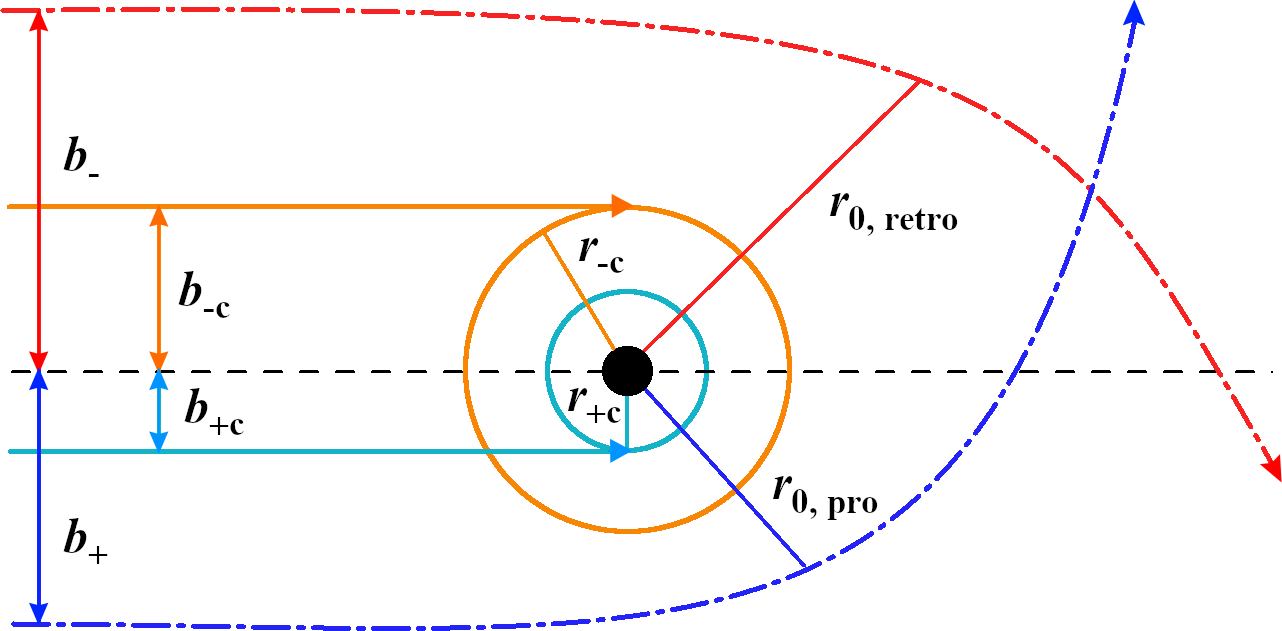}
\end{center}
\caption[Overhead view of the prograde and retrograde orbits (for the equatorial plane) around a Kerr black hole.]{Overhead view of the prograde and retrograde orbits on the equatorial plane of a Kerr black hole. Blue and red dashed-dotted lines indicate closest approach for prograde and retrograde cases respectively. Cyan and orange solid curves mark the respective innermost photon orbits. The spin vector points out of the plane.}
\label{Kerr2}
\end{figure}

\paragraph{Comparison and Interpretation:}

The functional behavior of both $b_{\rm sc}$ and $r_{\rm sc}$ as a function of spin $a$ reveals essential aspects of the lensing geometry:

- As $a/M \to 1$, $b_{\rm sc}^{\rm (prograde)} \to 2M$, which represents the smallest allowed impact parameter before capture for co-rotating photons.
- In contrast, $b_{\rm sc}^{\rm (retrograde)}$ diverges as $a/M \to -1$, indicating increasing angular resistance to counter-rotating orbits.
- The difference $\Delta b_{\rm sc} = b_{-\rm c} - b_{+\rm c}$ serves as a quantitative measure of asymmetry induced by rotation.

These features are not merely coordinate effects but manifest in physical observables such as the black hole shadow, caustic structure in gravitational lensing, and differential time delays in null geodesics, all of which inherit this spin-induced asymmetry.

\paragraph{Numerical Profiles:}

A quantitative summary is best appreciated through the plots of $b_{\rm sc}/M$ and $r_{\rm sc}/M$ versus $a/M$, reproduced in Figs.~\ref{Kerrplot_1} and \ref{Kerrplot_2}. These highlight the continuous transition from Schwarzschild geometry ($a=0$) to the extremal Kerr regime ($a \to \pm M$), delineating the physical extremities of photon capture and deflection.

\begin{figure}[hbt!]
\begin{center}
\includegraphics[scale=0.4]{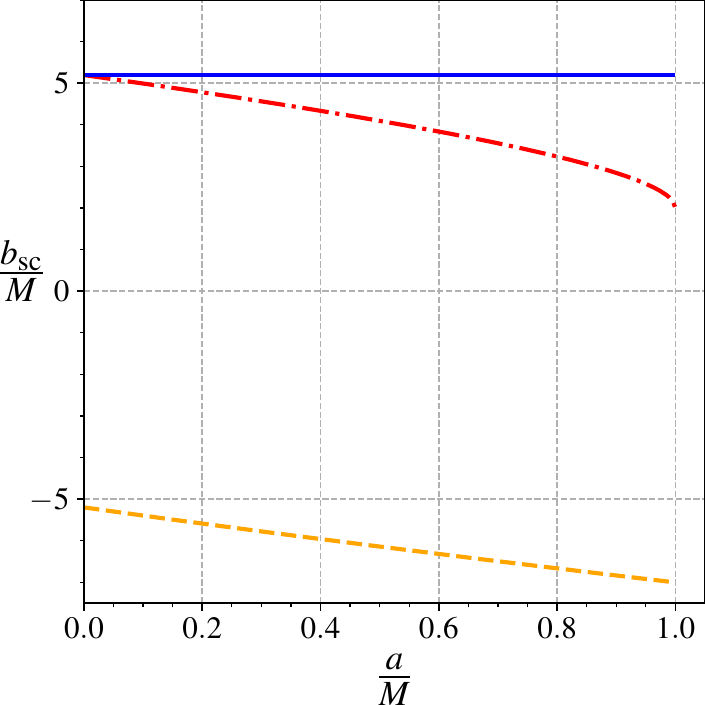}
\end{center}
\caption[Variation of $b_{\rm sc}/M$ with $a/M$ for prograde, retrograde and Schwarzschild cases.]{Variation of the normalized \emph{signed} critical impact parameter $b_{\rm sc}/M$ as a function of $a/M$. {\color{black} In the supplied plot, the blue solid horizontal curve is the positive Schwarzschild value $3\sqrt3$; the red dash--dotted curve is the prograde signed branch, which decreases to $2$; and the orange dashed curve is the retrograde signed branch, which reaches $-7$ at $a=M$.  Its physical retrograde impact-parameter magnitude is the absolute value of the orange branch.}}
\label{Kerrplot_1}
\end{figure}

\begin{figure}[hbt!]
\begin{center}
\includegraphics[scale=0.4]{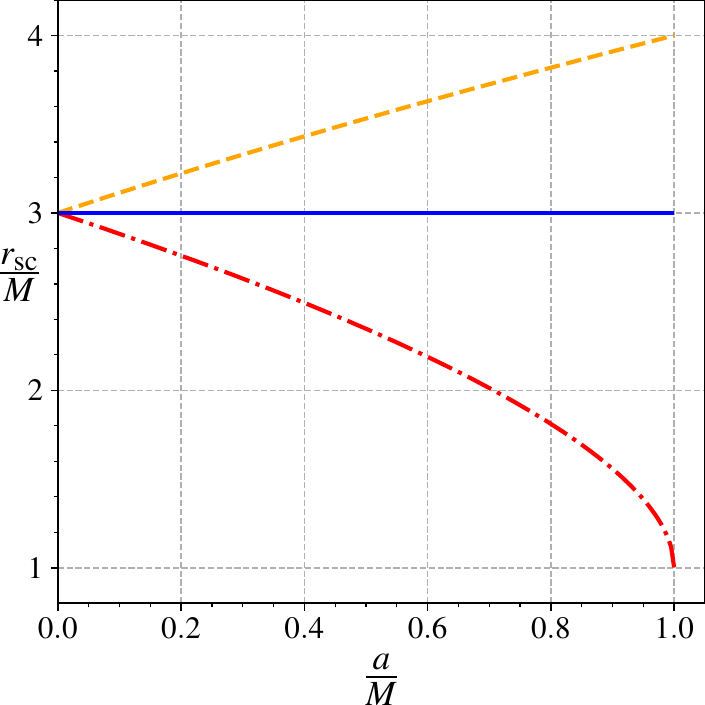}
\end{center}
\caption[Variation of $r_{\rm sc}/M$ with $a/M$ for prograde, retrograde and Schwarzschild cases.]{Variation of the normalized equatorial circular-photon radius $r_{\rm sc}/M$ versus black-hole spin $a/M$. {\color{black} In the supplied plot, the blue solid horizontal curve is Schwarzschild, $r_{\rm sc}/M=3$; the red dash--dotted curve is the prograde branch, which decreases to $1$; and the orange dashed curve is the retrograde branch, which increases to $4$.} Prograde motion allows tighter photon orbits, while retrograde motion pushes them outward.}
\label{Kerrplot_2}
\end{figure}

\paragraph{Significance:}

The expressions in ~\eqref{e3.38_re} and \eqref{e3.40_re}, derived via trigonometric resolution of the Kerr photon orbit condition, encapsulate the full influence of spin on null trajectories. Unlike perturbative or numerical treatments, this approach provides exact analytical control and offers direct insights into how frame dragging modulates light propagation.

The formalism developed by Iyer and Hansen \cite{Iyer1} is particularly advantageous in this regard. Their parametrization not only delivers precision but also encodes directional asymmetry implicitly through the sign parameter $s$, eliminating the need for separate case-by-case derivations. As a result, it stands as a robust tool for lensing analysis in spinning spacetimes.

\section{Rindler--Ishak Method}
\label{sec:RindlerIshakMethod}

Gravitational lensing theory always involves two logically distinct operations. First, one must determine the trajectory of a light ray in a given spacetime. Second, one must translate that trajectory into an angle that can be interpreted as what an observer measures. In asymptotically flat spacetimes these two operations are often blended together. The orbit is computed and the deflection is then read off by comparing asymptotic directions at infinity. This strategy is not conceptually clean when the spacetime is not asymptotically flat, or when the relevant measurement is performed at large but finite radius. Schwarzschild de Sitter geometry provides the simplest setting where this issue becomes visible.

Rindler and Ishak highlighted that, in Schwarzschild de Sitter spacetime, the differential equation governing the coordinate orbit of a null geodesic can be written in a form that contains no explicit cosmological constant\cite{Rindler:2007zz}. If one stops at that step, one may be tempted to claim that $\Lambda$ does not affect bending. The missing ingredient is the measurement prescription. A bending angle is not a property of a curve drawn on an $(r,\phi)$ diagram. It is a statement about the angle between directions in the observer's local spatial geometry. That local geometry is determined by the spatial part of the metric, and in Schwarzschild de Sitter it depends explicitly on $\Lambda$. The Rindler Ishak method is an invariant way to compute the local intersection angle between the photon direction and a chosen reference direction in the orbital surface. It therefore isolates where background fields enter the lensing signal.
\begin{figure}[t!]
\begin{center}\includegraphics[scale=0.75, angle=0]{Rindler-Ishak.png}
\end{center}
\caption{
Geometric definition of the Rindler--Ishak local angle on the orbital plane.
A trajectory (red curve) propagates from the source $S$ to the observer $O$ past the lens at $L$, with impact parameter $b$ defined with respect to the reference straight line.
At the observation point, the radial coordinate direction (along $\phi=\mathrm{const}$) is compared with the instantaneous direction of motion of the ray or particle.
Their invariant intersection angle is $\psi$, while $\phi$ denotes the corresponding coordinate angle on the orbital plane.
The one sided deflection used in the Rindler--Ishak prescription is $\epsilon=\psi-\phi$, evaluated at the receiver location.
}
\label{rindler}
\end{figure}

We consider a general static and spherically symmetric spacetime of the form
\begin{equation}
	ds^{2}
	=
	-f(r)\,dt^{2}
	+\frac{1}{g(r)}\,dr^{2}
	+r^{2}\left(d\theta^{2}+\sin^{2}\theta\,d\phi^{2}\right),
	\label{RI_metric}
\end{equation}
where $f(r)$ and $g(r)$ are assumed positive in the region where the photon trajectory and the observer are defined.

Spherical symmetry implies that the motion can be restricted to a plane through the origin.
Without loss of generality we choose the equatorial plane,
\begin{equation}
	\theta=\frac{\pi}{2}.
	\label{eq:RI_equatorial}
\end{equation}
On a constant time slice, the induced two dimensional spatial geometry on this orbital surface is
\begin{equation}
	dl^{2}=\frac{1}{g(r)}\,dr^{2}+r^{2}d\phi^{2}.
	\label{eq:RI_orbital_metric}
\end{equation}
 \ref{eq:RI_orbital_metric} provides the intrinsic inner product on the orbital plane and therefore fixes how local angles are defined.
This is the geometric ingredient that enters the Rindler and Ishak prescription.

Null geodesics follow from the standard variational principle applied to the Lagrangian
\begin{equation}
\mathcal{L}=g_{\mu\nu}\dot{x}^{\mu}\dot{x}^{\nu},
\label{RI_Lagrangian}
\end{equation}
together with the null condition
\begin{equation}
\mathcal{L}=0.
\label{RI_null}
\end{equation}
Because the metric coefficients in \ref{RI_metric} are independent of $t$ and $\phi$, there are two conserved quantities along any geodesic. The first corresponds to time translation symmetry and is interpreted as the photon energy parameter,
\begin{equation}
E=f(r)\,\dot{t}.
\label{RI_energy}
\end{equation}
The second corresponds to axial symmetry and is interpreted as the conserved angular momentum parameter,
\begin{equation}
\ell=r^{2}\dot{\phi}.
\label{RI_angmom}
\end{equation}
Substituting \ref{RI_energy} and \ref{RI_angmom} into the null condition \ref{RI_null} yields the radial first integral
\begin{equation}
\dot{r}^{2}=g(r)\left(\frac{E^{2}}{f(r)}-\frac{\ell^{2}}{r^{2}}\right).
\label{RI_rdot}
\end{equation}
Eliminating the affine parameter in favor of $\phi$ gives
\begin{equation}
\left(\frac{dr}{d\phi}\right)^{2}
=
\frac{r^{4}}{\ell^{2}}\,g(r)\left(\frac{E^{2}}{f(r)}-\frac{\ell^{2}}{r^{2}}\right).
\label{RI_dr_dphi_first}
\end{equation}
Introducing $u(\phi)=1/r(\phi)$ and differentiating leads to an orbital equation of the form
\begin{equation}
\frac{d^{2}u}{d\phi^{2}}
=
\frac{1}{2}\frac{d}{du}\!\left(\frac{g(u)}{f(u)}\right)\frac{E^{2}}{\ell^{2}}
-\left(
u\,g(u)+\frac{1}{2}u^{2}\frac{dg(u)}{du}
\right),
\label{RI_orbit_general}
\end{equation}
which determines the coordinate shape of the ray once appropriate boundary data are chosen.

In weak field lensing one rarely needs an exact solution of \ref{RI_orbit_general}. Instead, one constructs a controlled approximation by perturbing around the flat space trajectory. In flat space, where $f(r)=g(r)=1$, the orbit equation reduces to
\begin{equation}
\frac{d^{2}u}{d\phi^{2}}+u=0,
\label{RI_flat_orbit}
\end{equation}
and the straight line solution with closest approach at $\phi=\pi/2$ is
\begin{equation}
u_{0}(\phi)=\frac{\sin\phi}{R},
\qquad
r_{0}(\phi)=\frac{R}{\sin\phi}.
\label{RI_flat_solution}
\end{equation}
The parameter $R$ labels the reference line in the auxiliary flat geometry. It is a convenient expansion label. It is not automatically equal to the true closest approach of the curved orbit, and this distinction becomes important beyond leading order. The perturbative scheme proceeds by inserting $u_{0}(\phi)$ into the source terms in \ref{RI_orbit_general}, solving the resulting inhomogeneous linear equation, and iterating if higher order accuracy is required.

The Rindler Ishak contribution begins at the next logical step, where the coordinate orbit is turned into a measurable angle. The orbital surface at constant $t$ and $\theta=\pi/2$ inherits a two dimensional spatial metric from \ref{RI_metric},
\begin{equation}
dl^{2}=\frac{1}{g(r)}\,dr^{2}+r^{2}d\phi^{2}.
\label{RI_orbital_metric}
\end{equation}
The physical direction of the photon, as seen by a static observer, is encoded in the tangent to the orbit. Introduce the coordinate slope
\begin{equation}
\mathcal{A}(r,\phi)\equiv \frac{dr}{d\phi}.
\label{RI_A_def}
\end{equation}
The point is that $\mathcal{A}$ alone has no invariant meaning. The same curve can have different coordinate slopes under reparametrizations, yet the measured angle must be unique. Rindler and Ishak therefore define the local angle $\psi$ as an invariant angle between two directions in the Riemannian geometry \ref{RI_orbital_metric}. One direction is the ray direction, represented in $(r,\phi)$ components as $d^{i}=(\mathcal{A},1)$. The second direction is taken to be the radial coordinate direction along $\phi=\mathrm{const}$, represented as $\delta^{i}=(1,0)$. The local angle between these directions is fixed by the metric inner product,
\begin{equation}
\cos\psi
=
\frac{g_{ij}d^{i}\delta^{j}}
{\sqrt{g_{ij}d^{i}d^{j}}\sqrt{g_{ij}\delta^{i}\delta^{j}}},
\label{RI_cospsi}
\end{equation}
where $g_{ij}$ now denotes the components of the two metric in \ref{RI_orbital_metric}. Substituting the explicit vectors and simplifying yields a practical expression in terms of $\mathcal{A}$ and the metric function $g(r)$,
\begin{equation}
\tan\psi
=
\frac{r\sqrt{g(r)}}{\left|\mathcal{A}(r,\phi)\right|}.
\label{RI_tanpsi}
\end{equation}
\ref{RI_tanpsi} is the operational statement of the method. Once $r(\phi)$ is computed from the geodesic equation, one differentiates to obtain $\mathcal{A}$, and then evaluates $\psi$ using the spatial metric of the orbital surface.

To define a bending angle one compares this measured direction to a reference direction. Rindler and Ishak define the one sided bending at a point $(r,\phi)$ by
\begin{equation}
\epsilon(r,\phi)=\psi(r,\phi)-\phi,
\label{RI_eps_def}
\end{equation}
which measures how much the photon direction differs from the coordinate line $\phi=\mathrm{const}$ at that location. 
In asymptotically flat lensing this local description reduces to the standard deflection extracted from asymptotes. In non asymptotically flat spacetimes it remains meaningful because it is built from local geometric data. {\color{black} The quantity $\epsilon$ is a local, one-sided comparison and is not by itself a universal total deflection angle.  A reported lensing angle must specify whether it is evaluated at the source or receiver, the observer family, the reference radial direction, and how the two endpoint contributions are combined.  In an asymptotically flat limit with the usual source--lens--observer geometry, that complete prescription can be matched to the standard asymptotic deflection.} In practice it is evaluated at the observer position, which is at large but finite radius in many astrophysical settings.

\subsection{Application to Schwarzschild de Sitter spacetime}
\label{subsec:RI_SdS}

The Schwarzschild de Sitter case illustrates why this definition matters. The metric is
\begin{equation}
ds^{2}
=
-\alpha(r)\,dt^{2}
+\frac{1}{\alpha(r)}\,dr^{2}
+r^{2}\Big(d\theta^{2}+\sin^{2}\theta\,d\phi^{2}\Big),
\label{RI_SdS_metric}
\end{equation}
with
\begin{equation}
\alpha(r)=1-\frac{2m}{r}-\frac{\Lambda r^{2}}{3}.
\label{RI_alpha}
\end{equation}
Here $f(r)=g(r)=\alpha(r)$. One finds that the orbital equation for $u(\phi)=1/r(\phi)$ can be written as
\begin{equation}
\frac{d^{2}u}{d\phi^{2}}+u=3mu^{2},
\label{RI_SdS_orbit}
\end{equation}
which contains no explicit $\Lambda$. The physical bending, however, is determined by the local measurement \ref{RI_tanpsi}, and that measurement depends explicitly on $g(r)=\alpha(r)$, hence on $\Lambda$. This is the core message. A background field can disappear from the coordinate orbit equation yet remain present in the physical angle because the spatial geometry used to measure angles is modified.

In the weak field regime, one may use the standard first order orbit
\begin{equation}
u(\phi)
=
\frac{\sin\phi}{R}
+
\frac{3m}{2R^{2}}
\left(1+\frac{1}{3}\cos2\phi\right),
\label{RI_Sch_firstorbit}
\end{equation}
compute $\mathcal{A}=dr/d\phi$ from \ref{RI_A_def}, and evaluate $\psi$ through \ref{RI_tanpsi} with $g(r)=\alpha(r)$.
Expanding for small angles yields a bending that contains an explicit correction proportional to $\Lambda$.
The sign indicates that positive $\Lambda$ reduces focusing, which is consistent with the interpretation that the large scale geometry counters attractive convergence in the local spatial measurement.

From the perspective of lensing in more general backgrounds, the Rindler Ishak method is best viewed as a measurement map. The geodesic equation provides the coordinate curve $r(\phi)$. The invariant spatial metric on the orbital surface provides the correct inner product to convert a coordinate slope into a local direction angle. This separation is valuable whenever the asymptotic structure is nontrivial, whenever sources and observers are at finite distance, or whenever one studies how background fields modify bending through the spatial geometry rather than through the orbit equation alone.
}

\subsection{Application to Weyl conformal gravity: sign of the linear potential term from the invariant angle}
\label{subsec:RI_Weyl}

A second instructive setting is conformal Weyl gravity, where the static spherical exterior geometry is described by the Mannheim--Kazanas solution\cite{1989ApJ...342..635M}. In units $G=c=1$ the line element may be written as
\begin{equation}
	d\tau^{2}
	=
	-B(r)\,dt^{2}
	+\frac{1}{B(r)}\,dr^{2}
	+r^{2}\left(d\theta^{2}+\sin^{2}\theta\,d\phi^{2}\right),
	\label{eq:Weyl_metric}
\end{equation}
with
\begin{equation}
	B(r)=\alpha-\frac{2M}{r}+\gamma r-k r^{2},
	\qquad
	\alpha=\sqrt{1-6M\gamma}.
	\label{eq:Weyl_B}
\end{equation}
The parameters $\gamma$ and $k$ arise from the conformal theory and are commonly discussed in connection with galactic scale phenomenology. 

If one applies asymptotic angle comparison in a naive manner, one is led to a bending formula of the schematic form
\begin{equation}
	2\epsilon=\frac{4M}{r_{0}}-\gamma r_{0},
	\label{eq:Weyl_conventional}
\end{equation}
where $r_{0}$ is the closest approach radius. 

For $\gamma>0$ this expression suggests a reduction of bending relative to Schwarzschild.
However, the Mannheim--Kazanas geometry is not asymptotically flat in the strict sense required to interpret \ref{eq:Weyl_conventional} as an observable deflection, and the sign inferred from that asymptotic reasoning is not reliable. 

The Rindler and Ishak prescription resolves this tension by defining bending through the invariant local angle on the orbital two geometry. On the equatorial plane $\theta=\pi/2$, the measurement map is
\begin{equation}
	\tan\psi=\frac{\sqrt{B(r)}\,r}{|A|},
	\qquad
	A(r,\phi)=\frac{dr}{d\phi},
	\label{eq:Weyl_RI_angle}
\end{equation}
and the one sided bending is $\epsilon=\psi-\phi$.
A convenient evaluation point is $\phi=0$, where $\epsilon=\psi$.

Introducing $u(\phi)=1/r(\phi)$, the orbit equation implied by \ref{eq:Weyl_metric} takes the form
\begin{equation}
	\frac{d^{2}u}{d\phi^{2}}=-\alpha u+3Mu^{2}-\frac{\gamma}{2},
	\label{eq:Weyl_orbit}
\end{equation}
while the parameter $k$ drops out of the orbit equation itself.

A perturbative expansion in $M$ yields a consistent weak field orbit and therefore an explicit $A(r,\phi)$.

The essential physical point is that the measured angle \ref{eq:Weyl_RI_angle} depends on $B(r)$ evaluated along the trajectory, so both $\gamma$ and $k$ can enter the observable bending through the spatial geometry even if $k$ does not appear in \ref{eq:Weyl_orbit}. 

Carrying out the Rindler--Ishak evaluation gives a deflection structure of the schematic form
\begin{equation}
	2\psi
	=
	\frac{4M}{r_{0}}
	-\frac{k r_{0}^{3}}{2M}
	+\frac{15M^{2}\gamma}{r_{0}},
	\label{eq:Weyl_deflection_r0}
\end{equation}
and, upon identifying $k=\Lambda/3$ and rewriting in terms of an impact parameter $b$ at weak field level,
\begin{equation}
	2\psi
	\simeq
	\frac{4M}{b}
	+\frac{2Mb\Lambda}{3}
	+\frac{15M^{2}\gamma}{b}.
	\label{eq:Weyl_deflection_b}
\end{equation}
A notable outcome is that the $\gamma$ contribution enters with a positive sign for $\gamma>0$ in the Rindler--Ishak definition, curing the sign issue associated with \ref{eq:Weyl_conventional} and highlighting the role of the invariant local angle in non asymptotically flat geometries. 

\subsection{Application to 4D Einstein--Gauss--Bonnet--de Sitter black holes}
\label{subsec:RI_EGBdS}

Higher curvature modifications provide a complementary arena to test the Rindler--Ishak prescription because they can modify both the orbit dynamics and the spatial measurement geometry.
In the four dimensional Einstein--Gauss--Bonnet construction, the static spherical line element is
\begin{equation}
	ds^{2}=-f(r)\,dt^{2}+\frac{dr^{2}}{f(r)}+r^{2}\left(d\theta^{2}+\sin^{2}\theta\,d\phi^{2}\right),
	\label{eq:EGB_metric}
\end{equation}
with a metric function that depends on the Gauss--Bonnet coupling $\alpha$ and the cosmological constant $\Lambda$,
\begin{equation}
	f(r)=1+\frac{r^{2}}{2\alpha}
	\left(
	1 \pm \sqrt{1+\frac{8\alpha M}{r^{3}}+\frac{4\alpha\Lambda}{3}}
	\right).
	\label{eq:EGB_f}
\end{equation}
The negative branch approaches the Schwarzschild--de Sitter behaviour at large radii. 

Restricting to $\theta=\pi/2$ and using $u(\phi)=1/r(\phi)$, one finds that the null orbit equation in this framework can acquire explicit mixed contributions involving $\alpha$ and $\Lambda$.
Keeping the leading terms to first order in $\alpha$ and $\Lambda$ in the weak field regime yields the schematic structure
\begin{equation}
	\frac{d^{2}u}{d\phi^{2}}+u \approx 3Mu^{2}-2M\alpha\Lambda\,u^{2}-12\alpha M^{2}u^{5}.
	\label{eq:EGB_orbit}
\end{equation}
This is an instructive contrast with general relativistic Schwarzschild--de Sitter, where $\Lambda$ can be absent from the orbit equation.
Here, the theory itself moves cosmological information into the orbit dynamics through $\alpha\Lambda$ couplings. 

The perturbative solution is obtained by expanding about the straight line reference orbit
\begin{equation}
	u_{0}(\phi)=\frac{\sin\phi}{R},
	\label{eq:EGB_u0}
\end{equation}
and iterating corrections,
\begin{equation}
	u(\phi)=u_{0}(\phi)+\delta u_{1}(\phi)+\delta u_{2}(\phi)+\delta u_{3}(\phi)+\cdots.
	\label{eq:EGB_u_expand}
\end{equation}
Once $u(\phi)$ is known to the required order, one computes $r(\phi)$ and $dr/d\phi$ and then applies the Rindler--Ishak local angle map on the orbital two geometry,
\begin{equation}
	dl^{2}=\frac{dr^{2}}{f(r)}+r^{2}d\phi^{2},
	\label{eq:EGB_orbital_metric}
\end{equation}
which yields
\begin{equation}
	\tan\psi=\frac{r\sqrt{f(r)}}{\left|dr/d\phi\right|},
	\label{eq:EGB_tanpsi}
\end{equation}
and $\epsilon=\psi-\phi$.

The resulting weak field deflection, expressed in terms of the auxiliary orbit parameter $R$, takes the form
\begin{flalign}
	2\epsilon \approx
	\Bigg[
	\frac{4M}{R}
	+\frac{15\pi M^{2}}{4R^{2}}
	+\frac{177M^{3}}{4R^{3}}
	-\frac{\Lambda R^{3}}{6M}
	-\frac{71R\Lambda M}{96}
	+\frac{5\pi\Lambda R^{2}}{32}
	+\frac{25\pi\Lambda M^{2}}{512}
	\Bigg]\nonumber\\
	-\alpha
	\Bigg[
	\frac{15\pi M^{2}}{4R^{4}}
	+\frac{9\Lambda M}{4R}
	+\frac{5\pi\Lambda}{32}
	+\frac{R^{3}\Lambda^{2}}{18M}
	+\frac{5\pi R^{2}\Lambda^{2}}{96}
	\Bigg].
	\label{eq:EGB_deflection_R}
\end{flalign}
The sign structure shows that a positive $\alpha$ diminishes the deflection in this approximation, consistent with the intuition that the Gauss--Bonnet sector counteracts focusing.

For comparison across conventions, it is useful to relate $R$ to the true closest approach radius $r_{0}=r(\pi/2)$.
In the limit $\alpha\to 0$ and $\Lambda\to 0$, the perturbative orbit implies
\begin{equation}
	\frac{1}{r_{0}}=\frac{1}{R}+\frac{M}{R^{2}}+\frac{3M^{2}}{16R^{3}}+\frac{27M^{3}}{4R^{4}}+\cdots,
	\label{eq:EGB_R_to_r0}
\end{equation}
which is essential for consistent term by term comparisons between results expressed in $R$ and those expressed in $r_{0}$

\subsection{Charged massive particle bending in a charged galactic wormhole}
\label{subsec:RI_CGW_CMP}

In several astrophysical situations, the probe is not a photon but a massive particle, and the probe may carry charge.
In such cases the motion is not governed solely by spacetime geodesics, because the Lorentz force couples the dynamics to electromagnetic fields\cite{2026IJGMM..2350151H}.
A convenient way to keep the framework geometric is to recast the spatial motion at fixed energy into a Jacobi metric, so that the trajectory becomes a geodesic of an effective Riemannian metric on the spatial slice.
Once this effective orbital geometry is specified, the Rindler--Ishak definition of a locally measured direction angle can be applied.

The charged galactic wormhole model is represented by a static spherically symmetric line element of the form\cite{PhysRevD.63.064014}
\begin{equation}
	ds^{2}
	=
	-\left(1+\frac{Q^{2}}{r^{2}}\right)dt^{2}
	+
	\left(1-\frac{b(r)}{r}+\frac{Q^{2}}{r^{2}}\right)dr^{2}
	+
	r^{2}\left(d\theta^{2}+\sin^{2}\theta\,d\phi^{2}\right),
	\label{eq:CGW_metric}
\end{equation}
with wormhole charge $Q$ and shape function $b(r)$.
The latter is modeled in terms of a galactic scale profile with parameters such as $r_{s}$ and $\rho_{s}$. 

For a charged massive particle of mass $m$ and charge $q$ moving in a static spacetime with electrostatic potential $A_{t}(r)$, the Jacobi metric formulation yields an effective spatial metric of the form
\begin{equation}
	J_{ij}
	=
	\left[(E+qA_{t})^{2}-m^{2}g_{tt}\right]\frac{g_{ij}}{g_{tt}},
	\label{eq:Jacobi_metric_form}
\end{equation}
where $E$ is the conserved energy.
Specializing to equatorial motion $\theta=\pi/2$ and using a generic static spherical decomposition, one obtains a first integral for the orbit in terms of $U(\phi)=1/r(\phi)$,
\begin{equation}
	\left(\frac{dU}{d\phi}\right)^{2}
	=
	\frac{C^{2}U^{4}}{A(r)B(r)L^{2}}
	\left[
	(E+qA_{t})^{2}-\left(m^{2}+\frac{L^{2}}{C(r)}\right)A(r)
	\right],
	\label{eq:CMP_orbit_first_integral}
\end{equation}
where $L$ is the conserved angular momentum. 

The Rindler--Ishak step is then implemented on the orbital two geometry at fixed time and $\theta=\pi/2$ by defining the invariant angle $\psi$ between the direction of motion and the radial direction.
Writing $\gamma=dr/d\phi$, one arrives at an operational expression
\begin{equation}
	\tan\psi=\frac{\sqrt{g_{\phi\phi}/g_{rr}}}{|\gamma|},
	\label{eq:CMP_tanpsi}
\end{equation}
and defines the one sided deflection by
\begin{equation}
	\alpha=\psi(\phi)-\phi.
	\label{eq:CMP_alpha_one_sided}
\end{equation}
A finite distance receiver at angular location $\phi=\phi_{b}$ is treated by evaluating the orbit at $\phi_{b}$ and substituting into the invariant angle formula, leading to a compact expression for $\alpha_{\mathrm{RI}}$ in terms of $U(\phi)$ and $dU/d\phi$. 

From a review article standpoint, this example serves as a bridge between photon lensing and more general deflection phenomena.
The invariant angle construction remains the measurement map, while the dynamical input is generalized from null geodesics to Jacobi geodesics whose form depends on $(E+qA_{t})$.
This makes explicit why the sign of $q$ can qualitatively alter deflection behaviour even in a fixed background geometry, because the orbit itself changes once electromagnetic forces are included. 

\section{Bending Angle of Light and the Gauss-Bonnet Theorem}{\label{s4}} 
A common approach in the literature while defining the bending angle of particle trajectories is to assume the source to be located at infinity. However, in reality, lensing objects lie at finite distances from the source, and in non-asymptotically flat spacetimes, such an assumption becomes untenable. Hence, a more accurate definition of the bending angle must incorporate finite-distance corrections. In \cite{Gibbons_2008}, the Gauss-Bonnet theorem was applied to the optical (or Fermat) metric of spherically symmetric spacetimes to determine light deflection. Two domains were considered: 
\begin{itemize}
\item[$\mathscr{D}_1$] enclosed by two light rays with the lensing object at the centre—linking multiple images to the topology of spacetime.
\item[$\mathscr{D}_2$] bounded by a non-geodesic circular arc and a light ray—used for computing the asymptotic deflection angle as the surface integral of Gaussian curvature.
\end{itemize}
In this formulation, both observer and source were assumed to be in an asymptotically Euclidean region. Extending this, \cite{Ishihara_2016} and \cite{Ono_2017} incorporated finite-distance corrections to evaluate the bending angle in non-asymptotically flat spacetimes. Their procedure involves:
\begin{itemize}
\item[(a)] Expressing the metric in Randers-Finsler form to extract the Fermat metric.
\item[(b)] Restricting to the equatorial plane to obtain a two-dimensional Riemannian manifold for the particle trajectory.
\item[(c)] Defining the integral region bounded by: 
\begin{itemize}
\item[(i)] the particle trajectory, 
\item[(ii)] radial curves through the lensing object to the source and observer, and 
\item[(iii)] an auxiliary circular arc centred on the lens.
\end{itemize}
\item[(d)] Applying the Gauss-Bonnet theorem to relate the deflection angle to integrals of Gaussian and geodesic curvature.
\item[(e)] Showing the equivalence of this formulation to one involving only jump angles and azimuthal coordinate separation.
\end{itemize}
This framework \cite{Ono_2017} provides an elegant alternative that avoids direct curvature integration. Subsequent studies \cite{Ishihara_2017, Ono_2018, Avgun_2018, Ono_2019, Ono_2019b, Avgun_2019} have extended the method to various spacetime geometries, including strong deflection regimes. The following sub-sections briefly review the Gauss-Bonnet theorem and describe the \emph{modus operandi} of \cite{Ishihara_2016} and \cite{Ono_2017} for both spherically and axially symmetric spacetimes.

\subsection{The Gauss-Bonnet Theorem}
\begin{figure}[hbt!]
\begin{center}
\includegraphics[scale=0.2]{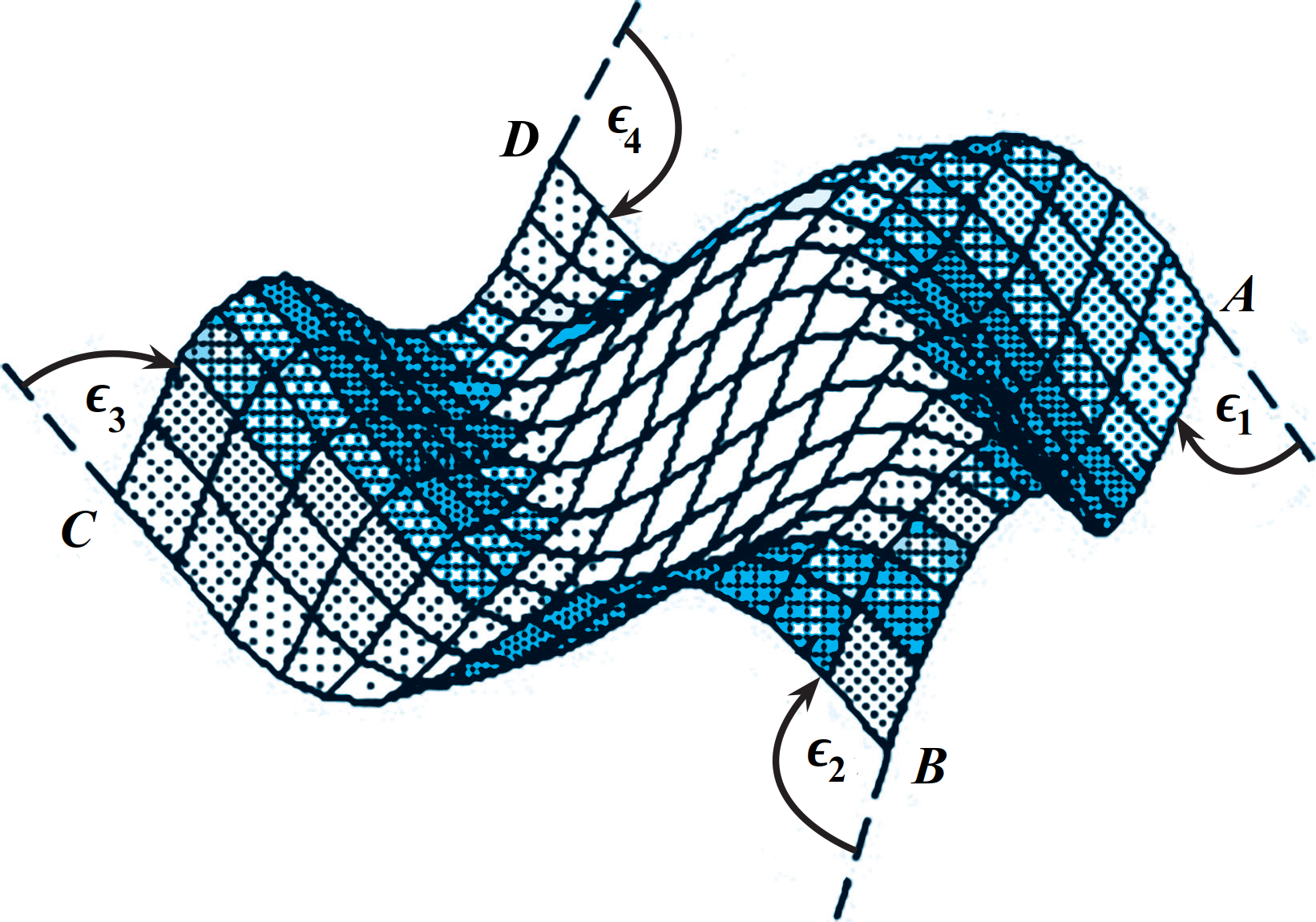}
\end{center}
\caption{\textbf{A patch of surface created by the intersection of several curves thereby  subtending the exterior angle $\epsilon_i$ at each point of intersection}}
\label{fig1}
\end{figure}

{\label{s4.1}}
The inter-relationship between the topology of a surface to its intrinsic differential geometry is one of the many areas of interest while studying the motion of particles in General Relativity. The Gauss-Bonnet theorem (\cite{Carmo}, \cite{John_Oprea2019-02-06}) provides us with an extremely potent tool to serve this very purpose. In fact it is one of the most profound theorems of differential geometry that links local and global geometry. Let us consider there exists a patch of surface $\mathscr{D}$ (\ref{fig1}) enclosed by a set of $n$ boundary curves denoted by $\partial\mathscr{D}_i$. Now, if $\epsilon_i$ denote the exterior angles formed where the boundary curves $\partial\mathscr{D}_i$ meet and they have the geodesic curvature $\kappa_g(s_i)$ with $s_i$ representing a point on $\partial\mathscr{D}_i$ then the mathematical formulation of the theorem reads

\begin{equation}
\iint\limits_\mathscr{D} K {\rm d}S + \sum_{i=1}^{n}\int\limits_{\partial\mathscr{D}_i} \kappa_g {\rm d}\ell + \sum_{i=1}^{n} \epsilon_i = 2\pi\chi(\mathscr{D})
\label{e25}
\end{equation}
where $K$, ${\rm d}S$ and ${\rm d}\ell$ are the Gaussian curvature, area element of $\mathscr{D}$ and line element of $\partial\mathscr{D}_i$ respectively. The term $\chi(\mathscr{D})$ is called the Euler characteristic number of $\mathscr{D}$. {\color{black} Compactness and connectedness do not fix the Euler characteristic: for example, a sphere has $\chi=2$, a disk has $\chi=1$, and an annulus or torus has $\chi=0$.  The lensing domains used below have $\chi=1$ only when they are topological disks, with no additional excisions, holes, or self-intersections.}. It should be noted that the sign of the line element $\ell$ should be chosen such that its compatibility with the orientation of the surface is maintained. 
\subsection{Gravitational Lens, Photon trajectory and Geometry}{\label{s4.2}}
The geometrical depiction of the manifold as chalked out by the presence of a lensing object, source of light, viewer and the photon trajectory lies at the heart of the current topic. Let us, therefore, briefly spend some time describing the geometry of the said manifold that provides the integrable region as required by the Gauss-Bonnet theorem.

\begin{figure}[hbt!]
\begin{center}
\includegraphics[scale=0.3]{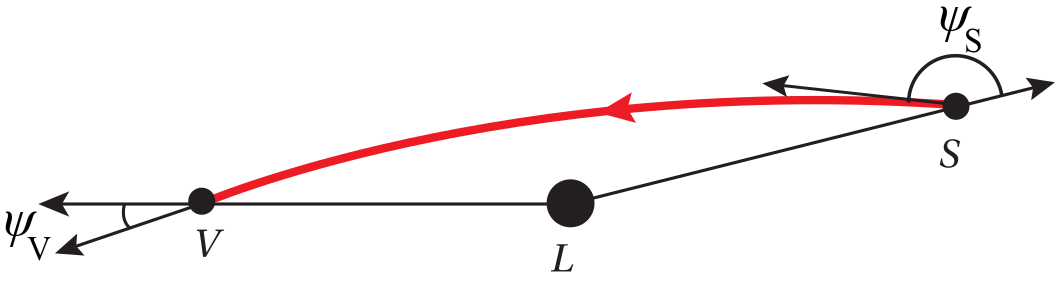}
\end{center}
\caption{\textbf{Two dimensional Riemannian manifold $^2\mathscr{R}$ where \textbf{LS}, \textbf{LV} \& \textbf{SV} are the spatial projections of the geodesics in the equatorial plane}}
\label{fig2a}
\end{figure}

\par Let us consider Fig.~\ref{fig2a}, which depicts the triangle $^{\rm V}\nabla_{\rm L}^{\rm S}$ formed by the photon trajectory between the source (\textbf{S}) and viewer (\textbf{V}) after deflection by the lensing object (\textbf{L}), together with the radial geodesics from \textbf{L} through \textbf{S} and \textbf{V}. This region is described by the Fermat metric extracted from the spacetime metric under consideration. Since we are interested in trajectories confined to the equatorial plane, Fig.~\ref{fig2a} effectively represents a two-dimensional Riemannian manifold $^2\mathscr{R}$, where \textbf{LS}, \textbf{LV}, and \textbf{SV} denote spatial projections of the corresponding geodesics. The photon path in this plane may possess a nonzero geodesic curvature depending on the underlying metric, though it always corresponds to a null geodesic in four-dimensional spacetime.

\par Let $\psi_{\rm V}$ and $\psi_{\rm S}$ be the angles between the photon trajectory and the radial direction at the viewer and source, respectively. If $\phi_{\rm V}$ and $\phi_{\rm S}$ denote their longitudes, then $\phi_{\rm VS} = \phi_{\rm V} - \phi_{\rm S}$ gives the coordinate separation angle. The internal angle at the lens \textbf{L} is $\psi_{\rm L}$. When applying the Gauss–Bonnet theorem to this surface, note that $\psi_{\rm V}$ corresponds to the exterior angle at \textbf{V} (equal to its interior angle), while $\psi_{\rm S}$ is the exterior angle at \textbf{S} (its interior angle being $\pi - \psi_{\rm S}$). Denoting the interior angles of the triangle by $\theta_i \, (i=1,2,3)$, we obtain
\begin{align*}
\sum_{i=1}^3 \theta_i 
&= \psi_{\rm V} + \psi_{\rm L} + \left(\pi - \psi_{\rm S}\right) 
= \psi_{\rm V} + \psi_{\rm L} - \psi_{\rm S} + \pi 
= \alpha_\psi + \pi, \\
\Rightarrow \alpha_\psi 
&= \sum_{i=1}^3 \theta_i - \pi, \numberthis \label{e26}
\end{align*}
where
\begin{align*}
\alpha_\psi = \psi_{\rm V} + \psi_{\rm L} - \psi_{\rm S}. 
\numberthis \label{e27}
\end{align*}
If the surface $^{\rm V}\nabla_{\rm L}^{\rm S}$ is flat, then $\sum_i \theta_i = \pi$ and hence $\alpha_\psi = 0$. Thus, $\alpha_\psi$ naturally quantifies the deviation from Euclidean geometry, forming the basis—through the Gauss–Bonnet theorem—for a consistent definition of the deflection angle discussed next.

\subsection{Mathematical Formulation of Deflection Angle from the Gauss-Bonnet Theorem}{\label{s4.3}}
In our attempt to determine the deflection angle with the help of Gauss-Bonnet theorem the choice of an appropriate integrable region is of critical importance. As will become evident during the course of this subsection a suitable choice for our purposes is quite tricky. The two regions that lend themselves readily for the purposes of our investigation are the embedded triangle $^{\rm V}\nabla_{\rm L}^{\rm S}$ and the embedded quadrilateral $^\infty_{\rm V}\square^\infty_{\rm S}$. 
\subsubsection{Gauss-Bonnet theorem in the region $^{\rm V}\nabla_{\rm L}^{\rm S}$}{\label{s4.3.1}}
\par Applying the Gauss–Bonnet theorem to the domain represented by the embedded triangle $^{\rm V}\nabla_{\rm L}^{\rm S}$, we can express Eq.~\ref{e25} as
\begin{equation}
\iint\limits_{^{\rm V}\nabla_{\rm L}^{\rm S}} K\,{\rm d}S 
+ \sum_{i=1}^{3}\int\limits_{\partial\mathscr{D}_i} \kappa_g\,{\rm d}\ell 
+ \sum_{i=1}^{3} \epsilon_i 
= 2\pi\chi(^{\rm V}\nabla_{\rm L}^{\rm S}).
\label{e28}
\end{equation}
Since the exterior and interior angles satisfy 
$\displaystyle{\sum_i \epsilon_i + \sum_i \theta_i} = 3\pi$, combining this with ~\ref{e26} yields
\begin{align*}
 \alpha_\psi 
= \iint\limits_{^{\rm V}\nabla_{\rm L}^{\rm S}} K\,{\rm d}S 
+ \int_{\rm S}^{\rm V}\!\kappa_{g(\rm SV)}\,{\rm d}\ell
+ 2\pi - 2\pi\chi(^{\rm V}\nabla_{\rm L}^{\rm S}),
\numberthis
\label{e29}
\end{align*}
where the geodesic curvatures along the radial geodesics \textbf{LS} and \textbf{VL} vanish, while that of \textbf{SV} may be nonzero.

\par Two issues arise in ~\ref{e29}. First, $\psi_{\rm L}$—the interior angle at the lens \textbf{L}—becomes ill-defined when the lens represents a singular object such as a black hole. Second, since \textbf{L} may correspond to a singularity, the region of $^{\rm V}\nabla_{\rm L}^{\rm S}$ intersecting the horizon is not smooth, invalidating the assumption $\chi(^{\rm V}\nabla_{\rm L}^{\rm S}) = 1$. Consequently, ~\ref{e29} does not yield a workable expression for $\alpha_\psi$. To overcome these difficulties, one must redefine the integration domain to a smooth manifold excluding the singular point, thereby eliminating the dependence on $\psi_{\rm L}$.

\subsubsection{Gauss-Bonnet theorem in the region $^\infty_{\rm V}\square^\infty_{\rm S}$}{\label{s4.3.2}}

\begin{figure}[hbt!]
\begin{center}
\includegraphics[scale=0.3]{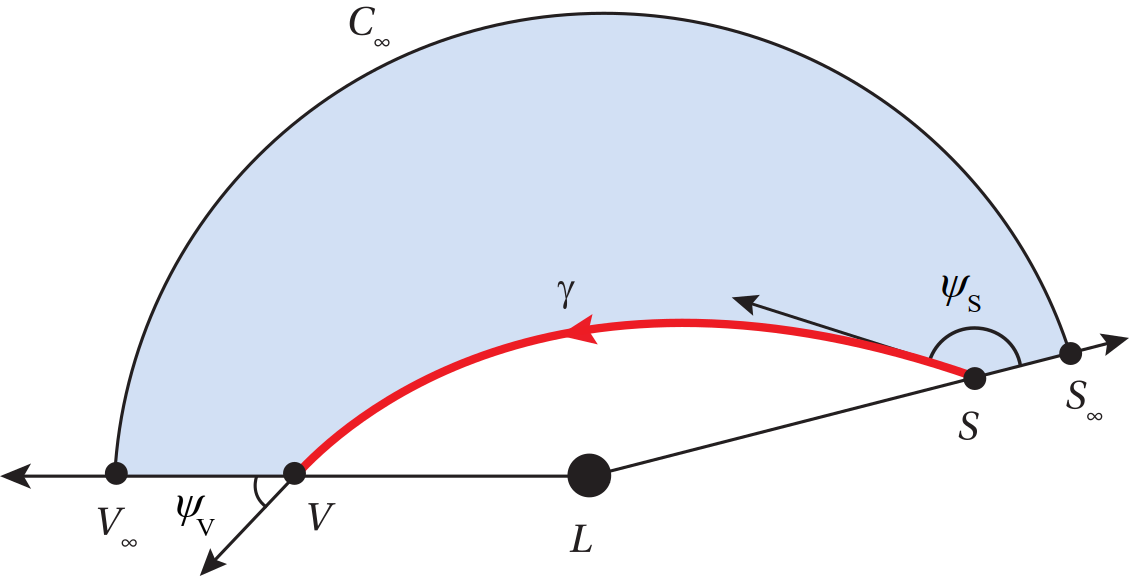}
\end{center}
\caption{\textbf{Extracting the integrable surface region depicted as the quadrilateral $^\infty_{\rm V}\square_{\rm S}^\infty$. The said region is indicated by the portion shaded in blue}}
\label{fig3}
\end{figure}

\par First let us consider another embedded triangle \ref{fig3} $^\infty\nabla_{\rm L}^\infty$ which encloses the embedded triangle $^{\rm V}\nabla_{\rm L}^{\rm S}$. The extremal boundary of the triangle $^\infty\nabla_{\rm L}^\infty$ is a circular arc of radius $r_c$ with \textbf{L} at its centre and intersecting with the projections of the radial geodesics through the viewer and source. In fact these points of intersections might very well be taken as the location for the viewer and source located at infinity(for $r_c\rightarrow \infty$). {\color{black}If asymptotic Euclidean behaviour is an independent property of the optical geometry, then at $r_c\rightarrow\infty$ the auxiliary arc satisfies $\kappa_c\rightarrow1/r_c$ and ${\rm d}\ell\rightarrow r_c{\rm d}\phi$.  Sending a coordinate radius to infinity does not itself establish asymptotic flatness; the metric fall-off must be checked.}\footnote{Since for $r_c\rightarrow\infty$ the spacetime can be considered asymptotically flat we can express the metric describing the line element involved with the integral of geodesic curvature as ${\rm d}\ell^2 = {\rm d}r^2 + r^2\left({\rm d}\theta^2 + \sin^2\theta{\rm d}\phi^2\right)$. For the equatorial plane $\theta=\dfrac{\pi}{2}$ which allows us to express ${\rm d}\ell$ as $r_c{\rm d}\phi$}. 
\par Let us now apply the Gauss-Bonnet theorem to the integrable region given by the embedded quadrilateral $^\infty_{\rm V}\square_{\rm S}^\infty$. The exterior angles at the vertices R, ${\rm V_\infty}$, ${\rm S_\infty}$ \& S are $\psi_{\rm V}$, $\dfrac{\pi}{2}$, $\dfrac{\pi}{2}$ \& $\pi- \psi_{\rm S}$ respectively. Also, geodesic curvatures associated with the radial geodesics will vanish of course. And since the embedded square $^\infty_{\rm V}\square_{\rm S}^\infty = ^\infty\nabla_{\rm L}^\infty - ^{\rm V}\nabla_{\rm L}^{\rm S}$ we do not have to take into consideration whether the lensing object is a black hole singularity or not, thereby allowing us to write $\chi(^\infty_{\rm V}\square_{\rm S}^\infty)=1$. 
\begin{subequations}\label{e30}
\begin{align}
& \iint\limits_{^\infty_{\rm V}\square_{\rm S}^\infty} K{\rm d}S + \sum_{i=1}^4 \int\limits_{\partial\mathscr{D}_i}\kappa_g {\rm d}\ell + \sum_{i=1}^4 \epsilon_i = 2\pi\chi(^\infty_{\rm V}\square_{\rm S}^\infty) \notag\\
\Rightarrow & \iint\limits_{^\infty_{\rm V}\square_{\rm S}^\infty} K{\rm d}S + \phi_{\rm VS} + \int\limits_{\rm V}^{\rm S} \kappa_{g}^{(\rm VS)} {\rm d}\ell + \psi_{\rm V} - \psi_{\rm S} = 0 \notag\\
 \therefore  \alpha_{\rm D} &= -\iint\limits_{^\infty_{\rm V}\square_{\rm S}^\infty} K{\rm d}S + \int\limits_{\rm S}^{\rm V} \kappa_{g}^{(\rm VS)} {\rm d}\ell \label{e30a}\\
 & =  \psi_{\rm V} - \psi_{\rm S} +  \phi_{\rm VS}  \label{e30b}
\end{align}
\end{subequations}

\ref{e30a} \& \ref{e30b} actually depict the generic expression for the deflection angle $\alpha_{\rm D}$ of a photon. A quick check of the formula can be performed by testing it out in the asymptotically flat case, where the expression for deflection angle should take the same form as the standard form generally given in textbooks. In the conventional approach for deriving the expression of bending angle, as \textbf{S} \& \textbf{V} are taken to be located at infinity we can safely assume that $\psi_{\rm S}=\pi$ and $\psi_{\rm V}=0$. This effectively reduces \ref{e30b} to the form $\alpha_{\rm D}^{\rm AF}=\phi_{\rm VS}-\pi$ which is nothing but the traditional textbook definition of deflection angle. A more rigorous and detailed discussion on the justification for accepting \ref{e30b} as a consistent definition of the bending angle maybe found in \cite{Takizawa_2020}.
\par However, the definition of bending angle involving the integrals of the Gaussian curvature and the geodesic curvature as expressed in \ref{e30a} demands the assumption of asymptotic flatness$(r\rightarrow\infty)$. \cite{Takizawa_2020}, followed by \cite{Huang_2024} have addressed this issue in detail. \cite{Huang_2024} have in fact firmly established a generalized GW formalism of determining the deflection angle which does not require any form of asymptotic flatness.     
\par Depending on the spacetime we are working with we can either use the form involving the integrals of the Gaussian curvature and geodesic curvature or the form involving the jump angles and coordinate angular separation between the source and observer. Although, in general the second form nearly always turns up complicated elliptic integrals. In spite of that a thorough consistency check between the results obtained by both the forms have been presented in \cite{Ono_2017}. Using the results as derived in this section we can now utilise them for computing the deflection angle for different types of spacetimes. {\color{black} Huang, Cao and L\"u generalized the curvature construction so that no outer boundary at spatial infinity is required, and Takahashi \emph{et al.} proved its equivalence to the OIA endpoint-angle form under the hypotheses of their construction \cite{Huang_2024,takahashi2023equivalence}.  These results remove an asymptotic-flatness requirement for that finite domain, but they do not remove the need to specify an admissible topology, a continuous ray branch, endpoint orientations, and a well-defined stationary optical geometry.} 
\begin{equation}
\alpha_{\rm D} =  \psi_{\rm V} - \psi_{\rm S} +  \phi_{\rm VS} = \int\limits_{\rm S}^{\rm V} \left(\dfrac{{\rm d}\psi}{{\rm d}\phi} + 1 \right){\rm d}\phi
\label{e31}
\end{equation}
{\color{black} For clarity, the generalized Huang--Cao--L\"u integrand can be written explicitly.  On the equatorial optical surface let
\begin{equation*}
{\rm d}\ell^2=\gamma_{rr}(r){\rm d}r^2+\gamma_{\phi\phi}(r){\rm d}\phi^2,
\qquad
\beta=\beta_\phi(r){\rm d}\phi,
\qquad
\gamma=\gamma_{rr}\gamma_{\phi\phi}.
\end{equation*}
Along the photon orbit $r=r_\gamma(\phi)$ define
\begin{align*}
\mathcal H(r)
&=-\frac{1}{2\sqrt{\gamma}}\frac{{\rm d}\gamma_{\phi\phi}}{{\rm d}r},\\
\mathcal T(r)
&=-\frac{{\rm d}\beta_\phi}{{\rm d}r}
\sqrt{\frac{1}{\gamma_{\phi\phi}}\left(\frac{{\rm d}r}{{\rm d}\phi}\right)^2
+\frac{1}{\gamma_{rr}}}.
\end{align*}
Then, with the same orientation conventions at the source and receiver,
\begin{equation}
\alpha_{\rm HC}
=\int_{\phi_{\rm S}}^{\phi_{\rm V}}
\left[1+\mathcal H\!\left(r_\gamma(\phi)\right)
+\mathcal T\!\left(r_\gamma(\phi)\right)\right]{\rm d}\phi .
\label{eq:HC_generalized}
\end{equation}
On each non-radial, continuously oriented branch, their relation
$\mathcal H+\mathcal T={\rm d}\psi/{\rm d}\phi$ reduces \ref{eq:HC_generalized} to \ref{e31}.  A trajectory with turning points or multiple windings must be divided into monotone branches and the endpoint angles must be followed continuously; an inverse trigonometric principal value alone does not retain this information.}
One of the most critical points to note in this context is that provided the photon trajectory remains fixed the deflection angle as evaluated by the Gauss-Bonnet based approach is not dependent on the domains of integration.
\par In the following section we mainly devote ourselves to reviewing the development of mathematical techniques essential for determining the bending angle of light rays in---
\begin{itemize}
\item[•] Spherically symmetric spacetimes
\item[•] Axisymmetric spacetimes
\end{itemize} 
using the formalism developed here. In particular the determination of the angles $\psi$ and $\phi$ are the main objectives in \ref{s4}. On the other hand determination of the bending angle by evaluating integrals of $K$ and $\kappa_g$ forms the main subject matter of \ref{s5} .

\subsection{Implications for Direct Astronomical Observations}
\begin{figure}[hbt!]
\begin{center}
\includegraphics[scale=0.187]{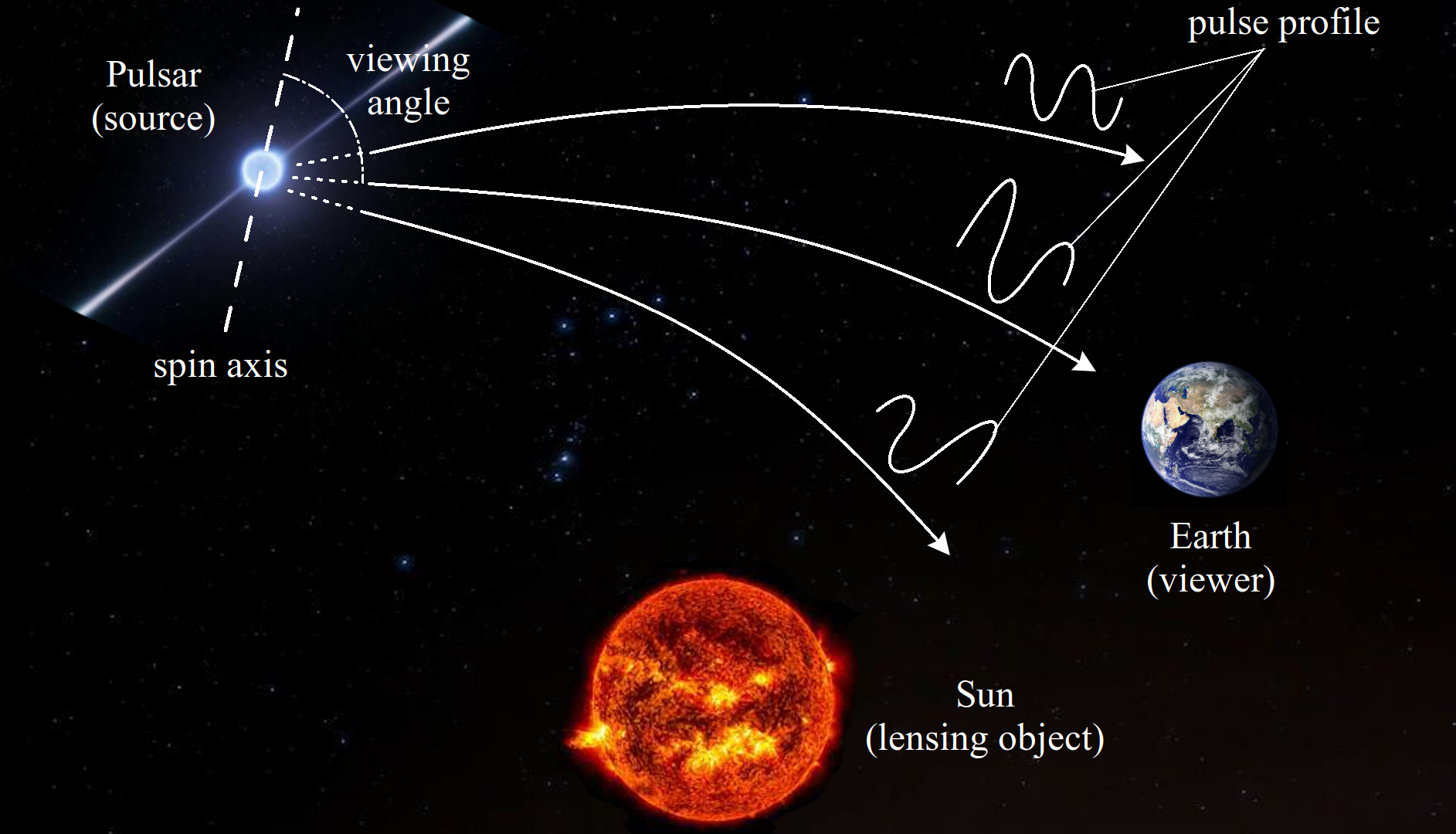}
\end{center}
\caption[Illustrating the geometric layout for verifying the finite distance corrections to the bending angle of light. The source is taken as a Pulsar emitting radio signals at periodic intervals.]{The figure presents a scenario where we have the Sun as the lensing object, Earth as the location of the viewer and a pulsar as the source. The pulsar emits radio signals at periodic intervals in a specific anisotropic manner. The pulse profile associated with the emissions allows us to determine the radiation direction at the actual location of the pulsar. The relative positions of the Earth, Sun \& Pulsar may be determined with the help of the ephemeris which in turn allows us to determine $\phi_{\rm VS}$ and $\psi_{\rm S}$. The angle $\psi_{\rm V}$ can be ascertained by observing the pulsar as described in the text. This illustrates a scenario wherein the accuracy of \ref{e30b} describing $\alpha_{D}$ can be tested.}
\label{pulsar1}
\end{figure}

{\label{s4.4}}
In terms of actual observations that can be carried out with the assistance of modern astronomical apparatus \ref{e30b} is immensely relevant. Given that the location of the source and the observer are known the angles described in \ref{e30b} can be easily determined in principle. Considering that the Very Long Baseline Interferometry(VLBI) is able to provide extremely precise radio observations the relative positions of a pulsar, the Earth and the lensing object Sun(say) can be determined from the ephemeris \cite{{Madison_2013}, {Wang_2017}, {Deller_2019}, {Liu_2023},  {Ding_2024}, {Pearlman_2024}}. Since there is a multitude of pulsars whose spin period, pulse signal, pulse profiles etc features are well documented they are ideal as potential source candidates for verifying weak lensing predictions \cite{{Walker_1996}, {Lai:2004ue}, {Hobill2008}}. As such, we can consider an arbitrary pulsar emitting radio signals with a constant period in an anisotropic manner as a source. Given these considerations we can easily determine the following: --
\begin{itemize}
\item[\textbf{(a)}] The angle $\psi_{\rm V}$ at the Earth between the solar directon and the pulsar direction.
\item[\textbf{(b)}] The angle $\phi_{\rm VS}$.
\item[\textbf{(c)}] In conjunction with information regarding the direction of pulse radiation and pulsar position the angle $\psi_{\rm S}$ can be easily determined. Since the viewing angle of the pulsar as seen by the viewer is accessible from the pulse profiles the direction of radiating pulses that reach the viewer can also be determined in principle. It is worth noting in this context that as a consequence of the Earth's motion around the Sun the viewing keeps continually changing with respect to time as illustrated in \ref{pulsar1}. 
\end{itemize}  
\par Determination of the angle $\psi_{\rm S}$ can be a very involved process and merits a somewhat more elaborate description. The pulse profile and shape are dependent on the viewing angle with respect to the spin axis of the pulsar. This provides us with a strong motivation for considering pulsars whose spin axis is known to us from some astronomical observations. Furthermore, a significant issue in this context is that the spin axis of any isolated pulsar remains unchanged with respect to time. As the Earth keeps changing position due to its motion around the Sun the viewer records different viewing angles subtended by the pulsar with passage of time. As a result of this the corresponding pulse shape also keeps changing. Now by recording these changes in pulse shape the intrinsic direction of the radio emissions can be determined,at least in principle. This in turn allows us to determine the angle between the spin axis and the direction of the emitted light from the pulsar to the viewer. An added benefit of this approach is that we can glean information like the radial direction from the lensing object or the intrinsic position of any such known pulsar from the ephemeris. While from a technological point of view such measurements are still subject to handicaps the scientific principle behind measuring $\psi_{\rm S}$ is quiet sound. Thus all the angles $\psi_{\rm V}$, $\psi_{\rm S}$ \& $\phi_{\rm VS}$ can be determined from astronomical observations thereby determining $\alpha_{\rm D}$ as given by \ref{e30b}. Thus it can be claimed that the theoretical framework developed here has a strong potential for experimental verification based on astronomical observations. 

\color{black}
\section{Generalized Expression of Bending Angle in Axisymmetric Spacetime by OIA Formalism}{\label{s5}}

Let us first consider the case of stationary axisymmetric spacetime followed by the spherically symmetric case. The generic form of the metric in the axisymmetric case in polar coordinates is given as 
\begin{align*}
{\rm d}s^2 &= g_{\mu\nu}{\rm d}x^\mu{\rm d}x^\nu \\
& =-A\left(r,\theta \right){\rm d}t^2 - 2H\left(r,\theta \right){\rm d}t{\rm d}\phi + B\left(r,\theta \right){\rm d}r^2 + C\left(r,\theta \right){\rm d}\theta^2 + D\left(r,\theta \right){\rm d}\phi^2 \numberthis
\label{e32}
\end{align*} 
For a photon orbit restricted to the equatorial plane, we work with the assumption that a local reflection symmetry with respect to $\theta = \dfrac{\pi}{2}$ is observed. Furthermore, $A\left(r,\dfrac{\pi}{2}  \right),\; B\left(r,\dfrac{\pi}{2}  \right),\; D\left(r,\dfrac{\pi}{2}  \right) > 0 $ is also an implicit assumption (\cite{Ono_2017}, \cite{Huang_2024}). As outlined in \ref{s4} in order to employ the OIA method we first need to reduce the metric to the Randers-Finsler form. 
\subsection{Extracting the Fermat Metric from the Randers-Finsler Form}{\label{s5.1}}
The Randers-Finsler \cite{{Chanda_2019}, {Werner_2012}, {Gibbons_2009}} form of any metric is an invaluable tool in the arsenal of any researcher investigating the geometrodynamics of any given spacetime. In general it is structured as 
\begin{equation}
{\rm d}s^2 = -h\left({\rm d}t - \omega_i{\rm d}x^i  \right)^2 + h^{-1} \xi_{ij}{\rm d}x^i{\rm d}x^j
\label{e33}
\end{equation}
Detailed accounts of this is available in the literature---\cite{Frolov_2013}, \cite{Landau1980}. With a suitable rescaling \ref{e33} can be written as 
\begin{equation}
{\rm d}S^2 = -\left({\rm d}t - \beta_i{\rm d}x^i \right)^2
+\gamma_{ij}^{\rm (F)}{\rm d}x^i{\rm d}x^j .
\label{e34}
\end{equation}
where ${\rm d}s^2 = h{\rm d}S^2$ and $\gamma_{ij}^{\rm (F)}$ is the spatial metric \cite{Asada_2000} commonly referred to as the optical metric or Fermat metric. Note in spite of its appearance the Fermat metric is not the same as the induced metric encountered during $3+1$ splitting of a 4-d metric. The Fermat metric \& $\beta_i$ can be expressed in terms of the components of the given spacetime metric as 
\begin{equation}
\gamma_{ij}^{\rm (F)} = \dfrac{g_{ij}}{h} + \dfrac{g_{ti}}{h}\dfrac{g_{tj}}{h},\;\;\;\;\;\; \beta_i=\dfrac{g_{ti}}{h}
\label{e35}
\end{equation}
{\color{black} The one-form $\beta=\beta_i{\rm d}x^i$ supplies the stationary Randers term.  The local force in the projected ray equation is controlled by the two-form ${\rm d}\beta$; an exact part of $\beta$ changes the travel-time functional only by an endpoint term.}. Invoking the null condition ${\rm d}s^2=0$ we can write from \ref{e34}
\begin{equation}
{\rm d}t = \sqrt{\gamma_{ij}^{\rm (F)}{\rm d}x^i{\rm d}x^j} + \beta_i{\rm d}x^i
\label{e36}
\end{equation}
$\gamma_{ij}^{\rm (F)}$ is a Riemannian metric whereas $\beta_i{\rm d}x^i$ represents an one-form. {\color{black} Staticity permits $\beta_i=0$ in a chart adapted to the hypersurface-orthogonal timelike Killing field.  Spherical symmetry alone does not force a shift one-form to vanish in every coordinate chart, and axisymmetry alone does not require a nonzero shift; a static axisymmetric metric is a counterexample.}. A significant observation can be inferred from this, namely that $\gamma_{ij}^{\rm (F)}$ which is a Riemannian metric is sufficient for structuring the optical geometry for spherically symmetric cases. Whereas, for axisymmetric cases the inclusion of a non-zero $\beta_i$ ensures that the optical geometry is described by a Randers-Finsler manifold \cite{Bao_D_2012-10-03} which essentially reduces to Riemannian geometry \textit{iff} $\beta_i=0$. In fact one of the reasons we favour the terminology Fermat metric over optical metric as a descriptor of $\gamma_{ij}^{\rm (F)}$ is specifically because it does not universally lay out the optical geometry.
   
\par The axisymmetric metric expressed by \ref{e32} can be reduced to the Randers-Finsler form as shown below. 
\begin{align}
{\rm d}s^2
&=-A\left({\rm d}t+\frac{H}{A}{\rm d}\phi\right)^2
+B\,{\rm d}r^2+C\,{\rm d}\theta^2
+\left(D+\frac{H^2}{A}\right){\rm d}\phi^2 \notag\\
&=A\left[
-\left({\rm d}t+\frac{H}{A}{\rm d}\phi\right)^2
+\frac{B}{A}{\rm d}r^2+\frac{C}{A}{\rm d}\theta^2
+\frac{AD+H^2}{A^2}{\rm d}\phi^2
\right].
\label{e37}
\end{align}
A comparison with the form presented in \ref{e34} with \ref{e37} allows us to write
\begin{equation}
\gamma_{ij}^{\rm (F)}{\rm d}x^i{\rm d}x^j = \dfrac{B}{A}{\rm d}r^2 + \dfrac{C}{A}{\rm d}\theta^2 + \dfrac{AD + H^2}{A^2}{\rm d}\phi^2
\label{e38}
\end{equation}
\begin{equation}
\beta_i{\rm d}x^i = -\dfrac{H}{A}{\rm d}\phi
\label{e39}
\end{equation}
The steps leading upto \ref{e37} is undoubtedly a pedestrian approach \emph{en route} to deriving the Fermat metric and we have simply provided it as a easy way for someone new to the literature to verify the accuracy of \ref{e35}. In any case it is needless to say that preference must be given to \ref{e35} while deriving $\gamma_{ij}^{\rm (F)}$ \& $\beta_i$. Thus the matrix form of $\gamma_{ij}^{\rm (F)}$, for axisymmetric spacetime is
\begin{equation}
\gamma_{ij}^{\rm (F)}=
\renewcommand\arraystretch{1.62}
\begin{pmatrix}
\dfrac{B}{A}  &  0  &  0\\
0  &  \dfrac{C}{A}  &  0\\
0  &  0  &  \dfrac{AD + H^2}{A^2}
\end{pmatrix}
\label{e40}
\end{equation}

\subsubsection{Fermat metric --- Why?}{\label{s5.1.1}}
Determining the jump angles at the viewer and source locations is central
to the methodology of GW-OIA formalism as is evident from our previous discussions in \ref{s4.2} \& \ref{s4.3}. Now, in order to accomplish this we need to determine two unit vectors namely ---
\begin{itemize}
\item[(a)] the \textbf{unit tangential vector} $\tau^i$ (say) along the photon trajectory
\item[(b)] the \textbf{unit radial vector} $\rho^i$ (say) along the outgoing direction 
\end{itemize}  
This is exactly where the Fermat metric comes in. The Fermat metric is critical for deriving a working expression of these two unit vectors. The beauty and simplicity of the Fermat metric lies in the fact that by the very nature of its definition it represents a Riemannian space($^3\mathscr{R}$) wherein the trajectory of a photon(or any massless particle for that matter) can be represented by a simple spatial curve. In fact in case of any stationary spacetime the arc length along the photon trajectory, as described by the Fermat metric, is directly linked with the time associated with the time-like Killing vector. Now, if we take $\ell_{\rm ph}$ as the arc length along the photon trajectory then we can structure its definition with the aid of the Fermat metric as
\begin{equation}
{\rm d}\ell_{\rm ph}^2 = \gamma_{ij}^{\rm (F)}{\rm d}x^i{\rm d}x^j
\label{e41}
\end{equation}
provided of course that $\gamma^{ij(F)}$ obeys the condition
\begin{equation}
\gamma^{ij{\rm (F)}}\gamma_{jk}^{\rm (F)} = \delta^i_k
\label{e42}
\end{equation}
\par {\color{black} The following distinction is essential: the spatial length that Asada and Kasai show to be affine up to a constant scale is not the Fermat length defined in \ref{e41}, but a conformally related spatial length \cite{Asada_2000}.  The Fermat length remains suitable for normalizing the OIA spatial tangent; it is simply not, in general, a four-dimensional affine parameter.}

\subsubsection{Affine parameter and the Asada--Kasai versus Fermat spatial metrics}{\label{s5.1.2}}

{\color{black} Let us assume $c=1$ and stipulate a stationary region where $h=-g_{00}>0$.  The decomposition used by Asada and Kasai is
\begin{equation*}
{\rm d}s^2=-h\left({\rm d}t-w_i{\rm d}x^i\right)^2
+h^{-1}\gamma^{\rm(AK)}_{ij}{\rm d}x^i{\rm d}x^j ,
\end{equation*}
with
\begin{equation*}
w_i=-\frac{g_{0i}}{g_{00}}=\frac{g_{0i}}{h},
\qquad
\gamma^{\rm(AK)}_{ij}
=-g_{00}\left(g_{ij}-\frac{g_{0i}g_{0j}}{g_{00}}\right)
=h\,g_{ij}+g_{0i}g_{0j}.
\end{equation*}
Define ${\rm d}\ell_{\rm AK}^2=\gamma^{\rm(AK)}_{ij}{\rm d}x^i{\rm d}x^j$.  By contrast, the Fermat metric in \ref{e35} is
\begin{equation*}
\gamma^{\rm(F)}_{ij}
=\frac{g_{ij}}{h}+\frac{g_{0i}g_{0j}}{h^2}
=\frac{\gamma^{\rm(AK)}_{ij}}{h^2},
\qquad
{\rm d}\ell_{\rm ph}=\frac{{\rm d}\ell_{\rm AK}}{h}.
\end{equation*}
The position-dependent factor $h^2$ is the point at which the two spatial metrics must not be confused.

Let $k^\mu={\rm d}x^\mu/{\rm d}\lambda$ be the tangent to an affinely parameterized null geodesic.  Stationarity gives the conserved covariant component
$k_0=g_{0\mu}k^\mu=-h(k^0-w_i k^i)$.  Applying the null condition to the decomposed metric gives
\begin{align}
0&=-h(k^0-w_i k^i)^2
+h^{-1}\gamma^{\rm(AK)}_{ij}k^ik^j,\notag\\
(k_0)^2&=\gamma^{\rm(AK)}_{ij}k^ik^j
=\left(\frac{{\rm d}\ell_{\rm AK}}{{\rm d}\lambda}\right)^2 .
\label{e43}
\end{align}
Hence ${\rm d}\ell_{\rm AK}/{\rm d}\lambda=|k_0|$ is constant, so $\ell_{\rm AK}$ is affine up to a constant scale and origin.  For the Fermat length, however,
\begin{equation}
\frac{{\rm d}\ell_{\rm ph}}{{\rm d}\lambda}
=\frac{|k_0|}{h\!\left(x(\lambda)\right)},
\label{e44}
\end{equation}
which is not constant unless $h$ is constant along the ray.  For example, in Schwarzschild spacetime $h=f(r)=1-2M/r$ and $k_0=-E$, so
\[
\frac{{\rm d}\ell_{\rm ph}}{{\rm d}\lambda}=\frac{E}{f(r)};
\]
it varies as the photon moves through the gravitational redshift factor.

Note that the full null contraction $g_{\mu\nu}k^\mu k^\nu=0$ does not imply $(k_0)^2=0$, nor does it imply that the positive spatial norm vanishes.  It relates those two nonzero terms as in \ref{e43}.  Confusing $\gamma^{\rm(AK)}_{ij}$ with $\gamma^{\rm(F)}_{ij}$ moves a factor $h^2$ and incorrectly promotes the Fermat arclength to an affine parameter.

The OIA construction does not require that promotion.  It needs only the unit spatial tangent
\begin{equation}
e^i\equiv\frac{{\rm d}x^i}{{\rm d}\ell_{\rm ph}},
\qquad
\gamma^{\rm(F)}_{ij}e^ie^j=1,
\label{e45}
\end{equation}
which follows directly from the definition \ref{e41}.  In the subsequent formulas the  symbol $\tau^i$ is to be understood as this normalized spatial tangent $e^i$, not as the spatial components of a four-dimensional affinely parameterized tangent.}
\subsection{Determination of the Angles $\psi_{\rm V}$, $\psi_{\rm S}$ \& $\phi_{\rm VS}$}{\label{s5.2}}
The angle $\phi_{\rm VS}$ is the one ever-present term in any mathematical formulation of the deflection angle and something which has been dealt with extensively in \ref{s4}. The only difference in this case is that we are no longer entertaining the assumption of an infinite separation between the source and viewer.   How the consideration of finite distance affects the representation of $\phi_{\rm VS}$ is discussed in \ref{s5.2.4} In order to find the angles $\psi_{\rm V}$ \& $\psi_{\rm S}$ we first need to define the unit tangent vector and unit radial vector as depicted in the \ref{fig2a} \& \ref{fig3}.
\subsubsection{Defining the unit tangent vector and unit radial vector}{\label{s5.2.1}}
From \ref{e45} we can extract the components of the unit tangent vector as 
\begin{equation}
\tau^i = \left(\dfrac{{\rm d}r}{{\rm d}\ell_{\rm ph}}, 0, \dfrac{{\rm d}\phi}{{\rm d}\ell_{\rm ph}}   \right)
\label{e46}
\end{equation}  
where there is no contribution from the $\theta$ part as we restrict ourselves to the equatorial plane wherein $\theta=\dfrac{\pi}{2}$. However, instead of keeping $\tau^i$ in the form expressed in \ref{e46} we can restructure it to our benefit as 
\begin{align*}
& \tau^i = \left(\dfrac{{\rm d}r}{{\rm d}\phi}\dfrac{{\rm d}\phi}{{\rm d}\ell_{\rm ph}}, 0, \dfrac{{\rm d}\phi}{{\rm d}\ell_{\rm ph}}   \right) \\
\Rightarrow & \tau^i = \dfrac{1}{\zeta} \left(\dfrac{{\rm d}r}{{\rm d}\phi}, 0, 1   \right) \numberthis
\label{e47}
\end{align*}
where we can easily determine the value of $\dfrac{1}{\zeta}$ by enforcing the condition $\gamma_{ij}^{\rm (F)}\tau^i\tau^j = 1$ and using \ref{e38} in conjunction with \ref{e47}. Although, since it will be helpful to express our values involving the impact parameter we first derive the orbit equations using the properties of killing vectors as shown in \ref{s5.2.2}. [The derivation of the final form of unit tangential vector is resumed in \ref{s5.2.3}.]

\par The unit radial vector is simply defined as an unit tangent vector along the radial line on the equatorial plane in $^3\mathscr{R}$. Mathematically speaking the unit radial vector may be represented by 
\begin{equation}
\rho^i = \left(\rho^r, 0, 0  \right)
\label{e48}
\end{equation} 
Now given the specifications of the Fermat metric the unit radial vector should abide by the stipulation 
\begin{equation}
\gamma_{ij}^{\rm (F)}\rho^i \rho^j = 1
\label{e49}
\end{equation}
which allows us to write
\begin{align*}
& \gamma_{ij}^{\rm (F)}\rho^i \rho^j = 1 \\
\Rightarrow & \gamma_{rr}^{\rm (F)} \left(\rho^r  \right)^2 + \gamma_{\theta\theta}^{\rm (F)} \left(\rho^\theta  \right)^2 + \gamma_{\phi\phi}^{\rm (F)} \left(\rho^\phi  \right)^2 = 1 \\
\Rightarrow & \gamma_{rr}^{\rm (F)} \left(\rho^r  \right)^2 = 1 \\
\Rightarrow & \rho^r = \dfrac{1}{\sqrt{\gamma_{rr}^{\rm (F)}}} \\
& \therefore \rho^i = \left(\dfrac{1}{\sqrt{\gamma_{rr}^{\rm (F)}}}, 0, 0  \right) \numberthis\label{e50}
\end{align*}
\subsubsection{Orbit equation in terms of impact parameter}{\label{s5.2.2}}
Energy $E$ and azimuthal component of angular momentum $J$ are conserved along the geodesics(since the metric has no explicit dependence on $t$ and $\phi$). If we express the killing vectors with $\Xi_{(t)}$ \& $\Xi_{(\phi)}$ then we can utilise it to construct two equations of motion in the spacetime \ref{e32} involving $E$ and $J$ as follows
\begin{equation}
E = -\Xi_{(t)\mu}p^\mu = -g_{tt}\dot{t} - g_{t\phi}\dot{\phi} = A\dot{t} + H\dot{\phi}
\label{e51}
\end{equation}
and
\begin{equation}
J = \Xi_{(\phi)\mu}p^\mu=p_\phi
=g_{\phi t}\dot{t}+g_{\phi\phi}\dot{\phi}
=-H\dot{t}+D\dot{\phi}.
\label{e52}
\end{equation}
One of the other equation motion involving the radial part can be constructed from the null condition $p^\mu p_\mu = 0$ as applied to the spacetime metric \ref{e32}. Now, it is well known that $u\cdot u = g_{\alpha\beta}\dfrac{{\rm d}x^\alpha}{{\rm d}\lambda}\dfrac{{\rm d}x^\beta}{{\rm d}\lambda}$ which coupled with the null condition as applied to \ref{e32} allows us to write
\begin{equation}
0 = -A\dot{t}^2 + B\dot{r}^2 + C\dot{\theta}^2 + D\dot{\phi}^2 - 2H\dot{t}\dot{\phi}
\label{e53}
\end{equation} 
For the equatorial plane this reduces to the form
\begin{equation}
0 = -A\dot{t}^2 + B\dot{r}^2 + D\dot{\phi}^2 - 2H\dot{t}\dot{\phi}
\label{e54}
\end{equation} 
$\lambda$ represents an affine parameter and $\cdot$ represents any derivative w.r.t to the affine parameter.

\par The impact parameter is defined as $b=\dfrac{J}{E}$ as usual. We can now use this to express the impact parameter in terms of the components of the metric \ref{e32}.
\begin{align*}
b&=\dfrac{J}{E}=\dfrac{-H\dot{t} + D\dot{\phi}}{A\dot{t} + H\dot{\phi}} = \dfrac{-H + D\dfrac{{\rm d}\phi}{{\rm d}\lambda}\dfrac{{\rm d}\lambda}{{\rm d}t}}{A + H\dfrac{{\rm d}\phi}{{\rm d}\lambda}\dfrac{{\rm d}\lambda}{{\rm d}t}}= \dfrac{-H + D \dfrac{{\rm d}\phi}{{\rm d}t}}{A + H \dfrac{{\rm d}\phi}{{\rm d}t}}\numberthis\label{e55}
\end{align*}
\par Let us now resume our treatment of the \ref{e54} to mould it into a more suitable form
\begin{align*}
\left(\dfrac{{\rm d}r}{{\rm d}\phi} \right)^2 = \dfrac{E\dot{t} - J\dot{\phi}}{B\dot{\phi}^2}
\numberthis\label{e56}
\end{align*}
Before proceeding further we need to express $\dot{t}$ and $\dot{\phi}$ in terms of $E$, $J$ and the components of the spacetime metric. A very simple algebraic manipulation of \ref{e51} \& \ref{e52} allows us to write 
\begin{equation}
\dot{\phi} = \dfrac{EH + AJ}{AD + H^2}
\label{e57}
\end{equation}
and
\begin{equation}
\dot{t} = \dfrac{ED - HJ}{AD + H^2}
\label{e58}
\end{equation}
Substituting the expressions in \ref{e57} \& \ref{e58} in \ref{e56} we get the radial equation in terms of the impact parameter and components of the spacetime metric as follows
\begin{align*}
\left(\dfrac{{\rm d}r}{{\rm d}\phi} \right)^2 = \dfrac{\left(AD + H^2   \right) \left(D - 2 Hb - Ab^2   \right)}{B\left(H + Ab  \right)^2}
\numberthis\label{e59}
\end{align*}
\ref{e59} in its present form can be immediately plugged into \ref{e50} to get the expression for the unit tangential vector discussed in \ref{s5.2.1}.
\par In terms of reciprocal coordinate $u=\dfrac{1}{r}$ we have
\begin{equation}
\left(\dfrac{{\rm d}u}{{\rm d}\phi}  \right)^2 = u^4 \dfrac{\left(AD + H^2   \right) \left(D - 2 Hb - Ab^2   \right)}{B\left(H + Ab  \right)^2} \equiv F(u)
\label{e61}
\end{equation}
\subsubsection{Deriving the expression of the angles $\psi_{\rm V}$ \& $\psi_{\rm S}$}{\label{s5.2.3}}
A careful study of the \ref{fig2a} establishes that the angles $\psi_{\rm V}$ and $\psi_{\rm S}$ are subtended between the unit tangential and unit radial vectors at the points \textbf{V} and \textbf{S} respectively. Now any angle $\psi$ measured from the radial direction to the tangent drawn on the spatial curve representing the photon trajectory can be mathematically expressed as
\begin{equation}
\cos\psi = \gamma_{ij}^{\rm (F)}\tau^i\rho^j
\label{e62}
\end{equation}
The only existing component of unit radial vector has already been derived in \ref{e50}. The expression for the unit tangential vector \ref{e47} is now engineered to its final form as shown below. First we evaluate the expression for $\dfrac{1}{\zeta}$ by utilising $\gamma_{ij}^{\rm (F)}\tau^i\tau^j = 1$,
\begin{align*}
 \dfrac{1}{\zeta} = \dfrac{A\left(H + Ab \right)}{AD + H^2}
\numberthis\label{e63}
\end{align*}  
Therefore, the components of the unit tangent vector can now be written as
\begin{equation}
\tau^i = \dfrac{A\left(H + Ab \right)}{AD + H^2} \left(\dfrac{{\rm d}r}{{\rm d}\phi}, 0, 1   \right)
\label{e64}
\end{equation}

Now, finally we can substitute the unit tangential and radial vectors given by \ref{e64} \& \ref{e50} respectively in \ref{e62}.
\begin{align*}
\cos\psi & = \gamma_{rr}^{\rm (F)}\tau^r \rho^r \\
& = \gamma_{rr}^{\rm (F)} \dfrac{A\left(H + Ab \right)}{AD + H^2} \dfrac{{\rm d}r}{{\rm d}\phi}\dfrac{1}{\sqrt{\gamma_{rr}^{\rm (F)}}} \\
& = \sqrt{\gamma_{rr}^{\rm (F)}} \dfrac{A\left(H + Ab \right)}{AD + H^2} \dfrac{{\rm d}r}{{\rm d}\phi}
\numberthis\label{e65}
\end{align*} 
However, there is a dependency on $\dfrac{{\rm d}r}{{\rm d}\phi}$ in \ref{e65}, which does not lend itself to yield a simple user friendly expression of $\psi$. This issue can be easily resolved by transforming the expression in terms of $\sin\psi$ so that $\psi$ can be expressed only in terms of local dependencies. Squaring \ref{e65} and substituting the expression of $\left(\dfrac{{\rm d}r}{{\rm d}\phi}\right)^2$ from \ref{e59} and $\gamma_{rr}^{\rm (F)}$ from \ref{e40} we get
\begin{equation}
\psi = \sin^{-1} \left(\dfrac{H + Ab}{\sqrt{AD + H^2}}\right)
\label{e67}
\end{equation}
Now, since we have restricted ourselves to the equatorial plane, $A$, $D$ \& $H$ are all functions of $r$ or $\dfrac{1}{u}$. Clearly at the points \textbf{V} \& \textbf{S}, $r$ assumes the value $r_{\rm V}=\dfrac{1}{u_{\rm V}}$ and $r_{\rm S}=\dfrac{1}{u_{\rm S}}$ respectively. Therefore, the expressions for $\psi_{\rm V}$ \& $\psi_{\rm S}$ can be written as 
\begin{equation}
\psi_{\rm V} = \sin^{-1}\left[ \dfrac{H\left(\dfrac{1}{u_{\rm V}} \right) + A\left(\dfrac{1}{u_{\rm V}} \right)b}{\sqrt{A\left(\dfrac{1}{u_{\rm V}} \right)D\left(\dfrac{1}{u_{\rm V}} \right) + H\left(\dfrac{1}{u_{\rm V}} \right)^2}}\right]
\label{e68}
\end{equation}
and
\begin{equation}
\psi_{\rm S} = \pi - \sin^{-1}\left[ \dfrac{H\left(\dfrac{1}{u_{\rm S}} \right) + A\left(\dfrac{1}{u_{\rm S}} \right)b}{\sqrt{A\left(\dfrac{1}{u_{\rm S}} \right)D\left(\dfrac{1}{u_{\rm S}} \right) + H\left(\dfrac{1}{u_{\rm S}} \right)^2}}\right]
\label{e69}
\end{equation}
respectively. 
\par The expressions given by \ref{e68} \& \ref{e69} are in their respective forms as $\psi_{\rm V}$ is acute and $\psi_{\rm S}$ is obtuse(\ref{fig3}). {\color{black} The principal value of $\sin^{-1}$ does not retain a quadrant or orientation.  Equations~\ref{e68}--\ref{e69} therefore require the assumed single-turning-point geometry, a continuous choice of tangent direction, and a sign convention for $H+Ab$.  If $b$ is signed, if the ray winds, or if $H+Ab$ changes sign, the orbit must be divided into continuous branches and the angle is most safely reconstructed from both sine and cosine (equivalently by an $\operatorname{atan2}$ prescription).}. 
\subsubsection{Deriving the expression of the angle $\phi_{\rm VS}$}{\label{s5.2.4}}
The azimuthal coordinate separation $\phi_{\rm VS}$ is simply expressed as
\begin{equation}
\phi_{\rm VS} = \int\limits_{\rm S}^{\rm V} {\rm d}\phi 
\label{e70}
\end{equation}
Now we have already derived $\psi_{\rm V}$ \& $\psi_{\rm S}$ in terms of $u$ we are motivated to express $\phi_{\rm VS}$ in the same fashion. To this end we can utilise \ref{e61} to express the integral in \ref{e70} in terms of $u$. The only tricky bit in this context is determining the limits of integration after transforming the integral. In the conventional case we were working with the definition describing a photon arriving from infinity to the distance of closest approach $r_0=\dfrac{1}{u_0}$ before traveling back to infinity. This would automatically mandate the lower limit of $u$ as $0$. However, since the very essence of the Gauss-Bonnet theorem-inspired approach relies on the realistic premise that the source and viewer are at finite distances, the limits should be determined accordingly. As such we simply segment the integral into two distinct portions namely
\begin{itemize}
\item[•] the photon trajectory from the source(\textbf{S}) to the turning point marking the distance of closest approach ($r_0$)
\item[•] the photon trajectory from the turning point to the viewer(\textbf{V})
\end{itemize}  
Therefore, \ref{e70} can can now be expressed as 
\begin{equation}
\phi_{\rm VS} = \int\limits_{u_{\rm S}}^{u_0} \dfrac{1}{\sqrt{F(u)}} {\rm d}u + \int\limits_{u_{\rm V}}^{u_0} \dfrac{1}{\sqrt{F(u)}} {\rm d}u
\label{e71}
\end{equation}
\par Substituting \ref{e68}, \ref{e69} \& \ref{e71} in \ref{e30} we can now express the bending angle of a photon trajectory in axisymmetric spacetimes as 
\begin{align}
\alpha_{\rm D}={}&
\sin^{-1}\!\frac{H(1/u_{\rm V})+A(1/u_{\rm V})b}
{\sqrt{A(1/u_{\rm V})D(1/u_{\rm V})+H(1/u_{\rm V})^2}}
-\pi
+\sin^{-1}\!\frac{H(1/u_{\rm S})+A(1/u_{\rm S})b}
{\sqrt{A(1/u_{\rm S})D(1/u_{\rm S})+H(1/u_{\rm S})^2}}\notag\\
&+\int_{u_{\rm S}}^{u_0}\frac{{\rm d}u}{\sqrt{F(u)}}
+\int_{u_{\rm V}}^{u_0}\frac{{\rm d}u}{\sqrt{F(u)}} .
\label{e72}
\end{align}
This follows immediately from $\alpha_{\rm D}=\psi_{\rm V}-\psi_{\rm S}+\phi_{\rm VS}$ and $\psi_{\rm S}=\pi-\sin^{-1}(\cdots)$.
Using this generic expression we can now evaluate the bending angle for any axisymmetric spacetime wherein the source and viewer are located at finite distances. We now apply \ref{e72} to the specific case of Kerr metric \cite{Teukolsky_2015}. 
\subsection{Applying OIA Formalism to Kerr metric}{\label{5.3}}
The Kerr metric can be expressed in terms of Boyer-Lindquist coordinates as 
\begin{equation}
{\rm d}s^2 = -\left(1 - \dfrac{2Mr}{\Sigma}  \right){\rm d}t^2 - \dfrac{4aMr\sin^2\theta}{\Sigma}{\rm d}\phi{\rm d}t + \dfrac{\Sigma}{\Delta}{\rm d}r^2 + \Sigma{\rm d}\theta^2 + \left(r^2 + a^2 + \dfrac{2a^2Mr\sin^2\theta}{\Sigma}  \right)\sin^2\theta{\rm d}\phi^2
\label{e73}
\end{equation} 
where 
\begin{equation}
\Sigma = r^2 + a^2\cos^2\theta
\label{e74}
\end{equation}
and 
\begin{equation}
\Delta = r^2 + a^2 - 2Mr
\label{e75}
\end{equation}
\subsubsection{Evaluating $\psi_{\rm V}$ \& $\psi_{\rm S}$ for Kerr spacetime}{\label{s5.3.1}}
Substituting the respective values of $A$, $D$ \& $H$ for the Kerr case in \ref{e67} and assuming $\theta=\dfrac{\pi}{2}$ (since we are only considering the equatorial case) we can express $\sin\psi$ as
\begin{align*}
\sin\psi &= \dfrac{H + Ab}{\sqrt{AD + H^2}} = \dfrac{\dfrac{2aMr}{r^2} + \left(1 - \dfrac{2Mr}{r^2}  \right)b}{\sqrt{\left(1 - \dfrac{2Mr}{r^2}  \right) \left(r^2 + a^2 + \dfrac{2a^2Mr}{r^2}  \right) + \dfrac{4a^2M^2r^2}{r^4}}}\\
& = \dfrac{b + \dfrac{2M}{r}\left(a - b  \right)}{\sqrt{r^2 + a^2 + \dfrac{2a^2M}{r} - 2Mr - \dfrac{2a^2M}{r} - \dfrac{4a^2M}{r^2} + \dfrac{4a^2M}{r^2}}} = \dfrac{b + \dfrac{2M}{r}\left(a - b  \right)}{\sqrt{r^2 + a^2 - 2Mr}} \\
& = \dfrac{b}{r} \times \left(1 - \dfrac{2M}{r} + \dfrac{2aM}{br}   \right) \times \left(1 - \dfrac{2M}{r} + \dfrac{a^2}{r^2}   \right)^{-\frac{1}{2}} \\
& = \dfrac{b}{r} \times \left(1 - \dfrac{2M}{r} + \dfrac{2aM}{br}   \right) \times \left[1 + \dfrac{M}{r} - \dfrac{a^2}{2r^2}  + \bigO\left\lbrace\left(\dfrac{2M}{r} - \dfrac{a^2}{r^2}  \right)^2\right\rbrace \right] \\
& = \dfrac{b}{r} \times \left[1 + \dfrac{M}{r} - \dfrac{a^2}{2r^2} - \dfrac{2M}{r} - \dfrac{2M^2}{r^2} + \dfrac{Ma^2}{r^3} + \dfrac{2aM}{br} + \dfrac{2aM^2}{br^2} - \dfrac{a^3M}{br^3} +  \bigO\left\lbrace\left(\dfrac{2M}{r} - \dfrac{a^2}{r^2}  \right)^2\right\rbrace \right] \\
& = \dfrac{b}{r} \times \left[1 - \dfrac{M}{r} + \dfrac{2aM}{br} \right] + \bigO\left(\dfrac{M^2}{r^2}, \dfrac{a^2}{r^2}, \dfrac{aM^2}{r^3}  \right) \\
& = bu \times \left(1 - Mu + \dfrac{2aM}{b}u  \right) + \bigO\left(M^2u^2, a^2u^2, aM^2u^3  \right)
\numberthis
\label{e76}
\end{align*}
In the above expression, we only retain terms of the order $\dfrac{M}{r}$ in the leading term and terms upto $\dfrac{M^2}{r^3}$ in the expansion while completely dismissing terms involving $r^{-4}$ and beyond. We can now solve for $\psi$ as
\begin{align*}
\psi &= \sin^{-1}\left( bu  - bMu^2 + 2aMu^2  \right) + \bigO\left(M^2u^2, a^2u^2, aM^2u^3  \right) \\
& = \sin^{-1}\left( bu \right) - \frac{bMu^2}{\sqrt{1-b^2u^2}} + \frac{2aMu^2}{\sqrt{1-b^2u^2}} + \bigO\left(M^2u^2, a^2u^2, aM^2u^3  \right) \numberthis
\label{e77}
\end{align*}
We have applied Taylor expansion\footnote{$f(x\pm h) = f(x) \pm hf^\prime(x) + \frac{h^2}{2}f^{\prime\prime}(x) \pm \frac{h^3}{6}f^{\prime\prime}(x) + .\; .\; . \; . $} to the $\sin^{-1}$ term with $bu$ as the variable $x$ and $\left(-bMu^2 + 2aMu^2\right)$ as $h$ where $h << x$. Terms are only retained upto the first order. From the generic expression of $\psi$ in Kerr spacetime we can write the expressions of the terms $\psi_{\rm V}$ and $\psi_{\rm S}$ as 
\begin{equation}
\psi_{\rm V} = \sin^{-1}\left( bu_{\rm V}\right) - \frac{bMu_{\rm V}^2}{\sqrt{1-b^2u_{\rm V}^2}} + \frac{2aMu_{\rm V}^2}{\sqrt{1-b^2u_{\rm V}^2}} + \bigO\left(M^2u_{\rm V}^2, a^2u_{\rm V}^2, aM^2u_{\rm V}^3  \right) \numberthis
\label{e78}
\end{equation}
and
\begin{align*}
\pi - \psi_{\rm S} = \sin^{-1}\left( bu_{\rm S}\right) - \frac{bMu_{\rm S}^2}{\sqrt{1-b^2u_{\rm S}^2}} + \frac{2aMu_{\rm S}^2}{\sqrt{1-b^2u_{\rm S}^2}} + \bigO\left(M^2u_{\rm S}^2, a^2u_{\rm S}^2, aM^2u_{\rm S}^3  \right) \\
\Rightarrow \psi_{\rm S} = \pi - \sin^{-1}\left( bu_{\rm S}\right) + \frac{bMu_{\rm S}^2}{\sqrt{1-b^2u_{\rm S}^2}} - \frac{2aMu_{\rm S}^2}{\sqrt{1-b^2u_{\rm S}^2}} + \bigO\left(M^2u_{\rm S}^2, a^2u_{\rm S}^2, aM^2u_{\rm S}^3  \right)
\numberthis
\label{e79}
\end{align*}
\subsubsection{Evaluating $\phi_{\rm VS}$ for Kerr spacetime}{\label{s5.3.2}}
The angle $\phi_{\rm VS}$ for the Kerr spacetime can be derived from the expression presented in \ref{e71}. But first we need to evaluate $F(u)$ by substituting the appropriate values of $A$, $B$, $D$ \& $H$ for the Kerr case in \ref{e61}. Let us start off with the form in \ref{e59} before we derive the expression of $F(u)$.   
\begin{align}
\left(\dfrac{{\rm d}r}{{\rm d}\phi}\right)^2 &= \dfrac{AD + H^2}{B} \dfrac{D - 2Hb -Ab^2}{\left(H + Ab\right)^2} \notag
\\
& = {\dfrac{\left(1 - \dfrac{2Mr}{r^2}  \right) \left(r^2 + a^2 + \dfrac{2a^2Mr}{r^2}   \right) + \left(\dfrac{2aMr}{r^2}  \right)^2}{\dfrac{r^2}{r^2 - 2Mr + a^2}}} \notag
\\
     & \qquad\qquad\qquad\qquad\qquad\qquad\qquad \meqbox{A}{{} \times \dfrac{\left(r^2 + a^2 + \dfrac{2a^2Mr}{r^2}   \right) - \dfrac{4aMr}{r^2}b - \left(1 - \dfrac{2Mr}{r^2} \right)b^2}{\left[\dfrac{2amr}{r^2} + \left(1-\dfrac{2Mr}{r^2}  \right)b  \right]^2}} \notag
\\
& \simeq \dfrac{b^2\left[\dfrac{r}{b}\left(\dfrac{r}{b}-\dfrac{2M}{b}  \right)  \right]^2 \left[-\dfrac{4aM}{b^2} + \dfrac{2M}{b} - \dfrac{r}{b} + \dfrac{r^3}{b^3}   \right]}{\dfrac{r}{b}\left(\dfrac{2aM}{b^2} + \dfrac{r}{b} - \dfrac{2M}{b}   \right)^2} \notag \\
       &\qquad\qquad\qquad \meqbox{B}{{}  [\because\text{Neglecting terms}\; \geq a^2\; \text{for slow rotation approximation}}] \notag \\
& = \dfrac{b^2\left[x\left(x-\dfrac{2M}{b}  \right)  \right]^2 \left[-\dfrac{4aM}{b^2} + \dfrac{2M}{b} - x + x^3   \right]}{x\left(x + \dfrac{2aM}{b^2} - \dfrac{2M}{b} \right)^2}  \notag \\ 
& \qquad\qquad\qquad \meqbox{B}{{}  [\because\text{Substituting}\; \dfrac{r}{b} = x\;}] \notag \\
& \simeq \dfrac{\dfrac{b^2x^2}{x^3}\left(x^2 - \dfrac{4M}{b} x   \right) \left(-\dfrac{4aM}{b^2} + \dfrac{2M}{b} - x + x^3   \right)}{\left[1 - \dfrac{2M}{bx}\left(1 - \dfrac{a}{b}   \right)   \right]^2} \notag \\ 
&\qquad\qquad\qquad \meqbox{B}{{}  [\because\text{Neglecting terms}\; \geq M^2\;}] \notag \\
& = b^2\left(x -\dfrac{4M}{b}  \right) \left(-\dfrac{4aM}{b^2} + \dfrac{2M}{b} - x + x^3 \right) \left(1 + \dfrac{4M}{bx} - \dfrac{4aM}{b^2x}  \right)  \notag \\
& = b^2\dfrac{r^4}{b^4} - b^2 \dfrac{r^2}{b^2} + b^2\dfrac{2M}{b}\dfrac{r}{b} - b^2\dfrac{4aM}{b^2}\dfrac{r^3}{b^3} \notag \\
&\qquad\qquad\qquad \meqbox{B}{{}  [\because\text{Substituting back}\; x=\dfrac{r}{b}}] \notag \\
& = \dfrac{r^4}{b^2} - r^2 + 2Mr - \dfrac{4aM}{b^3}r^3 + \bigO (a^2) \numberthis \label{e80}
\end{align}
Now, using \ref{e80} and substituting in \ref{e71} we can write,
\begin{align*}
\left(\dfrac{{\rm d}u}{{\rm d}\phi}  \right)^2 = F(u) &= u^4\left(\dfrac{{\rm d}r}{{\rm d}\phi}  \right)^2 \\
& = u^4\left[\dfrac{r^4}{b^2} - r^2 + 2Mr - \dfrac{4aM}{b^3}r^3  \right] + \bigO (a^2u^4) \\
\Rightarrow F(u) & = \dfrac{1}{b^2} - u^2 + 2Mu^3 - \dfrac{4aM}{b^3}u + \bigO (a^2u^4) \numberthis
\label{e81}
\end{align*} 
Now, if we consider the distance of closest approach $r=r_0=\dfrac{1}{u_0}$ and $\phi=\dfrac{\pi}{2}$ then $\left(\dfrac{{\rm d}u}{{\rm d}\phi}\right)_{u=u_0,\;\phi=\frac{\pi}{2}} = 0$. We can employ this condition to find the expression of $b$ from \ref{e81} as shown below
\begin{align}
& \dfrac{1}{b^2} - u_0^2 + 2Mu_0^3 - \dfrac{4aM}{b^3}u_0 = 0 \notag\\
\Rightarrow  &\dfrac{1}{b^2} = u_0^2\left[1 - \left(2Mu_0 - \dfrac{4aM}{b^3u_0}  \right)\right]\notag\\
\Rightarrow & b = \dfrac{1}{u_0} \left[1 - \left(2Mu_0 - \dfrac{4aM}{b^3u_0}  \right)  \right]^{-\frac{1}{2}} \notag\\
 &  \simeq \dfrac{1}{u_0}\left[1 -\left(-\dfrac{1}{2}  \right) \left(2Mu_0 - \dfrac{4aM}{b^3u_0}  \right)  \right] \notag\\
&\qquad\qquad\qquad \meqbox{C}{{}  [\because\text{Expanding binomially and neglecting the higher order terms}}] \notag \\
& \simeq \dfrac{1}{u_0} + M - 2aMu_0 \notag\\
\Rightarrow & b = \dfrac{1}{u_0} + M - 2aMu_0 + \bigO(M^2, a^2) \numberthis
\label{e83}
\end{align}  
\par Let us now express $\dfrac{1}{\sqrt{F(u)}}$ in a form more conducive to performing the integration. Starting with \ref{e81} and substituting the value of $\dfrac{1}{b^2}$ we get,
\begin{align}
F(u) & = \dfrac{1}{b^2} - u^2 + 2Mu^3 - \dfrac{4aM}{b^3}u + \bigO (a^2u^4)\notag\\
     & = u_0^2 - 2Mu_0^3 + \dfrac{4aM}{b^3}u_0 - u^2 + 2Mu^3 - \dfrac{4aM}{b^3}u \notag\\
     & = \left(u_0^2 - u^2  \right)\left[1 - \left\lbrace 2M\left(\dfrac{u_0^3 - u^3}{u_0^2 - u^2}\right) - \dfrac{4aM}{b^3}\left(\dfrac{u_0 - u}{u_0^2 - u^2}\right) \right\rbrace  \right]\notag\\
     \Rightarrow \dfrac{1}{\sqrt{F(u)}} & = \dfrac{1}{\sqrt{u_0^2 - u^2}}\left[1 - \left\lbrace  2M\left( \dfrac{u_0^3 - u^3}{u_0^2 - u^2}\right) - \dfrac{4aM}{b^3}\left( \dfrac{u_0 - u}{u_0^2 - u^2}\right) \right\rbrace \right]^{-\frac{1}{2}}\notag\\
     & = \dfrac{1}{\sqrt{u_0^2 - u^2}} + M\dfrac{\left( u_0^3 - u^3 \right)}{\left( u_0^2 - u^2 \right)^{\frac{3}{2}}} - 2aM\dfrac{\left( u_0 - u \right)}{\left( u_0^2 - u^2 \right)^{\frac{3}{2}}}\times \dfrac{1}{\dfrac{1}{u_0^3}\left(1 + Mu_0 - 2aMu_0^2  \right)^3}\notag\\
     &\qquad\qquad\qquad \meqbox{C}{{}  [\because\text{Substituting}\;b=\dfrac{1}{u_0} + M - 2aMu_0 }] \notag \\
     & \simeq \dfrac{1}{\sqrt{u_0^2 - u^2}} + M\dfrac{\left( u_0^3 - u^3 \right)}{\left( u_0^2 - u^2 \right)^{\frac{3}{2}}} - 2aM\dfrac{u_0^3\left( u_0 - u \right)}{\left( u_0^2 - u^2 \right)^{\frac{3}{2}}}\label{e84}
\end{align}  
All terms $\geq M^2$ have been neglected while deriving \ref{e84}.
\par We can now substitute this in \ref{e71} to derive the final expression for $\phi_{\rm VS}$ in the case of Kerr metric
\begin{align}
\phi_{\rm VS} & = \int\limits_{u_{\rm S}}^{u_0} \dfrac{1}{\sqrt{F(u)}} {\rm d}u + \int\limits_{u_{\rm V}}^{u_0} \dfrac{1}{\sqrt{F(u)}} {\rm d}u \notag\\
	& = \int\limits_{u_{\rm S}}^{u_0}\dfrac{1}{\sqrt{u_0^2 - u^2}}{\rm d}u + \int\limits_{u_{\rm S}}^{u_0}M\dfrac{\left( u_0^3 - u^3 \right)}{\left( u_0^2 - u^2 \right)^{\frac{3}{2}}}{\rm d}u - \int\limits_{u_{\rm S}}^{u_0}2aM\dfrac{u_0^3\left( u_0 - u \right)}{\left( u_0^2 - u^2 \right)^{\frac{3}{2}}}{\rm d}u\notag\\
	 &\qquad\qquad\qquad \meqbox{C}{ + \int\limits_{u_{\rm V}}^{u_0}\dfrac{1}{\sqrt{u_0^2 - u^2}}{\rm d}u + \int\limits_{u_{\rm V}}^{u_0}M\dfrac{\left( u_0^3 - u^3 \right)}{\left( u_0^2 - u^2 \right)^{\frac{3}{2}}}{\rm d}u - \int\limits_{u_{\rm V}}^{u_0}2aM\dfrac{u_0^3\left( u_0 - u \right)}{\left( u_0^2 - u^2 \right)^{\frac{3}{2}}}{\rm d}u } 
\label{e85}
\end{align}
\par Substituting the values of the integrals back in \ref{e85} to evaluate $\phi_{\rm VS}$ 
\begin{align}
\phi_{\rm VS} & = \int\limits_{u_{\rm S}}^{u_0}\dfrac{1}{\sqrt{u_0^2 - u^2}}{\rm d}u + \int\limits_{u_{\rm S}}^{u_0}M\dfrac{\left( u_0^3 - u^3 \right)}{\left( u_0^2 - u^2 \right)^{\frac{3}{2}}}{\rm d}u - \int\limits_{u_{\rm S}}^{u_0}2aM\dfrac{u_0^3\left( u_0 - u \right)}{\left( u_0^2 - u^2 \right)^{\frac{3}{2}}}{\rm d}u\notag\\
	 &\qquad\qquad\qquad \meqbox{C}{ + \int\limits_{u_{\rm V}}^{u_0}\dfrac{1}{\sqrt{u_0^2 - u^2}}{\rm d}u + \int\limits_{u_{\rm V}}^{u_0}M\dfrac{\left( u_0^3 - u^3 \right)}{\left( u_0^2 - u^2 \right)^{\frac{3}{2}}}{\rm d}u - \int\limits_{u_{\rm V}}^{u_0}2aM\dfrac{u_0^3\left( u_0 - u \right)}{\left( u_0^2 - u^2 \right)^{\frac{3}{2}}}{\rm d}u } \notag\\
	 &= \left[\dfrac{\pi}{2} - \sin^{-1}\left(\frac{u_{\rm S}}{u_0}\right) + M \dfrac{\left(2u_0 + u_{\rm S}  \right)\left(u_0 - u_{\rm S}  \right)}{\sqrt{u_0^2 - u_{\rm S}^2} } - 2aMu_0^2\dfrac{u_0 - u_{\rm S}}{\sqrt{u_0^2 - u_{\rm S}^2}} \right]\notag\\
	 &\qquad\qquad \meqbox{b}{ + \left[\dfrac{\pi}{2} - \sin^{-1}\left(\frac{u_{\rm V}}{u_0}\right) + M \dfrac{\left(2u_0 + u_{\rm V}  \right)\left(u_0 - u_{\rm V}  \right)}{\sqrt{u_0^2 - u_{\rm V}^2} } - 2aMu_0^2\dfrac{u_0 - u_{\rm V}}{\sqrt{u_0^2 - u_{\rm V}^2}} \right] }\notag\\
	 &= \left[\dfrac{\pi}{2} - \sin^{-1}\left(\dfrac{u_{\rm S}}{u_0}\right) + \dfrac{Mu_0^2\left(2 - \dfrac{u_{\rm S}}{u_0} - \dfrac{u_{\rm S}^2}{u_0^2}  \right)}{u_0\sqrt{1 - \dfrac{u_{\rm S}^2}{u_0^2}}}  - \dfrac{2aMu_0^2\times u_0\left(1 - \dfrac{u_{\rm S}}{u_0} \right)}{u_0\sqrt{1 - \dfrac{u_{\rm S}^2}{u_0^2}}} \right]\notag\\
	 &\qquad\qquad \meqbox{v}{ + \left[\dfrac{\pi}{2} - \sin^{-1}\left(\dfrac{u_{\rm V}}{u_0}\right) + \dfrac{Mu_0^2\left(2 - \dfrac{u_{\rm V}}{u_0} - \dfrac{u_{\rm V}^2}{u_0^2}  \right)}{u_0\sqrt{1 - \dfrac{u_{\rm V}^2}{u_0^2}}}  - \dfrac{2aMu_0^2\times u_0\left(1 - \dfrac{u_{\rm V}}{u_0} \right)}{u_0\sqrt{1 - \dfrac{u_{\rm V}^2}{u_0^2}}} \right]}	 
\label{e94}
\end{align}
In the terms involving $\sin^{-1}$ we make the substitution $\dfrac{1}{u_0} = b - M + 2aMu_0$ from \ref{e83} and expand it in a Taylor series. As for the remaining terms we make the same substitution and neglect terms $\geq M^2$,
\begin{align}
\phi_{\rm VS} & = \left[\dfrac{\pi}{2} - \sin^{-1}\left\lbrace u_{\rm S} \left(b - M + 2aMu_0  \right) \right\rbrace + \dfrac{Mu_0^2\left(2 - \dfrac{u_{\rm S}}{u_0} - \dfrac{u_{\rm S}^2}{u_0^2}  \right)}{u_0\sqrt{1 - \dfrac{u_{\rm S}^2}{u_0^2}}}  - \dfrac{2aMu_0^2\times u_0\left(1 - \dfrac{u_{\rm S}}{u_0} \right)}{u_0\sqrt{1 - \dfrac{u_{\rm S}^2}{u_0^2}}} \right]\notag\\
	 & \meqbox{Z}{ + \left[\dfrac{\pi}{2} - \sin^{-1}\left\lbrace u_{\rm V} \left(b - M + 2aMu_0  \right) \right\rbrace + \dfrac{Mu_0^2\left(2 - \dfrac{u_{\rm V}}{u_0} - \dfrac{u_{\rm V}^2}{u_0^2}  \right)}{u_0\sqrt{1 - \dfrac{u_{\rm V}^2}{u_0^2}}}  - \dfrac{2aMu_0^2\times u_0\left(1 - \dfrac{u_{\rm V}}{u_0} \right)}{u_0\sqrt{1 - \dfrac{u_{\rm V}^2}{u_0^2}}} \right]}\notag\\
	 &= \dfrac{\pi}{2} - \sin^{-1}\left(bu_{\rm S} \right) + \dfrac{Mu_{\rm S}}{\sqrt{1 - b^2u_{\rm S}^2}} - \dfrac{2aM}{b}\dfrac{u_{\rm S}}{\sqrt{1 - b^2u_{\rm S}^2}} - \dfrac{Mu_{\rm S}}{\sqrt{1 - b^2u_{\rm S}^2}} + \dfrac{M}{b}\dfrac{\left(2 - b^2u_{\rm S}^2 \right)}{\sqrt{1 - b^2u_{\rm S}^2}} \notag\\
	 & \meqbox{Z}{  - \dfrac{2aM}{b^2}\dfrac{1}{\sqrt{1 - b^2u_{\rm S}^2}} + \dfrac{2aM}{b}\dfrac{u_{\rm S}}{\sqrt{1 - b^2u_{\rm S}^2}}} \notag\\
	  & + \dfrac{\pi}{2} - \sin^{-1}\left(bu_{\rm V} \right) + \dfrac{Mu_{\rm V}}{\sqrt{1 - b^2u_{\rm V}^2}} - \dfrac{2aM}{b}\dfrac{u_{\rm V}}{\sqrt{1 - b^2u_{\rm V}^2}} - \dfrac{Mu_{\rm V}}{\sqrt{1 - b^2u_{\rm V}^2}} + \dfrac{M}{b}\dfrac{\left(2 - b^2u_{\rm V}^2 \right)}{\sqrt{1 - b^2u_{\rm V}^2}} \notag\\
	 & \meqbox{Z}{  - \dfrac{2aM}{b^2}\dfrac{1}{\sqrt{1 - b^2u_{\rm V}^2}} + \dfrac{2aM}{b}\dfrac{u_{\rm V}}{\sqrt{1 - b^2u_{\rm V}^2}}} \notag\\
	 &= \pi - \sin^{-1}\left(bu_{\rm S} \right) - \sin^{-1}\left(bu_{\rm V} \right) + \dfrac{M}{b}\left[\dfrac{2 - b^2u_{\rm S}^2}{\sqrt{1 - b^2u_{\rm S}^2}} + \dfrac{2 - b^2u_{\rm V}^2}{\sqrt{1 - b^2u_{\rm V}^2}}\right]\notag\\
	 & \meqbox{Z}{ - \dfrac{2aM}{b^2}\left[\dfrac{1}{\sqrt{1 - b^2u_{\rm S}^2}} + \dfrac{1}{\sqrt{1 - b^2u_{\rm V}^2}} \right]}
\label{e95}
\end{align}
\subsubsection{Bending angle of a photon trajectory in Kerr metric from OIA formalism}{\label{s5.3.3}}
\par Finally we can now substitute the values of $\psi_{\rm V}$ (from \ref{e78}), $\psi_{\rm S}$ (from \ref{e79}) \& $\phi_{\rm VS}$ (from \ref{e95}) into \ref{e30b} to find the expression for deflection angle of a photon trajectory in Kerr spacetime
\begin{align}
\alpha_{\rm D_{Kerr}}^{\rm pro} &= \psi_{\rm V} - \psi_{\rm S} + \phi_{\rm VS}\notag\\
	&= \sin^{-1}\left(bu_{\rm V} \right) - \dfrac{bMu_{\rm V}^2}{\sqrt{1 - b^2u_{\rm V}^2}} + \dfrac{2aMu_{\rm V}^2}{\sqrt{1 - b^2u_{\rm V}^2}} -\pi + \sin^{-1}\left(bu_{\rm S} \right) - \dfrac{bMu_{\rm S}^2}{\sqrt{1 - b^2u_{\rm S}^2}} + \dfrac{2aMu_{\rm S}^2}{\sqrt{1 - b^2u_{\rm S}^2}} \notag\\
	&\qquad \meqbox{n}{+\pi - \sin^{-1}\left(bu_{\rm S} \right) - \sin^{-1}\left(bu_{\rm V} \right) + \dfrac{2M}{b\sqrt{1 - b^2u_{\rm S}^2}} - \dfrac{bMu_{\rm S}^2}{\sqrt{1 - b^2u_{\rm S}^2}} + \dfrac{2M}{b\sqrt{1 - b^2u_{\rm V}^2}} - \dfrac{bMu_{\rm V}^2}{\sqrt{1 - b^2u_{\rm V}^2}} }\notag\\
	&\qquad \meqbox{g}{-\dfrac{2aM}{b^2}\dfrac{1}{\sqrt{1 - b^2u_{\rm S}^2}} -\dfrac{2aM}{b^2}\dfrac{1}{\sqrt{1 - b^2u_{\rm V}^2}} }\notag\\
	&= -\dfrac{2bMu_{\rm S}^2}{\sqrt{1 - b^2u_{\rm S}^2}} -\dfrac{2bMu_{\rm V}^2}{\sqrt{1 - b^2u_{\rm V}^2}} + \dfrac{2M}{b}\dfrac{1}{\sqrt{1 - b^2u_{\rm S}^2}} + \dfrac{2M}{b}\dfrac{1}{\sqrt{1 - b^2u_{\rm V}^2}}\notag\\
	&\qquad\qquad \meqbox{u}{+ \dfrac{2aMu_{\rm S}^2}{\sqrt{1 - b^2u_{\rm S}^2}} + \dfrac{2aMu_{\rm V}^2}{\sqrt{1 - b^2u_{\rm V}^2}} - \dfrac{2aM}{b^2}\dfrac{1}{\sqrt{1 - b^2u_{\rm S}^2}} - \dfrac{2aM}{b^2}\dfrac{1}{\sqrt{1 - b^2u_{\rm V}^2}} }\notag\\
	&= \dfrac{2M}{b}\dfrac{1}{\sqrt{1 - b^2u_{\rm S}^2}}\left( 1 - b^2u_{\rm S}^2 \right) + \dfrac{2M}{b}\dfrac{1}{\sqrt{1 - b^2u_{\rm V}^2}}\left( 1 - b^2u_{\rm V}^2 \right)\notag\\
	&\qquad\qquad \meqbox{u}{ - \dfrac{2aM}{b^2\sqrt{1 - b^2u_{\rm S}^2}}\left(1 - b^2u_{\rm S}^2 \right) - \dfrac{2aM}{b^2\sqrt{1 - b^2u_{\rm V}^2}}\left(1 - b^2u_{\rm V}^2 \right)}\notag\\
	&= \dfrac{2M}{b}\left(\sqrt{1 - b^2u_{\rm S}^2} + \sqrt{1 - b^2u_{\rm V}^2} \right) - \dfrac{2aM}{b^2}\left(\sqrt{1 - b^2u_{\rm S}^2} + \sqrt{1 - b^2u_{\rm V}^2} \right)\notag\\
	\Rightarrow \alpha_{\rm D_{Kerr}}^{\rm pro} &= \left(\sqrt{1 - b^2u_{\rm S}^2} + \sqrt{1 - b^2u_{\rm V}^2} \right)\left(\dfrac{2M}{b} - \dfrac{2aM}{b^2}  \right) + \bigO\left(\dfrac{M^2}{b^2} \right)
\label{e96}
\end{align}  
\ref{e96} is actually the expression for the prograde case of the photon orbit in Kerr spacetime. For the retrograde case we simply invert the sign of the term linear in $a$ giving us the orbit expression as
\begin{equation}
 \alpha_{\rm D_{Kerr}}^{\rm retro} = \left(\sqrt{1 - b^2u_{\rm S}^2} + \sqrt{1 - b^2u_{\rm V}^2} \right)\left(\dfrac{2M}{b} + \dfrac{2aM}{b^2}  \right) + \bigO\left(\dfrac{M^2}{b^2} \right)
\label{e97}
\end{equation}
Note that for the asymptotic case $r_{\rm S}\; \text{\&}\; r_{\rm V}\rightarrow\infty$ or $u_{\rm S}\; \text{\&}\; u_{\rm V}\rightarrow 0$ the deflection angle for a photon trajectory reduces to the form already present in the existing literature \cite{Chandrasekhar}, \cite{Epstein} \& \cite{Ibanez}.
\begin{subequations}{\label{e98}}
\begin{align}
\alpha_{\rm D_{Kerr}}^{\rm pro} \rightarrow \dfrac{4M}{b} - \dfrac{4aM}{b^2} + \bigO\left(\dfrac{M^2}{b^2} \right)\label{e98a}\\
\alpha_{\rm D_{Kerr}}^{\rm retro} \rightarrow \dfrac{4M}{b} + \dfrac{4aM}{b^2} + \bigO\left(\dfrac{M^2}{b^2} \right)
\label{e98b}
\end{align}
\end{subequations} 
These same results are obtained by working out the expression presented in \ref{e30a}. This is basically an extension of the formalism initially developed in \cite{Gibbons_2008} and improved on by \cite{Ono_2017}. \ref{s6} presents the detailed exposition of the technique while illustrating the equivalence of the two separate formalism.  
\section{Generalized Expression of Bending Angle in Axisymmetric Spacetime by GW-OIA Formalism}{\label{s6}} 

\par As mentioned earlier in \ref{s4.3} the bending angle of a photon trajectory can be evaluated in terms of the surface integral of the Gaussian curvature $K$ and line integral of the geodesic curvature $\kappa_g$. Although we have discussed some critical geometric aspects related to photon trajectory \& Fermat metric in \ref{s4.2} \& \ref{s5.1} any significant discussion on Gaussian and geodesic curvature has not been presented so far. However, in order to fully appreciate the Gibbons-Werner---Ono-Ishihara-Asada formalism it is essential to have a concrete idea of these quantities.

\subsection{Gaussian curvature \& Geodesic curvature}{\label{s6.1}}
The Gaussian curvature and the geodesic curvature are perhaps the most critical parameters associated with any optical geometry responsible fr describing the photon trajectory in any spacetime. A very concise but succinct definition of these two quantities may be given as :--
\begin{itemize}
\item \textbf{Gaussian curvature}-- It is the intrinsic curvature of a two dimensional surface. This gives us a measure of the departure of the surface from being intrinsically flat.
\item \textbf{Geodesic curvature}-- It gives us a measure of the departure of any one-dimensional curve on a two dimensional curved surface from being a geodesic.    
\end{itemize}
Therefore, it automatically follows that the Gaussian curvature of an intrinsically flat surface must be zero and the geodesic curvature of any geodesic curve must be zero. It is worth noting that both the Gaussian curvature as well as the geodesic curvature strictly relies on the intrinsic metric of the two dimensional surface, \emph{i.e.,} there is no bearing on the embedding of this surface in higher dimensional spacetime.   
\par {\color{black} For a regular curve $\gamma$ lying on a surface embedded in a three-dimensional Riemannian space, the ambient curvature vector splits into a component tangent to the surface and a component normal to it.  Their signed magnitudes are the geodesic curvature $\kappa_g$ and normal curvature $\kappa_n$, respectively.  The latter depends both on the embedding and on the tangent direction of the chosen curve.  The \emph{total curvature of the embedded curve}, not of the surface, satisfies $\kappa^2=\kappa_g^2+\kappa_n^2$.}  

\par
\begin{figure}[hbt!]
\begin{center}
\includegraphics[scale=0.2]{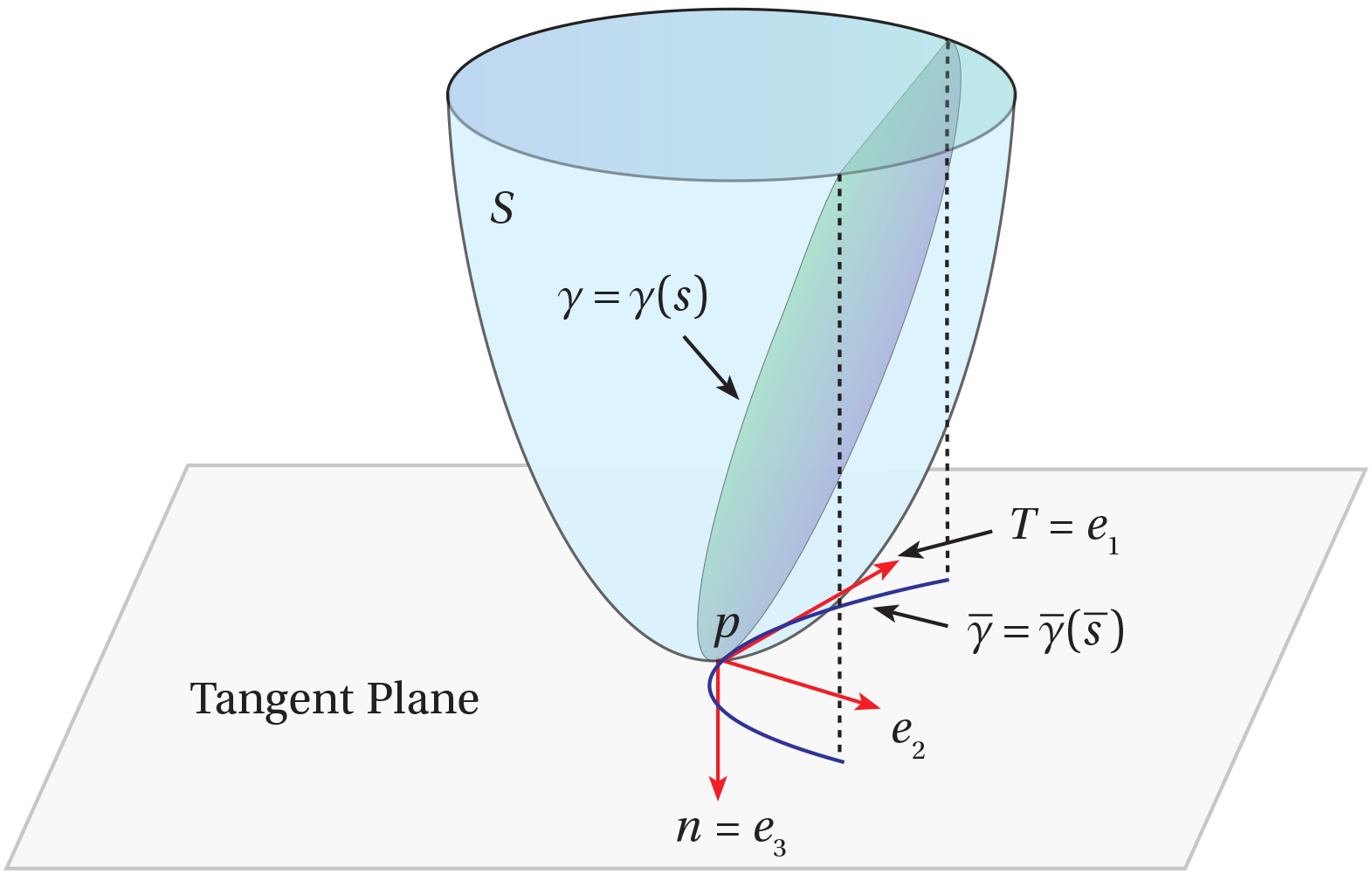}
\end{center}
\caption{\textbf{Embedding of a three dimensional space with associated curves and surfaces relevant to the geometry}}
\label{fig5}
\end{figure}

 The embedding of a surface $S$ in three dimensional Euclidean space is depicted in \ref{fig5}. The basic features and characteristics of these quantities are tabulated in \ref{t1}
	
\pagebreak
\begin{table}[hbt!]
\centering
\caption{\textbf{Characteristics of different types of curvatures associated with a curve on a 2D surface}}
\label{t1}
\small
\begin{tabular}{|@{}P{.1\linewidth}|c|l|l|l|}
\hline
  & \bf Symbol & \bf \makecell{Definition} & \bf \makecell{Dependency} & \makecell{\bf Null curvature criteria} \\
\hline  
\bf \qline{Gaussian curvature}  & $K$ & \Mline{It is the intrinsic curvature of a two dimensional surface.} & \Nline{Independent of the embedding of surface $S$ into higher dimensional space. Intrinsic property.} & \mline{{\color{black}$K=0$ on a neighbourhood implies local intrinsic flatness there; global flatness also requires global information.}} \\
\hline  
\bf \qline{Geodesic curvature}  & $\kappa_g$ & \Mline{The curvature of a curve $\gamma$ in a two dimensional surface $S$.} & \Nline{Independent of the embedding of surface $S$ into higher dimensional space. Intrinsic property.} & \mline{$\kappa_g = 0$ implies that the curve $\gamma$ on the surface $S$ is a geodesic.} \\
\hline  
\bf \qline{Normal curvature}   & $\kappa_n$ & \Mline{{\color{black}Normal component of the ambient curvature of $\gamma$, fixed by the second fundamental form in the tangent direction.}} & \Nline{Dependent on the embedding of surface $S$ into higher dimensional space and on the tangent direction. Extrinsic property.} & \mline{{\color{black}$\kappa_n=0$ identifies an asymptotic direction, when one exists; it does not imply that the whole surface is flat.}} \\
\hline  
\bf \qline{Total curvature}  & $\kappa $ & \Mline{Curvature of a curve $\gamma$ when embedded in a higher dimensional space, with $\kappa^2=\kappa_g^2+\kappa_n^2$.} & \Nline{Dependent on the ambient geometry and embedding.} &  \mline{{\color{black}$\kappa=0$ means that $\gamma$ is locally straight in the ambient space; a surface geodesic need only have $\kappa_g=0$ and may have $\kappa_n\neq0$.}} \\
\hline 
\end{tabular}
\end{table}
	
\subsection{Solving the Orbit Equation to Evaluate $r_{\rm VS}$}{\label{s6.2}}
In order to solve the surface integral of the Gaussian curvature we first need to solve the orbit equation derived in \ref{s5.2.2} as \ref{e59}. We simply have to substitute the appropriate values of $A$, $B$, $D$ \& $H$ to find the solution of the orbit equation.

\begin{align}
\left(\dfrac{{\rm d}r}{{\rm d}\phi}  \right)^2 & = \dfrac{AD + H^2}{B} \times \dfrac{D - 2Hb - Ab^2}{\left(H + Ab   \right)^2}\notag\\
& = \dfrac{\left(1 - \dfrac{2Mr}{r^2}  \right)  \left(r^2 + a^2 + \dfrac{2a^2Mr}{r^2}   \right) + \left(\dfrac{2aMr}{r^2}  \right)^2}{\dfrac{r^2}{r^2 - 2Mr + a^2}}\notag\\
&\qquad\qquad\qquad\qquad\qquad \meqbox{h}\times \dfrac{\left( r^2 + a^2 + \dfrac{2a^2Mr}{r^2} \right) - \dfrac{4aMrb}{r^2} - \left(1 - \dfrac{2Mr}{r^2}  \right)b^2}{\left[\dfrac{2aMr}{r^2} + \left(1 - \dfrac{2Mr}{r^2}   \right)b   \right]^2}\notag\\
& = \dfrac{\left(r^2 + a^2 - 2Mr\right)^2}{r^2} \times \dfrac{r^2 + a^2 + \dfrac{2a^2M}{r} - \dfrac{4abM}{r} - b^2 + \dfrac{2Mb^2}{r}}{\dfrac{b^4}{r^2}\left(\dfrac{2aM}{b^2} + \dfrac{r}{b} - \dfrac{2M}{b}  \right)^2}\notag\\
& = \dfrac{\left(r^2 + a^2 - 2Mr\right)^2 \times \left( r^2 + a^2 + \dfrac{2a^2M}{r} - \dfrac{4abM}{r} - b^2 + \dfrac{2Mb^2}{r}\right)}{b^4\left(\dfrac{2aM}{b^2} + \dfrac{r}{b} - \dfrac{2M}{b}  \right)^2}\notag\\
& = \dfrac{b^4 \left(\dfrac{r^2}{b^2} + \dfrac{a^2}{b^2} - \dfrac{2Mr}{b^2}  \right)^2 \times \dfrac{b^3}{r}\left(\dfrac{r^3}{b^3} + \dfrac{a^2r}{b^3} + \dfrac{2a^2M}{b^3} - \dfrac{4aM}{b^2} - \dfrac{r}{b} + \dfrac{2M}{b}   \right)}{b^4 \left(\dfrac{2aM}{b^2} + \dfrac{r}{b} - \dfrac{2M}{b}   \right)^2}\notag\\
& = \dfrac{b^2\left[\dfrac{r}{b}\left(\dfrac{r}{b} - \dfrac{2M}{b}     \right)   \right]^2 \left[- \dfrac{4aM}{b^2} + \dfrac{2M}{b} - \dfrac{r}{b} + \dfrac{r^3}{b^3}  \right] }{\dfrac{r}{b}\left(\dfrac{2aM}{b^2} + \dfrac{r}{b} - \dfrac{2M}{b}   \right)^2}
\label{e99}
\end{align}      
	Substituting $\dfrac{r}{b} = x$ in \ref{e99} we get,
	
	\begin{align}
\left(\dfrac{{\rm d}r}{{\rm d}\phi}  \right)^2	& =\dfrac{b^2\left[x\left(x - \dfrac{2M}{b}  \right)   \right]^2 \left[-\dfrac{4aM}{b^2} + \dfrac{2M}{b} - x + x^3  \right]}{x\left[x + \dfrac{2aM}{b^2} - \dfrac{2M}{b}  \right]^2}\notag\\
	& = b^2\left(x - \dfrac{4M}{b}  \right) \left(- \dfrac{4aM}{b^2} + \dfrac{2M}{b} -x + x^3 \right)\left(1 + \dfrac{4M}{bx} - \dfrac{4aM}{b^2x} \right)\notag\\
	& = b^2\left(- \dfrac{4aM}{b^2} + \dfrac{2M}{b} -x + x^3  \right)\left(x + \dfrac{4M}{b} - \dfrac{4aM}{b^2} - \dfrac{4M}{b}  \right)\notag\\
	& \simeq b^2\left(-\dfrac{4aMx}{b^2} + \dfrac{2Mx}{b} - x^2 + \dfrac{4aMx}{b^2} + x^4 - \dfrac{4aMx^3}{b^2} \right)\notag\\
	& = b^2 \dfrac{r^4}{b^4} - b^2 \dfrac{r^2}{b^2} + b^2\dfrac{2M}{b}\dfrac{r}{b} - b^2 \dfrac{4aM}{b^2}\dfrac{r^3}{b^3} \qquad\qquad\qquad\qquad {\rm [Substituting} \;x=\dfrac{r}{b}{\rm ]} \notag\\
	& = \dfrac{r^4}{b^2} - r^2 + 2Mr - \dfrac{4r^3aM}{b^3} + \bigO\left(a^2 \right)\label{e100}
	\end{align}
	
	We have neglected any terms involving $M^2$. In terms of the reciprocal coordinates we can write \ref{e100} as
	\begin{align}
	\left(\dfrac{{\rm d}u}{{\rm d}\phi}  \right)^2 = F(u) & = u^4\left(\dfrac{{\rm d}r}{{\rm d}\phi}  \right)^2\notag\\
	& = u^4 \left[\dfrac{r^4}{b^2} - r^2 + 2Mr - \dfrac{4r^3aM}{b^3}   \right] + \bigO\left(a^2u^4  \right)\notag\\
	& = \dfrac{1}{b^2} - u^2 + 2Mu^3 - \dfrac{4aMu}{b^3} + \bigO\left(a^2u^4  \right)\label{e101}
	\end{align}
 \par The solution of the \ref{e101} can be solved iteratively by taking \ref{e101} and truncating it. We proceed by finding the zeroth order solution first followed by linear order solution so on and so forth.
 \par The zeroth order solution is found by solving the equation,
 
 \begin{align}
 &\left(\dfrac{{\rm d}u}{{\rm d}\phi} \right)^2  = \dfrac{1}{b^2} - u^2 + \bigO\left(Mu^3,\;aMu^4,\; a^2u^4  \right)\notag\\
 &\Rightarrow {\rm d}u = \sqrt{\dfrac{1}{b^2} - u^2} {\rm d}\phi \notag\\
 &\Rightarrow \dfrac{b}{\sqrt{1 - b^2u^2}} {\rm d}u = {\rm d}\phi \notag\\
 &\Rightarrow \sin^{-1}\left(bu \right) = \phi + C \qquad\qquad\qquad\qquad {\rm [Integrating\; both\; sides] }\label{e102}
\end{align} 	
	 Since the distance of closest approach in the present case is given by $r = r_0 = \dfrac{1}{u_0}, \; \phi = \dfrac{\pi}{2}$ we can employ the boundary condition $\left(\dfrac{{\rm d}u}{{\rm d}\phi}\right)_{\phi = \frac{\pi}{2}} = 0$. This in turn yields the value of the constant of integration as 0. Hence, the zeroth order solution is 
	 
	 \begin{align}
	 u_0= \dfrac{\sin \phi}{b} \label{e103}
	 \end{align}
	 
Similarly, we can assume that the linear order solution with $M$ is in the form

\begin{align}
u &= u_0(\phi) + u_1(\phi)M \notag\\
\Rightarrow u &= \dfrac{\sin \phi}{b} + u_1(\phi)M \label{e104}
\end{align}

 In order to find $u_1(\phi)$ we simply need to substitute \ref{e104} in \ref{e101} with the terms bearing a linear dependency on $M$. This leads upto the desired solution as follows,
 
 \begin{align}
&  \left(\dfrac{{\rm d}u}{{\rm d}\phi} \right)^2  = \dfrac{1}{b^2} - u^2 + 2Mu^3 + \bigO\left(aMu^4, a^2u^4 \right)\notag\\
&  \Rightarrow\left[\dfrac{{\rm d}}{{\rm d}\phi} \left(\dfrac{\sin \phi}{b} + u_1(\phi)M  \right)\right]^2 = \dfrac{1}{b^2} - \dfrac{\sin^2\phi}{b^2} + u_1^2(\phi)M^2 + 2 \dfrac{\sin \phi}{b}u_1(\phi)M \notag\\
 & \qquad\qquad\qquad\qquad\qquad\qquad\qquad\qquad \meqbox{h} + 2M\left[\dfrac{\sin^3\phi}{b^3} + u_1^3(\phi)M^3 + \dfrac{3M}{b}\sin\phi u_1(\phi)\left(\dfrac{\sin \phi}{b} + u_1(\phi)M \right)  \right]\notag\\
& \Rightarrow \left[\dfrac{\cos\phi}{b} + u_1^\prime(\phi)M\right]^2  \simeq \dfrac{1}{b^2} - \dfrac{\sin^2\phi}{b^2} - \dfrac{2M}{b}\sin\phi u_1(\phi) + \dfrac{2M}{b^3}\sin^3\phi\notag\\
& \Rightarrow \dfrac{\cos^2\phi}{b^2} + \dfrac{2M}{b}\cos\phi u_1^\prime(\phi) = \left( \dfrac{1}{b^2} - \dfrac{\sin^2\phi}{b^2}   \right) - \dfrac{2M}{b}\sin\phi u_1(\phi) + \dfrac{2M}{b^3}\sin^3\phi\notag\\
& \Rightarrow \dfrac{2M}{b}\cos\phi u_1^\prime(\phi) = -\dfrac{2M}{b}\sin\phi\left[u_1(\phi) - \dfrac{\sin^2\phi}{b^2}  \right]\notag\\
&\Rightarrow \cos \phi \dfrac{{\rm d}u_1(\phi)}{{\rm d}\phi} = - \sin\phi\left[u_1(\phi) - \dfrac{\sin^2\phi}{b^2}   \right]\notag\\
& \Rightarrow \dfrac{{\rm d}u_1(\phi)}{{\rm d}\phi} + \tan\phi u_1(\phi) - \dfrac{1}{b^2}\tan\phi \sin^2 \phi = 0 \label{e105}
\end{align}  
We can easily solve the differential equation in \ref{e105} by employing the boundary condition $\dfrac{{\rm d}u_1}{{\rm d}\phi}=0$ at $\phi=\dfrac{\pi}{2}$. The equation yields the linear order solution of $u$ as 

\begin{align}
u_1(\phi) = \dfrac{1}{b^2}\left(1 + \cos^2\phi  \right)\label{e106}
\end{align} 
  	
Therefore, the solution with $a$ can be said to be 
\begin{align}
u=\dfrac{\sin\phi}{b} + \dfrac{M}{b^2}\left(1 + \cos^2\phi  \right) + u_2(\phi)a
\label{e107}
\end{align} 

However, it is evident from \ref{e101} that there are no terms which bear a linear dependency on $a$. As such we can conclude that $u_2(\phi)=0$. Finally, considering the term involving $aM$ we assume the solution to be 

\begin{align}
u= \dfrac{\sin\phi}{b} + \dfrac{M}{b^2}\left(1 + \cos^2\phi  \right) + u_3(\phi)aM
\label{e108}
\end{align}  	
  	 
 Let us now substitute \ref{e108} in \ref{e101} to find $u_3(\phi)$.
 \begin{align}
& \left(\dfrac{{\rm d}u}{{\rm d}\phi}  \right)^2 = \dfrac{1}{b^2} - u^2 + 2Mu^3 - \dfrac{4uaM}{b^3} + \bigO\left(a^2u^4  \right)\notag\\
& \Rightarrow \left[\dfrac{{\rm d}}{{\rm d}\phi}\left\lbrace \dfrac{\sin\phi}{b} + \dfrac{M}{b^2}\left(1 + \cos^2\phi  \right) + u_3(\phi)aM   \right\rbrace  \right]^2 \notag\\ & \qquad\qquad \meqbox{h} = \dfrac{1}{b^2} - \dfrac{\sin^2\phi}{b^2} - M^2\left[\dfrac{1}{b^2}\left(1 + \cos^2\phi \right) + u_3(\phi)a  \right]^2 - \dfrac{2\sin\phi M}{b}\left[\dfrac{1}{b^2}\left(1 + \cos^2\phi \right) + u_3(\phi)a  \right]\notag\\
& \qquad\qquad\qquad \meqbox{h} + 2M\dfrac{\sin^3\phi}{b^3} - \dfrac{4aM}{b^3}\dfrac{\sin\phi}{b}   \notag\\
& \Rightarrow \left[\dfrac{\cos\phi}{b} - \dfrac{2M}{b^2}\sin\phi\cos\phi + aMu_3^\prime (\phi)  \right]^2  \notag\\ & \qquad\qquad \meqbox{h} \simeq  \dfrac{1}{b^2} - \dfrac{\sin^2\phi}{b^2} - \dfrac{2M}{b^3}\left(1 + \cos^2\phi  \right)\sin\phi - \dfrac{2aM}{b}\sin\phi u_3(\phi) + \dfrac{2M}{b^3}\sin^3\phi - \dfrac{4aM}{b^4}\sin\phi \notag\\
& \Rightarrow \dfrac{\cos^2\phi}{b^2} + \dfrac{2aM}{b}\cos\phi u_3^\prime (\phi) - \dfrac{4M}{b^3}\sin\phi\cos^2\phi \notag\\
& \qquad\qquad \meqbox{h} = \dfrac{1}{b^2} - \dfrac{\sin^2\phi}{b^2} - \dfrac{2M}{b^3}\sin\phi - \dfrac{2M}{b^3}\sin\phi\cos^2\phi - \dfrac{2aM}{b}\sin\phi u_3(\phi) + \dfrac{2M}{b^3}\sin^3\phi - \dfrac{4aM}{b^4}\sin\phi \notag\\
& \Rightarrow u_3^{\prime}(\phi) + \tan \phi u_3(\phi) = -\dfrac{2}{b^3}\tan \phi \notag\\
& \Rightarrow \dfrac{{\rm d}}{{\rm d}\phi} \left[u_3\left(\phi \right)\sec \phi   \right] = - \dfrac{2}{b^3} \sec \phi \tan \phi \notag\\
& \Rightarrow u_3\left(\phi \right) \sec \phi = -\dfrac{2}{b^3}\sec \phi + C_3
\label{e109}
\end{align}  

Applying the boundary condition we find the value of the integration constant to be zero. Thus, the solution of $u_3$ is

\begin{align}
u_3 = -\dfrac{2}{b^3}
\label{e110}
\end{align}

The complete solution of $u$ is therefore given as,

\begin{align}
u & = u_0 + Mu_1 + au_2 + aMu_3 \notag\\
& = \dfrac{\sin \phi}{b} + \dfrac{M}{b^2}\left(1 + \cos^2 \phi   \right) - \dfrac{2aM}{b^3} + \bigO \left(\dfrac{M^2}{b^3},\; \dfrac{a^2}{b^3}    \right)
\label{e111}
\end{align} 	 

We can further manipulate \ref{e111} with the help of Taylor expansion to get a solution of $\phi$.

\begin{align}
ub & = \sin \phi + \dfrac{M}{b} \left(1 + 1 - \sin^2 \phi   \right) - \dfrac{2aM}{b^2} \notag\\
& = \sin \phi + \dfrac{2M}{b} - \dfrac{M}{b}\sin^2 \phi - \dfrac{2aM}{b^2} \notag\\
\Rightarrow  \phi & = \arcsin \left(bu   \right) - \dfrac{2M}{b}\dfrac{{\rm d}}{{\rm du}}\left[\arcsin \left(bu   \right)   \right] + \dfrac{M}{b} b^2u^2 \dfrac{{\rm d}}{{\rm d}u}\left[\arcsin \left(bu   \right)   \right] + \dfrac{2aM}{b^2} \dfrac{{\rm d}}{{\rm d}u}\left[\arcsin \left(bu   \right)  \right] \notag\\
& = \arcsin \left(bu  \right) - \dfrac{2M}{b}\dfrac{1}{\sqrt{1-b^2u^2}} + \dfrac{Mb^2u^2}{b\sqrt{1 - b^2u^2}} + \dfrac{2aM}{b^2\sqrt{1 - b^2u^2}} 
\label{e112}
\end{align}   

The domain of $\phi$ can be chosen to lie in the range $-\pi \leq \phi < \pi$ without compromising generality. With consideration to the source location the angular coordinate $\phi_{\rm S}$ will be contained in the domain $-\dfrac{\pi}{2} \leq \phi_{\rm S} < \dfrac{\pi}{2}$	whereas for the viewer location the angular coordinate $\phi_{\rm V}$ will abide by the constraint $|\phi_{\rm V}|>\dfrac{\pi}{2}$. Therefore, within the specific domains the solution of $\phi$ will be,

\begin{align}
\phi &= \arcsin(bu) + \dfrac{-2 + b^2u^2}{b\sqrt{1 - b^2u^2}}M + \dfrac{2aM}{b^2\sqrt{1-b^2u^2}} + \bigO\left(\dfrac{M^2}{b^3},\; \dfrac{a^2}{b^3}  \right) \;\;\;\;\;\;\;\;\;\;\;\; \left[|\phi|<\dfrac{\pi}{2}  \right] \notag\\
&= \pi - \arcsin(bu) - \dfrac{-2+b^2u^2}{b\sqrt{1-b^2u^2}}M - \dfrac{2aM}{b^2\sqrt{1-b^2u^2}} + \bigO\left(\dfrac{M^2}{b^3},\; \dfrac{a^2}{b^3}   \right) \;\;\;\;\;\;\;\;\;\;\;\; \left[\dfrac{\pi}{2}<|\phi|    \right]
\label{e113}
\end{align}  
	\par It is worth noting in this context that $|bu|<1$ since the square roots involved in \ref{e113} cannot be imaginary or equate to zero, and of course both $b$ and $u$ should be positive. As such $bu$ should abide by the condition $0 < bu < 1$. We can now proceed with evaluating the surface integral of the Gaussian curvature which requires the values $\phi_{\rm V}$ and $\phi_{\rm S}$ as the limits of the integral.

\subsection{Surface Integral of the Gaussian Curvature for Kerr metric (in equatorial plane)}{\label{s6.3}}	
	\par For the purpose of the present discussion the mathematical formulation of the Gaussian curvature can be defined in terms of the 2-dimensional Riemann tensor and the Fermat metric as follows,
	
{\color{black}For a two-dimensional metric $\gamma_{ab}$, Gaussian curvature is the scalar
	\begin{equation}
	K=\frac{R_{r\phi r\phi}}{\det(\gamma_{ab})}.
    \label{e114_n}
	\end{equation}
	For the diagonal, axisymmetric line element
	${\rm d}\ell^2=E(r){\rm d}r^2+G(r){\rm d}\phi^2$, this becomes
	\begin{equation*}
	K=-\frac{1}{\sqrt{E G}}\frac{{\rm d}}{{\rm d}r}
	\left[\frac{1}{2\sqrt{E G}}\frac{{\rm d}G}{{\rm d}r}\right].
	\end{equation*}
	Division by the tensor $\gamma_{ij}^{\rm(F)}$ is undefined; the determinant is required.	}
	
	\begin{align}
		K &= \dfrac{1}{\sqrt{\det \gamma_{ij}^{(\rm F)}}} \left[\dfrac{\partial}{\partial \phi} \left(\dfrac{\sqrt{\det \gamma_{ij}^{(\rm F)}}}{\gamma_{rr}^{(\rm F)}}\Gamma^\phi_{rr}   \right) - \dfrac{\partial}{\partial r} \left(\dfrac{\sqrt{\det \gamma_{ij}^{(\rm F)}}}{\gamma_{rr}^{(\rm F)}}\Gamma^\phi_{r\phi}   \right)    \right]
    \label{e114}
	\end{align}	
	\par The terms $R_{r \phi r \phi}$ and $\Gamma_{jk}^{i}$ can be expressed in terms of the Fermat metric $\gamma_{ij}^{(\rm F)}$ as defined on the equatorial plane. The surface element is defined in terms of the Fermat metric as ${\rm d}S = \sqrt{\det \gamma_{ij}^{(\rm F)}}{\rm d}r{\rm d}\phi$. This helps us to express the surface integral of the Gaussian curvature as required by \ref{e30a} in the form,	
	\begin{align}
	\iint\limits_{^\infty_{\rm V}\square_{\rm S}^\infty} K{\rm d}S = \int\limits_{\phi_S}^{\phi_V}\int\limits_{r_{VS}}^{\infty} K\sqrt{\det \gamma_{ij}^{(\rm F)}}{\rm d}r{\rm d}\phi \label{e115}
	\end{align}	 
	where $r_{VS}$ represents the solution of the orbit equation.
	
	Evaluating the Christoffel symbols we find that $\Gamma_{rr}^\phi = 0$ and $\Gamma_{r\phi}^\phi=\dfrac{1}{2}\times \dfrac{A^2}{AD+H^2}\times \dfrac{\partial}{\partial r}\left(\dfrac{AD+H^2}{A^2}  \right)$. Upon substituting the these values in \ref{e114} we find that 
	{\color{black} \begin{equation}
	K =-\sqrt{\dfrac{A^3}{B\left(AD + H^2  \right)}}\,
	\dfrac{\partial}{\partial r}\left[\dfrac{1}{2} \sqrt{\dfrac{A^3}{B\left(AD + H^2  \right)}}\,
	\dfrac{\partial}{\partial r}\left(\dfrac{AD + H^2}{A^2}  \right)  \right].
	\label{e116}
	\end{equation}
	The overall minus sign is required by \ref{e114}; with it, the exact Kerr result in \ref{e117} has the stated weak-field limit $K=-2M/r^3+\cdots$.}
	
	A simple but lengthy bit of algebra is required to give the expression for the Gaussian curvature as 
	\begin{align}
	K = \dfrac{M \left[ -6r\left(a^2 + M^2  \right) + 6a^2M + 7Mr^2 - 2r^3   \right]}{r^5\left(r - 2M\right)} 
	\label{e117}
	\end{align}
	Under the weak-field and slow rotation approximations \ref{e117} can be further simplified to the form 
	\begin{align}
	K = -\dfrac{2M}{r^3} + \bigO\left(\dfrac{M^2}{r^4}, \dfrac{a^2M}{r^5}   \right)\label{e118}
	\end{align}
	
	\par The surface element of the of the Fermat metric ${\rm d}S = \sqrt{\det \gamma_{ij}^{(\rm F)}}{\rm d}r{\rm d}\phi$ can be similarly simplified as follows
	
	\begin{align}
	{\rm d}S & = \sqrt{\det \gamma_{ij}^{(\rm F)}}{\rm d}r{\rm d}\phi \notag\\
	& = \sqrt{\dfrac{B\left(AD + H^2  \right)}{A^3}}{\rm d}r{\rm d}\phi \notag\\
	& = \sqrt{\dfrac{\dfrac{r^2}{r^2-2Mr+a^2}\times \left(1-\dfrac{2M}{r}  \right)\left(r^2+a^2+\dfrac{2a^2M}{r}   \right) + \left(\dfrac{2aM}{r}  \right)^2}{\left(1-\dfrac{2M}{r}  \right)^3}}{\rm d}r{\rm d}\phi\notag\\
	& = r\times\left(1+\dfrac{3}{2}\times\dfrac{2M}{r}  \right){\rm d}r{\rm d}\phi\notag\\
	\Rightarrow {\rm d}S & = \left[r + 3M + \bigO \left(\dfrac{M^2}{r}  \right)\right] {\rm d}r{\rm d}\phi \label{e119}
	\end{align}
	where we have neglected the terms involving $a^2$ \& $M^2$.
		
Now, we can finally proceed with evaluating the surface integral of the Gaussian curvature on the equatorial plane, by substituting the values of $r_{\rm VS}$, $\phi_{\rm V}$ \& $\phi_{\rm S}$ at the appropriate vacancies.
 
	\begin{align}
	\iint\limits_{^\infty_{\rm V}\square_{\rm S}^\infty} K{\rm d}S &= \int\limits_{\phi_{\rm S}}^{\phi_{\rm V}}\int\limits_{r_{\rm VS}}^{\infty} K\sqrt{\det \gamma_{ij}^{(\rm F)}}{\rm d}r{\rm d}\phi \notag\\
	&= \int\limits_{\phi_{\rm S}}^{\phi_{\rm V}}\int\limits_{r_{\rm VS}}^{\infty} -\dfrac{2M}{r^3} \left(r + 3M   \right)  {\rm d}r{\rm d}\phi + \bigO\left(\dfrac{M^2}{b^2},\; \dfrac{aM^2}{b^3}, \; \dfrac{a^2M}{b^3} \right)  \notag\\
	&= -2M \int\limits_{\phi_{\rm S}}^{\phi_{\rm V}} \int\limits_{0}^{\frac{\sin \phi}{b} + \frac{M}{b^2}\left(1 + \cos^2\phi   \right) - \frac{2aM}{b^3}}  {\rm d}u{\rm d}\phi + \bigO\left(\dfrac{M^2}{b^2},\; \dfrac{aM^2}{b^3},\; \dfrac{a^2M}{b^3}   \right) \notag\\
	&= -2M \int\limits_{\phi_{\rm S}}^{\phi_{\rm V}}\dfrac{\sin \phi}{b}{\rm d}\phi + \bigO\left(\dfrac{M^2}{b^2},\; \dfrac{aM^2}{b^3},\; \dfrac{a^2M}{b^3}   \right)\;\;\;\;\;\;\;\;\;\;\;\;\;\;\;\;\; [\because {\rm Neglecting\;\; terms\;\; involving\;\;} M^2 ] \notag\\
	&= -\dfrac{2M}{b}\left[\sqrt{1-b^2u^2_{\rm S}} +  \sqrt{1-b^2u^2_{\rm V}} \right]+ \bigO\left(\dfrac{M^2}{b^2},\; \dfrac{aM^2}{b^3},\; \dfrac{a^2M}{b^3}   \right)
	\label{e120}
	\end{align}
	
\subsection{Evaluating the Path Integral of Geodesic Curvature (in equatorial plane)  }{\label{s6.4}}
In order to establish the final form of the deflection angle we now need to evaluate the path integral of the geodesic curvature. In this regard, the geodesic curvature \cite{Carmo}, \cite{John_Oprea2019-02-06} can be vectorially expressed as

\begin{align}
\kappa_g \equiv \vec{T}^{\prime}\cdot (\vec{T} \times \vec{N}) \label{e121}
\end{align}
where,
\par $\vec{T}$ -- unit tangent vector for the curve after reparameterizing the curve using its arc length.
\par $\vec{T}^{\prime}$ -- derivative of $T$ w.r.t the parameter
\par $\vec{N}$ -- unit normal to the surface under consideration

{\color{black} 
%
%
%

In what follows we abbreviate the Riemannian part of the Fermat metric as
\[
\gamma_{ij}\equiv\gamma^{\rm(F)}_{ij},
\]
and write
\[
\beta=\beta_i\,{\rm d}x^i
\]
for the stationary Randers one-form.  The discussion is local to a spatial
region in which $\gamma_{ij}$ is positive definite and the corresponding
stationary Fermat--Randers structure is regular.  For an ordinary Randers
metric
\[
F(x,y)=\sqrt{\gamma_{ij}(x)y^iy^j}+\beta_i(x)y^i,
\]
the familiar strong-convexity condition is
\[
\gamma^{ij}\beta_i\beta_j<1.
\]
This condition guarantees that the Randers norm is positive and strongly
convex on the non-zero tangent bundle; see, for example,
\cite{Gibbons_2009}.

Let $x^i(\sigma)$ be a regular trial spatial curve joining the spatial
positions of the source and viewer.  We choose $\sigma$ to be an arbitrary
\emph{orientation-preserving} monotone parameter and write
\[
\dot x^i\equiv\frac{{\rm d}x^i}{{\rm d}\sigma}.
\]
The restriction to an orientation-preserving reparameterization is relevant
because a general Randers metric is non-reversible: reversing the tangent
vector need not leave its Finsler norm unchanged.

From Eq.~\ref{e36}, the null condition gives the stationary Fermat relation
\[
{\rm d}t
=
\sqrt{\gamma_{ij}\,{\rm d}x^i{\rm d}x^j}
+\beta_i\,{\rm d}x^i.
\]
Substituting
\[
{\rm d}x^i=\dot x^i\,{\rm d}\sigma
\]
and integrating along the trial curve gives the coordinate travel-time
functional
\begin{equation}
T[x]=\int_{\sigma_{\rm S}}^{\sigma_{\rm V}}
\left[
\sqrt{\gamma_{ij}(x)\dot x^i\dot x^j}
+\beta_i(x)\dot x^i
\right]{\rm d}\sigma .
\label{e122}
\end{equation}
The relativistic Fermat principle states that the spatial projection of a
light ray is a stationary curve of the corresponding arrival-time/Fermat
functional; ``stationary'' is the appropriate variational statement and
does not in general mean that the travel time is a global minimum
\cite{Caponio_2011}.

It is useful at this stage to define
\begin{equation}
\mathcal L(x,\dot x)=v+\beta_i\dot x^i,
\qquad
v\equiv\sqrt{\gamma_{ij}\dot x^i\dot x^j}.
\label{e123}
\end{equation}
The quantity $v$ has a direct geometrical meaning.  If
\[
{\rm d}\ell_{\rm ph}^2
=
\gamma_{ij}\,{\rm d}x^i{\rm d}x^j
\]
denotes Riemannian arclength with respect to $\gamma_{ij}$, then
\[
\left(\frac{{\rm d}\ell_{\rm ph}}{{\rm d}\sigma}\right)^2
=
\gamma_{ij}
\frac{{\rm d}x^i}{{\rm d}\sigma}
\frac{{\rm d}x^j}{{\rm d}\sigma}
=
\gamma_{ij}\dot x^i\dot x^j
=
v^2.
\]
For an orientation-preserving parametrization,
\[
\boxed{
v=\frac{{\rm d}\ell_{\rm ph}}{{\rm d}\sigma}>0 .
}
\]
Thus $v$ is simply the speed of the trial curve measured by the Riemannian
metric $\gamma_{ij}$.

There is also an important variational reason for retaining an arbitrary
parameter at this stage.  For any $\lambda>0$,
\[
\begin{aligned}
\mathcal L(x,\lambda\dot x)
&=
\sqrt{\gamma_{ij}
      (\lambda\dot x^i)(\lambda\dot x^j)}
+\beta_i(\lambda\dot x^i)
\\
&=
\lambda
\sqrt{\gamma_{ij}\dot x^i\dot x^j}
+\lambda\beta_i\dot x^i
\\
&=
\lambda\mathcal L(x,\dot x).
\end{aligned}
\]
Hence $\mathcal L$ is positively homogeneous of degree one in the tangent
vector.  Consequently, under an orientation-preserving change of parameter
$\sigma=\sigma(\tilde\sigma)$,
\[
\frac{{\rm d}\sigma}{{\rm d}\tilde\sigma}>0,
\]
we have
\[
\mathcal L
\left(
x,\frac{{\rm d}x}{{\rm d}\tilde\sigma}
\right)
{\rm d}\tilde\sigma
=
\mathcal L
\left(
x,
\dot x\,\frac{{\rm d}\sigma}{{\rm d}\tilde\sigma}
\right)
{\rm d}\tilde\sigma
=
\mathcal L(x,\dot x)\,{\rm d}\sigma .
\]
The functional $T[x]$ therefore depends on the oriented spatial curve and
not on the particular increasing parameter used to describe it.  This
allows us to carry out the variation with a general $\sigma$ and only
afterwards choose the convenient Riemannian-arclength parametrization.

We now derive the two partial derivatives entering the Euler--Lagrange
equation.  First,
\[
v=
\left(\gamma_{ij}\dot x^i\dot x^j\right)^{1/2}.
\]
When differentiating with respect to $x^k$, the components $\dot x^i$ are
held fixed.  Therefore,
\begin{align*}
\frac{\partial v}{\partial x^k}
&=
\frac{1}{2}
\left(\gamma_{ij}\dot x^i\dot x^j\right)^{-1/2}
\frac{\partial}{\partial x^k}
\left(\gamma_{ij}\dot x^i\dot x^j\right)
\\
&=
\frac{1}{2v}\,
\gamma_{ij,k}\dot x^i\dot x^j,
\end{align*}
where
\[
\gamma_{ij,k}\equiv\partial_k\gamma_{ij}.
\]
Similarly,
\[
\frac{\partial}{\partial x^k}
\left(\beta_i\dot x^i\right)
=
\beta_{i,k}\dot x^i.
\]
Hence
\begin{equation}
\frac{\partial\mathcal L}{\partial x^k}
=
\frac{1}{2v}\gamma_{ij,k}\dot x^i\dot x^j
+\beta_{i,k}\dot x^i .
\label{e124}
\end{equation}

Next consider the velocity derivative.  Since $\gamma_{ij}$ is symmetric,
\begin{align*}
\frac{\partial v}{\partial\dot x^k}
&=
\frac{1}{2v}
\frac{\partial}{\partial\dot x^k}
\left(\gamma_{ij}\dot x^i\dot x^j\right)
\\
&=
\frac{1}{2v}
\gamma_{ij}
\left(
\delta^i{}_k\dot x^j
+\dot x^i\delta^j{}_k
\right)
\\
&=
\frac{1}{2v}
\left(
\gamma_{kj}\dot x^j
+\gamma_{ik}\dot x^i
\right)
\\
&=
\frac{\gamma_{kj}\dot x^j}{v}.
\end{align*}
Moreover,
\[
\frac{\partial}{\partial\dot x^k}
\left(\beta_i\dot x^i\right)
=
\beta_i\delta^i{}_k
=
\beta_k.
\]
Therefore,
\begin{equation}
\frac{\partial\mathcal L}{\partial\dot x^k}
=
\frac{\gamma_{kj}\dot x^j}{v}+\beta_k .
\label{e125}
\end{equation}

The Euler--Lagrange equations are
\[
\frac{{\rm d}}{{\rm d}\sigma}
\left(
\frac{\partial\mathcal L}{\partial\dot x^k}
\right)
-
\frac{\partial\mathcal L}{\partial x^k}
=0.
\]
Substituting Eqs.~\eqref{e124} and \eqref{e125} gives
\[
0=
\frac{{\rm d}}{{\rm d}\sigma}
\left(
\frac{\gamma_{ki}\dot x^i}{v}
+\beta_k
\right)
-
\frac{1}{2v}\gamma_{ij,k}\dot x^i\dot x^j
-
\beta_{i,k}\dot x^i .
\]
We expand each derivative explicitly.  First,
\begin{align*}
\frac{{\rm d}}{{\rm d}\sigma}
\left(
\frac{\gamma_{ki}\dot x^i}{v}
\right)
={}&
\frac{1}{v}
\frac{{\rm d}}{{\rm d}\sigma}
\left(\gamma_{ki}\dot x^i\right)
-
\frac{\gamma_{ki}\dot x^i}{v^2}
\frac{{\rm d}v}{{\rm d}\sigma}
\\
={}&
\frac{\gamma_{ki}}{v}\ddot x^i
+
\frac{\gamma_{ki,j}}{v}\dot x^i\dot x^j
-
\frac{\gamma_{ki}\dot x^i}{v^2}
\frac{{\rm d}v}{{\rm d}\sigma},
\end{align*}
where
\[
\ddot x^i\equiv\frac{{\rm d}^2x^i}{{\rm d}\sigma^2}.
\]
For the one-form term,
\[
\frac{{\rm d}\beta_k}{{\rm d}\sigma}
=
\beta_{k,j}\dot x^j.
\]
The Euler--Lagrange equation before fixing the parameter is consequently
\begin{align*}
0={}&
\frac{\gamma_{ki}}{v}\ddot x^i
+
\frac{\gamma_{ki,j}}{v}\dot x^i\dot x^j
-
\frac{\gamma_{ki}\dot x^i}{v^2}
\frac{{\rm d}v}{{\rm d}\sigma}
\\
&\quad
-\frac{1}{2v}\gamma_{ij,k}\dot x^i\dot x^j
+
\left(\beta_{k,j}-\beta_{j,k}\right)\dot x^j .
\end{align*}
In obtaining the final term we have merely renamed the dummy index in
$\beta_{i,k}\dot x^i$.

Only now do we make the convenient parameter choice
\[
\sigma=\ell_{\rm ph}.
\]
Because
\[
v=\frac{{\rm d}\ell_{\rm ph}}{{\rm d}\sigma},
\]
this choice gives
\[
v=1,
\qquad
\frac{{\rm d}v}{{\rm d}\ell_{\rm ph}}=0.
\]
Define the unit tangent vector of the projected spatial curve by
\[
e^i\equiv
\frac{{\rm d}x^i}{{\rm d}\ell_{\rm ph}}.
\]
Its normalization follows directly:
\[
\gamma_{ij}e^ie^j
=
\gamma_{ij}
\frac{{\rm d}x^i}{{\rm d}\ell_{\rm ph}}
\frac{{\rm d}x^j}{{\rm d}\ell_{\rm ph}}
=1.
\]
Moreover,
\[
\frac{{\rm d}e^i}{{\rm d}\ell_{\rm ph}}
=
\frac{{\rm d}^2x^i}{{\rm d}\ell_{\rm ph}^2}.
\]
The Euler--Lagrange equation therefore reduces to
\begin{align}
0={}&
\gamma_{ki}\frac{{\rm d}e^i}{{\rm d}\ell_{\rm ph}}
+\gamma_{ki,j}e^ie^j
-\frac12\gamma_{ij,k}e^ie^j
+(\beta_{k,j}-\beta_{j,k})e^j .
\label{e126}
\end{align}

It is important to clarify the role of $\ell_{\rm ph}$ here.
It is being used solely as the \emph{Riemannian arclength parameter of the
spatially projected curve with respect to $\gamma_{ij}$}.  No identification
of $\ell_{\rm ph}$ with a four-dimensional affine parameter is required in
this derivation.  The stationary-spacetime Fermat correspondence determines
the projected light ray as a Fermat/Randers geodesic up to an appropriate
reparameterization; choosing a convenient constant Riemannian-speed
parameter is therefore a spatial parametrization choice
\cite{Caponio_2011}.

We next show explicitly how the first three terms of Eq.~\eqref{e126}
combine into a covariant acceleration.  Let
\[
\Gamma^i{}_{mn}
=
\frac12\gamma^{ip}
\left(
\gamma_{pm,n}
+\gamma_{pn,m}
-\gamma_{mn,p}
\right)
\]
be the Levi--Civita connection of $\gamma_{ij}$.  The covariant derivative
of $e^i$ along the curve is
\[
\frac{D e^i}{{\rm d}\ell_{\rm ph}}
\equiv
\frac{{\rm d}e^i}{{\rm d}\ell_{\rm ph}}
+
\Gamma^i{}_{mn}e^me^n .
\]
Lowering the free index with $\gamma_{ki}$ gives
\begin{align*}
\gamma_{ki}\frac{D e^i}{{\rm d}\ell_{\rm ph}}
={}&
\gamma_{ki}\frac{{\rm d}e^i}{{\rm d}\ell_{\rm ph}}
+
\gamma_{ki}\Gamma^i{}_{mn}e^me^n
\\
={}&
\gamma_{ki}\frac{{\rm d}e^i}{{\rm d}\ell_{\rm ph}}
+
\frac12
\left(
\gamma_{km,n}
+\gamma_{kn,m}
-\gamma_{mn,k}
\right)e^me^n .
\end{align*}
Since $e^me^n$ is symmetric under the interchange $m\leftrightarrow n$,
the first two terms inside the parentheses give equal contributions:
\[
\frac12
\left(
\gamma_{km,n}+\gamma_{kn,m}
\right)e^me^n
=
\gamma_{km,n}e^me^n .
\]
Relabelling dummy indices therefore yields
\[
\boxed{
\gamma_{ki}\frac{D e^i}{{\rm d}\ell_{\rm ph}}
=
\gamma_{ki}\frac{{\rm d}e^i}{{\rm d}\ell_{\rm ph}}
+\gamma_{ki,j}e^ie^j
-\frac12\gamma_{ij,k}e^ie^j .
}
\]
Thus the first three terms of Eq.~\eqref{e126} are precisely the
Levi--Civita covariant acceleration associated with the Riemannian metric
$\gamma_{ij}$.  Equation~\eqref{e126} consequently becomes
\begin{equation}
\gamma_{ki}\frac{D e^i}{{\rm d}\ell_{\rm ph}}
=
(\beta_{j,k}-\beta_{k,j})e^j .
\label{e127}
\end{equation}

Multiplying by the inverse metric $\gamma^{mk}$ gives
\[
\frac{D e^m}{{\rm d}\ell_{\rm ph}}
=
\gamma^{mk}
(\beta_{j,k}-\beta_{k,j})e^j .
\]
Relabelling dummy and free indices gives the equivalent form
\begin{equation}
\frac{D e^i}{{\rm d}\ell_{\rm ph}}
=
\gamma^{ij}\left(\beta_{k,j}-\beta_{j,k}\right)e^k
\equiv a^i .
\label{e128}
\end{equation}
This is the standard projected-ray equation used in the stationary
Fermat/Randers description; in particular, it is the form employed in the
Ono--Ishihara--Asada treatment of stationary axisymmetric lensing
\cite{Ono_2017}.

The geometrical content of Eq.~\eqref{e128} becomes especially transparent
if we introduce the two-form
\[
\mathcal B\equiv{\rm d}\beta.
\]
For a one-form $\beta=\beta_i{\rm d}x^i$ its components are
\[
\mathcal B_{jk}
=
({\rm d}\beta)_{jk}
=
\partial_j\beta_k-\partial_k\beta_j
=
\beta_{k,j}-\beta_{j,k}.
\]
Hence
\[
\boxed{
a^i
=
\gamma^{ij}\mathcal B_{jk}e^k.
}
\]
Because the Levi--Civita connection is torsion free,
\[
\Gamma^m{}_{jk}=\Gamma^m{}_{kj},
\]
the same antisymmetric tensor may equally well be written using covariant
derivatives:
\begin{align*}
\nabla_j\beta_k-\nabla_k\beta_j
&=
\left(
\partial_j\beta_k-\Gamma^m{}_{jk}\beta_m
\right)
-
\left(
\partial_k\beta_j-\Gamma^m{}_{kj}\beta_m
\right)
\\
&=
\partial_j\beta_k-\partial_k\beta_j
\\
&=
\mathcal B_{jk}.
\end{align*}
Equation~\eqref{e128} may therefore also be written as
\[
a^i
=
\gamma^{ij}
\left(
\nabla_j\beta_k-\nabla_k\beta_j
\right)e^k,
\]
which makes direct contact with the covariant notation commonly used in
the stationary-lensing literature \cite{Ono_2017}.

There is an immediate and useful consistency check.  Since
$\mathcal B_{jk}$ is antisymmetric,
\[
\mathcal B_{jk}=-\mathcal B_{kj},
\]
whereas $e^je^k$ is symmetric.  Contracting the acceleration with the
unit tangent therefore gives
\begin{align*}
\gamma_{ij}e^i a^j
&=
\gamma_{ij}e^i
\gamma^{jm}\mathcal B_{mk}e^k
\\
&=
e^m\mathcal B_{mk}e^k
\\
&=0.
\end{align*}
Thus
\[
\boxed{
\gamma_{ij}e^i a^j=0 .
}
\]
The effective acceleration is everywhere orthogonal, with respect to
$\gamma_{ij}$, to the unit tangent of the projected curve.  Equivalently,
\[
\frac{{\rm d}}{{\rm d}\ell_{\rm ph}}
\left(\gamma_{ij}e^ie^j\right)
=
2\gamma_{ij}e^i
\frac{D e^j}{{\rm d}\ell_{\rm ph}}
=
2\gamma_{ij}e^i a^j
=
0.
\]
Hence the force term changes the \emph{direction} of the unit tangent
relative to the Riemannian geometry without changing its
$\gamma$-norm.  This is precisely the mathematical structure of a magnetic
flow on a Riemannian manifold: the Randers one-form plays the role of a
magnetic potential and ${\rm d}\beta$ plays the role of the associated
magnetic two-form \cite{Gibbons_2009}.

The word ``force'' in this statement must be interpreted geometrically.
The photon is not being acted upon by an additional non-gravitational
force in the four-dimensional spacetime.  The four-dimensional light ray
is still a null geodesic.  Rather, after projecting that null geodesic onto
the Riemannian spatial geometry described by $\gamma_{ij}$, the projected
curve need not be a Riemannian geodesic of $\gamma_{ij}$.  The departure
from a $\gamma_{ij}$-geodesic is encoded by the antisymmetric two-form
${\rm d}\beta$.  This is the Lorentz-force-type or magnetic-flow
interpretation of stationary Randers geometry discussed by
Gibbons--Herdeiro--Warnick--Werner and, in the gravitational-lensing
setting, by Ono--Ishihara--Asada
\cite{Gibbons_2009,Ono_2017}.

The static or locally curl-free limit provides another useful check.
If
\[
{\rm d}\beta=0
\]
in the region under consideration, then
\[
a^i=0
\]
and Eq.~\eqref{e128} reduces to
\[
\frac{D e^i}{{\rm d}\ell_{\rm ph}}=0.
\]
The projected light ray is then a geodesic of the Riemannian metric
$\gamma_{ij}$.  Conversely, a non-zero ${\rm d}\beta$ produces a
non-vanishing covariant acceleration relative to that Riemannian metric.
For stationary rotating geometries this term is the geometrical origin of
the gravitomagnetic contribution to the projected ray equation discussed
in the OIA formalism \cite{Ono_2017}.

Finally, Eq.~\eqref{e128} shows that the local trajectory depends on
$\beta$ only through its exterior derivative.  To demonstrate this
explicitly, consider the transformation
\[
\beta\longrightarrow\beta'=\beta+{\rm d}\chi,
\qquad
\beta'_i=\beta_i+\partial_i\chi ,
\]
where $\chi(x)$ is a smooth scalar function.  The corresponding Fermat
functional is
\begin{align*}
T'[x]
&=
\int_{\sigma_{\rm S}}^{\sigma_{\rm V}}
\left[
v+
\left(
\beta_i+\partial_i\chi
\right)\dot x^i
\right]{\rm d}\sigma
\\
&=
T[x]
+
\int_{\sigma_{\rm S}}^{\sigma_{\rm V}}
\partial_i\chi\,
\frac{{\rm d}x^i}{{\rm d}\sigma}
\,{\rm d}\sigma
\\
&=
T[x]
+
\int_{\sigma_{\rm S}}^{\sigma_{\rm V}}
\frac{{\rm d}\chi(x(\sigma))}{{\rm d}\sigma}
\,{\rm d}\sigma
\\
&=
T[x]+\chi_{\rm V}-\chi_{\rm S}.
\end{align*}
For fixed endpoints, the additional term has zero first variation and
therefore does not alter the local Euler--Lagrange equation.  Equivalently,
\[
{\rm d}\beta'
=
{\rm d}\beta+{\rm d}^2\chi
=
{\rm d}\beta,
\]
because ${\rm d}^2=0$.  This exact-one-form freedom of Randers data is
standard in stationary optical geometry
\cite{Gibbons_2009}.

The precise local statement is therefore not that the magnitude of
$\beta_i$ itself supplies the non-geodesic term.  Rather, ${\rm d}\beta$ is the local gauge-invariant two-form that determines the
departure of the projected ray from a geodesic of $\gamma_{ij}$.

A change $\beta\rightarrow\beta+{\rm d}\chi$ can change the numerical value
of the Fermat functional by the endpoint contribution
$\chi_{\rm V}-\chi_{\rm S}$, but it leaves ${\rm d}\beta$ and hence the
local projected trajectory unchanged.
```
}

The \ref{e121} can be tensorially expressed as 
\begin{align}
\kappa_g = \varepsilon_{ijk}N^ia^je^k
\label{e129}
\end{align}
where $\vec{T}$ and $\vec{T^{\prime}}$ are in correspondence with $e^k$ and $a^j$.
 
 \par The geodesic curvature can therefore be expressed as shown below. We simply need to substitute $\gamma_{ij}e^ie^j=1$ and $\gamma_{ij}e^iN^j=0$ at the appropriate junctures.
\begin{align}
\kappa_g &= \varepsilon_{ijk}N^i\gamma^{jl}\left(\beta_{n;l} - \beta_{l;n} \right)e^n e^k \notag\\
&= \gamma^jaN^ie^ke^b\varepsilon_{ijk}\varepsilon_{sab}\varepsilon^{sml}\beta_{l;m}\notag\\
&= N_ie_ke^b\left(\delta^i_s \delta^k_b - \delta^i_b \delta^k_s   \right)\varepsilon^{sml}\beta_{l;m}\notag\\
&= - \varepsilon^{ijk}N_i\beta_{j;k} 
\label{e130}
\end{align}
 The unit vector normal to the equatorial plane is $N_p = \dfrac{1}{\sqrt{\gamma^{\theta\theta}}}\delta^\theta_p$. We have chosen the upward direction over here without compromising generality. Since, we are concerned with the equatorial plane the term $\varepsilon^{\theta xy}\beta_{y;x}$ can be expressed as $ = - \dfrac{1}{\sqrt{\gamma}}\beta_{\phi,r}$ since $\varepsilon^{\theta r \phi} = -\dfrac{1}{\sqrt{\gamma}}$ and $\beta_{r,\phi}$ vanishes on account of axisymmetry. Now, substituting all of this back into \ref{e130} we get the expression for geodesic curvature on the equatorial plane as
 
 \begin{align}
 \kappa_g = -\dfrac{1}{\sqrt{\gamma \gamma^{\theta \theta}}}\beta_{\phi,r}
 \label{e131}
 \end{align}
  As for the line element ${\rm d}\ell_{\rm ph}$ we can easily define it for the equatorial plane$\left(\theta = \dfrac{\pi}{2}   \right)$ with the help of ${\rm d}\ell_{\rm ph}^2 \equiv \gamma_{ij}{\rm d}x^i {\rm d}x^j$ as
  
  \begin{align}
  {\rm d}\ell_{\rm ph} = \sqrt{\gamma_{rr}\left(\dfrac{{\rm d}r}{{\rm d}\phi}   \right)^2 + \gamma_{\phi \phi}}{\rm d}\phi
  \label{e132}
  \end{align}
 {\color{black} \par Recalling that for the Kerr metric the Fermat line element can be expressed as
  \begin{align}
  {\rm d}\ell_{\rm ph}^2
  =\gamma_{ij}^{\rm(F)}{\rm d}x^i{\rm d}x^j
  ={}&\dfrac{\Sigma^2}{\Delta\left(\Sigma-2Mr\right)}{\rm d}r^2
  +\dfrac{\Sigma^2}{\Sigma-2Mr}{\rm d}\theta^2\notag\\
  &+\left(r^2+a^2+
  \dfrac{2a^2Mr\sin^2\theta}{\Sigma-2Mr}\right)
  \dfrac{\Sigma\sin^2\theta}{\Sigma-2Mr}{\rm d}\phi^2 .
  \label{e133}
  \end{align}}
  
  Using this we can get the required expression for $\gamma^{\theta\theta}$ as 
  \begin{align}
  \gamma^{\theta\theta} = \dfrac{\Sigma - 2Mr}{\Sigma^2} = \dfrac{r^2 + a^2\cos^2\theta - 2Mr}{\left(r^2 + a^2\cos^2\theta  \right)^2} = \dfrac{r^2 - 2Mr}{r^4} = \dfrac{1}{r^2} - \dfrac{2M}{r^3}
  \label{e134}
  \end{align}
  
  and that of $\gamma$ as
  \begin{align}
  \gamma &= \dfrac{\Sigma^2}{\Delta \left(\Sigma - 2Mr   \right)}\dfrac{\Sigma^2}{\Sigma - 2Mr} \left(r^2 + a^2 + \dfrac{2a^2Mr\sin^2\theta}{\Sigma - 2Mr}   \right)\dfrac{\Sigma \sin^2\theta}{\Sigma - 2Mr} \notag\\
  &= \dfrac{\left(r^2 + a^2\cos^2\theta  \right)^2}{\Delta\left(r^2 + a^2\cos^2\theta - 2Mr  \right)} \dfrac{\left(r^2 + a^2\cos^2\theta   \right)^2}{r^2 + a^2\cos^2\theta - 2Mr} \left(r^2 + a^2 + \dfrac{2a^2 Mr\sin^2\theta}{r^2 + a^2\cos^2 \theta - 2Mr}   \right)\dfrac{\left(r^2 + a^2\cos^2\theta \right)\sin^2\theta}{\left(r^2 + a^2\cos^2\theta - 2Mr  \right)} \notag\\
  &\simeq \dfrac{r^4}{\left(r^2 - 2Mr + a^2 \right)\left(r^2 - 2Mr  \right)} \dfrac{r^4}{r^2 - 2Mr} \left(r^2 + a^2 + \dfrac{2a^2Mr}{r^2 - 2Mr} \right)\dfrac{r^2}{r^2 - 2Mr } \notag\\
  & = \dfrac{r^{10}}{\left(r^2 - 2Mr  \right)^3 \left(r^2 - 2Mr + a^2  \right)}\left(r^2 + a^2 + \dfrac{2a^2Mr}{r^2 - 2Mr}    \right)
  \label{e135}
  \end{align}
  Therefore, $\gamma\gamma^{\theta\theta}$ assumes the form,
  \begin{align}
  \gamma\gamma^{\theta\theta} &= \dfrac{r^2 -2Mr}{r^4}\dfrac{r^{10}}{\left(r^2 - 2Mr  \right)^3 \left(r^2 - 2Mr + a^2  \right)}\left(r^2 + a^2 + \dfrac{2a^2Mr}{r^2 - 2Mr}  \right) \notag\\
  &= \dfrac{r^6}{\left(r^2 - 2Mr  \right)^2\left(r^2 - 2Mr + a^2  \right)}\left(r^2 + a^2 + \dfrac{2a^2Mr}{r^2 - 2Mr}  \right)\notag\\
 &= \dfrac{r^5}{\left(r - 2M \right)^3\left(r^2 - 2Mr + a^2   \right)}\left(r^2 - 2Mr + a^2   \right)\notag\\
  &= \dfrac{r^5}{\left(r - 2M  \right)^3}
  \label{e136}
  \end{align}  
 On the other hand, $\beta_\phi$ in the equatorial plane will reduce down to,
 \begin{align}
 \beta_{\phi} = -\dfrac{2aMr\sin^2\theta}{r^2 + a^2\cos^2\theta - 2Mr} = -\dfrac{2aM}{r - 2M}
 \label{e137}
\end{align} 
Hence, $\beta_{\phi,\;r}$ becomes
\begin{align}
\beta_{\phi,\;r} = - \dfrac{\partial}{\partial r}\dfrac{2aM}{r - 2M} = \dfrac{2aM}{\left(r - 2M  \right)^2}
\label{e138}
\end{align}
Finally, we can plug in all these expressions to find the final expression of the geodesic curvature,
\begin{align}
\kappa_g &= - \dfrac{1}{\sqrt{\gamma\gamma^{\theta\theta}}}\beta_{\phi ,\;r} \notag\\
&= - \dfrac{1}{\sqrt{\dfrac{r^5}{\left(r - 2M  \right)^3}}}\dfrac{2aM}{\left(r - 2M  \right)^2}\notag\\
&= - \dfrac{2aM}{r^3}\left(1 - \dfrac{2M}{r}  \right)^{-\frac{1}{2}} \notag\\
&= -\dfrac{2aM}{r^3} + \bigO\left(\dfrac{aM^2}{r^4}  \right)
\label{e139}
\end{align} 
We have neglected any terms involving $a^nM$ over here for $n\geq 2$. For the line element ${\rm d}\ell_{\rm ph}$ in \ref{e132} we can similarly substitute the expressions of $\gamma_{rr}$ to reduce it to the form
\begin{align}
{\rm d}\ell_{\rm ph} = \left[\dfrac{b}{\sin^2\phi} + \bigO(M)  \right]{\rm d}\phi
\label{e140}
\end{align}
Now, we can finally amalgamate the two expressions presented in \ref{e139} \& \ref{e140} to get the expression of the path integral of the geodesic curvature as follows
\begin{align}
\int\limits_{V}^{S} \kappa_g {\rm d}\ell_{\rm ph} &= \int\limits_{S}^{V}\dfrac{2aM}{r^3}{\rm d}\ell_{\rm ph} + \bigO\left(\dfrac{aM^2}{r^4}  \right)\notag\\
&= \int\limits_{\phi_{\rm S}}^{\phi_{\rm V}}2aM\times \dfrac{\sin^3\phi}{b^3}\times\dfrac{b}{\sin^2\phi}{\rm d}\phi + \bigO\left(\dfrac{aM^2}{r^4}  \right)\notag\\
&=\dfrac{2aM}{b^2} \int\limits_{\phi_{\rm S}}^{\phi_{\rm V}}\sin\phi{\rm d}\phi + \bigO\left(\dfrac{aM^2}{r^4}  \right)\notag\\
&=\dfrac{2aM}{b^2}
\left[
\sqrt{1-b^2u_{\rm S}^2}+\sqrt{1-b^2u_{\rm V}^2}
\right]
+\bigO\left(\dfrac{aM^2}{b^3}\right).
\label{e141}
\end{align}

\par Tracing back to \ref{e30a} we can see that the angle of deviation as evaluated from the expressions involving the summation of the surface integral of Gaussian curvature and the line integral of geodesic curvature is 
\begin{align}
\alpha_{\rm D_{Kerr}}^{\rm pro}
&= -\iint\limits_{^\infty_{\rm V}\square_{\rm S}^\infty} K{\rm d}S
   + \int\limits_{S}^{V} \kappa_g {\rm d}\ell_{\rm ph}\notag\\
&= \left[
\sqrt{1-b^2u_{\rm S}^2}+\sqrt{1-b^2u_{\rm V}^2}
\right]
\left(\dfrac{2M}{b}-\dfrac{2aM}{b^2}\right)
+\bigO\left(\dfrac{M^2}{b^2}\right).
\label{e142}
\end{align}
which agrees exactly with \ref{e96}. And of course tweaking the sign of the term containing $a$ gives the expression for the retrograde trajectory for the Kerr case as before which is the same as \ref{e97}. {\color{black} Agreement in Kerr is a useful consistency check, but the general equivalence, with its hypotheses and orientation conventions, is the result proved in \cite{takahashi2023equivalence}; the Kerr calculation illustrates that theorem.}  

\section{Generalized Expression of Bending Angle in Static Spherically Symmetric Spacetime by OIA Formalism}{\label{s7}}

{\color{black} The preceding Kerr calculation has checked the agreement of the two forms in one stationary axisymmetric example; the cited general proof supplies the broader equivalence under its stated assumptions \cite{takahashi2023equivalence}.}. Furthermore, the fact that the results reduce to the more familiar expressions of deflection angle under the consideration of $r_{\rm S},\;r_{\rm SV}\rightarrow \infty$ gives validation to the GW-OIA formalism. For, the sake of thoroughness we can now step down to the case of spherically symmetric spacetime and derive the more generic expression of deflection angle in that context. For the current study we restrict ourselves to the Schwarzschild metric only.

\par The generic representation of a spherically symmetric spacetime is expressed as
\begin{align}
{\rm d}s^2 &= g_{ij}{\rm d}x^i {\rm d}x^j \notag\\
		&= -A(r){\rm d}t^2 + B(r){\rm d}r^2 + r^2\left({\rm d}\theta^2 + \sin^2\theta{\rm d}\phi^2   \right) 
\label{e143}
\end{align}    	
As usual we restrict ourselves to studying the photon orbit in the equatorial plane and evaluate the impact parameter as done earlier in \ref{s5.2.2}. Under the specified conditions the energy $E$ and the azimuthal component of angular momentum $J$ are conserved along the geodesics and hence can be expressed as
\begin{equation}
E = -g_{tt}\dot{t}-g_{t\phi}\dot{\phi}=A(r)\dot t .
\label{e144}
\end{equation}
and 
\begin{align}
J = g_{\phi t}\dot{t} + g_{\phi \phi}\dot{\phi} = r^2\dot{\phi}
\label{e145}
\end{align}
	Therefore, the impact parameter of the photon trajectory under the influence of a Schwarzschild metric is 
	
	\begin{align}
	b=\dfrac{J}{E}=\dfrac{r^2}{A(r)}\dfrac{{\rm d} \phi}{{\rm d} t} = \dfrac{r^3}{r-2M}\dfrac{{\rm d} \phi}{{\rm d}t}
	\label{e146}
	\end{align}

From, the mathematical framework already developed in \ref{s5.1} as \ref{e38} \& \ref{e39} we can derive the Fermat metric for the Schwarzschild case as follows
\begin{align}
\gamma_{ij}^{(\rm F)} {\rm d}x^i {\rm d}x^j &= \dfrac{\dfrac{r}{r-2M}}{\dfrac{r-2M}{r}}{\rm d}r^2 + \dfrac{r^2}{\dfrac{r-2M}{r}}{\rm d}\theta^2 + \dfrac{r^2\sin^2\theta}{\dfrac{r-2M}{r}}{\rm d}\phi^2 \notag\\
&= \dfrac{r^2}{\left(r-2M  \right)^2}{\rm d}r^2 + \dfrac{r^3}{r-2M}{\rm d}\theta^2 + \dfrac{r^3 \sin^2\theta}{r-2M}{\rm d}\phi^2
\label{e147}
\end{align}  
As there are no cross terms in the Schwarzschild metric we have $\beta_i{\rm d}x^i=0 \Rightarrow \beta_i = 0$.

\par Now, starting with the Schwarzschild metric in the form expressed below we can derive the equation for the photon trajectory by setting ${\rm d}s^2 = 0$. 
\begin{align}
{\rm d}s^2=-\left(1-\dfrac{2M}{r}   \right){\rm d}t^2 + \dfrac{{\rm d}r^2}{\left(1 - \dfrac{2M}{r}  \right)} + r^2{\rm d}\theta^2 + r^2\sin^2\theta {\rm d}\phi^2
\label{e148}
\end{align}
Therefore,
\begin{align}
& -A\dot{t}^2 + B\dot{r}^2 + D\dot{\phi}^2 =0 \notag\\
\Rightarrow & \dot{r}^2 = \dfrac{A}{B}\dot{t}^2 - \dfrac{D}{B}\dot{\phi}^2 \notag\\
\Rightarrow & \left(\dfrac{{\rm d}r}{{\rm d}\ell_{\rm ph}}  \right)^2 = \dfrac{A}{B}\left(\dfrac{{\rm d}t}{{\rm d}\ell_{\rm ph}}  \right)^2 - \dfrac{D}{B}\left(\dfrac{{\rm d}\phi}{{\rm d}\ell_{\rm ph}}   \right)^2 \notag\\
\Rightarrow & \left(\dfrac{{\rm d}r}{{\rm d}\phi}  \right)^2 = \dfrac{A}{B} \left(\dfrac{{\rm d}t}{{\rm d}\ell_{\rm ph}}  \right)^2\left(\dfrac{{\rm d}\ell_{\rm ph}}{{\rm d}\phi}  \right)^2 - \dfrac{D}{B} \notag\\
\Rightarrow & \left(\dfrac{{\rm d}r}{{\rm d}\phi}  \right)^2 + \dfrac{r^2\sin^2\theta}{B} = \dfrac{A}{B}\dfrac{1}{\left(\dfrac{{\rm d}\phi}{{\rm d}t}   \right)^2} = \dfrac{A}{B}\dfrac{1}{\left(\dfrac{Ab}{r^2}  \right)^2} = \dfrac{r^4}{ABb^2} \notag\\
\Rightarrow & \left(\dfrac{{\rm d}r}{{\rm d}\phi}  \right)^2 = \dfrac{r^4}{b^2AB} - \dfrac{r^2}{B}
\label{e149}
\end{align}
This expression will come in useful while evaluating the value of $\phi_{\rm VS}$. As has been illustrated in the case of Kerr metric earlier determining the deflection angle by evaluating the surface integral of the Gaussian curvature and the line integral of the geodesic curvature is comparatively tedious. As such we proceed with the OIA formalism. 
\par The unit tangential vector along the photon trajectory is given by $\tau^i = \dfrac{1}{\zeta}\left(\dfrac{{\rm d}r}{{\rm d}\phi},\; 0,\; 1   \right)$. Now, developing from the condition $\gamma_{ij}^{\rm (F)}\tau^i\tau^j = 1$ we get the following equation,
\begin{align}
&\gamma_{ij}^{\rm (F)}\tau^i\tau^j=1 \notag\\
\Rightarrow & \gamma_{rr}\tau^r \tau^r + \gamma_{\phi\phi}\tau^\phi \tau^\phi = 1 \notag\\ 
\Rightarrow & \dfrac{r^2}{\left(r-2M \right)^2}\left(\dfrac{1}{\zeta}\dfrac{{\rm d}r}{{\rm d}\phi}  \right)^2 + \dfrac{r^3\sin^2 \theta}{r-2M}\dfrac{1}{\zeta^2} = 1 \notag\\
\Rightarrow & \dfrac{B}{A}\left(\dfrac{1}{\zeta}\dfrac{{\rm d}r}{{\rm d}\phi}  \right)^2 + \dfrac{r^2}{A}\dfrac{1}{\zeta^2} = 1 \notag\\
\Rightarrow & \dfrac{1}{\zeta^2}\left[\dfrac{B}{A}\left(\dfrac{{\rm d}r}{{\rm d}\phi}  \right)^2 + \dfrac{r^2}{A}  \right] = 1
\label{e150}
\end{align}  
Substituting the expression for $\left(\dfrac{{\rm d}r}{{\rm d}\phi}  \right)^2$ from \ref{e149} in \ref{e150} we get,
\begin{align}
&\dfrac{1}{\zeta^2}\left[\dfrac{B}{A}\left(\dfrac{r^4}{b^2AB} - \dfrac{r^2}{B}  \right) + \dfrac{r^2}{A} \right] = 1 \notag\\
\Rightarrow & \dfrac{1}{\zeta^2}\left[\dfrac{r^4}{b^2A^2}   \right] = 1 \notag\\
\Rightarrow & \dfrac{1}{\zeta} = \dfrac{bA}{r^2}
\label{e151}
\end{align}
Similarly, in reference to the unit radial vector $\rho^i = \left(\rho^r,\; 0,\; 0  \right)$ along the photon trajectory we can proceed from the condition $\gamma_{ij}^{\rm (F)}\rho^i\rho^j=1$ to derive the expression for $\rho^r$
\begin{align}
& \gamma_{ij}^{\rm (F)}\rho^i\rho^j=1 \notag\\
\Rightarrow & \gamma_{rr}\rho^r\rho^r + \gamma_{\theta\theta}\rho^\theta \rho^\theta + \gamma_{\phi \phi} \rho^\phi \rho^\phi = 1 \notag\\
\Rightarrow & \gamma_{rr}\left(\rho^r  \right)^2 = 1 \notag\\
\Rightarrow & \rho^r = \dfrac{1}{\sqrt{\gamma_{rr}}}
\label{e152}
\end{align}
Now, in order to evaluate the jump angles $\psi_{\rm V}$ \& $\psi_{\rm S}$ we need to determine the expression for $\sin \psi$. This is easily done by working through the mathematical formulation for $\cos \psi$
\begin{align}
 \cos \psi &= \gamma_{ij}^{\rm (F)}\rho^i \tau^j \notag\\
 & = \gamma_{rr}\rho^r \tau^r \notag\\
 & = \gamma_{rr}\dfrac{1}{\sqrt{\gamma_{rr}}}\dfrac{1}{\zeta}\dfrac{{\rm d}r}{{\rm d}\phi} \notag\\
 & = \sqrt{\gamma_{rr}}\dfrac{bA}{r^2}\dfrac{{\rm d}r}{{\rm d}\phi} \notag\\
\Rightarrow \cos^2\psi &= \gamma_{rr} \dfrac{b^2A^2}{r^4}\left(\dfrac{r^4}{b^2AB} - \dfrac{r^2}{B}  \right) \notag\\
 &= \gamma_{rr} \dfrac{b^2A^2}{r^2}\dfrac{1}{B}\left(\dfrac{r^2}{b^2A} - 1 \right) \notag\\
 &= \dfrac{B}{A}\dfrac{b^2A^2}{r^2B}\left(\dfrac{r^2}{b^2A} - 1  \right) \notag\\
 &= 1 - \dfrac{b^2A}{r^2} \notag\\
\Rightarrow 1-\cos^2\psi &= \dfrac{b^2A}{r^2} \notag\\
\Rightarrow \sin\psi &= \dfrac{b\sqrt{A}}{r} = \dfrac{b}{r}\left(1-\dfrac{2M}{r}  \right)^{\frac{1}{2}} = bu - bMu^2 + \bigO \left( M^2 u^3 \right)
\label{e153}
\end{align}
Expanding in Taylor's series we can now express the required expressions for $\psi_{\rm V}$ \& $\psi_{\rm S}$ as
\begin{align}
\psi_{\rm V} = \sin^{-1}\left(bu_{\rm V}  \right) - \dfrac{bMu_{\rm V}^2}{\sqrt{1-b^2u_{\rm V}^2}}
\label{e154}
\end{align}
and
\begin{align}
\pi - \psi_{\rm S} = \sin^{-1}\left(bu_{\rm S}  \right) - \dfrac{bMu_{\rm S}^2}{\sqrt{1-b^2u_{\rm S}^2}}
\label{e155}
\end{align}
The remaining contribution to the deflection angle comes from the term $\phi_{\rm VS}$. We can easily achieve this by restructuring \ref{e149} in terms of reciprocal coordinates.
\begin{align}
\left(\dfrac{{\rm d}u}{{\rm d}\phi}  \right)^2 = \dfrac{1}{b^2AB} - \dfrac{u^2}{B} = \dfrac{1}{b^2} - u^2 + 2Mu^3 = F(u) \notag\\
\label{e156}
\end{align}
For the distance of closest approach $u=u_0$ and $\dfrac{{\rm d}u}{{\rm d}\phi}=0$ which for the Schwarzschild case leads to 
\begin{align}
&0=\dfrac{1}{b^2} - u_0^2 + 2Mu_0^3 \notag\\
\Rightarrow & \dfrac{1}{b^2} = u_0^2\left[1-2Mu_0  \right] \notag\\
\Rightarrow & b = \dfrac{1}{u_0}\left(1-2Mu_0  \right)^{-\frac{1}{2}} = \dfrac{1}{u_0}+M 
\label{e157}
\end{align}
Substituting the expression for the impact parameter in terms of the distance of closest approach in \ref{e156}
\begin{align}
F(u) &= \dfrac{1}{b^2} - u^2 + 2Mu^3 \notag\\
& = u_0^2 - 2Mu_0^3 - u^2 + 2Mu^3 \notag\\
& = \left(u_0^2 - u^2\right)\left[1 - 2M\left(\dfrac{u_0^3 - u^3}{u_0^2 - u^2}   \right)  \right] \notag\\
\Rightarrow \dfrac{1}{\sqrt{F(u)}} &= \dfrac{1}{\sqrt{u_0^2 - u^2}}\left[1 - 2M\left(\dfrac{u_0^3 - u^3}{u_0^2 - u^2}   \right)  \right]^{-\frac{1}{2}}\notag\\
&= \dfrac{1}{\sqrt{u_0^2 - u^2}} + M \dfrac{u_0^3 - u^3}{\left(u_0^2 - u^2  \right)^{\frac{3}{2}}}
\label{e158}
\end{align} 
The expression for $\dfrac{1}{\sqrt{F(u)}}$ when substituted in \ref{e71} for the Schwarzschild case yields the integrals that have already been calculated earlier. As such, we can write down the expression for $\phi_{\rm VS}$ pertaining to Schwarzschild case as
\begin{align}
\phi_{\rm VS} = \pi - \sin^{-1}\left(bu_{\rm V}  \right) - \sin^{-1}\left(bu_{\rm S}  \right) + \dfrac{M}{b}\dfrac{\left(2-b^2u_{\rm V}^2  \right)}{\sqrt{1-b^2u_{\rm V}^2}} + \dfrac{M}{b}\dfrac{\left(2-b^2u_{\rm S}^2  \right)}{\sqrt{1-b^2u_{\rm S}^2}}
\label{e159}
\end{align}
Finally we can combine \ref{e154}, \ref{e155} \& \ref{e159} to get the expression for deflection angle of a photon in Schwarzschild spacetime as
\begin{align}
\alpha_{\rm D}^{\rm Sch} &= \psi_{\rm V} - \psi_{\rm S} + \phi_{\rm VS} \notag\\
&= \dfrac{2M}{b}\left[\sqrt{1-b^2u_{\rm V}^2} + \sqrt{1-b^2u_{\rm S}^2}  \right]
\label{e160}
\end{align}
As usual under the limits $u_{\rm V}\rightarrow 0$ \& $u_{\rm S}\rightarrow 0$ \ref{e160} reduces to the familiar form of $\alpha_{\rm D}^{\rm Sch} \rightarrow \dfrac{4M}{b}$.

\begingroup
\providecommand{\dd}{}
\providecommand{\Lag}{}
\providecommand{\Ham}{}
\providecommand{\Order}{}
\providecommand{\sg}{}
\providecommand{\eg}{}
\providecommand{\RN}{}
\providecommand{\EH}{}
\providecommand{\ID}{}
\providecommand{\Hay}{}
\renewcommand{\dd}{\mathrm d}
\renewcommand{\Lag}{\mathcal L}
\renewcommand{\Ham}{\mathcal H}
\renewcommand{\Order}{\mathcal O}
\renewcommand{\sg}{\mathrm{SG}}
\renewcommand{\eg}{\mathrm{EG}}
\renewcommand{\RN}{\mathrm{RN}}
\renewcommand{\EH}{\mathrm{EH}}
\renewcommand{\ID}{\mathrm{ID}}
\renewcommand{\Hay}{\mathrm H}

\makeatletter
\@ifundefined{labelformat}{}{%
  \labelformat{section}{#1}%
  \labelformat{subsection}{#1}%
  \labelformat{subsubsection}{#1}%
  \labelformat{subsubsubsection}{#1}%
  \labelformat{equation}{#1}%
  \labelformat{figure}{#1}%
  \labelformat{subfigure}{\thefigure#1}%
  \labelformat{table}{#1}%
  \labelformat{appendix}{#1}%
}
\makeatother

\section{Nonlinear electrodynamics in finite-distance light deflection}
\label{sec:NEDChapter}

Nonlinear electrodynamics (NED) enters a finite-distance light-deflection calculation through both the background geometry and the propagation of electromagnetic disturbances. The nonlinear field contributes to the stress tensor that sources $g_{\mu\nu}$, while its characteristic equations determine the electromagnetic characteristic cones. In the geometrical-optics limit, the two polarization modes are described by the quadratic factors of the electromagnetic Fresnel polynomial \cite{Novello2000,DeLorenci2000,ObukhovRubilar2002}.

Background null geodesics provide the photon trajectories when the characteristic cone of the relevant polarization coincides with the background null cone. This coincidence holds in Maxwell electrodynamics, but need not hold for each polarization branch in NED. A finite-distance prediction therefore combines three ingredients: the background geometry, the electromagnetic characteristics, and the observer construction:
\begin{equation}
\boxed{\begin{aligned}
\text{Einstein--NED field equations}&\ \Longrightarrow\ g_{\mu\nu},
&\quad&\text{background geometry},\\
\text{electromagnetic principal symbol}&\ \Longrightarrow\
\bar g_{(a)}^{\mu\nu},
&&\text{polarization characteristics},\\
\text{observer tetrad and endpoints}&\ \Longrightarrow\
\beta_a,\,\alpha_a,\,\omega_a,
&&\text{measured quantities}.
\end{aligned}}
\label{eq:threeLayers}
\end{equation}
Here $a$ labels a polarization branch.  A generic one-invariant theory
$\Lag(F)$ contains an ordinary factor proportional to
$g^{\mu\nu}k_\mu k_\nu$ and an extraordinary factor involving $\Lag_F$ and
$\Lag_{FF}$.  A two-invariant theory $\Lag(F,G)$ generally has two deformed
factors.  Thus the phrase ``the effective metric of $\Lag(F)$'' is incomplete
unless it explicitly means the extraordinary branch.

Related gravitational-optics studies provide useful points of comparison, while
describing a different physical mechanism.  Roy and Sen applied a material-medium
refractive-index construction to Kerr light bending
\cite{RoySenKerrMaterial2015} and to
Reissner--Nordstr\"om and Janis--Newman--Winicour geometries (with
electromagnetic and scalar charges, respectively)
\cite{RoySenChargedMedium2017}.  More recent implementations address
non-equatorial Kerr--Newman deflection
\cite{RoyKalaGhoshKerrNewman2025} and the rotating charged Kerr--Sen solution
of heterotic string theory \cite{RoyKalaSinghaKerrSen2025}.
These works encode the influence of a prescribed background geometry through an
equivalent optical medium; they do not, solely by using a refractive index,
establish the polarization-dependent characteristic cones of a nonlinear
electromagnetic field theory.  In particular, the cited material-medium
calculations should not be read as finite-endpoint NED predictions without
the separate propagation and observer constructions developed here.

The examples below illustrate complementary aspects of this hierarchy.  The
Dymnikova solution separates an RN-like background from a low-order
extraordinary-cone correction; the fully regular $h$-family exhibits
strong-field sensitivity despite unusually high far-field powers; a
quadratic $\Lag(F,G)$ model displays branch-resolved birefringence; and the
Hayward geometry clarifies what can be inferred from a regular background
metric alone
\cite{dePaulaDymnikova2023,dePaulaAzreg2025,
dePaulaBirefringence2026,dePaulaHayward2023}.  Table~\ref{tab:NEDAnalysisHierarchy}
summarizes the sequence from the action to an observable interpretation.

\begin{table}[H]
\centering
\caption{Logical hierarchy of an NED lensing calculation.  Each stage supplies
a distinct mathematical input to the construction and interpretation of an
observable.}
\label{tab:NEDAnalysisHierarchy}
\small
\begin{tabularx}{\textwidth}{
>{\raggedright\arraybackslash}p{0.15\textwidth}
>{\raggedright\arraybackslash}p{0.23\textwidth}
>{\raggedright\arraybackslash}p{0.24\textwidth}
>{\raggedright\arraybackslash}X}
\toprule
Stage & Central question & Invariant input or output & Role in the optical analysis \\
\midrule
Action translation
& Are the invariant, matter-action sign, and field normalization consistent?
& Complete normalized action $S[g,A]$
& Keeps the stress tensor and QED matching within one convention.\\
Background branch
& Which spacetime and exterior domain solve the Einstein--NED equations?
& $g_{\mu\nu}$, horizons, regularity, and parameter domain
& Identifies the physically admissible exterior of the formal solution.\\
Principal symbol
& Which polarizations are electromagnetic characteristics?
& Fresnel polynomial and co-metrics $\bar g_{(a)}^{\mu\nu}$
& Distinguishes electromagnetic polarizations from background-null probes.\\
Hamiltonian reduction
& What controls turning points, light rings, and orbit integrals?
& $H_a(x,p)$, $U_a^2=\bar A_a/\bar C_a$, and $b=L/E$
& Relates the characteristic co-metric consistently to the orbit equations.\\
Observer map
& Which local direction or frequency is actually measured?
& Background tetrad acting on $p_\mu$, $K^\mu$, and $u^\mu$
& Specifies whether phase, energy transport, or frequency is measured.\\
Inference
& What additional assumptions connect the optical quantity to data?
& Endpoint geometry, mass--distance scale, emission, plasma, and likelihood model
& Separates an optical prediction from astrophysical parameter inference.\\
\bottomrule
\end{tabularx}
\end{table}

\subsection{Where NED enters a lensing calculation}

\subsubsection{Action normalization and background field equations}

We use signature $(-,+,+,+)$, units $G=c=1$, and invariants
\begin{equation}
 F\equiv F_{\mu\nu}F^{\mu\nu},
 \qquad
 G\equiv F_{\mu\nu}{}^{\star}\!F^{\mu\nu},
 \qquad
 {}^{\star}\!F^{\mu\nu}\equiv
 \frac12\epsilon^{\mu\nu\rho\sigma}F_{\rho\sigma}.
\label{eq:invariants}
\end{equation}
The action convention for this section is
\begin{equation}
 S=\frac1{16\pi}\int\dd^4x\sqrt{-g}\,[R-\Lag(F,G)].
\label{eq:action}
\end{equation}
Source papers may instead define $\mathcal F=F/4$, $\mathcal G=G/4$, absorb
constants into the matter action, or reverse the metric signature.  Such
choices must be converted at the level of the complete action.  If
\begin{equation}
 S=S_{\rm EH}-\frac{\varepsilon}{4\pi}
 \int\sqrt{-g}\,L_{\rm src}(\mathcal F,\mathcal G)\,\dd^4x,
 \qquad \varepsilon=\pm1,
\end{equation}
then comparison with Eq.~\eqref{eq:action} gives
\begin{equation}
 \boxed{\Lag(F,G)=4\varepsilon L_{\rm src}(F/4,G/4)},
 \quad
 \Lag_{FF}=\frac{\varepsilon}{4}(L_{\rm src})_{\mathcal F\mathcal F},
 \quad
 \Lag_{GG}=\frac{\varepsilon}{4}(L_{\rm src})_{\mathcal G\mathcal G}.
\label{eq:conventionConversion}
\end{equation}
The invariant definition, matter-action sign, field normalization, and
nonlinear coefficients must therefore move together.  An overall rescaling of
a characteristic co-metric preserves its cone; an unaccompanied rescaling of
the matter action changes the stress tensor and is not the same freedom.

At fixed metric,
\begin{equation}
 \delta F=2F^{\mu\nu}\delta F_{\mu\nu},\qquad
 \delta G=2{}^{\star}\!F^{\mu\nu}\delta F_{\mu\nu},\qquad
 \delta F_{\mu\nu}=2\nabla_{[\mu}\delta A_{\nu]}.
\end{equation}
Variation with respect to $A_\mu$ gives
\begin{equation}
 \boxed{\nabla_\mu\!\left(\Lag_FF^{\mu\nu}
 +\Lag_G{}^{\star}\!F^{\mu\nu}\right)=0},
 \qquad
 \boxed{\nabla_\mu{}^{\star}\!F^{\mu\nu}=0}.
\label{eq:NEDfield}
\end{equation}
Using
\begin{equation}
 \delta F=2F_{\mu\alpha}F_\nu{}^\alpha\delta g^{\mu\nu},
 \qquad
 \delta G=\frac12Gg_{\mu\nu}\delta g^{\mu\nu},
\end{equation}
the metric variation gives
\begin{equation}
 \boxed{G_{\mu\nu}=2\Lag_FF_{\mu\alpha}F_\nu{}^\alpha
 +\frac12(G\Lag_G-\Lag)g_{\mu\nu}.}
\label{eq:EinsteinNED}
\end{equation}
The Maxwell correspondence is $\Lag\to F$, $\Lag_F\to1$, and vanishing
higher derivatives.

\subsubsection{Dual descriptions and branch control}

For electrically charged regular solutions it is often useful to introduce
\begin{equation}
 P_{\mu\nu}\equiv\Lag_FF_{\mu\nu},\qquad
 P\equiv P_{\mu\nu}P^{\mu\nu},\qquad
 \Ham(P)\equiv2F\Lag_F-\Lag(F).
\label{eq:Pframework}
\end{equation}
Since $P=F\Lag_F^2$,
\begin{equation}
 \frac{\dd\Ham}{\dd F}=\Lag_F+2F\Lag_{FF},\qquad
 \frac{\dd P}{\dd F}=\Lag_F(\Lag_F+2F\Lag_{FF}).
\end{equation}
On an invertible Legendre branch,
\begin{equation}
 \boxed{\Ham_P=\frac1{\Lag_F},\qquad
 F_{\mu\nu}=\Ham_PP_{\mu\nu},\qquad
 \Lag=2P\Ham_P-\Ham.}
\label{eq:Legendre}
\end{equation}
The $F$ and $P$ frameworks are therefore branch-dependent descriptions of the
same constitutive theory, not interchangeable formulas at points where the
Legendre map degenerates.

\subsubsection{The principal symbol and its polarization factors}

Let $k_\mu$ be normal to a wavefront.  For a one-invariant theory the Hadamard
system separates according to the scalar discontinuity amplitude
$\xi=F^{\mu\nu}f_{\mu\nu}$.  The sector $\xi\ne0$ gives the extraordinary
characteristic equation \cite{Novello2000,DeLorenci2000}
\begin{equation}
 \boxed{\bar g^{\mu\nu}k_\mu k_\nu=0,\qquad
 \bar g^{\mu\nu}=\Lag_Fg^{\mu\nu}
 -4\Lag_{FF}F^\mu{}_{\alpha}F^{\alpha\nu}.}
\label{eq:effectiveF}
\end{equation}
The separate sector $\xi=0$, for $\Lag_F\ne0$ and nontrivial polarization,
gives the ordinary factor \cite{Novello2000,ObukhovRubilar2002}
\begin{equation}
 \boxed{g^{\mu\nu}k_\mu k_\nu=0.}
\label{eq:ordinaryCone}
\end{equation}
In a generic nonlinear $\Lag(F)$ theory, the two Fresnel factors describe an ordinary polarization on the background cone and an extraordinary polarization governed by Eq.~\eqref{eq:effectiveF}. The factors coincide in the Maxwell limit. Each characteristic cone is unchanged by a nonzero conformal rescaling of its co-metric.

In the $P$ framework the extraordinary factor can be written
\cite{dePaulaDymnikova2023}
\begin{equation}
 \boxed{g_{\rm eff}^{\mu\nu}=\frac1{\Ham_P}g^{\mu\nu}
 +4\frac{\Ham_{PP}}{F_P}P^\mu{}_{\sigma}P^{\sigma\nu},
 \qquad
 F_P=\Ham_P(\Ham_P+2P\Ham_{PP}).}
\label{eq:effectiveP}
\end{equation}

Equations~\eqref{eq:effectiveF} and \eqref{eq:effectiveP} define
contravariant characteristic \emph{co-metrics}.  A covariant line element
requires a matrix inverse.  For example,
\begin{equation}
 \mathcal G^{\mu\nu}=\operatorname{diag}\!\left(
 -\frac{X}{f},Xf,\frac{Y}{r^2},\frac{Y}{r^2\sin^2\theta}\right)
 \Longrightarrow
 \mathcal G_{\mu\nu}=\operatorname{diag}\!\left(
 -\frac{f}{X},\frac1{Xf},\frac{r^2}{Y},
 \frac{r^2\sin^2\theta}{Y}\right).
\label{eq:cometricInverse}
\end{equation}
The turning-point test makes the consequence independent of any preferred
line-element representative.  With $E=-p_t$, $L=p_\phi$, and $p_r=0$, the
co-metric itself gives
\begin{equation}
 2H=-\frac{X}{f}E^2+\frac{Y}{r^2}L^2=0
 \quad\Longrightarrow\quad
 \boxed{\frac{E^2}{L^2}=\frac{fY}{Xr^2}.}
\label{eq:directCometricTurningPoint}
\end{equation}
Copying the coefficients of $\mathcal G^{\mu\nu}$ unchanged into $\dd s^2$
instead gives $E^2/L^2=fX/(Yr^2)$.  The two expressions are reciprocal in
$W=Y/X$, and their ratio is $W^2(r)$.  Unless $W^2$ is constant, no conformal
rescaling relates them.  This is therefore an index-placement error, not a
choice of signature, notation, or conformal gauge.  The distinction is
relevant to the Einstein--Bronnikov shadow construction of
Ref.~\cite{Allahyari2020} and the rational-NED shadow construction of
Ref.~\cite{Kruglov2020}, for which the characteristic-Hamiltonian potential
differs from the potential obtained by copying the co-metric coefficients into
a line element.  The implications are discussed in
Section~\ref{sec:NEDLiteratureConsistency}.

For $\Lag(F,G)$, eliminating two scalar discontinuity amplitudes produces a
quartic Fresnel polynomial.  When it factorizes,
\begin{equation}
 \mathcal P_+(k)\mathcal P_-(k)=0,
 \qquad
 \mathcal P_\pm(k)=\bar g_\pm^{\mu\nu}k_\mu k_\nu.
\label{eq:FresnelFactor}
\end{equation}
The factors are the two electromagnetic polarizations; the background cone is
not a third mode \cite{DeLorenci2000,dePaulaBirefringence2026}.

Nondegeneracy and Lorentzian signature of each factor are necessary local
tests, but they are not a complete well-posedness result.  For the factorized
two-cone Pleba\'nski class studied by Abalos \emph{et al.}, symmetric
hyperbolicity holds if and only if the interiors of the two timelike covector
cones intersect \cite{Abalos2015}.  This common-co-cone statement is local;
global causality and dynamical stability require additional analysis.  The
logical order is therefore: invertibility, Lorentzian signature, common
timelike-covector interior, and finally global questions.  The discussion
below is restricted to parameter ranges in which the relevant characteristic
factors are nondegenerate and Lorentzian before their light rings or observer
maps are evaluated.

\subsection{From characteristic covectors to finite-distance observables}

\subsubsection{Hamiltonian turning points and critical curves}

Write a static spherical representative of one photon branch as
\begin{equation}
 \dd\bar s^2=-\bar A(r)\dd t^2+\bar B(r)\dd r^2
 +\bar C(r)\dd\Omega^2.
\label{eq:genericEffective}
\end{equation}
The overbar is omitted when the background metric itself governs the chosen
trajectory.  In the equatorial plane the branch Hamiltonian is
\begin{equation}
 H_a=\frac12\left[-\frac{E^2}{\bar A}
 +\frac{p_r^2}{\bar B}+\frac{L^2}{\bar C}\right]=0,
 \qquad E=-p_t,\qquad L=p_\phi.
\label{eq:genericHamiltonian}
\end{equation}
The angular potential
\begin{equation}
 U_a^2(r)\equiv\frac{\bar A(r)}{\bar C(r)}
\label{eq:angularPotential}
\end{equation}
is invariant under a common conformal rescaling of the branch metric.  At a
circular photon orbit $r=r_l^{(a)}$,
\begin{equation}
 \boxed{\left.\frac{\dd U_a^2}{\dd r}\right|_{r_l^{(a)}}=0,
 \qquad
 (b_l^{(a)})^2\equiv\frac{L^2}{E^2}
 =\left.\frac{\bar C}{\bar A}\right|_{r_l^{(a)}}.}
\label{eq:genericLR}
\end{equation}
The orbit is unstable when $U_a^2$ has a local maximum.  This Hamiltonian
construction avoids assigning physical meaning to a particular conformal
representative and makes the co-metric/metric index placement explicit.

Solving Eq.~\eqref{eq:genericHamiltonian} for the orbit gives
\begin{equation}
 \boxed{\frac{\dd\phi}{\dd r}
 =\frac{b\sqrt{\bar A\bar B}}
 {\sqrt{\bar C}\sqrt{\bar C-b^2\bar A}}.}
\label{eq:genericOrbit}
\end{equation}
For asymptotically flat endpoints at infinity,
\begin{equation}
 \Theta_a(b)=2\int_{r_0}^{\infty}
 \left|\frac{\dd\phi}{\dd r}\right|\dd r-\pi,
 \qquad b^2=\frac{\bar C(r_0)}{\bar A(r_0)}.
\label{eq:directDeflection}
\end{equation}
At finite source and receiver distance, the infinite limits are replaced by
endpoint integrals and local angles.  The conserved ratio $b=L/E$ belongs to
the selected Hamiltonian; using the same symbol for two branches compares
coefficient structure but does not assert identical turning points.

\subsubsection{Two local directions: phase normal and energy ray}

The photon Hamiltonian is effective, but a material observer is timelike with
respect to the background spacetime
\begin{equation}
 \dd s^2=-A(r)\dd t^2+B(r)\dd r^2+C(r)\dd\Omega^2.
\label{eq:backgroundGeneric}
\end{equation}
A static observer at $r_o$ carries the orthonormal frame
\begin{equation}
 e_{\hat t}=A_o^{-1/2}\partial_t,\qquad
 e_{\hat r}=B_o^{-1/2}\partial_r,\qquad
 e_{\hat\phi}=C_o^{-1/2}\partial_\phi.
\label{eq:staticTetrad}
\end{equation}
The eikonal phase $S$ defines a wave covector $p_\mu=\nabla_\mu S$.  For
\begin{equation}
 H(x,p)=\frac12\bar g^{\mu\nu}p_\mu p_\nu=0,
 \qquad
 K^\mu\equiv\frac{\dd x^\mu}{\dd\lambda}
 =\frac{\partial H}{\partial p_\mu}=\bar g^{\mu\nu}p_\nu,
\label{eq:phaseRay}
\end{equation}
$p_\mu$ is normal to a phase front whereas $K^\mu$ is tangent to the
Hamiltonian ray.  They need not determine the same spatial direction in the
background tetrad \cite{Perlick2000,Oancea2020}.

The covector projections are
\begin{equation}
 p_{\hat r}=\frac{p_r}{\sqrt{B_o}},
 \qquad p_{\hat\phi}=\frac{L}{\sqrt{C_o}}.
\end{equation}
Defining $\tan\beta_k=|p_{\hat\phi}/p_{\hat r}|$ and using the effective null
condition gives
\begin{equation}
 \boxed{\tan^2\beta_{k,o}=
 \frac{b^2B_o}
 {C_o\bar B_o(\bar A_o^{-1}-b^2\bar C_o^{-1})}.}
\label{eq:generalFiniteAngle}
\end{equation}
Projecting the spatial ray tangent instead gives
\begin{equation}
 \boxed{\tan\beta_{v,o}=
 \frac{C_o\bar B_o}{B_o\bar C_o}\tan\beta_{k,o}.}
\label{eq:genericRayPhaseAngle}
\end{equation}
When the two metrics coincide or are conformally related,
$\beta_v=\beta_k$ and
\begin{equation}
 \sin^2\beta_o=\frac{b^2A_o}{C_o}.
\label{eq:standardFiniteAngle}
\end{equation}
At finite distance, a result must say whether it reports the phase-normal
angle or the ray-arrival angle.  For an asymptotically Maxwellian branch the
constitutive ratio in Eq.~\eqref{eq:genericRayPhaseAngle} tends to unity, so
both conventions approach the same $b_l^{(a)}$ in the distant-observer limit.

Frequency requires a separate invariant statement.  An observer of
four-velocity $u^\mu$ measures
\begin{equation}
 \boxed{\omega=-p_\mu u^\mu.}
\label{eq:measuredFrequency}
\end{equation}
Lowering the ray tangent $K^\mu$ with the background metric does not in general
recover $p_\mu$; doing so mixes two geometries and changes the observer map.

\subsubsection{Optical curvature and endpoint geometry}

The equatorial null condition of Eq.~\eqref{eq:genericEffective} defines the
two-dimensional optical metric
\begin{equation}
 \dd t^2=\frac{\bar B}{\bar A}\dd r^2
 +\frac{\bar C}{\bar A}\dd\phi^2
 =\dd r_*^2+v^2(r_*)\dd\phi^2,
 \quad
 \dd r_*=\sqrt{\frac{\bar B}{\bar A}}\dd r,
 \quad
 v=\sqrt{\frac{\bar C}{\bar A}}.
\label{eq:opticalMetric}
\end{equation}
Its Gaussian curvature is
\begin{equation}
 K=-\frac1v\frac{\dd^2v}{\dd r_*^2}.
\label{eq:GaussianCurvature}
\end{equation}
For a static spherical branch, the projected light ray is a geodesic of this
optical metric.  The finite-distance Gauss--Bonnet relation can therefore be
written \cite{Gibbons_2008,Ishihara_2016,Ono_2017}
\begin{equation}
 \alpha_a=\Psi_R^{(a)}-\Psi_S^{(a)}+\phi_{RS}^{(a)}
 =-\iint_{D_a}K_a\,\dd S_a,
\label{eq:GBfinite}
\end{equation}
with a consistent domain, orientation, and endpoint convention.  In a
birefringent theory there are two optical metrics, two curvatures, and two
deflection angles.  The method is not changed; the geometric input is branch
dependent.
The background metric $g_{\mu\nu}$ determines horizons, curvature invariants, and timelike emitter motion. The selected characteristic co-metric $\bar g_{(a)}^{\mu\nu}$ and its Hamiltonian determine photon characteristics, light rings, and weak bending. Local directions are obtained by projecting $p_\mu$ or $K^\mu$ onto a background tetrad, while the measured frequency is given by Eq.~\eqref{eq:measuredFrequency}. These roles remain distinct whenever the characteristic and background metrics are not conformally related.

\subsection{Representative NED models and optical regimes}

\subsubsection{Regular background with a nontrivial cone: the Dymnikova spacetime}

The Dymnikova solution is most naturally written in the $P$ framework.  Its
structural function is \cite{Dymnikova2004,dePaulaDymnikova2023}
\begin{equation}
 \Ham(P)=\frac{P}{(1+\alpha\sqrt{-P})^2},
 \qquad P<0.
\label{eq:IDH}
\end{equation}
For a purely radial electric field, the first integral of the field equation is
\begin{equation}
 P=-\frac{2Q^2}{r^4}.
\label{eq:IDP}
\end{equation}
Defining
\begin{equation}
 z\equiv\frac{\pi Q^2}{8M},
\label{eq:IDz}
\end{equation}
regularity at the centre and ADM normalization give
\begin{equation}
 \boxed{f_{\ID}(r)=1-\frac{4M}{\pi r}
 \left[\arctan\!\left(\frac r z\right)
 -\frac{rz}{r^2+z^2}\right].}
\label{eq:IDf}
\end{equation}
The two asymptotic regimes are
\begin{align}
 f_{\ID}(r)&=1-\frac{2M}{r}+\frac{Q^2}{r^2}
 -\frac{2Q^2z^2}{3r^4}+\Order(r^{-6}),
 &&r\rightarrow\infty,
\label{eq:IDfar}\\
 f_{\ID}(r)&=1-\frac{Q^2}{3z^4}r^2+\Order(r^4),
 &&r\rightarrow0.
\label{eq:IDcore}
\end{align}
The first line establishes an RN-like exterior; the second establishes a
finite de Sitter-type core.

The constitutive derivatives entering the extraordinary co-metric are
\begin{equation}
 \Ham_P=\frac{r^6}{(r^2+z^2)^3},
 \qquad
 F_P=\frac{r^{12}(r^2-2z^2)}{(r^2+z^2)^7}.
\label{eq:IDHPFP}
\end{equation}
The zero of $F_P$ defines a characteristic boundary
\begin{equation}
 r_{\rm eff}=\sqrt2\,z.
\label{eq:IDreff}
\end{equation}
On the black-hole branch used in Ref.~\cite{dePaulaDymnikova2023}, this radius
lies below the outer horizon; the exterior extraordinary cone is evaluated on
the nondegenerate Lorentzian branch.

The extraordinary and ordinary angular potentials are
\begin{equation}
 U_{\ID,\eg}^2(r)=\frac{F_Pf_{\ID}}{\Ham_P^2r^2},
 \qquad
 U_{\ID,\sg}^2(r)=\frac{f_{\ID}}{r^2}.
\label{eq:IDpotentials}
\end{equation}
The extraordinary critical impact parameter is
\begin{equation}
 b_l^{\ID}=\frac{r_l\Ham_P(r_l)}
 {\sqrt{F_P(r_l)f_{\ID}(r_l)}}.
\label{eq:IDbc}
\end{equation}
Projecting the extraordinary wave covector on a static background tetrad gives
the phase-normal screen radius reported in the source paper,
\begin{equation}
 r_s^{\ID}(r_o)=r_ob_l^{\ID}
 \sqrt{\frac{(r_o^2+z^2)f_o}
 {r_o^2(r_o^2+z^2)+3(b_l^{\ID})^2z^2f_o}}.
\label{eq:IDfiniteShadow}
\end{equation}
It approaches $b_l^{\ID}$ as $r_o\to\infty$.  The Hamiltonian-ray angle follows
from Eq.~\eqref{eq:genericRayPhaseAngle} and generally differs at finite
$r_o$.  At extremality the extraordinary critical size is about $9.29\%$
larger than the ordinary critical size in the same spacetime
\cite{dePaulaDymnikova2023}.  This is a strong-field effect of the
characteristic cone, not merely of the regular background.

The weak series separates the two layers even more sharply.  The ordinary
branch agrees with RN through $b^{-3}$ because the first non-RN metric term in
Eq.~\eqref{eq:IDfar} is $r^{-4}$.  The extraordinary branch changes already at
$b^{-2}$ because $\Ham_P^2/F_P$ differs from unity at $r^{-2}$:
\begin{align}
 \Theta_{\ID,\sg}
 &=\frac{4M}{b}+\frac{3\pi(5M^2-Q^2)}{4b^2}
 +\frac{16M(8M^2-3Q^2)}{3b^3}+\Order(b^{-4}),
\label{eq:IDweakSG}\\
 \Theta_{\ID,\eg}
 &=\Theta_{\ID,\sg}
 +\frac{6Mz^3}{Q^2b^2}+\frac{16Mz^2}{b^3}+\Order(b^{-4}).
\label{eq:IDweakEG}
\end{align}
The conserved $b=L/E$ is defined branch by branch.  The important result is
the order at which propagation enters, not an identification of the two
turning-point maps.

Figure~\ref{fig:DymnikovaObservables} summarizes the principal strong-field
observables.  The perimetral light-ring radius in panel (a) is
$\varrho_l=r_l\sqrt{\Ham_P(r_l)}$ for the extraordinary geometry.  The
critical impact parameter in panel (b) is the asymptotic shadow radius, and
panel (c) displays that radius directly as a circular screen edge.  The
comparison keeps the normalized position within each model's black-hole
branch fixed: $q=Q/Q_{\rm ext}$ is formed with the Dymnikova extremal charge
for the Dymnikova curves and with $Q_{\rm ext}=M$ for RN.

The two curves in panel (b) cross at the numerically determined value
$q=0.865986$.  That equality is observable-specific: it does not make the two
light-ring radii or radial potentials equal.  Panel (d) also shows why a
background-null calculation would miss the dominant effect near extremality.
At $q=1$, $b_l^{\eg}/b_l^{\sg}=1.09290$ in the same Dymnikova spacetime even
though both branches share the lapse in Eq.~\eqref{eq:IDf}.

\begin{figure}[!tbp]
\centering
\includegraphics[width=0.98\textwidth]{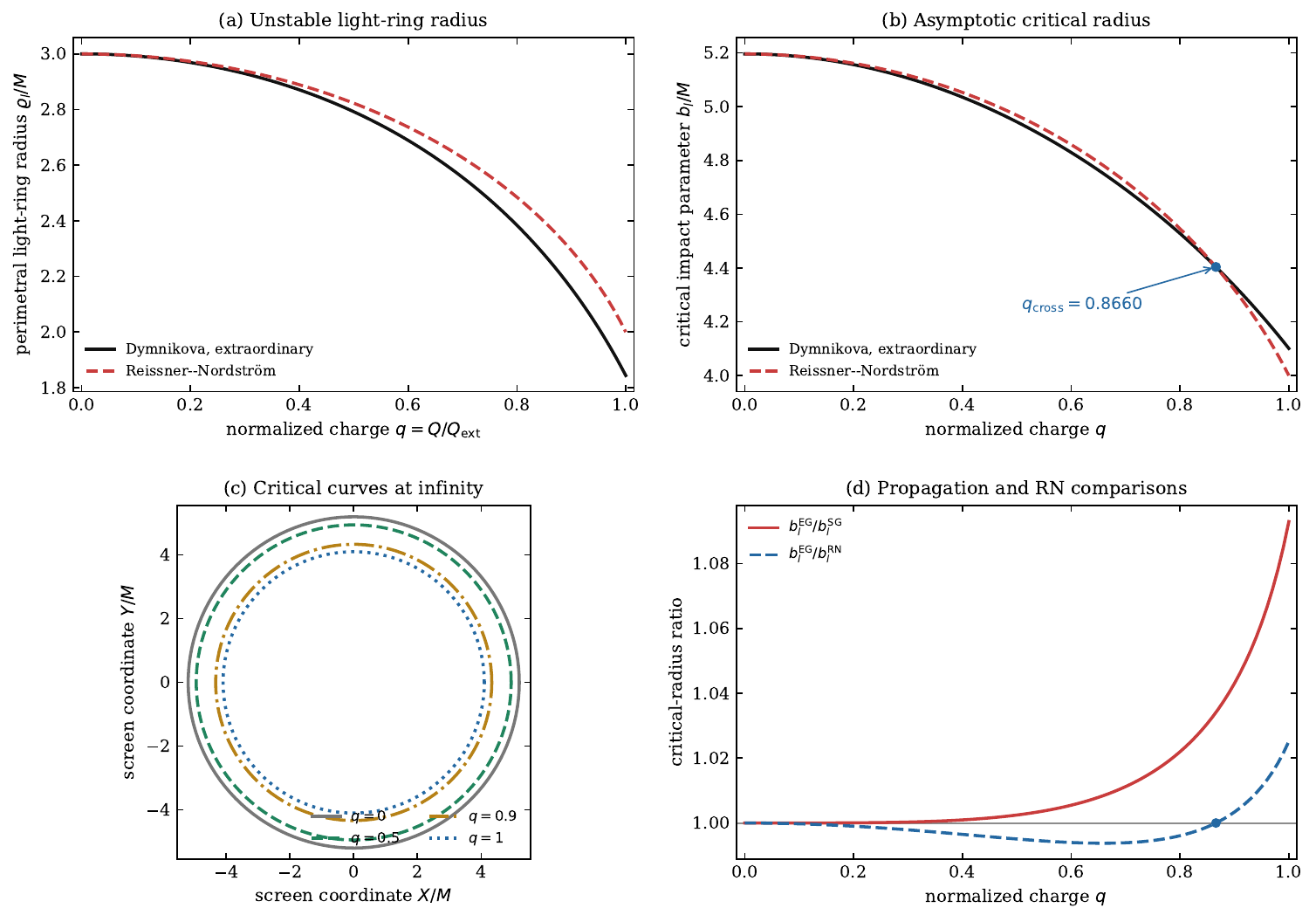}
\caption{Dymnikova light-ring and shadow observables.  (a) Extraordinary
perimetral light-ring radius compared with RN.  (b) Critical impact
parameters, with the nonzero Dymnikova/RN crossing marked.  (c) Asymptotic
extraordinary critical curves for representative normalized charges.  (d)
Extraordinary-to-ordinary and extraordinary-to-RN critical-radius ratios.}
\label{fig:DymnikovaObservables}
\end{figure}

\subsubsection{Background-metric optics in the Hayward spacetime}

For the magnetic interpretation of the static Hayward geometry used in
Refs.~\cite{Hayward2006,dePaulaHayward2023},
\begin{equation}
 F=\frac{2Q^2}{r^4},
 \qquad
 \Lag(F)=\frac{12M}{|Q|Q^2}
 \frac{(Q^2F/2)^{3/2}}
 {[1+(Q^2F/2)^{3/4}]^2}.
\label{eq:HayL}
\end{equation}
For $Q>0$, substitution of $F(r)$ gives
\begin{equation}
 \Lag(r)=\frac{12MQ^3}{(r^3+Q^3)^2}.
\end{equation}
Writing $f=1-2m(r)/r$, the $t$--$t$ Einstein equation gives
\begin{equation}
 m'(r)=\frac{r^2\Lag(r)}4
 =\frac{3MQ^3r^2}{(r^3+Q^3)^2},
\end{equation}
and $m(0)=0$ yields
\begin{equation}
 \boxed{m(r)=\frac{Mr^3}{r^3+Q^3},
 \qquad
 f_{\Hay}(r)=1-\frac{2Mr^2}{r^3+Q^3}.}
\label{eq:Hayf}
\end{equation}
The core and far field are
\begin{align}
 f_{\Hay}(r)&=1-\frac{2M}{Q^3}r^2+\Order(r^5),
 &&r\rightarrow0,\\
 f_{\Hay}(r)&=1-\frac{2M}{r}
 +\frac{2MQ^3}{r^4}+\Order(r^{-7}),
 &&r\rightarrow\infty.
\label{eq:Hayfar}
\end{align}
The displayed model does not recover $\Lag\sim F$ in the weak-field limit;
instead $\Lag\propto F^{3/2}$.  The optical quantities considered here are
constructed from null geodesics of the background Hayward metric and therefore
characterize neutral massless probes.  A polarization-resolved NED photon
interpretation would additionally require the electromagnetic characteristic
co-metrics of the underlying theory.

For those background null geodesics,
\begin{equation}
 U_{\Hay}^2=\frac{f_{\Hay}}{r^2},
 \qquad
 (r_c^3+Q^3)^2=3Mr_c^5,
 \qquad
 b_c=\frac{r_c}{\sqrt{f_{\Hay}(r_c)}}.
\label{eq:HayLR}
\end{equation}
The extremal parameters are
\begin{equation}
 r_{\rm ext}=\frac{4M}{3},
 \qquad
 Q_{\rm ext}=\frac{2^{5/3}}{3}M.
\label{eq:HayExt}
\end{equation}
The weak-deflection result displayed in Ref.~\cite{dePaulaHayward2023} is
\begin{equation}
 \boxed{\Theta_{\Hay}(b)=\frac{4M}{b}
 +\frac{15\pi M^2}{4b^2}+\Order(b^{-3}).}
\label{eq:Hayweak}
\end{equation}
It agrees with Schwarzschild through the calculated order, whereas RN has an
additional $-3\pi Q^2/(4b^2)$ term.  The Hayward parameter is not optically
irrelevant; its first displayed metric correction is simply too high in the
inverse-radius expansion to affect Eq.~\eqref{eq:Hayweak}.

The distinction between a weak-field coincidence and a strong-field change is
visible in Fig.~\ref{fig:HaywardObservables}.  At
$\alpha=Q/Q_{\rm ext}=0.8$, the background-null radial equation gives
$r_h/M=1.815996$, $r_c/M=2.848459$, and $b_c/M=5.068549$.  Rays with
$b<b_c$ cross the horizon, the critical ray approaches the unstable circular
orbit, and rays with $b>b_c$ return to infinity.  The corresponding critical
circles contract as $\alpha$ increases even though the $b^{-2}$ coefficient in
Eq.~\eqref{eq:Hayweak} remains Schwarzschild-like.

Accordingly, the curves describe geodesics of the Hayward background rather
than trajectories of a specified electromagnetic polarization.  Extending the
analysis to NED photons requires the characteristic polynomial associated with
Eq.~\eqref{eq:HayL}, its admissible polarization branches, and the corresponding
observer projection.

\begin{figure}[!tbp]
\centering
\includegraphics[width=0.98\textwidth]{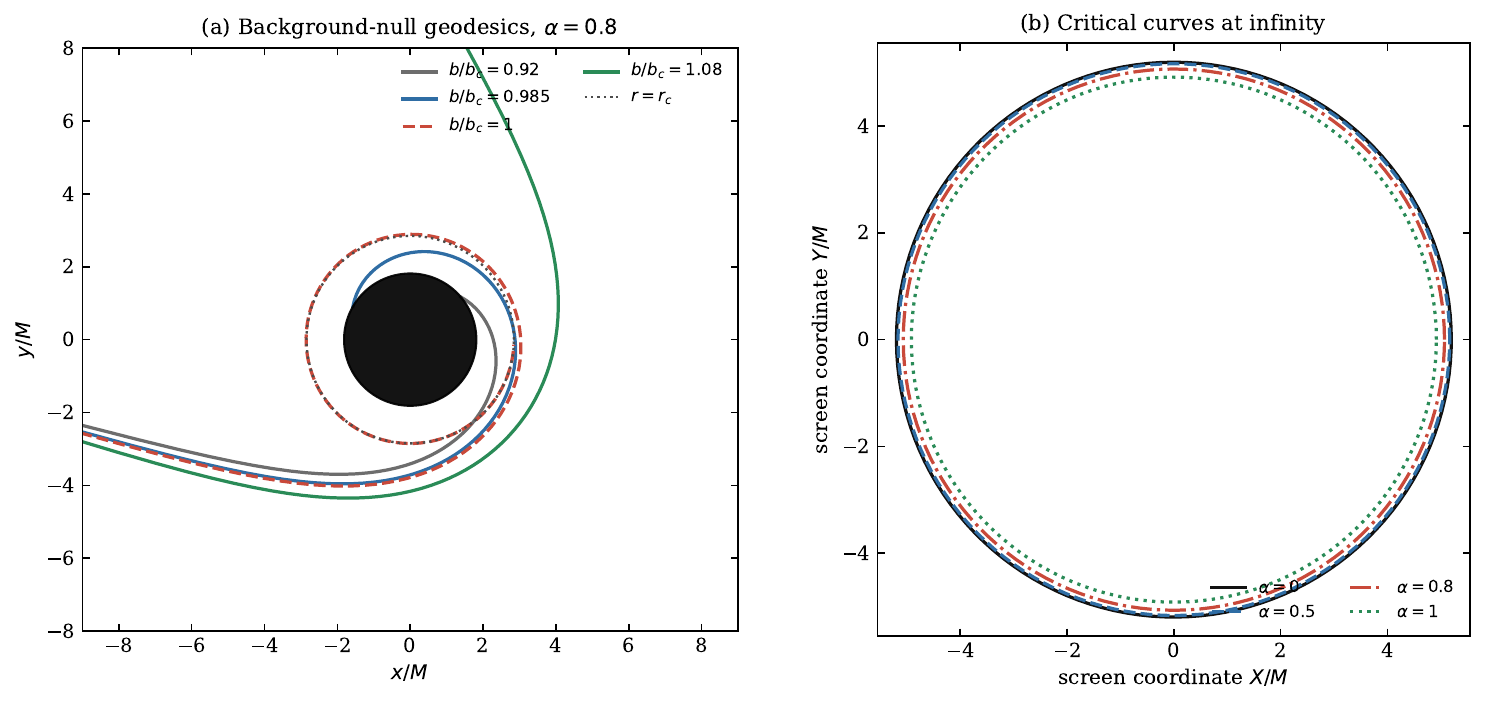}
\caption{Hayward background-metric optics.  (a) Equatorial null geodesics at
$\alpha=0.8$; the black disk ends at the outer horizon and the dotted circle
is the unstable orbit $r=r_c$.  (b) Asymptotic critical curves for four
charges.  Both panels refer to null propagation in the background geometry.}
\label{fig:HaywardObservables}
\end{figure}

\subsubsection{Suppressed far-field corrections and frequency shifts: the
\texorpdfstring{$h$}{h}-family}

The family of Ref.~\cite{dePaulaAzreg2025} is specified directly in the $F$
framework by
\begin{equation}
 \Lag_F(x)=1+\frac{h}{x^4}+\frac1{x^8},
 \qquad x\equiv\frac rK.
\label{eq:hLF}
\end{equation}
For a radial electric field, Eq.~\eqref{eq:NEDfield} integrates to
\begin{equation}
 r^2\Lag_FE=Q,
 \qquad
 E(x)=\frac{Qx^6}{K^2(x^8+hx^4+1)},
\end{equation}
and hence
\begin{equation}
 F(x)=-\frac{2Q^2x^{12}}
 {K^4(x^8+hx^4+1)^2}.
\label{eq:hF}
\end{equation}
The constitutive combination controlling the radial characteristic sector is
\begin{equation}
 \Phi\equiv\Lag_F+2F\Lag_{FF}
 =\frac{(x^8+hx^4+1)^2}
 {x^8(x^8-hx^4-3)}.
\label{eq:hPhi}
\end{equation}
The chosen extraordinary branch satisfies
\begin{equation}
 r>r_{\rm CR}=K\left(\frac{h+\sqrt{h^2+12}}2\right)^{1/4}.
\label{eq:hCR}
\end{equation}
For the black holes considered in the source, $r_{\rm CR}$ lies below the
outer horizon.

The two regular members used there have
\begin{equation}
 K_0=\frac{\sqrt{\sqrt2+1}\,\pi Q^2}
 {6\,2^{1/4}M},
 \qquad
 K_2=\frac{\pi Q^2}{4\sqrt2M}.
\label{eq:hK}
\end{equation}
Their exact metric functions can be written
\begin{equation}
 f_0(x)=1-\frac{36\sqrt2(\sqrt2-1)M^2}{\pi^2Q^2}p_0(x),
 \qquad
 f_2(x)=1-\frac{32M^2}{\pi^2Q^2}p_2(x),
\label{eq:hmetrics}
\end{equation}
where the elementary functions $p_0,p_2$ are given in
Ref.~\cite{dePaulaAzreg2025}.  Their far fields, more useful for the lensing
comparison, are
\begin{align}
 f_0(r)&=1-\frac{2M}{r}+\frac{Q^2}{r^2}
 -\frac{(17+12\sqrt2)\pi^8Q^{18}}
 {181398528M^8r^{10}}+\Order(r^{-18}),
\label{eq:h0far}\\
 f_2(r)&=1-\frac{2M}{r}+\frac{Q^2}{r^2}
 -\frac{\pi^4Q^{10}}{5120M^4r^6}+\Order(r^{-10}).
\label{eq:h2far}
\end{align}
Both backgrounds are regular at the centre and RN-like to unusually high
order in the exterior.

Matrix inversion of the electric $\Lag(F)$ co-metric gives the extraordinary
line element
\begin{equation}
 \boxed{\dd\bar s^2=-\frac{f}{\Phi}\dd t^2
 +\frac{\dd r^2}{f\Phi}+\frac{r^2}{\Lag_F}\dd\Omega^2.}
\label{eq:hEffectiveLine}
\end{equation}
The corresponding angular potential and critical impact parameter are
\begin{equation}
 U_h^2=\frac{f\Lag_F}{r^2\Phi},
 \qquad
 b_l=r_l\sqrt{\frac{\Phi_l}{f_l(\Lag_F)_l}},
\label{eq:hLR}
\end{equation}
where
\begin{equation}
 f_l\left[2-r_l\frac{(\Lag_F)'_l}{(\Lag_F)_l}
 +r_l\frac{\Phi_l'}{\Phi_l}\right]-r_lf_l'=0.
\end{equation}
The finite-position phase-normal screen radius is
\begin{equation}
 \boxed{r_s(r_o)=r_ob_l\sqrt{\frac{f_o\Phi_o}
 {r_o^2\Phi_o+b_l^2f_o[\Phi_o-(\Lag_F)_o]}}.}
\label{eq:hFiniteShadow}
\end{equation}

The far-field photon factor is
\begin{equation}
 \frac{\Phi}{\Lag_F}=
 \begin{cases}
 1+4x^{-8}+\Order(x^{-16}),&h=0,\\
 1+4x^{-4}+\Order(x^{-8}),&h=2.
 \end{cases}
\label{eq:hPhotonFar}
\end{equation}
Neither the background corrections in Eqs.~\eqref{eq:h0far}--\eqref{eq:h2far}
nor these photon factors enter through $b^{-3}$.  Both ordinary and
extraordinary branches therefore give
\begin{equation}
 \boxed{\Theta_h(b)=\frac{4M}{b}
 +\frac{3\pi(5M^2-Q^2)}{4b^2}
 +\frac{16M(8M^2-3Q^2)}{3b^3}+\Order(b^{-4}).}
\label{eq:hWeak}
\end{equation}
This is an asymptotic-order agreement, not equality of the two cones.  Their
strong-field critical sizes remain different.

Redshift probes the observer map rather than only the ray path.  For
tangential emission by a massive circular geodesic, define
\begin{equation}
 \chi=\frac{rf'}{2f}.
\end{equation}
Using $\omega=-p_\mu u^\mu$ with the extraordinary characteristic equation
gives
\begin{equation}
 \boxed{1+z_\pm=\frac{1\mp\sqrt{\chi\Phi/\Lag_F}}
 {\sqrt f\sqrt{1-\chi}},
 \qquad
 z_{\rm grav}=f^{-1/2}-1.}
\label{eq:hCorrectRedshift}
\end{equation}
\paragraph{Observer-frequency relation.}
Equation~\eqref{eq:hCorrectRedshift} follows from the invariant contraction
$\omega=-p_\mu u^\mu$: it has no overall factor of $\Phi$, and its directional
factor is $\Phi/\Lag_F$.  In NED, $p_\mu$ and
$g_{\mu\nu}K^\nu$ are generally different covectors, so background-metric
lowering of the effective-ray tangent does not give the phase covector used in
the frequency measurement.  This distinction affects the frequency shift but
not the background solution, the characteristic cone, or the ray trajectories
obtained from that cone.
A physical stellar emitter must additionally satisfy $2f-rf'>0$ and radial
stability; a limiting null circular orbit is not a stable material orbit.

The shadow and frequency effects are displayed together in
Fig.~\ref{fig:hFamilyObservables}.  In the $h=2$ model the extraordinary
asymptotic shadow is larger than
the background-null critical curve for every nonzero charge and reaches
$\bar r_s/r_s=1.013463$ at extremality.  When the comparison is instead made
with RN at equal normalized charge, the $h=2$ curve crosses unity at
$q=0.971611$, while the $h=0$ extraordinary curve remains below the RN curve
throughout the black-hole interval.  The two comparisons hold different
geometries fixed and therefore cannot be interchanged.

The lower panels use Eq.~\eqref{eq:hCorrectRedshift} at $Q/M=0.79$.  The near
coincidence of the $h=0$ and RN curves follows from the very high-order tails
in Eqs.~\eqref{eq:h0far} and \eqref{eq:hPhotonFar}.  The
$h=2$ difference is concentrated near the smallest allowed circular-emitter
radius and decays rapidly outward.  The curves include the mathematically
defined but radially unstable timelike circular branch; only the portion
outside the appropriate marginally stable orbit can describe a long-lived
stellar emitter.

\begin{figure}[!tbp]
\centering
\includegraphics[width=0.98\textwidth]{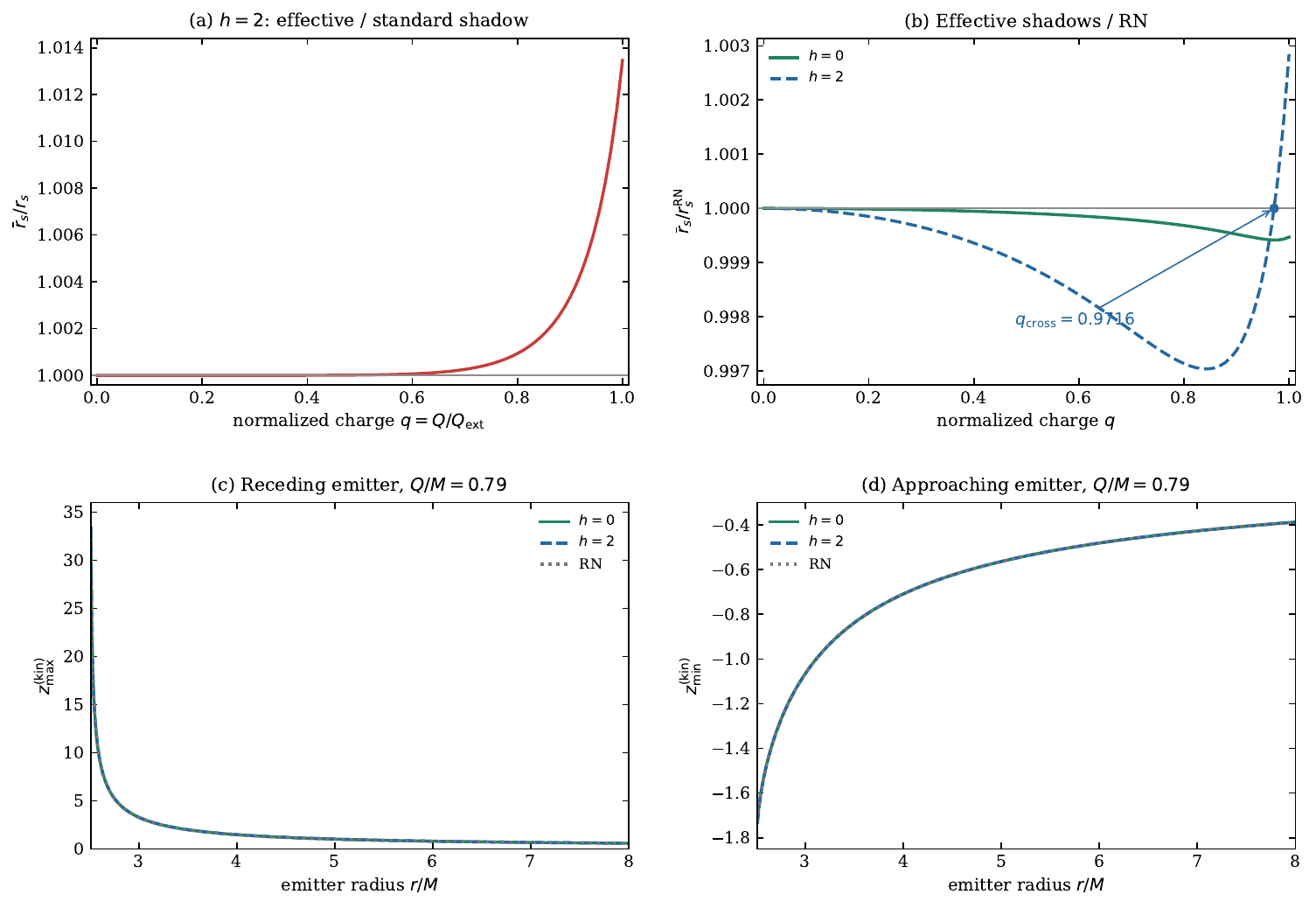}
\caption{Optical observables for the fully regular $h$-family.  (a)
Extraordinary-to-ordinary asymptotic critical-radius ratio for $h=2$.  (b)
Extraordinary-to-RN ratios for $h=0$ and $h=2$, with each charge normalized by
the relevant extremal charge.  (c,d) Kinematic redshift and blueshift extrema
for $Q/M=0.79$, with RN included as the far-field reference.  The frequency
shifts use $\omega=-p_\mu u^\mu$ and Eq.~\eqref{eq:hCorrectRedshift}.}
\label{fig:hFamilyObservables}
\end{figure}

\subsubsection{Branch-resolved birefringence in a quadratic two-invariant model}

The model used in Ref.~\cite{dePaulaBirefringence2026} is
\begin{equation}
 \Lag(F,G)=F-\mu\left(F^2+\frac74G^2\right).
\label{eq:EHL}
\end{equation}
Its coefficient cannot be imported from QED without matching conventions.  In
rationalized Heaviside--Lorentz units the one-loop matter density is
\begin{equation}
 L_{\rm EH}=-\mathcal F
 +\frac{8\alpha^2}{45m_e^4}\mathcal F^2
 +\frac{14\alpha^2}{45m_e^4}\mathcal G^2+\cdots.
\end{equation}
Matching the Maxwell term to Eq.~\eqref{eq:action}, including the field
rescaling $F_{\mu\nu}^{\rm M}=\sqrt{4\pi}\,F_{\mu\nu}^{\rm HL}$, gives
\begin{equation}
 \boxed{\mu_{\rm QED}=\frac{\alpha^2}{90\pi m_e^4}}
\label{eq:muQED}
\end{equation}
before conversion to a geometrized length scale
\cite{HeisenbergEuler1936,Dunne2004}.  The illustrative value
$\mu/M^2=0.05$ below is therefore a parameter of the classical quadratic
model, not a controlled QED prediction.

For a magnetic charge, $F=2Q^2/r^4$ and $G=0$, and the background metric is
\begin{equation}
 \boxed{f_{\EH}(r)=1-\frac{2M}{r}+\frac{Q^2}{r^2}
 -\frac{2\mu Q^4}{5r^6}.}
\label{eq:EHf}
\end{equation}
Unlike the preceding regular backgrounds, this geometry remains singular at
the centre.  Its role here is birefringence, not regularity.

At a degenerate horizon set $\rho=r_{\rm ext}/M$, $q=Q/M$, and
$\widehat\mu=\mu/M^2$.  Solving $f=f'=0$ gives the correlated curve
\begin{equation}
 \boxed{q^2=\frac{\rho(5-3\rho)}2,
 \qquad
 \widehat\mu=\frac{5\rho^3(1-\rho)}{(5-3\rho)^2}.}
\label{eq:EHcorrelatedExtremality}
\end{equation}
Its cusp is at $\rho=5/6$, with $q^2=25/24$ and
$\widehat\mu=25/324$.  These maxima are coordinate projections of one curve,
not independent bounds whose Cartesian product defines the horizon domain.

At $G=0$, the two polarization sectors of the principal symbol decouple.  For
the polynomial in Eq.~\eqref{eq:EHL}, direct substitution gives, up to separate
nonzero conformal factors,
\begin{align}
 \bar g_+^{\mu\nu}
 &=\left(\Lag_F-2F\Lag_{GG}\right)g^{\mu\nu}
   -4\Lag_{GG}F^\mu{}_{\lambda}F^{\lambda\nu}
 =\left(1+5\mu F\right)g^{\mu\nu}
   +14\mu F^\mu{}_{\lambda}F^{\lambda\nu},\nonumber\\
 \bar g_-^{\mu\nu}
 &=\Lag_Fg^{\mu\nu}
   -4\Lag_{FF}F^\mu{}_{\lambda}F^{\lambda\nu}
 =\left(1-2\mu F\right)g^{\mu\nu}
   +8\mu F^\mu{}_{\lambda}F^{\lambda\nu}.
\label{eq:EHExactFactors}
\end{align}
These are the two co-metric representatives displayed in
Ref.~\cite{dePaulaBirefringence2026}.  Their covariant inverses may be written
\begin{equation}
 \dd\bar s_\pm^2=\frac1{G_1^\pm}
 \left(-f\dd t^2+\frac{\dd r^2}{f}\right)
 +\frac{r^2}{G_2^\pm}\dd\Omega^2,
\label{eq:EHmetrics}
\end{equation}
where
\begin{align}
 G_1^+&=1+\frac{10\mu Q^2}{r^4},
 &G_2^+&=1-\frac{4\mu Q^2}{r^4},\nonumber\\
 G_1^-&=1-\frac{4\mu Q^2}{r^4},
 &G_2^-&=1-\frac{12\mu Q^2}{r^4}.
\label{eq:EHG}
\end{align}
There are two logically different readings.  If Eq.~\eqref{eq:EHL} is
postulated as a standalone classical polynomial theory, the decoupled
co-metrics in Eq.~\eqref{eq:EHExactFactors} and their rational matrix inverses
are algebraically exact for this $G=0$ background.  If Eq.~\eqref{eq:EHL} is
instead read as the Euler--Heisenberg action truncated at first order in
$\mu$, as in the source paper, observables must be expanded consistently
through $\Order(\mu)$, with $|\mu F|\ll1$ and a separate derivative/frequency
window.  Solving the rational inverse factors without re-expanding them is then
a useful partial resummation of the displayed model, not a higher-order QED
prediction.

The first characteristic-degeneracy radii are
\begin{equation}
 r_{\rm deg}^+=(4\mu Q^2)^{1/4},
 \qquad
 r_{\rm deg}^-=(12\mu Q^2)^{1/4}.
\label{eq:EHsignature}
\end{equation}
The exterior calculation requires the event horizon to lie outside the
relevant radius; the two-cone system also requires the common-co-cone test.

For each polarization,
\begin{equation}
 U_\pm^2=\frac{G_2^\pm f}{G_1^\pm r^2},
\label{eq:EHPotential}
\end{equation}
and Eq.~\eqref{eq:genericLR} becomes
\begin{equation}
 f\left[2-r\frac{(G_2^\pm)'}{G_2^\pm}
 +r\frac{(G_1^\pm)'}{G_1^\pm}\right]-rf'=0,
 \qquad
 b_l^\pm=r_l^\pm\sqrt{\frac{G_1^\pm(r_l^\pm)}
 {G_2^\pm(r_l^\pm)f(r_l^\pm)}}.
\label{eq:EHLR}
\end{equation}
One background spacetime thus has two polarization-dependent unstable light
rings and two critical impact parameters.

For a static observer, the phase-normal convention of the source gives
\begin{align}
 \sin\beta_l^+&=\frac{b_l^+\sqrt{f_o}}
 {\sqrt{r_o^2+A_+(r_o)(b_l^+)^2f_o}},
 &A_+(r)&=\frac{14\mu Q^2}{r^4+10\mu Q^2},\nonumber\\
 \sin\beta_l^-&=\frac{b_l^-\sqrt{f_o}}
 {\sqrt{r_o^2+A_-(r_o)(b_l^-)^2f_o}},
 &A_-(r)&=\frac{8\mu Q^2}{r^4-4\mu Q^2}.
\label{eq:EHfinite}
\end{align}
The phase-normal screen radii are $r_s^\pm=r_o\sin\beta_l^\pm$.  They approach
$b_l^\pm$ at infinity, whereas the finite-distance ray angles follow from
Eq.~\eqref{eq:genericRayPhaseAngle}.  At $M=Q=1$ and $\mu=0.05M^2$, the
critical impact parameters are
\begin{equation}
 b_l^+=4.0827996M,
 \qquad
 b_l^-=4.0510941M,
 \qquad
 b_l^{\sg}=4.0024777M.
\label{eq:EHnumbers}
\end{equation}
Here $b_l^{\sg}$ is a background-metric test-ray value, not a third photon
polarization.

The first-order dependence on $\mu$ can be obtained directly at the stationary
point.  Put
$r_0=(3M+\sqrt{9M^2-8Q^2})/2$, $b_0=r_0/\sqrt{f_0(r_0)}$, and
$c_+=14$, $c_-=8$.  Then
\begin{equation}
 \boxed{\left.\frac{\dd b_l^\pm}{\dd\mu}\right|_{\mu=0}
 =\frac{Q^4b_0^3}{5r_0^8}
 +\frac{c_\pm Q^2b_0}{2r_0^4}.}
\label{eq:EHimpactCorrection}
\end{equation}
At $M=Q=1$ this gives $1.80$ for the plus mode and $1.05$ for the minus
mode.  Equations~(77)--(79) of Ref.~\cite{dePaulaBirefringence2026} agree with
the plus value but imply $2.55$ for the minus mode.  The discrepancy also
persists for nonextremal charge-to-mass ratios and is confined to the symbolic
first-order minus-branch expression.  The nonperturbative roots of the
characteristic factors used in Eq.~\eqref{eq:EHnumbers}, including the
numerical Sgr~A$^*$ limits reported in that reference, do not depend on this
first-order coefficient.

Despite the strong-field splitting, the two polarization series and the
background test-ray series agree through $b^{-3}$:
\begin{equation}
 \boxed{\Theta_+(b)=\Theta_-(b)=\Theta_{\sg}(b)
 =\frac{4M}{b}+\frac{3\pi(5M^2-Q^2)}{4b^2}
 +\frac{16M(8M^2-3Q^2)}{3b^3}+\Order(b^{-4}).}
\label{eq:EHweak}
\end{equation}
The background correction in Eq.~\eqref{eq:EHf} starts at $r^{-6}$, whereas
the characteristic factors start at $r^{-4}$.  The first
polarization-dependent weak term can therefore occur only beyond the displayed
order.

\begin{figure}[!tbp]
\centering
\includegraphics[width=0.98\textwidth]{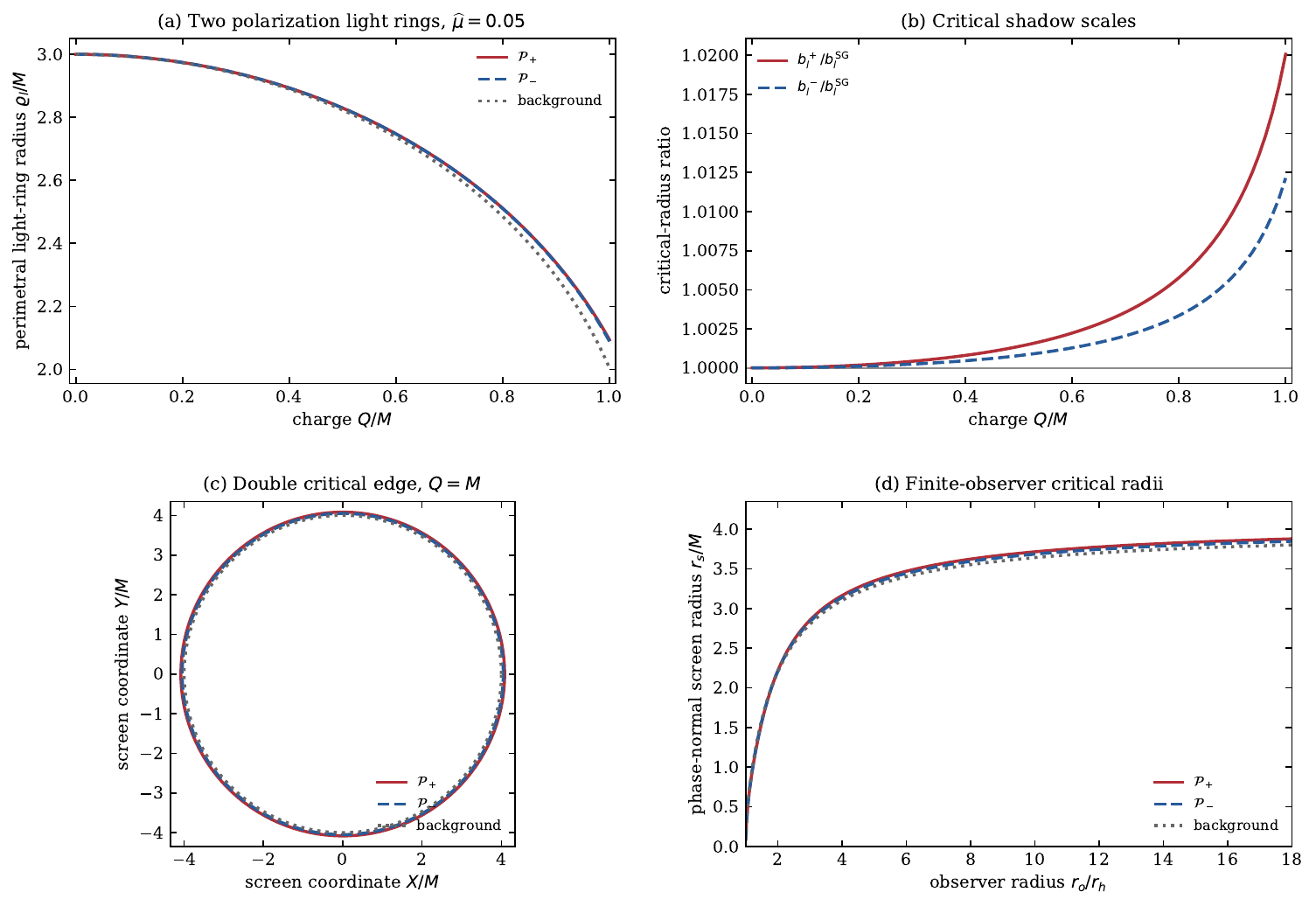}
\caption{Polarization-resolved observables for the quadratic model at
$\widehat\mu=0.05$.  (a) Perimetral radii of the two electromagnetic light rings and the
background-null reference orbit.  (b) Critical-radius ratios relative to that
reference.  (c) The two asymptotic electromagnetic critical edges at $Q=M$.
(d) Their phase-normal finite-observer screen radii.  The gray background
curve is a background-metric reference and is not a third polarization.  The sweep
$0\le Q/M\le1$ lies inside the correlated black-hole domain at this coupling.}
\label{fig:QuadraticObservables}
\end{figure}

Figure~\ref{fig:QuadraticObservables} shows the optical meaning of the
factorized Fresnel polynomial without converting it into a generic ``NED
deviation.''  Both electromagnetic branches move outward relative to the
background reference, but they do so by different amounts.  At $Q=M$ the
critical radii are those in
Eq.~\eqref{eq:EHnumbers}; panel (c) renders their separation as two circular
edges.  Panel (d) then applies the finite-position map
Eq.~\eqref{eq:EHfinite}: the screen radii vanish at the static-observer limit
near the horizon and approach their branch-specific $b_l^\pm$ values at large
$r_o$.

The plotted parameter line is admissible for the standalone-classical reading
of Eq.~\eqref{eq:EHL}.  At $Q=M$ the outer horizon is
$r_h/M=1.113965$, while $r_{\rm deg}^+/M=0.668740$ and
$r_{\rm deg}^-/M=0.880112$ both remain hidden.  Under the strict effective-
field-theory reading, the same curves are illustrative resummations of the
truncated model and only their expansion through the controlled order should
be used quantitatively.  This qualification affects the physical scale of the
splitting, not the existence of two algebraic polarization factors.

\subsection{Physical synthesis: which observable sees which nonlinear layer?}

The model-resolved figures above should not be read as four estimates of one
generic NED correction.  Each holds a different part of the problem fixed.
Figure~\ref{fig:DymnikovaObservables} compares the extraordinary and ordinary
cones of one regular electric spacetime and then compares the extraordinary
critical scale with RN.  Figure~\ref{fig:HaywardObservables} concerns the
background metric.  Figure~\ref{fig:hFamilyObservables} combines
a characteristic critical curve with a frequency measurement from a massive
emitter.  Figure~\ref{fig:QuadraticObservables} compares two electromagnetic
polarizations on the same background and then moves the observer away from
infinity.  The distinctions among these operations carry more physical
content than a single cross-model ranking.

\subsubsection{Far-field bending and strong-field capture are different
filters}

For any chosen reference geometry, it is useful to keep the two nonlinear
inputs conceptually separate:
\begin{equation}
 \delta_{\rm bg}(r):=f(r)-f_{\rm ref}(r),
 \qquad
 \delta_{\rm cone}^{(a)}(r):=
 \frac{U_a^2(r)}{U_{\rm bg}^2(r)}-1.
\label{eq:twoCorrections}
\end{equation}
The first term describes a change in the spacetime generated by the matter distribution, whereas the second describes a change in electromagnetic propagation on that background.

Their radial falloffs explain why a strong-field critical-curve plot can show
a clear effect even when a displayed weak-deflection series looks familiar.
In the Dymnikova solution the first non-RN term of the lapse is
$\Order(r^{-4})$, whereas the extraordinary cone changes the bending already
at $b^{-2}$ through Eq.~\eqref{eq:IDweakEG}.  The $h=0$ background and cone
corrections are delayed to $r^{-10}$ and $r^{-8}$, respectively, yet the
critical curve is still determined at $r=\Order(M)$.  For $h=2$ and for the
quadratic model, the first cone corrections scale as $r^{-4}$ and therefore
lie beyond the $b^{-3}$ series retained in Eqs.~\eqref{eq:hWeak} and
\eqref{eq:EHweak}.  Hayward has no Coulomb term and first departs from
Schwarzschild at $r^{-4}$, so Eq.~\eqref{eq:Hayweak} remains Schwarzschild-like
through $b^{-2}$ although Fig.~\ref{fig:HaywardObservables} shows a changing
capture separatrix.

This is a hierarchy of radial sensitivity, not a contradiction.  A
large-impact-parameter ray samples the far-field tail near its turning point.
A critical curve is fixed by a stationary point of $U_a^2$ in the strong
field.  The redshift of a circular emitter samples the lapse, orbital velocity,
and characteristic direction at the emitter radius.  A finite-observer shadow
adds a fourth radius.  Agreement in one channel therefore gives no theorem of
agreement in the others.

\subsubsection{The exterior branch must exist before it can cast a shadow}

Regularity of the background curvature does not certify the electromagnetic
initial-value problem.  The Dymnikova extraordinary branch contains the zero
of $F_P$ at $r_{\rm eff}=\sqrt2z$; the $h$-family contains the pole of $\Phi$
at $r_{\rm CR}$; and the quadratic model contains the separate degeneracy
radii $r_{\rm deg}^\pm$.  Before solving for a critical orbit, one must verify
the ordering
\begin{equation}
 r_{\rm boundary}^{(a)}<r_h<r_l^{(a)}<r_o
\label{eq:admissibleOrdering}
\end{equation}
for the branch and observer under consideration.  If there are two cones,
their individual Lorentzian signatures must be supplemented by the common
timelike-covector condition relevant to symmetric hyperbolicity
\cite{Abalos2015}.

The parameter ranges used in Figs.~\ref{fig:DymnikovaObservables},
\ref{fig:hFamilyObservables}, and \ref{fig:QuadraticObservables} satisfy
Eq.~\eqref{eq:admissibleOrdering}.  In the quadratic case at $Q=M$, for
example, both characteristic boundaries lie
below $r_h/M=1.113965$, while the two light rings lie near $2M$.  This is a
result of the selected parameter line, not an automatic consequence of the
action.  The Hayward figure has no electromagnetic characteristic-boundary
marker because its optical curves are restricted to background null
geodesics.

A consistent analysis therefore proceeds as follows:
\begin{enumerate}
 \item solve the background equations and identify the black-hole domain;
 \item derive and factorize the electromagnetic principal polynomial;
 \item locate zeros, poles, and sign changes of every characteristic factor;
 \item select a common admissible exterior and test the light-ring stability;
 \item project the phase covector or Hamiltonian ray on a declared observer
 tetrad; and
 \item only then compare the resulting quantity with another model or with
 data.
\end{enumerate}
This order prevents a formally correct extremum equation from being evaluated
on a branch that has lost its intended signature.

\subsubsection{Finite distance changes the observable, not the critical
orbit}

The asymptotic equalities $r_s=b_l$ used in the circular panels of
Figs.~\ref{fig:DymnikovaObservables}, \ref{fig:HaywardObservables}, and
\ref{fig:QuadraticObservables} are limiting statements.  At finite $r_o$, the
critical impact parameter remains fixed by the unstable orbit, but the local
screen radius also contains the lapse and characteristic coefficients at the
observer.  Panel (d) of Fig.~\ref{fig:QuadraticObservables} shows this directly:
the three curves share the expected near-horizon suppression, separate in the
intermediate region, and converge toward three distinct asymptotic critical
scales.

There is a second finite-distance distinction.  The phase normal is obtained
from the spatial projection of $p_\mu$, whereas the energy-transport direction
is obtained from the Hamiltonian tangent
$K^\mu=\bar g^{\mu\nu}p_\nu$.  Equation~\eqref{eq:genericRayPhaseAngle} relates
their local angles.  They agree for an asymptotically Maxwellian cone at
infinity, but not generically at a finite endpoint.  Thus an interferometric
phase construction, a wavefront-normal shadow convention, and an arrival
direction inferred from energy transport must state which vector was placed
on the screen.

The redshift in Fig.~\ref{fig:hFamilyObservables} is another observer
projection.  It depends on the invariant frequency $\omega=-p_\mu u^\mu$, not
on lowering $K^\mu$ with the background metric.  The near coincidence of the
$h=0$ and RN curves over most of the displayed range is therefore meaningful:
the remaining difference is suppressed by the model's far-field structure.

\subsubsection{Critical-size degeneracies are broken by other channels}

The nonzero crossing in panel (b) of Fig.~\ref{fig:DymnikovaObservables}
provides a concrete degeneracy: a Dymnikova extraordinary branch and an RN
black hole can have the same asymptotic $b_l$ at one normalized charge.  Panel
(b) of Fig.~\ref{fig:hFamilyObservables} provides another.  Conversely, the
quadratic model contains an internal discriminator, because the two
polarizations cannot both be replaced by the single gray background curve in
Fig.~\ref{fig:QuadraticObservables}.

These equalities and splittings are compactly described by
\begin{equation}
 \mathcal R_{\rm prop}^{(a)}=\frac{b_l^{(a)}}{b_l^{\rm bg}},
 \qquad
 \mathcal R_{\rm ref}^{(a)}=\frac{b_l^{(a)}}{b_l^{\rm ref}},
 \qquad
 \Delta_{\rm pol}=
 \frac{2(b_l^+-b_l^-)}{b_l^++b_l^-}.
\label{eq:observableRatios}
\end{equation}
The first isolates propagation on one spacetime, the second compares distinct
backgrounds, and the third is internal to one birefringent solution.  None is
a universal measure of ``distance from GR.''  Einstein gravity supplies every
background considered here; what differs is the matter model, the optical
branch, or the comparison spacetime.

One matching condition fixes one number.  Equality of $b_l$ does not imply
equality of $r_l$, $U_a(r)$, weak-deflection coefficients, finite-distance
screen radii, or redshifts.  A physically useful discrimination strategy must
therefore combine at least two channels, for example a critical size with a
polarization split, a finite-endpoint dependence, or a weak-field coefficient
that reaches the first model-dependent order.  This statement concerns model
identifiability; it is not a forecast that present instruments can resolve all
of those channels.

\subsubsection{Matched parameters and observational scope}

The figures use the dimensionless parameters
\begin{equation}
 q=\frac{Q}{Q_{\rm ext}(\text{model})},
 \qquad
 \widehat\mu=\frac{\mu}{M^2},
 \qquad
 R_o=\frac{r_o}{r_h}.
\label{eq:comparisonCoordinates}
\end{equation}
Equal $q$ compares positions within black-hole branches but generally means
different physical values of $Q/M$.  Equal $Q/M$ instead compares the same
asymptotic charge-to-mass coefficient but can place two models at different
distances from extremality.  In the quadratic theory even $Q_{\rm ext}$ varies
with $\widehat\mu$ along Eq.~\eqref{eq:EHcorrelatedExtremality}; independent
coordinate-wise maxima do not define a rectangular admissible domain.

For the same reason, the clean circular curves shown here are theoretical
capture edges, not synthetic interferometric images.  Translating them into a
constraint on a compact object additionally requires the mass--distance
calibration, emission and plasma transfer, instrument response, and a
likelihood definition of the measured diameter.  Their interpretation therefore
remains limited to the optical quantities defined above.

\subsection{Conventions and consistency across the literature}
\label{sec:NEDLiteratureConsistency}

Comparing NED lensing studies requires particular care with tensor placement,
action normalization, perturbative order, and the definition of a measured
frequency.  Table~\ref{tab:literatureConsistency} summarizes the principal
formula-level distinctions encountered in the examples above and identifies
which parts of the corresponding optical interpretation are affected.

\begin{table}[H]
\centering
\caption{Formula-level distinctions relevant to comparisons among NED optical
studies.}
\label{tab:literatureConsistency}
\footnotesize
\setlength{\tabcolsep}{4pt}
\renewcommand{\arraystretch}{0.96}
\begin{tabularx}{\textwidth}{
>{\raggedright\arraybackslash}p{0.19\textwidth}
>{\raggedright\arraybackslash}p{0.25\textwidth}
>{\raggedright\arraybackslash}p{0.23\textwidth}
>{\raggedright\arraybackslash}X}
\toprule
Topic & Governing relation & Literature context & Physical implication \\
\midrule
Characteristic co-metric and covariant line element in the
Einstein--Bronnikov and rational-NED constructions
\cite{Allahyari2020,Kruglov2020}
& Use $H=\tfrac12g_{\rm eff}^{\mu\nu}p_\mu p_\nu$ directly, or matrix-invert
$g_{\rm eff}^{\mu\nu}$ before writing a line element
& Copying the same coefficient placement into both tensors produces reciprocal
angular potentials that are not generally conformally related
& The background solutions are unchanged; the extraordinary light ring is
defined by the characteristic-Hamiltonian potential.\\
$h$-family observer-frequency map \cite{dePaulaAzreg2025}
& Evaluate $\omega=-p_\mu u^\mu$ from the characteristic covector
& The phase covector and the background-metric lowering of the ray tangent are
not generally equivalent
& The frequency map follows from the phase covector; the background and
characteristic ray trajectories are unaffected.\\
Rational-NED weak-field series \cite{Kruglov2020}
& Expand $-\mathcal F/(1+2\beta\mathcal F)$ as a geometric series
& The cubic coefficient is $-4\beta^2\mathcal F^3$, rather than the coefficient
shown in Eq.~(14)
& The subsequent quadratic matching and exact-model calculation are
unchanged.\\
Quadratic-model QED normalization \cite{HeisenbergEuler1936,Dunne2004}
& Convert the complete matter action, including field normalization, before
matching the $F^2$ and $G^2$ coefficients
& Gives $\mu_{\rm QED}=\alpha^2/(90\pi m_e^4)$ in the convention of
Eq.~\eqref{eq:action}
& Changes the physical scale and interpretation of $\mu$; it does not alter
the internally defined classical quadratic model.\\
Quadratic minus-branch first-order coefficient
\cite{dePaulaBirefringence2026}
& Expand $b_l=U(r_l)^{-1}$ at the stationary point
& Equation~\eqref{eq:EHimpactCorrection} gives $1.05$ at $M=Q=1$, whereas the
published symbolic expression gives $2.55$
& The difference is confined to the first-order minus coefficient; the plus
branch and the roots of the unexpanded characteristic factors are unchanged.\\
\bottomrule
\end{tabularx}
\end{table}

For the diagonal characteristic tensor of
Eq.~\eqref{eq:cometricInverse}, let $W=Y/X$.  The Hamiltonian gives
$U_{\rm H}^2=fW/r^2$, whereas copying the same coefficients into a covariant
line element gives $f/(Wr^2)$.  Their ratio $W^2(r)$ is generally not constant,
so the two potentials are not related by conformal freedom.  This distinction
applies to the Einstein--Bronnikov and rational-NED constructions cited above:
the background geometry is unchanged, while the extraordinary orbit is fixed
by the characteristic Hamiltonian.

The rational-NED expansion provides a separate algebraic point:
\begin{equation}
 -\frac{\mathcal F}{1+2\beta\mathcal F}
 =-\mathcal F+2\beta\mathcal F^2
 -4\beta^2\mathcal F^3+\Order(\mathcal F^4),
\label{eq:KruglovSeriesCorrection}
\end{equation}
whose cubic coefficient differs from the $-6\beta^2\mathcal F^3$ term shown in
Eq.~(14) of Ref.~\cite{Kruglov2020}.  Since the parameter matching is performed
at quadratic order and the subsequent calculation uses the exact rational
Lagrangian, this coefficient does not change those later results.

In the quadratic model, action-level QED matching and the first-order
minus-branch coefficient concern different questions.  The former determines
the physical interpretation of $\mu$; the latter concerns only the
perturbative expansion of one polarization.  As noted below
Eq.~\eqref{eq:EHimpactCorrection}, the unexpanded characteristic factors and
their numerical roots remain unchanged.  Consequently, an illustrative large
coupling should not be interpreted as a QED prediction, while the
nonperturbative optical curves of the classical model remain well defined in
their stated parameter domain.

\subsection{From optical calculations to observational claims}

The studies reviewed here relate analytic lensing quantities to shadow scales, ray-traced images, and redshifts. Comparing these predictions with observations requires an emission model, a specified observer geometry, and the appropriate calibration of the measured quantities.

The Dymnikova and $h$-family critical radii demonstrate how characteristic
geometry changes a strong-field scale.  A comparison with Sgr~A$^*$ or
M87$^*$ constrains a combination of metric parameters, photon geometry,
mass-to-distance calibration, and the observational definition of the quoted
diameter.  It does not by itself identify a unique NED Lagrangian.  The charge
values giving the largest theoretical shifts may also be near extremality and
larger than conventional astrophysical expectations for a net electric charge
\cite{dePaulaAzreg2025}.

The quadratic example provides a conceptually clean signature---two
polarization-dependent critical edges---but the benchmark $\mu$ is treated as
a free classical parameter and can be much larger than the QED value at the
relevant mass scale \cite{dePaulaBirefringence2026}.  Its plots demonstrate how
vacuum birefringence would modify compact-object optics; they are not a claim
that the displayed splitting is presently observable.

Backward ray tracing in the Dymnikova and quadratic studies finds the most
conspicuous differences near the critical curve
\cite{dePaulaDymnikova2023,dePaulaBirefringence2026}.  The detailed appearance
of a ray-traced image also depends on the emission model, plasma transfer, and
observer configuration, so it supplements rather than replaces the
branch-resolved optical quantities defined above.

Table~\ref{tab:observableInterpretation} summarizes the additional information
needed to relate an optical quantity to an observational conclusion.

\begin{table}[H]
\centering
\caption{Relation between theoretical optical quantities and their
observational interpretation in NED lensing.}
\label{tab:observableInterpretation}
\small
\begin{tabularx}{\textwidth}{
>{\raggedright\arraybackslash}p{0.23\textwidth}
>{\raggedright\arraybackslash}p{0.36\textwidth}
>{\raggedright\arraybackslash}X}
\toprule
Optical quantity & Additional ingredients before comparison with data &
Interpretation supported by the optical calculation \\
\midrule
Branch critical impact parameter $b_l^{(a)}$
& Mass--distance calibration, definition of the measured ring or shadow
diameter, emission morphology, plasma, and instrument response
& The selected characteristic branch predicts a capture separatrix and an
asymptotic angular scale.\\
Finite-position $\beta_k$ or $\beta_v$
& Observer tetrad, screen convention, source and receiver positions, and a
decision whether the measurement tracks phase or energy transport
& The branch predicts a local direction for the stated observer and endpoint
geometry.\\
Weak-deflection coefficient
& Lens equation, impact-parameter regime, finite endpoints, and enough orders
to reach the first nonlinear term allowed by the separate background and cone
expansions
& The coefficient tests a particular radial tail; lower-order benchmark
agreement is not a no-go result for strong-field deviations.\\
Polarization splitting $b_l^+-b_l^-$
& Polarimetric transfer through the source and plasma, bandwidth, angular
resolution, and a coupling inside the model's validity window
& Birefringence creates two branch-dependent critical scales in one
background; the benchmark is not automatically a QED forecast.\\
Emitter--observer redshift
& A physically allowed and stable emitter orbit, emission direction,
kinematic model, and the conserved phase covector
& The frequency map tests a background orbit and a characteristic branch
jointly; it is not determined by the ray curve alone.\\
\bottomrule
\end{tabularx}
\end{table}

\subsection{Conclusions}

NED changes a finite-distance light-deflection problem structurally.  The
Einstein equations determine a background spacetime, while the principal
electromagnetic equations determine polarization characteristics.  A
finite-distance construction must then distinguish the phase covector from the
Hamiltonian ray tangent before either is projected on an observer tetrad.
These steps cannot be collapsed into the phrase ``null geodesics of the
black-hole metric.''

The four representative models emphasize different aspects of this structure.
Dymnikova shows that the propagation layer can enter
weak bending before the metric's regularizing tail.  Hayward shows why central
regularity and a background light ring do not define a NED photon observable.
The $h$-family shows that high-order far-field suppression can coexist with
appreciable strong-field and redshift structure.  The quadratic two-invariant
model shows how one spacetime can carry two non-background polarization cones
and hence two critical scales.

Separate expansions of the background and propagation corrections
identify the leading inverse-radius terms that control weak-deflection sensitivity. Critical curves and redshifts provide complementary information because they weight the radial structure and characteristic geometry differently. These comparisons require a covariant effective line element constructed by inverting the characteristic co-metric. Background regularity, the Lorentzian signature of each characteristic branch, and the common timelike-covector condition for symmetric hyperbolicity must also be assessed separately.

The resulting hierarchy is
\begin{equation}
 \boxed{\Lag\ \longrightarrow\ g_{\mu\nu}\ \longrightarrow\
 \bar g_{(a)}^{\mu\nu}\ \longrightarrow\
 U_a(r),b_l^{(a)}\ \longrightarrow\
 \Psi_S^{(a)},\Psi_R^{(a)},\alpha_a\ \longrightarrow\
 \text{observable comparison}.}
\label{eq:finalWorkflow}
\end{equation}
Each calculation is restricted to the admissible domain of the chosen background solution and characteristic branch. The resulting light-ring curves, ray trajectories, critical edges, finite-observer angular scales, and redshift profiles show how the optical predictions depend on the compact-object geometry, endpoint positions, and observer.

\endgroup


\begingroup
\setlength{\emergencystretch}{2em}
\interdisplaylinepenalty=10000
\makeatletter

\@ifundefined{labelformat}{}{%
\labelformat{section}{#1}\labelformat{subsection}{#1}%
\labelformat{equation}{#1}\labelformat{figure}{#1}\labelformat{table}{#1}}
\makeatother
\section{Light deflection across methods and spacetimes}
\label{sec:c15}

The methods developed in this review offer several ways to answer the
same question: how much does a ray bend as it passes a gravitational lens?
The geodesic approach follows the ray, the invariant-angle approach relates
its direction to local measurements, and the Gauss--Bonnet approach expresses
the bending in terms of optical curvature. Their connection becomes most
useful when the lens departs from the simplest vacuum, spherical case.
Rotation changes the bending on opposite sides of the lens; charge changes
both the field and, in nonlinear electrodynamics, the propagation of light
through it. Close to capture, the instability of the photon orbit controls
how rapidly the angle grows.

We compare these effects through the deflection functions themselves.
Schwarzschild provides the mass baseline, followed by Kerr, Kottler, and
charged lenses with different photon characteristics. Within each case,
full orbit and curvature calculations are placed alongside the expansions
derived earlier in the review. The comparison then extends to three
consequences of those bending laws: the angular concentration of a scattered
beam, the frequency dependence introduced by plasma, and the spacing of
rays that execute successive turns around the lens. Together, these
examples show both where the familiar approximations work and which
physical distinctions survive when two lenses produce nearly the same
bending.

We use $G=c=1$. Numerical lengths are expressed in units of $M$ and times
in units of $M$; their dimensional counterparts are $GM/c^2$ and $GM/c^3$.
For vacuum and NED rays, $b=|L/E|$ denotes the impact-parameter magnitude;
the geometric impact parameter in plasma is denoted by $B$ below.
Angles are in radians unless otherwise indicated. For Kerr, $a=J/M>0$
and $s=+1$ ($s=-1$) denotes prograde (retrograde) motion.

\subsection{Schwarzschild: from Newtonian bending to optical curvature}
\label{subsec:c15sch}

\subsubsection{The mass contribution}

For distant sources and receivers, the Newtonian calculation in Section 4
gives $\alpha_{\rm N}=2M/b$ at leading order. The relativistic geodesic
calculation in Section 7 gives
\begin{equation}
 \alpha_{\rm geo}=2\int_{r_0}^{\infty}
 \frac{b\,dr}{r^2\sqrt{1-b^2f(r)/r^2}}-\pi,
 \qquad f(r)=1-\frac{2M}{r},\qquad
 b^2=\frac{r_0^2}{f(r_0)}.
 \label{eq:c15sch_exact}
\end{equation}
Its first two weak-deflection approximations are
\begin{equation}
 \alpha_{\rm geo}^{(1)}=\frac{4M}{b},\qquad
 \alpha_{\rm geo}^{(2)}=\frac{4M}{b}
       +\frac{15\pi M^2}{4b^2}.
 \label{eq:c15sch_series}
\end{equation}
The first relativistic term doubles the Newtonian prediction. In the
weak-field optical metric, spatial curvature supplies the contribution
missing from the Newtonian calculation, alongside the effect of
gravitational time dilation. The next term has a different meaning: it is
additional bending by the same mass, and its relative importance grows
as the ray passes closer to the lens.

The Gauss--Bonnet construction in Section 10 reaches the same first-order
result through the curvature of the optical surface
\cite{Gibbons_2008}. For
$dt^2=f^{-2}dr^2+r^2f^{-1}d\phi^2$, define
\begin{equation}
 dr_*=\frac{dr}{f},\qquad v=\frac{r}{\sqrt f},\qquad
 H(r)=\frac{dv}{dr_*}=\frac{1-3M/r}{\sqrt{1-2M/r}}.
 \label{eq:c15Hsch}
\end{equation}
Since $K=-v^{-1}d^2v/dr_*^2$, radial integration of the Gaussian curvature
over the exterior optical domain yields
\begin{equation}
 \alpha_{\rm GW}=\int_{\phi_S}^{\phi_R}[1-H(r(\phi))]\,d\phi.
 \label{eq:c15GW}
\end{equation}
Using $K=-2M/r^3+O(M^2/r^4)$ and the leading ray $r=b/\sin\phi$
gives $4M/b$. Using the full curvature and the full ray in
Eq.~\eqref{eq:c15GW} gives the complete angle in
Eq.~\eqref{eq:c15sch_exact}. The nonlinear growth of the angle is therefore also contained in the
curvature description; it enters through both the optical surface and the
shape of the ray.

At finite source and receiver radii, the OIA expression becomes
\begin{align}
 \alpha_{\rm OIA}&=\psi_S+\psi_R+\Phi-\pi,
 &\psi_i&=\arcsin\frac{b\sqrt{f(r_i)}}{r_i},\nonumber\\
 \Phi&=\sum_{i=S,R}\int_{r_0}^{r_i}
 \frac{b\,dr}{r^2\sqrt{1-b^2f/r^2}}.
 \label{eq:c15finite}
\end{align}
Here $\psi_S$ is the acute angle associated with the incoming ray; its
angle from the outward radial direction is $\pi-\psi_S$.
The OIA and finite-distance curvature calculations in Sections 11--13
give the common leading expression \cite{Ishihara_2016,Ono_2017}
\begin{equation}
 \alpha_{\rm OIA}^{(1)}=\alpha_{\rm GW}^{(1)}
 =\frac{2M}{b}\left(
 \sqrt{1-\frac{b^2}{r_S^2}}+
 \sqrt{1-\frac{b^2}{r_R^2}}\right).
 \label{eq:c15sch_finiteweak}
\end{equation}
Equation~\eqref{eq:c15GW}, evaluated along the corresponding finite ray,
also gives Eq.~\eqref{eq:c15finite} without a weak-field expansion.

\begin{figure}[!htb]
\centering
\includegraphics[width=\linewidth]{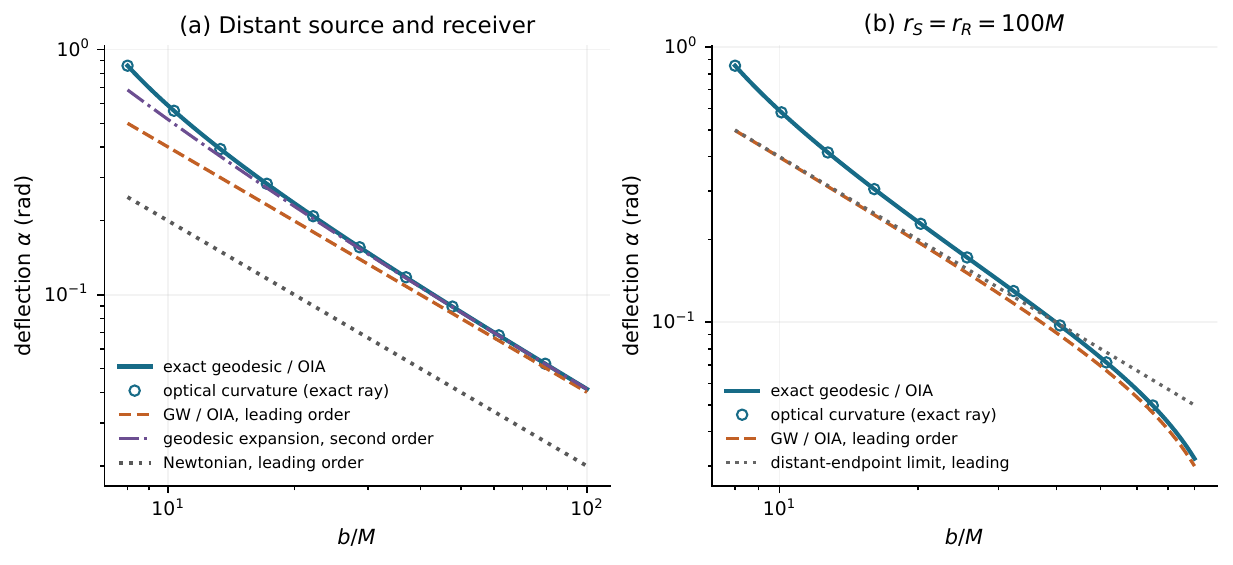}
\caption{Schwarzschild deflection from the methods developed in the review.
(a) Distant source and receiver: Newtonian bending, the leading GW/OIA
result, the second-order geodesic expansion, and the full geodesic integral.
Open circles are separate evaluations of the full optical-curvature
integral and coincide with the exact curve.
(b) The comparison at $r_S=r_R=100M$. The dotted curve is the
distant-endpoint leading expression, included to display the physical
change of observing geometry. All other curves and markers in this panel
use the stated finite radii.}
\label{fig:c15sch}
\end{figure}

\subsubsection{From weak bending to smaller impact parameters}

The difference between the approximations is substantial at the smaller
impact parameters in Fig.~\ref{fig:c15sch}(a). At $b=10M$, the full angle
is about $0.59040$ rad, compared with $0.4$ rad at first order and
$0.51781$ rad at second order. By $b=100M$, the three values have drawn
much closer: $0.0412225$, $0.04$, and $0.0411781$ rad, respectively.
The second-order term accounts for most of the additional bending there.
The curvature markers follow the full geodesic curve throughout, including
the region where the leading expression is visibly inadequate.

\begin{table}[!htb]
\centering\small
\caption{Schwarzschild deflection angles for representative points in
Fig.~\ref{fig:c15sch}. The second-order column is the asymptotic expression
in Eq.~\eqref{eq:c15sch_series}; it is used only for distant endpoints.}
\label{tab:c15sch}
\begin{tabular}{rrrrr}\toprule
$b/M$ & $R/M$ & Full angle & Leading OIA/GW & Asymptotic second order\\\midrule
10 & $\infty$ & 0.5903958 & 0.4000000 & 0.5178097 \\
30 & $\infty$ & 0.1482481 & 0.1333333 & 0.1464233 \\
100 & $\infty$ & 0.0412225 & 0.0400000 & 0.0411781 \\
30 & 100 & 0.1420779 & 0.1271919 & -- \\
80 & 100 & 0.0319356 & 0.0300000 & -- \\

\bottomrule\end{tabular}
\end{table}

A finite source--receiver separation introduces another scale, visible in
Fig.~\ref{fig:c15sch}(b). With $r_S=r_R=R$ fixed, a larger impact parameter
both weakens the gravitational interaction and reduces the angular span
covered by the ray between the endpoints. The latter effect gives the
factor $\sqrt{1-b^2/R^2}$ at leading order. It is small when $b\ll R$,
but becomes pronounced as the ray's closest approach moves toward the
source and receiver radii. Even a weak gravitational field can therefore
produce a noticeable departure from the distant-source formula.

\subsection{Kerr: mass bending and frame dragging}
\label{subsec:c15kerr}

\subsubsection{The two orbital orientations}

A rotating lens distinguishes rays passing with and against its spin.
For $\ell=sb$, the equatorial Kerr geodesic equations can be
written as \cite{Iyer1}
\begin{align}
 P&=r^2+a^2-a\ell,&
 \Delta&=r^2-2Mr+a^2,&
 \mathcal R&=P^2-\Delta(\ell-a)^2,\nonumber\\
 \left|\frac{d\phi}{dr}\right|
 &=\frac{|\ell-a+aP/\Delta|}{\sqrt{\mathcal R}}.
 \label{eq:c15kerr_orbit}
\end{align}
The largest exterior root of $\mathcal R$ gives the scattering turning
point. Twice the integral to infinity, minus $\pi$, gives the exact
asymptotic deflection on each side of the black hole.

The OIA and GW--OIA derivations in Sections 11 and 12 give
\begin{equation}
 \alpha_s^{\rm OIA}=\alpha_s^{\rm GW-OIA}
 =\left(\frac{2M}{b}-s\frac{2Ma}{b^2}\right)D+\cdots,
 \qquad
 D=\sqrt{1-\frac{b^2}{r_S^2}}+
   \sqrt{1-\frac{b^2}{r_R^2}}.
 \label{eq:c15kerr_leading}
\end{equation}
The first term arises from the mass contribution to optical curvature;
the second is the leading gravitomagnetic contribution. In the
GW--OIA representation, the latter includes the line integral of
geodesic curvature associated with the rotating Fermat geometry
\cite{Werner_2012,Ono_2017}. For distant endpoints, $D\to2$.
The complete asymptotic expansion through second order in $M/b$, at fixed
$a/M$, is
\begin{equation}
 \alpha_s^{(2)}=\frac{4M}{b}
 +\frac{15\pi M^2/4-s\,4Ma}{b^2}
 +O(M^3/b^3).
 \label{eq:c15kerr_second}
\end{equation}
Equations~\eqref{eq:c15kerr_leading} and \eqref{eq:c15kerr_second}
therefore retain the same leading spin term, while the latter also includes
the second-order monopole contribution.

For comparison with the full geodesic result, the generalized curvature
expression introduced in Section 10 is evaluated without expansion
\cite{Huang_2024,takahashi2023equivalence}. On the equatorial optical
surface, write
\begin{align}
 dt&=\sqrt{\gamma_{rr}dr^2+\gamma_{\phi\phi}d\phi^2}
       +\beta_\phi d\phi,\nonumber\\
 \gamma_{rr}&=\frac{r^2}{\Delta A},\qquad
 \gamma_{\phi\phi}=\frac{\Delta}{A^2},\qquad
 \beta_\phi=-\frac{2Ma}{r-2M},\qquad A=1-\frac{2M}{r}.
 \label{eq:c15kerr_fermat}
\end{align}
With the azimuth $\varphi=s\phi$ oriented along the ray,
\begin{align}
 H&=\frac{1}{2\sqrt{\gamma_{rr}\gamma_{\phi\phi}}}
       \frac{d\gamma_{\phi\phi}}{dr},\nonumber\\
 T_s&=-s\frac{d\beta_\phi}{dr}
 \sqrt{\frac{1}{\gamma_{\phi\phi}}
       \left(\frac{dr}{d\varphi}\right)^2+\frac{1}{\gamma_{rr}}},
 &\alpha_s^{\rm gen.GW}&=\int[1-H+T_s]\,d\varphi.
 \label{eq:c15kerr_HC}
\end{align}
The orientation factor in $T_s$ gives the change of sign under reversal
of orbital direction. The trajectories used here remain outside the
equatorial ergosurface, where this stationary optical representation is
regular.

\begin{figure}[!htb]
\centering
\includegraphics[width=\linewidth]{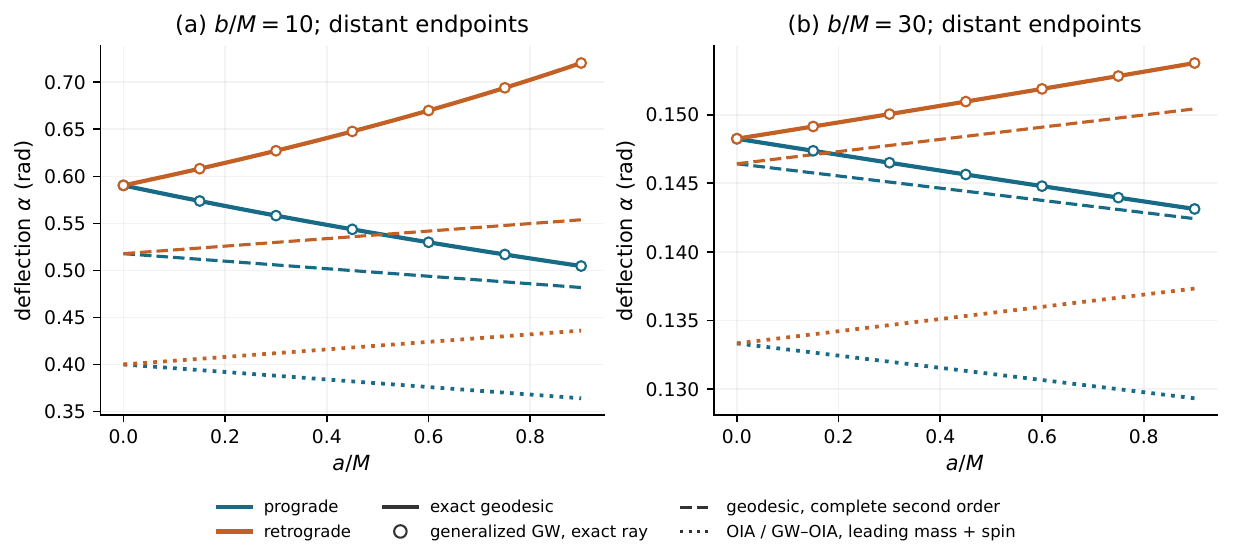}
\caption{Deflection versus Kerr spin at fixed impact parameter, with source
and receiver at infinity. Colour distinguishes orbital orientation and
line style distinguishes the expressions being evaluated. Circles are the
unexpanded generalized GW calculation. The dotted OIA/GW--OIA curves retain
leading mass and spin terms; the dashed geodesic expansion also retains
$15\pi M^2/(4b^2)$. Panels (a) and (b) compare the same methods at
$b=10M$ and $b=30M$, respectively.}
\label{fig:c15kerr}
\end{figure}

\subsubsection{Separating the mass and spin responses}

At fixed $|L/E|$, prograde rays bend less and retrograde rays bend more as
$a$ increases over the range shown in Fig.~\ref{fig:c15kerr}. The splitting
is much larger at $b=10M$ than at $b=30M$, as expected from the leading
$b^{-2}$ spin term. The full curves also show that the response ceases to
be linear in $a$ when the ray samples the stronger field.

It is useful to distinguish this splitting from the common upward shift
produced by the second-order mass term. For a Kerr black hole,
$|a|\le M$, so $15\pi M^2/(4b^2)$ is at least
$15\pi/16\simeq2.95$ times the magnitude of the leading spin correction
on either side. This explains why adding the complete second order moves
both curves appreciably even though the leading formula already contains
frame dragging. The common shift and the separation of the two curves
carry different physical information.

The even and odd combinations under orbital reversal make that distinction
explicit:
\begin{equation}
 E_b=\frac{\alpha_-+\alpha_+}{2},\qquad
 O_b=\frac{\alpha_--\alpha_+}{2}.
 \label{eq:c15parity}
\end{equation}
To the orders displayed above,
\begin{equation}
 E_b\longrightarrow\frac{4M}{b}+\frac{15\pi M^2}{4b^2},
 \qquad O_b\longrightarrow\frac{4Ma}{b^2}
 \label{eq:c15parityweak}
\end{equation}
at infinity. The leading finite-distance expressions instead contain
$E_b=2MD/b$ and $O_b=2MaD/b^2$, so that $bO_b/E_b\to a$ at that order.
The common distance factor cancels in this leading spin-to-mass ratio.
For an actual pair of images, however, the impact parameters need not be
equal: their relation to the source position must be supplied by the lens
equation before the asymmetry can be attributed to spin.

\subsection{Kottler: cosmological curvature and the measured direction}
\label{subsec:c15kottler}

Kottler spacetime brings the local-angle construction into sharper focus.
Its static region has no asymptotically Euclidean optical boundary, so the
bending must be related to the source and receiver within that region.
The orbit equation already hints at a subtlety:

\begin{equation}
 f_\Lambda(r)=1-\frac{2M}{r}-\frac{\Lambda r^2}{3},\qquad
 \left(\frac{du}{d\phi}\right)^2
 =\frac{1}{b^2}+\frac{\Lambda}{3}-u^2+2Mu^3.
 \label{eq:c15kottler_orbit}
\end{equation}
Differentiating gives $u''+u=3Mu^2$, with no explicit $\Lambda$ term.
Yet $\Lambda$ remains in the first integral that selects the orbit and in
the metric used to measure its direction.
The Rindler--Ishak angle of the outward ray measured by a static observer
at radius $R$ is \cite{Rindler:2007zz}
\begin{equation}
 \tan\psi_{\rm RI}=\frac{R\sqrt{f_\Lambda(R)}}{|dr/d\phi|_R},
 \qquad
 \sin\psi_{\rm RI}=\frac{b\sqrt{f_\Lambda(R)}}{R}.
 \label{eq:c15RI}
\end{equation}
For symmetric endpoints $r_S=r_R=R$, the corresponding total bending is
$\alpha=2\psi_{\rm RI}+\Phi-\pi$. Combining the two local directions with the swept azimuth puts the
Rindler--Ishak construction in the same form as the OIA total angle.

The finite-distance OIA expansion obtained by Ishihara \emph{et al.}
\cite{Ishihara_2016}, retaining the leading $M$, $\Lambda$, and mixed
$M\Lambda$ terms, is
\begin{align}
 \alpha_{\rm OIA}^{\rm weak}
 =&\ \frac{2M}{b}\sum_{i=S,R}\sqrt{1-b^2/r_i^2}
 -\frac{\Lambda b}{6}\sum_{i=S,R}r_i\sqrt{1-b^2/r_i^2}
 \nonumber\\
 &+\frac{M\Lambda b}{6}
       \sum_{i=S,R}\frac{1}{\sqrt{1-b^2/r_i^2}}+\cdots.
 \label{eq:c15kottler_series}
\end{align}
The omitted terms include higher mass orders and higher powers of
$\Lambda$ with their associated length factors. At fixed $b/r_i$, the
expansion is useful for $M/b\ll1$ and $|\Lambda|r_i^2\ll1$.
The unexpanded generalized GW expression is
\begin{equation}
 \alpha_{\rm gen.GW}=
 \int_{\phi_S}^{\phi_R}
 \left[1-\frac{1-3M/r}{\sqrt{f_\Lambda(r)}}\right]\,d\phi.
 \label{eq:c15kottler_HC}
\end{equation}
This finite-domain form uses the stationary optical geometry between the
endpoints and remains applicable inside the static patch.

\begin{figure}[!htb]
\centering
\includegraphics[width=\linewidth]{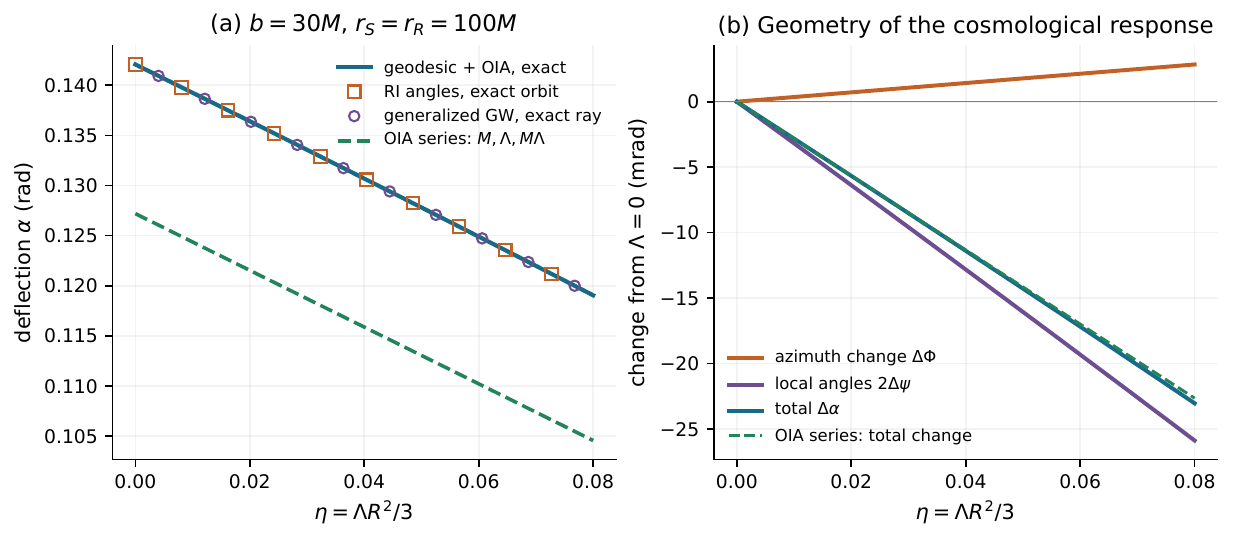}
\caption{Method comparison in Kottler spacetime at fixed $b=30M$ and
$r_S=r_R=R=100M$. The horizontal variable is $\eta=\Lambda R^2/3$.
(a) The full geodesic/OIA result, the RI angle reconstructed from the
coordinate slope, the generalized GW integral, and the weak OIA expression.
The exact local-angle and curvature markers coincide with the full curve.
(b) Changes relative to $\Lambda=0$: the azimuthal contribution,
the contribution from the two local angles, and their sum. The finite
values of $\eta$ illustrate the response of this static geometry.}
\label{fig:c15kottler}
\end{figure}

The full calculations in Fig.~\ref{fig:c15kottler}(a) give a decreasing
angle as $\Lambda$ increases. The weak OIA curve follows the same trend
from a lower baseline, since it omits the nonlinear mass terms seen in
the Schwarzschild comparison. Subtracting each curve's $\Lambda=0$
value exposes the cosmological response more directly.

At fixed $b$, the effective orbit parameter
$B=(b^{-2}+\Lambda/3)^{-1/2}$ decreases as $\Lambda$ increases. The
coordinate azimuthal span consequently increases in this example, while
the local angles decrease. At $\eta=0.08$, the two contributions to the
change in bending are approximately $+2.844$ and $-25.878$ milliradians;
their sum is $-23.034$ milliradians. The two effects oppose one another, with the change in local direction
dominating here. Thus the absence of $\Lambda$ from the differentiated
orbit equation does not remove it from the bending measured by static
observers. The finite-domain curvature integral includes both effects.

The result is specific to fixed $b$, fixed areal endpoint radii, and static
observers in Kottler spacetime. A cosmological lens model must also relate
these quantities to source redshift, angular-diameter distances, and the
receiver's motion \cite{c15LebedevLake2013}.

\subsection{Charge and nonlinear electrodynamics: geometry and propagation}
\label{subsec:c15ned}

\subsubsection{Background geometry and photon propagation}

For the nonlinear electromagnetic theories discussed in Section 14,
specifying the spacetime metric does not yet specify every photon path.
The characteristic cone of the chosen polarization also enters. In the
spherical examples, both ingredients can be written in the conformal
representative

\begin{equation}
 d\bar s^2=-f(r)dt^2+\frac{dr^2}{f(r)}
             +\frac{r^2}{W(r)}d\Omega^2,
 \qquad V(r)=\frac{f(r)W(r)}{r^2}.
 \label{eq:c15cone}
\end{equation}
The background-null case has $W=1$. For distant endpoints, the
Hamiltonian and optical-curvature expressions are
\begin{align}
 \alpha_{\rm H}
 &=2\int_{r_0}^{\infty}
      \frac{bW(r)\,dr}{r^2\sqrt{1-b^2V(r)}}-\pi,
 &b^2V(r_0)&=1,\nonumber\\
 \alpha_{\rm GW}
 &=2\int_{r_0}^{\infty}[1-H_W(r)]
      \frac{bW(r)\,dr}{r^2\sqrt{1-b^2V(r)}},\nonumber\\
 H_W&=\sqrt{\frac{f}{W}}
       \left[1-\frac r2\left(\frac{f'}f+\frac{W'}W\right)\right].
 \label{eq:c15ned_methods}
\end{align}
The second formula follows from
$v=r/\sqrt{fW}$ and $dr_*=dr/f$, just as in the Schwarzschild calculation.
Here $f$ describes the gravitational background and $W$ describes the
angular deformation of the selected photon cone. Their separate derivatives
in $H_W$ show how each enters the optical curvature. For the models below, $W\to1$ at infinity, where optical
and material-observer direction angles coincide.

\subsubsection{Charge at fixed mass: RN and Dymnikova lenses}

The RN metric has $f_{\rm RN}=1-2M/r+Q^2/r^2$ and $W=1$.
For the Dymnikova electric solution \cite{Dymnikova2004,dePaulaDymnikova2023},
\begin{align}
 z&=\frac{\pi Q^2}{8M},\qquad
 f_{\rm ID}=1-\frac{4M}{\pi r}
 \left[\arctan\frac rz-\frac{rz}{r^2+z^2}\right],\nonumber\\
 W_{\rm ID}^{\rm ord}&=1,\qquad
 W_{\rm ID}^{\rm ext}=\frac{r^2-2z^2}{r^2+z^2}.
 \label{eq:c15ID}
\end{align}
The ordinary branch samples the background null cone and the extraordinary
branch samples the deformed electromagnetic cone. These are the two
characteristic factors of the local $\mathcal L(F)$ theory
\cite{ObukhovRubilar2002}; they coincide in the Maxwell limit.
Regularity also ties the constitutive scale to the mass and charge:
the structural function in Section 14 has coupling
$\alpha_{\rm NED}=z^2/(\sqrt2\,|Q|)$ for $Q\ne0$.
Thus a fixed-mass ID charge sequence samples regular solutions with
different constitutive couplings \cite{dePaulaDymnikova2023}.
Their far-field structure is
\begin{equation}
 f_{\rm ID}=1-\frac{2M}{r}+\frac{Q^2}{r^2}
                -\frac{2Q^2z^2}{3r^4}+O(r^{-6}),\qquad
 W_{\rm ID}^{\rm ext}=1-\frac{3z^2}{r^2}+O(r^{-4}).
 \label{eq:c15IDtails}
\end{equation}
The departure of the photon cone from its Maxwell limit falls off as
$r^{-2}$, whereas the first difference between the two background metrics
falls off as $r^{-4}$. The propagation law can therefore affect the weak
angle before the regularizing part of the metric becomes visible.

Expanding Eq.~\eqref{eq:c15ned_methods}, or the equivalent optical
curvature expression to the same order, gives
\begin{align}
 \alpha_{\rm RN}
 &=\frac{4M}{b}+\frac{3\pi(5M^2-Q^2)}{4b^2}
        +\frac{16M(8M^2-3Q^2)}{3b^3}+O(b^{-4}),\nonumber\\
 \alpha_{\rm ID}^{\rm ext}
 &=\frac{4M}{b}+\frac{3\pi(5M^2-Q^2+z^2)}{4b^2}
        +\frac{16M(8M^2-3Q^2+3z^2)}{3b^3}+O(b^{-4}).
 \label{eq:c15charged_series}
\end{align}
The ordinary ID branch shares the displayed RN coefficients.
These are the weak-deflection expressions discussed in
Section 14; the full integrals extend their comparison to smaller impact
parameters.

\begin{figure}[!htb]
\centering
\includegraphics[width=\linewidth]{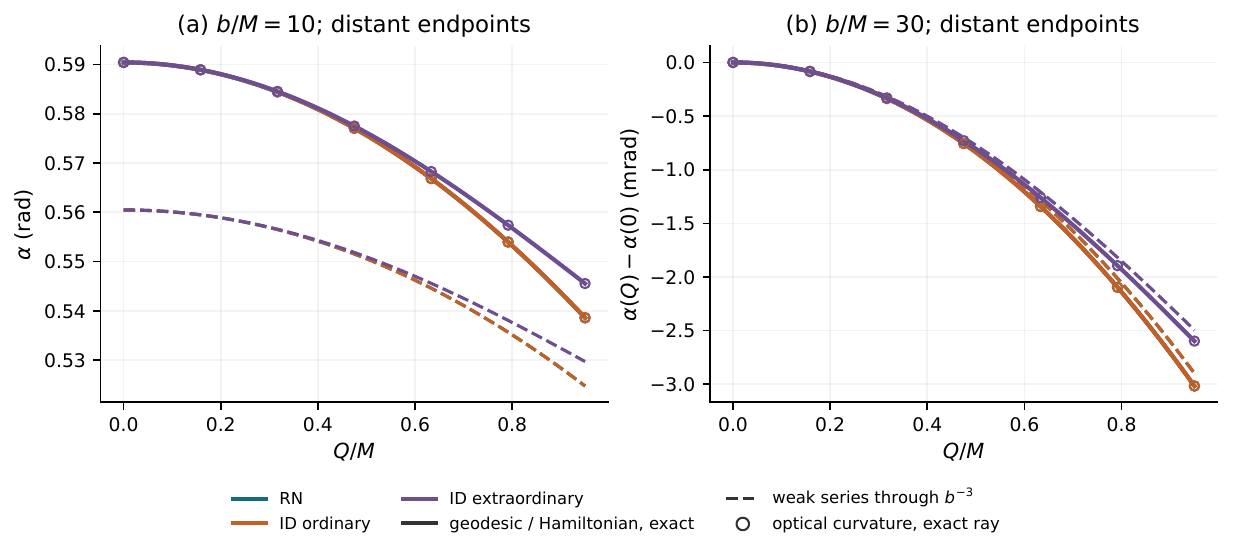}
\caption{Deflection as a function of physical charge at equal ADM mass.
Solid curves evaluate the exact geodesic or characteristic-Hamiltonian
integral; dashed curves use the weak series through $b^{-3}$; open circles
evaluate the exact optical-curvature integral. (a) Total deflection at
$b=10M$. (b) Charge-induced change at $b=30M$, subtracting the $Q=0$
value within each expression. The three colours distinguish RN, ordinary
Dymnikova, and extraordinary Dymnikova propagation. The RN and ordinary-ID
weak series coincide at the displayed order.}
\label{fig:c15charge}
\end{figure}

At the same $M$ and $b$, the charge term reduces RN bending relative to
Schwarzschild. The ordinary Dymnikova curve remains close to RN in
Fig.~\ref{fig:c15charge}, but the extraordinary mode bends more than either
of them. Its positive $z^2$ term partly offsets the negative $Q^2$ term.
This offset comes from the photon cone and is already present at orders
where the regularizing term in the background lapse has not entered.

The same falloff argument connects the other NED examples in Section 14.
Hayward background-null bending receives its first model-dependent metric
term at $b^{-4}$. The leading angular-cone term also appears at $b^{-4}$
in the $h=2$ and quadratic models, but only at $b^{-8}$ in the $h=0$
model. Dymnikova's extraordinary mode is more prominent in low-order weak
bending because its cone already differs at $r^{-2}$. Agreement through a
few weak orders can thus persist between lenses whose propagation laws
are quite different closer to the compact object.

\subsubsection{From the bending angle to the concentration of a beam}

The bending function also determines how a bundle of incident rays is
redistributed in angle. Consider a uniform beam arriving from infinity
and retain the direct branch, for which the deflection is the scattering
angle $0<\vartheta=\alpha(b)<\pi$. An incident annulus of area
$2\pi b\,|db|$ maps into the outgoing solid angle
$2\pi\sin\vartheta\,d\vartheta$. Conservation of ray number then gives
the direct-branch differential cross section
\begin{equation}
 \frac{d\sigma_0}{d\Omega}
 =\frac{b}{\sin\vartheta}\left|\frac{db}{d\vartheta}\right|
 =\frac{b}{\sin\vartheta\,|\alpha'(b)|}.
 \label{eq:c15crosssection}
\end{equation}
The complete geometrical cross section sums over the other winding
branches as well \cite{c15Oliveira2018}. Here the direct branch lets us
follow the connection between the weak bending function and the angular
concentration of a beam. This quantity is a scattering Jacobian; image
magnification additionally depends on the source--lens--observer geometry.

To compare lenses without the universal forward divergence obscuring
their differences, define
\begin{equation}
 \mathcal F(\vartheta)=
 \frac{\vartheta^3\sin\vartheta}{16M^2}
 \frac{d\sigma_0}{d\Omega}.
 \label{eq:c15focusing}
\end{equation}
For $\alpha=A_1/b+A_2/b^2+A_3/b^3+\cdots$, with $A_1=4M$,
inverting the series gives
\begin{align}
 b(\vartheta)
 &=\frac{A_1}{\vartheta}+\frac{A_2}{A_1}
   +\frac{A_1A_3-A_2^2}{A_1^3}\,\vartheta
   +O(\vartheta^2),\nonumber\\
 \mathcal F(\vartheta)
 &=1+\frac{A_2}{A_1^2}\,\vartheta+O(\vartheta^3).
 \label{eq:c15focusing_series}
\end{align}
There is no quadratic term in $\mathcal F$: the contribution from the
linear term of $b(\vartheta)$ cancels when $b$ is multiplied by
$|db/d\vartheta|$. Thus the next coefficient visible in the angle does
not automatically provide an independent next-order contribution to the
normalized angular concentration.

\begin{figure}[!htb]
\centering
\includegraphics[width=\linewidth]{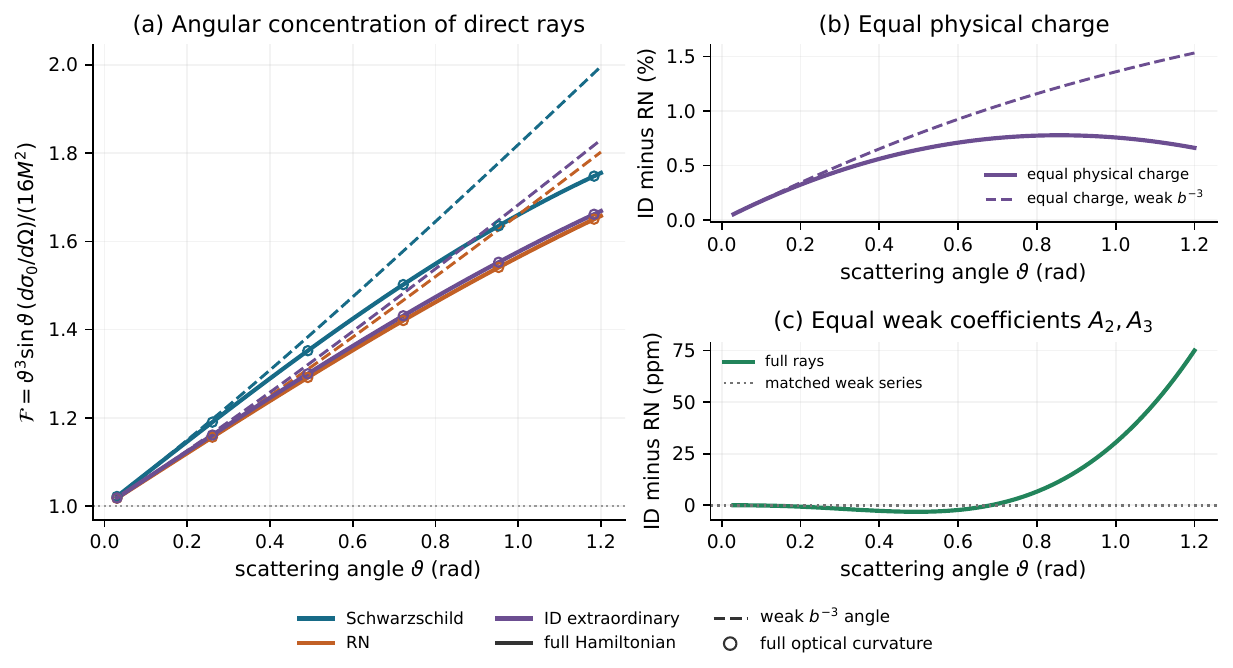}
\caption{Angular concentration of direct rays. (a) Schwarzschild, RN, and
extraordinary Dymnikova propagation, with $Q/M=0.95$ for both charged
models. Solid curves use the full bending function and its derivative;
dashed curves use the angle truncated at $b^{-3}$. Circles use separately
differentiated optical-curvature integrals. (b) The relative difference
$\mathcal F_{\rm ID}/\mathcal F_{\rm RN}-1$ at equal physical charge,
expressed as a percentage. (c) The same ratio after changing the RN charge
to match the two displayed weak coefficients, now expressed in parts per
million. The curves concern ray-number redistribution
on the direct branch, with distant incident and outgoing directions.}
\label{fig:c15scattering}
\end{figure}

Figure~\ref{fig:c15scattering}(a) shows how the increasing nonlinear
bending changes the angular concentration beyond its leading value
$\mathcal F=1$. At equal physical charge, the extraordinary Dymnikova
branch gives a larger concentration than RN over the plotted range.
Much of this difference can be absorbed into the weak coefficients:
choosing $Q_{\rm RN}^2=Q_{\rm ID}^2-z^2$ matches both $A_2$ and $A_3$.
For $Q_{\rm ID}=0.95M$, this requires $Q_{\rm RN}\simeq0.88142M$.
At $\vartheta=1.2$, the ID--RN contrast falls from about $0.663\%$ at
equal charge to $0.00750\%$ after the match. The residual even changes
sign at smaller angles. Differentiating the angle exposes the way
neighbouring rays separate, but it does not remove the close weak-field
resemblance of these two optical geometries.

\subsubsection{Polarization and the relation between bending and travel time}

The classical quadratic action
$\mathcal L(F,G)=F-\mu(F^2+7G^2/4)$ provides two electromagnetic
branches on one magnetic background \cite{dePaulaBirefringence2026}:
\begin{align}
 f&=1-\frac{2M}{r}+\frac{Q^2}{r^2}-\frac{2\mu Q^4}{5r^6},\nonumber\\
 W_+&=\frac{1-4\mu Q^2/r^4}{1+10\mu Q^2/r^4},
 &W_-&=\frac{1-12\mu Q^2/r^4}{1-4\mu Q^2/r^4}.
 \label{eq:c15quadratic}
\end{align}
The two modes share the RN series through $b^{-3}$, so the first
polarization difference lies beyond that common approximation. Its radial
scaling can be obtained directly from Fermat's principle. Write
$W_a=1-k_a/r^n+\cdots$, hold $f$ fixed, and use the weak reference ray
$r^2=b^2+x^2$, where $x$ is the longitudinal coordinate along the path.
Fermat's optical line element gives
\begin{equation}
 dt=\left[1+\frac{k_ab^2}
              {2(b^2+x^2)^{(n+2)/2}}\right]dx+\cdots.
 \label{eq:c15fermat_tail}
\end{equation}
Stationarity removes the first-order displacement of the path from the
travel-time variation when its spatial endpoints are fixed. The resulting
distant-endpoint delay and scattering response are
\begin{equation}
 \delta T_a=C_n k_ab^{1-n},\qquad
 \delta\alpha_a=-\frac{d\delta T_a}{db}
       =(n-1)C_n\frac{k_a}{b^n},\qquad
 C_n=\frac{\sqrt\pi\,\Gamma[(n+1)/2]}{2\Gamma[(n+2)/2]}.
 \label{eq:c15fermat_response}
\end{equation}
The same angular coefficient follows directly from the leading
optical-curvature integral. With $f=1$ at this order,
$H_W=1-(n-1)k_a/(2r^n)+\cdots$; substituting the straight ray into
Eq.~\eqref{eq:c15GW} gives
$\frac12(n-1)k_ab^{-n}\int_0^\pi\sin^n\phi\,d\phi$,
which equals Eq.~\eqref{eq:c15fermat_response}.

For Eq.~\eqref{eq:c15quadratic}, $n=4$, $k_+=14\mu Q^2$, and
$k_-=8\mu Q^2$. Thus
\begin{equation}
 \Delta\alpha_{\rm weak}=\frac{27\pi\mu Q^2}{8b^4},\qquad
 \Delta T_{\rm weak}=\frac{9\pi\mu Q^2}{8b^3},\qquad
 \frac{b\Delta\alpha_{\rm weak}}{\Delta T_{\rm weak}}=3,
 \label{eq:c15polarization_weak}
\end{equation}
where $\Delta\alpha=\alpha_+-\alpha_-$ and
$\Delta T=T_+-T_-$. The shared metric contribution cancels from this
leading polarization difference. The full characteristic integrals can now be compared with this leading
response without the much larger common mass contribution obscuring the
splitting.

\begin{figure}[!htb]
\centering
\includegraphics[width=\linewidth]{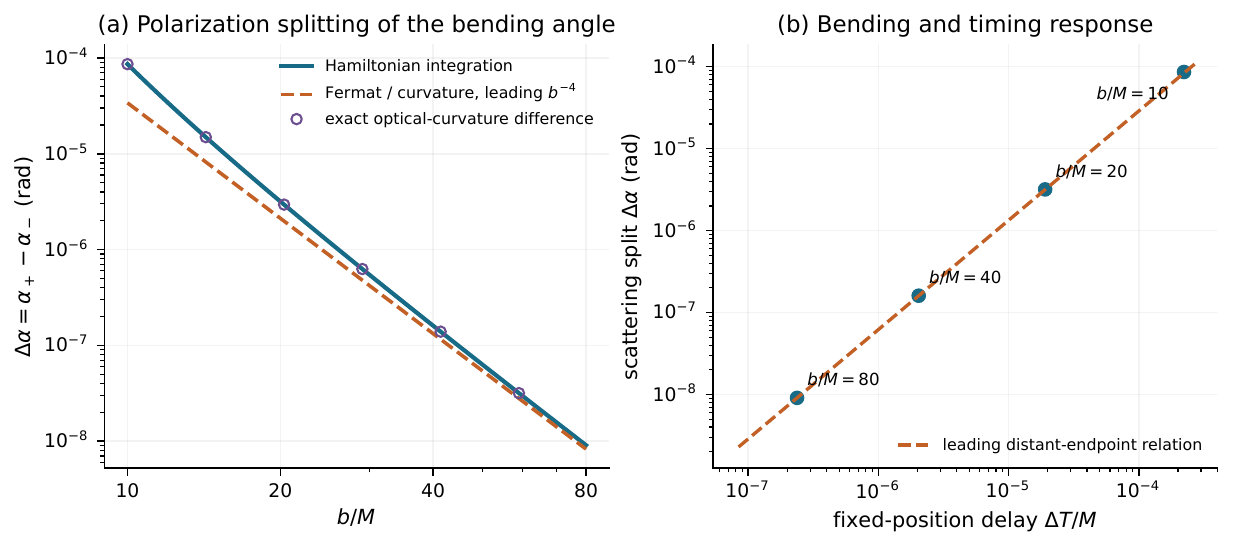}
\caption{Quadratic-NED polarization comparison at $Q/M=0.8$ and
$\mu/M^2=0.05$. (a) The difference of the full characteristic bending
angles, the independently integrated curvature difference, and the leading
$b^{-4}$ Fermat/curvature expression. (b) Scattering splitting paired with
the time difference between pulses connecting the same finite spatial
endpoints, at $R=5b_{\rm ref}$. Labels give the common reference scale
$b/M=b_{\rm ref}/M$; the two timing rays have separately solved impact
parameters. The dashed relation follows from the distant-endpoint leading
expressions in Eq.~\eqref{eq:c15polarization_weak}.}
\label{fig:c15polarization}
\end{figure}

As $b$ increases, the full splitting in Fig.~\ref{fig:c15polarization}(a)
approaches the predicted $b^{-4}$ behaviour. At smaller $b$, gravitational
bending carries the ray through a region where the cone deformation is
stronger, enhancing the separation of the two modes. A single curve based
on their common $b^{-3}$ series would miss this polarization dependence.

The timing comparison in Fig.~\ref{fig:c15polarization}(b) uses a reference
RN ray with impact parameter $b_{\rm ref}$ to set the source--receiver
angular separation $\Phi_{\rm ref}$. For each polarization, its own
impact parameter is obtained from $\Phi_a(b_a;R,R)=\Phi_{\rm ref}$, and
\begin{equation}
 T_a=2\int_{r_{0,a}}^R
 \frac{dr}{f\sqrt{1-b_a^2fW_a/r^2}},\qquad
 \Delta\tau_R=\sqrt{f(R)}\,\Delta T.
 \label{eq:c15timing}
\end{equation}
The factor $\sqrt{f(R)}$ converts the coordinate delay to the static
receiver's proper time. The two panels pair scattering at a common
asymptotic $b$ with arrival times between common positions; matching those
positions requires the separate values $b_+$ and $b_-$.

\begin{table}[!htb]
\centering\small
\caption{Bending and timing for the quadratic model in
Fig.~\ref{fig:c15polarization}. The delay is calculated between common
finite spatial endpoints. In the last column,
$\Delta T^{(1)}_\infty$ is the first-order cone response evaluated on the
full RN reference orbit with distant endpoints.}
\label{tab:c15pol}
\begin{tabular}{rrrr}\toprule
$b/M$ & $\Delta\alpha$ & $\Delta T/M$ & $b\Delta\alpha/\Delta T^{(1)}_\infty$\\\midrule
10 & $8.604\times10^{-5}$ & $2.224\times10^{-4}$ & 3.8669 \\
20 & $3.177\times10^{-6}$ & $1.907\times10^{-5}$ & 3.3317 \\
40 & $1.604\times10^{-7}$ & $2.037\times10^{-6}$ & 3.1492 \\
80 & $9.090\times10^{-9}$ & $2.368\times10^{-7}$ & 3.0711 \\

\bottomrule\end{tabular}
\end{table}

The reference-orbit response used in the last column is
\begin{equation}
 \Delta T^{(1)}_\infty=6\mu Q^2b^2
 \int_{r_{0,\rm RN}}^\infty
 \frac{dr}{r^6\sqrt{1-b^2f_{\rm RN}(r)/r^2}}.
 \label{eq:c15timing_response}
\end{equation}
It follows from the first variation of the optical angular coefficient,
$\delta\gamma_{\phi\phi}=6\mu Q^2/(f_{\rm RN}r^2)$, along the
unperturbed RN orbit. It retains the gravitational bending of that orbit
while linearizing the polarization difference in $\mu$.

The leading ratio $b\Delta\alpha/\Delta T\to n-1$ is especially useful:
it depends on the radial power of the cone correction but not on its
amplitude. The quadratic model therefore approaches $3$. At $b=10M$,
the ratio formed with the full reference-orbit response is about $3.87$.
A small constitutive correction can still be sampled by a substantially
bent ray, so weak polarization splitting does not by itself imply weak
gravitational bending. Here $\mu$ is the coupling of the stated classical
quadratic-NED model, with both characteristics regular in the exterior
region sampled by the rays.

\subsection{Frequency-dependent bending in a plasma}
\label{subsec:c15plasma}

The NED examples change the propagation law through the electromagnetic
action. A plasma introduces a different effect: dispersion. To isolate it,
consider Schwarzschild spacetime with a prescribed homogeneous, cold,
nonmagnetized test plasma, neglecting the medium's self-gravity and
absorption. A constant plasma frequency $\omega_p$ does not imply a
constant refractive index, because the frequency measured locally changes
with gravitational redshift. At a fixed frequency $E=\omega_\infty$,
\begin{equation}
 \chi=\frac{\omega_p^2}{\omega_\infty^2}<1,\qquad
 v_\infty=\sqrt{1-\chi},\qquad
 n^2(r)=1-\chi f(r),\qquad f=1-\frac{2M}{r}.
 \label{eq:c15plasma_index}
\end{equation}
The spatial rays can be described either by the dispersive Hamiltonian
$\mathcal H=\tfrac12(g^{\mu\nu}p_\mu p_\nu+\omega_p^2)=0$ or by a
frequency-dependent optical metric \cite{c15CrisnejoGallo2018}.

The geometric impact parameter is now
$B=|L|/\sqrt{E^2-\omega_p^2}$, so that $|L/E|=Bv_\infty$.
Keeping $B$ fixed compares rays with the same incident offset at infinity.
The Hamiltonian gives
\begin{align}
 F_\omega(r)&=1-f(r)
       \left(\chi+\frac{B^2v_\infty^2}{r^2}\right),
 &F_\omega(r_0)&=0,\nonumber\\
 \alpha_\omega^{\rm H}
 &=2\int_{r_0}^{\infty}
       \frac{Bv_\infty\,dr}{r^2\sqrt{F_\omega(r)}}-\pi.
 \label{eq:c15plasma_H}
\end{align}
For the optical calculation, use
\begin{equation}
 d\ell_\omega^2=n^2(r)
   \left(\frac{dr^2}{f^2}+\frac{r^2}{f}\,d\phi^2\right),\qquad
 H_\omega=\sqrt f\left(1+\frac{rn'}n-\frac{rf'}{2f}\right).
 \label{eq:c15plasma_optical}
\end{equation}
This metric determines the path at fixed frequency; in a dispersive
medium its arclength is not the group travel time. Radially integrating
its Gaussian curvature gives the independent expression
\begin{equation}
 \alpha_\omega^{\rm GW}
 =2\int_{r_0}^{\infty}(1-H_\omega)
       \frac{Bv_\infty\,dr}{r^2\sqrt{F_\omega(r)}}.
 \label{eq:c15plasma_GW}
\end{equation}
Both reduce to the vacuum formulas as $\chi\to0$.

The weak expansion through second order, obtainable from the orbit or
the optical curvature, is \cite{c15CrisnejoGallo2018,c15Crisnejo2019}
\begin{equation}
 \alpha_\omega^{(2)}
 =\frac{2M}{B}\left(1+\frac{1}{v_\infty^2}\right)
  +\frac{3\pi M^2}{4B^2}\left(1+\frac{4}{v_\infty^2}\right)
  +O(M^3/B^3).
 \label{eq:c15plasma_series}
\end{equation}
The expansion is taken at fixed $v_\infty>0$. As the frequency approaches
the propagation cutoff, the coefficients grow and a large $B/M$ alone
no longer ensures weak bending.

\begin{figure}[!htb]
\centering
\includegraphics[width=\linewidth]{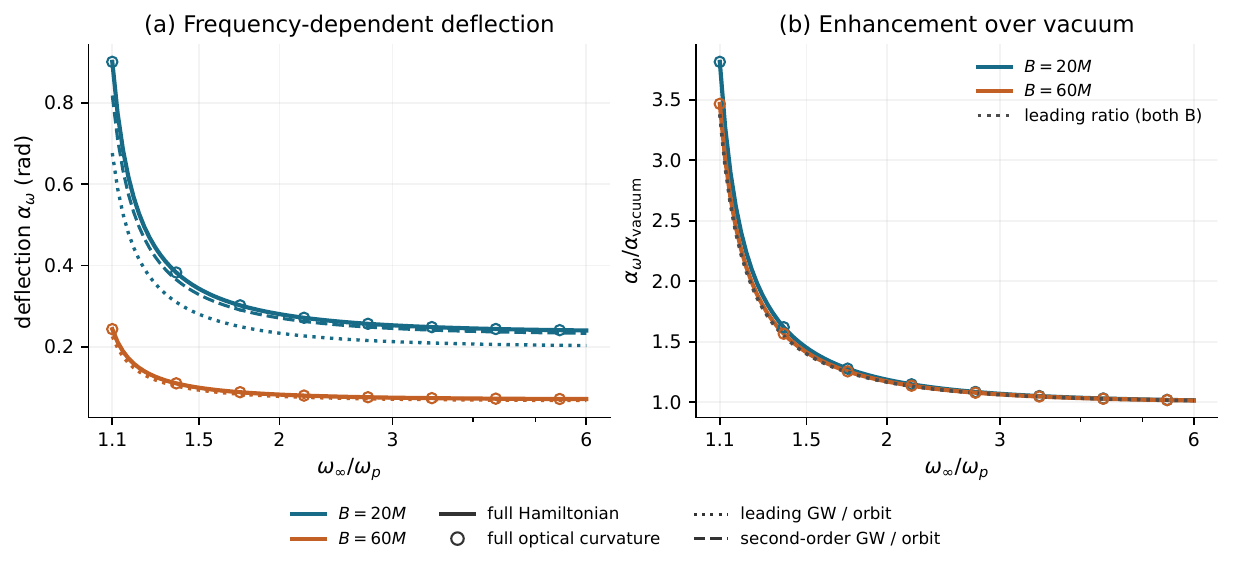}
\caption{Schwarzschild bending in a homogeneous test plasma at fixed
geometric impact parameters $B=20M$ and $60M$. (a) The full Hamiltonian
and optical-curvature results, compared with the leading and complete
second-order expressions. (b) Each full angle divided by its vacuum value
at the same $B$. The leading ratio is independent of $B$, while the full
enhancement retains a dependence on how close the ray passes to the lens.
The source and receiver are distant; all plotted frequencies propagate
through the exterior, and all plotted rays scatter.}
\label{fig:c15plasma}
\end{figure}

Lower-frequency rays bend more in Fig.~\ref{fig:c15plasma}, with the full
Hamiltonian and optical-curvature curves agreeing. The leading enhancement
over vacuum is $\tfrac12(1+v_\infty^{-2})$, common to both impact
parameters. The full calculation breaks this simple factorization:
at $\omega_\infty/\omega_p=1.1$, the enhancement is about $3.81$ for
$B=20M$ and $3.47$ for $B=60M$, compared with the leading value $3.38$.
Dispersion therefore changes more than the overall amplitude of a vacuum
bending curve; its coupling to the gravitational field depends on the
ray's closest approach.

This provides a useful distinction from the preceding polarization study.
The classical NED characteristics used there separate modes through their
radial structure, whereas the plasma comparison separates frequencies
within one isotropic medium. A spatially varying plasma density would add
ordinary refraction to the redshift-dependent effect isolated here, and
could change the sign of the plasma contribution. Frequency and
polarization comparisons consequently probe different aspects of the
propagation law \cite{c15Crisnejo2019}.

\subsection{From weak bending to rays near capture}
\label{subsec:c15strong}

\subsubsection{Lenses with the same capture scale}

Two lenses can have the same capture scale and still bend nearby rays
differently. To examine this possibility, we fix the ADM mass and the
critical impact parameter $b_c$, then determine the RN and extraordinary
Dymnikova charges separately from

\begin{equation}
 V'(r_c)=0,\qquad V''(r_c)<0,\qquad b_c=V(r_c)^{-1/2}.
 \label{eq:c15capture}
\end{equation}
Their apparent capture sizes are then equal for a distant observer at
the same distance, although their charges and optical potentials differ.

These lenses also have a close weak-field resemblance. Writing
$\alpha=4M/b+A_2/b^2+A_3/b^3+\cdots$, the two series in
Eq.~\eqref{eq:c15charged_series} depend on a single effective charge
combination:

\begin{equation}
 Q_{\rm opt}^2=Q_{\rm ID}^2-z^2,\qquad
 A_3=\frac{64M}{3\pi}A_2-\frac{112M^3}{3}.
 \label{eq:c15weak_degeneracy}
\end{equation}
An RN lens with $Q_{\rm RN}^2=Q_{\rm opt}^2$ reproduces both displayed
extraordinary-ID coefficients. The second and third weak coefficients consequently carry the same
parameter information in this comparison. The equal-capture solutions in
Table~\ref{tab:c15matched} lie close to that degeneracy, making their
near-critical rays a particularly useful place to look for a difference.

\begin{table}[!htb]
\centering\small
\caption{RN and extraordinary Dymnikova parameters at equal mass and
critical impact parameter. $A_2$ is the second weak-deflection coefficient;
$\bar a$ is the logarithmic strong-deflection coefficient.
All listed solutions have a nonextremal horizon and an admissible exterior
photon characteristic.}
\label{tab:c15matched}
\begin{tabular}{rlrrr}\toprule
$b_c/M$ & Model & $Q/M$ & $A_2/M^2$ & $\bar a$\\\midrule
4.15 & RN & 0.9581994 & 9.6176419 & 1.2907526 \\
4.15 & ID & 1.0520332 & 9.6182889 & 1.3030335 \\
4.30 & RN & 0.9072521 & 9.8415739 & 1.2104562 \\
4.30 & ID & 0.9826181 & 9.8447189 & 1.2167308 \\
4.60 & RN & 0.7716034 & 10.3781609 & 1.1084764 \\
4.60 & ID & 0.8136244 & 10.3804386 & 1.1097314 \\
4.90 & RN & 0.5649214 & 11.0290254 & 1.0443974 \\
4.90 & ID & 0.5800095 & 11.0294445 & 1.0445106 \\

\bottomrule\end{tabular}
\end{table}

\subsubsection{The growth of the angle near trapping}

Near capture, a ray spends an increasing azimuthal interval close to the
unstable circular orbit. Expanding the radial equation of
Eq.~\eqref{eq:c15cone} about the potential maximum gives
\begin{equation}
 \left(\frac{dr}{dt}\right)^2=f^2(1-b^2V),\qquad
 \lambda_t^2=-\frac{f_c^2V''_c}{2V_c},\qquad
 \Omega_c=\frac{1}{b_c}.
 \label{eq:c15instability}
\end{equation}
Here $\lambda_t$ is the rate at which a nearby ray separates radially
from the unstable orbit in time normalized at infinity
\cite{c15Cardoso2009}. The strong-deflection expansion is
\begin{equation}
 \alpha(b)=-\bar a\log\epsilon+\bar b+o(1),\qquad
 \epsilon=\frac{b}{b_c}-1,\qquad
 \bar a=\frac{\Omega_c}{\lambda_t}.
 \label{eq:c15strong_law}
\end{equation}
The logarithmic form follows from integrating through the neighbourhood
of a nondegenerate potential maximum \cite{c15Bozza2002}.
It replaces the expansion in inverse powers of $b$ when
$0<\epsilon\ll1$.
For rays that wind around the lens, the curvature integral uses the
unwrapped azimuth: successive passages contribute separately on the
covering optical surface.

\begin{figure}[!htb]
\centering
\includegraphics[width=\linewidth]{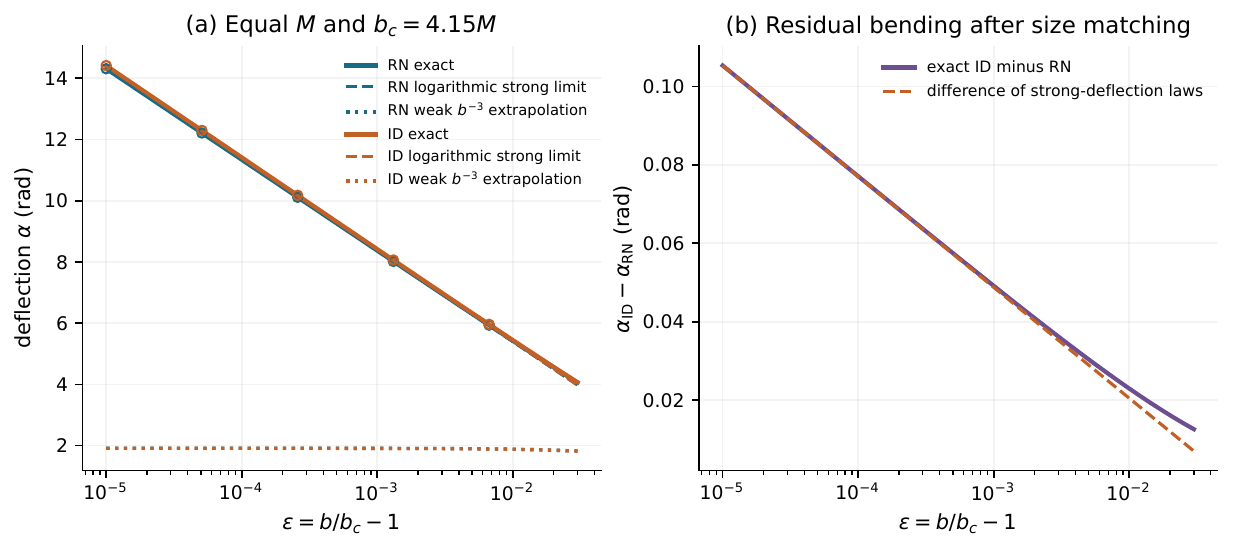}
\caption{Weak- and strong-deflection expressions for RN and extraordinary
Dymnikova lenses matched at $b_c=4.15M$.
(a) Full scattering integrals, generalized optical-curvature markers,
the logarithmic strong-deflection form, and an extrapolation of the weak
$b^{-3}$ series. The latter is shown outside its weak-field range to
display its failure to describe the near-capture growth. For the logarithmic
curves, $\bar a$ is calculated from the potential curvature and
$\bar b$ is extracted from the full scattering integrals near capture.
(b) The difference between the two full bending angles and the difference
between their strong-deflection expressions.}
\label{fig:c15strong}
\end{figure}

Close to capture, the full angles in Fig.~\ref{fig:c15strong} rise far
above the extrapolated weak series and approach the logarithmic law.
The RN and Dymnikova curves nevertheless separate, despite their common
$b_c$. At $b_c=4.15M$, their second weak coefficients differ by only about
$0.00673\%$, while their logarithmic slopes differ by about $0.951\%$.
That small slope difference is multiplied by an increasingly large
$|\log\epsilon|$ as the ray approaches capture.

To leading strong order,
\begin{equation}
 \alpha_{\rm ID}-\alpha_{\rm RN}
 =-(\bar a_{\rm ID}-\bar a_{\rm RN})\log\epsilon
   +(\bar b_{\rm ID}-\bar b_{\rm RN})+o(1).
 \label{eq:c15strong_difference}
\end{equation}
The distinction comes from the radial escape rate. Equal capture size
fixes the height of the optical potential, but leaves its curvature free
to differ. For this matched pair, the smaller Dymnikova escape rate lets
nearby rays spend longer around the circular orbit and accumulate more
azimuth before escaping.

The same escape rate governs the spacing of rays with different winding
numbers. Adding one full turn gives, at leading strong order,
\begin{equation}
 \frac{(b-b_c)_{N+1}}{(b-b_c)_N}\longrightarrow
 e^{-2\pi/\bar a},\qquad
 T_{\rm loop}\longrightarrow 2\pi b_c.
 \label{eq:c15accumulation}
\end{equation}
The spacing can therefore distinguish equal-size lenses even when their
leading additional loop times agree \cite{c15BozzaMancini2004}.

\subsubsection{The spacing of successive winding rays}

The spacing can be calculated directly rather than inferred only from
the logarithmic slope. Choose a residual scattering angle $\delta$ and
solve the full bending equation for the family
\begin{equation}
 \alpha(b_N)=2\pi N+\delta,\qquad
 \epsilon_N=\frac{b_N-b_c}{b_c}.
 \label{eq:c15winding_target}
\end{equation}
These rays have the same incoming and outgoing asymptotic directions,
but differ by $N$ complete turns. Inverting the strong-deflection law
gives
\begin{equation}
 \epsilon_N^{\rm log}
 =\exp\left(\frac{\bar b-\delta-2\pi N}{\bar a}\right).
 \label{eq:c15winding_log}
\end{equation}
Figure~\ref{fig:c15winding} compares this prediction with the roots of the
full scattering integrals for the same equal-capture pair used above.
The potential curvature supplies $\bar a$; $\bar b$ is the near-critical
constant already extracted for Fig.~\ref{fig:c15strong}.

\begin{figure}[!htb]
\centering
\includegraphics[width=\linewidth]{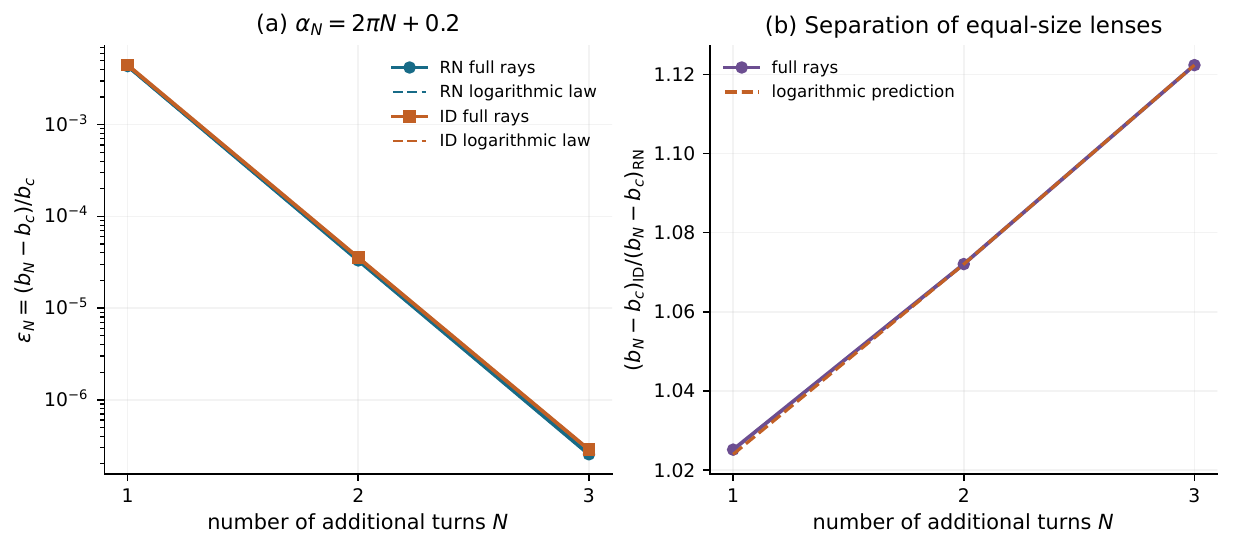}
\caption{Successive winding rays for RN and extraordinary Dymnikova lenses
with $b_c=4.15M$, at residual angle $\delta=0.2$ rad. (a) The impact-parameter
excess obtained by solving the full angle for $N=1,2,3$, together with the
inverted logarithmic expression. (b) The ratio of the two excesses. Lines
join the discrete winding orders to guide the eye. The ratio increases
while both absolute excesses become rapidly smaller.}
\label{fig:c15winding}
\end{figure}

The two sequences look almost identical on the logarithmic scale in
panel (a), but their ratio reveals a systematic separation. The Dymnikova
excess is about $2.51\%$ larger for one turn, $7.21\%$ larger for two,
and $12.24\%$ larger for three. A difference of less than one percent in
$\bar a$ has thus become a larger relative difference in the locations of
the winding rays. For a distant observer, $b_N\simeq D_{\rm OL}\theta_N$,
so the same ratio applies to angular offsets from the common critical
curve at leading order in $M/D_{\rm OL}$.

The amplification follows directly from
\begin{equation}
 \log\frac{\epsilon_N^{\rm ID}}{\epsilon_N^{\rm RN}}
 \simeq
 \frac{\bar b_{\rm ID}-\delta}{\bar a_{\rm ID}}
 -\frac{\bar b_{\rm RN}-\delta}{\bar a_{\rm RN}}
 +2\pi N\left(\frac{1}{\bar a_{\rm RN}}
              -\frac{1}{\bar a_{\rm ID}}\right).
 \label{eq:c15winding_ratio}
\end{equation}
It accompanies a rapid decrease in absolute scale: by $N=3$,
$\epsilon_N$ is only about $2.54\times10^{-7}$ for RN and
$2.85\times10^{-7}$ for ID. The comparison identifies a geometric way
to distinguish the lenses; resolving the corresponding images also
depends on angular resolution, source structure, and radiative transfer.
Equal capture size, equal leading loop time, and nearly equal weak bending
can therefore coexist with a distinct hierarchy of near-critical rays.

\subsection{Concluding discussion}
\label{subsec:c15conclusion}

The comparisons bring the approaches developed in this review into a
common physical picture. Following a ray, measuring its local direction,
and integrating optical curvature lead to the same full bending function
when they describe the same propagation law. Each route nevertheless
emphasizes a different part of the problem. The orbit equations make
turning points and capture explicit; invariant angles connect the orbit
to an observer; Fermat's principle relates bending to travel-time response;
and optical curvature shows how the geometry accumulates along the path.
Their weak and strong expansions are useful because they separate the
physical scales entering that common result.

Several distinctions recur across the examples. The nonlinear mass
contribution shifts both Kerr orientations, whereas frame dragging splits
them. A photon cone can change the angle at a lower weak order than a
regularizing term in the background metric. Polarization and plasma add
dependences that a vacuum bending curve cannot contain: the former
separates characteristic modes, while the latter introduces frequency
dependence even in a homogeneous medium. The relevant comparison is
therefore richer than the value of the angle at one impact parameter.
Its variation with orientation, polarization, and frequency identifies
which part of the lens or propagation law is being sampled.

The beam and winding studies sharpen this point. Angular concentration
involves the derivative of the bending function, yet it can retain the
same low-order parameter degeneracy as the angle itself. Near capture,
the logarithmic slope supplies different information: lenses with the
same critical size can produce appreciably different spacings of
successive winding rays. Far-field bending, capture size, and the approach
to trapping thus constrain different features of the optical geometry,
although some combinations of them remain correlated.

For applications, these bending functions must be placed in a lens
geometry to obtain image positions, magnifications, and delays. The
comparisons suggest where that next step is most informative: combine
rays that sample different impact parameters and, where available,
different orientations or propagation modes. Extending the analysis to
inclined rotating rays and realistic plasma profiles will test how much
of this separation survives in less symmetric settings. The enduring
strength of the methods reviewed here is that they connect such changes
in the observed light to identifiable features of the gravitational field
and of the medium through which it travels.

\clearpage
\endgroup

\section*{Acknowledgements}
Author BM thanks DST (GoI) for financial support during this work (vide reference no DST/WISE-PDF/PM-66/2023) under WISE Post-Doctoral Fellowship programme.

	\providecommand{\href}[2]{#2}\begingroup\raggedright\endgroup

	
\end{document}